\pdfoutput=1
\documentclass{icldt}
\usepackage{amsmath,amssymb,amsfonts,amsxtra,makeidx,graphics,graphicx,amsthm,epsfig,epstopdf,verbatim ,epstopdf,mathrsfs,multirow,placeins}
\usepackage[usenames,dvipsnames]{color}
\usepackage[all]{xy}
\newcommand{\bi}{\begin{itemize}}
\newcommand{\ei}{\end{itemize}}

\newcommand{\be}{\begin{equation}}
\newcommand{\ee}{\end{equation}}
\newcommand{\beq}{\begin{equation}}
\newcommand{\beql}[1]{\begin{equation}\label{#1}}
\newcommand{\eeq}{\end{equation}} 
\newcommand{\ba}{\begin{array}}
\newcommand{\ea}{\end{array}}
\newcommand{\bea}{\begin{eqnarray}}
\newcommand{\beal}[1]{\begin{eqnarray}\label{#1}}
\newcommand{\eea}{\end{eqnarray}}
\newcommand{\ben}{\begin{enumerate}}
\newcommand{\een}{\end{enumerate}}
\newcommand{\bean}{\begin{eqnarray*}}
\newcommand{\eean}{\end{eqnarray*}}
\newcommand{\eref}[1]{(\ref{#1})}

\newcommand{\tref}[1]{Table~\ref{#1}}
\newcommand{\fref}[1]{Figure~\ref{#1}}
\newcommand{\nn}{\nonumber}

\newcommand{\btab}[1]{\begin{tabular}{#1}}
\newcommand{\etab}{\end{tabular}}

\newcommand{\Diag}{\mbox{Diag}}
\newcommand{\tr}{\mathop{\rm Tr}}

\newcommand{\ra}{\rightarrow}

\newcommand{\cB}{{\cal B}}
\newcommand{\cC}{{\cal C}}
\newcommand{\cD}{{\cal D}}
\newcommand{\cE}{{\cal E}}
\newcommand{\cF}{{\cal F}}

\newcommand{\BP}{\mathbb{P}}
\newcommand{\BC}{\mathbb{C}}
\newcommand{\BR}{\mathbb{R}}
\newcommand{\BZ}{\mathbb{Z}}

\newcommand{\CN}{{\cal N}}

\newcommand{\ud}{\mathrm{d}}

\newcommand{\CP}{\mathbb P}

\newcommand{\CMm}{{\cal M}^{\mathrm{mes}}}

\newcommand{\BF}{\mathbb{F}}

\newcommand{\perm}{\mathrm{perm}}

\newcommand{\tmat}[1]{{\tiny \left(\begin{matrix} #1 \end{matrix}\right)}}

\newcommand{\f}{{\cal F}^{\flat}}

\newcommand{\firr}[1]{{}^{{\rm Irr}}\!{\cal F}^{\flat}_{#1}}

\newcommand{\bmi}{\begin{minipage}}
\newcommand{\emi}{\end{minipage}}
\newcommand{\Section}[1]{\section{#1} \setcounter{equation}{0}}
\newcommand{\conxc}{$\cal{C} \times \BC$ }
\newcommand{\CC}{\cal{C} }
\title{Brane Tilings, M2-branes and Orbifolds}
\author{John Paul Davey}
\date{August 1, 2011}
\department{Physics}
\supervisor{Professor Amihay Hanany}
\dedication{}

\begin{document}

\maketitle

\newpage
\mbox{}
\chapter*{Declaration}
I herewith certify that, to the best of my knowledge, all of the material in this dissertation which is not my own
work has been properly acknowledged.

\hfill John Paul Davey
\newpage
\mbox{}
\begin{abstract}
Brane Tilings represent one of the largest classes of superconformal theories with known gravity duals in 3+1 and also 2+1 dimensions. They provide a useful link between a large class of quiver gauge theories and their moduli spaces, which are the toric Calabi-Yau (CY) singularities.

This thesis includes a discussion of an algorithm that can be used to generate all brane tilings with any given number of superpotential terms. All tilings with at most 8 superpotential terms have been generated using an implementation of this method.

Orbifolds are a subject of central importance in string theory. It is widely known that there may be two or more orbifolds of a space by a finite group. Abelian Calabi-Yau orbifolds of the form $\BC^3 / \Gamma$ can be counted according to the size of the group $|\Gamma|$. Three methods of counting these orbifolds will be given.

A brane tiling together with a set of Chern Simons levels is sufficient to define a quiver Chern-Simons theory which describes the worldvolume theory of the M2-brane probe. A forward algorithm exists which allows us to easily compute the toric data associated to the moduli space of the quiver Chern-Simons theory from knowledge of the tiling and Chern-Simons levels. This forward algorithm will be discussed and illustrated with a few examples. It is possible that two different Chern-Simons theories have the same moduli-space. This effect, sometimes known as `toric duality' will be described further. We will explore how two Chern--Simons theories (corresponding to brane tilings) can be related to each other by the Higgs mechanism and how brane tilings (with CS levels) that correspond to 14 fano 3-folds have been constructed.

The idea of `child' and `parent' brane tilings will be introduced and we will discuss how it has been possible to count `children' using the symmetry of the `parent' tiling.
\end{abstract}

\makededication

\tableofcontents
\listoftables
\listoffigures
\chapter{Introduction and Outline}

Since Maxwell's formulation of electrodynamics almost 200 years ago, gauge theory has played a central role in our understanding of Physics. Through the 20th century, gauge theory has been developed and has culminated in the standard model of particle physics which has proved to be a phenomenally successful theory. The theory is capable of describing three of the four fundamental forces of nature amazingly well even when tested at the world's largest colliders, where physicists smash tiny particles together at colossal energies.

However we know that the standard model is not a fundamental theory of nature. One issue is that gravity is not described at all and so the model is useless as a tool for describing the universe in its infancy. A second problem is enormous difference between the Weak and the Planck scale which currently requires severe fine tuning of parameters in the model.

For decades some of the world's finest minds have tried but largely failed to find a theory that can supersede the standard model. String theory offers one promising avenue of research although it is not yet fully developed and many of its features are not well understood. It is not even clear whether the theory will ever be able to make a falsifiable prediction. Despite these issues, String Theory is currently our most developed quantum theory of gravity and has sparked developments in Geometry and also theoretical condensed matter.

The discovery of D-branes, which are explicit realisations of charged BPS states in superstring theory, is seen as being a remarkable advance \cite{Dai:1989ua,Polchinski:1995mt}. World-volume Lagrangians for D5-branes located at the fixed point of the orbifold $\mathbb{C}^2 / \mathbb{Z}_n$ were later constructed \cite{Gimon:1996rq,DouglasMoore96}. The massless spectrum of these worldvolume theories can be understood by making the crucial observation that if a point is an allowed endpoint for open strings, then all of its images under the orbifold group must also be allowed endpoints. This lead to the discovery that the field content of the worldvolume theory on the D5-brane is a Super Yang-Mills (SYM) theory with a matter content that can be displayed in a `quiver' diagram. 

\begin{figure}
\begin{center}
\includegraphics[totalheight=6cm]{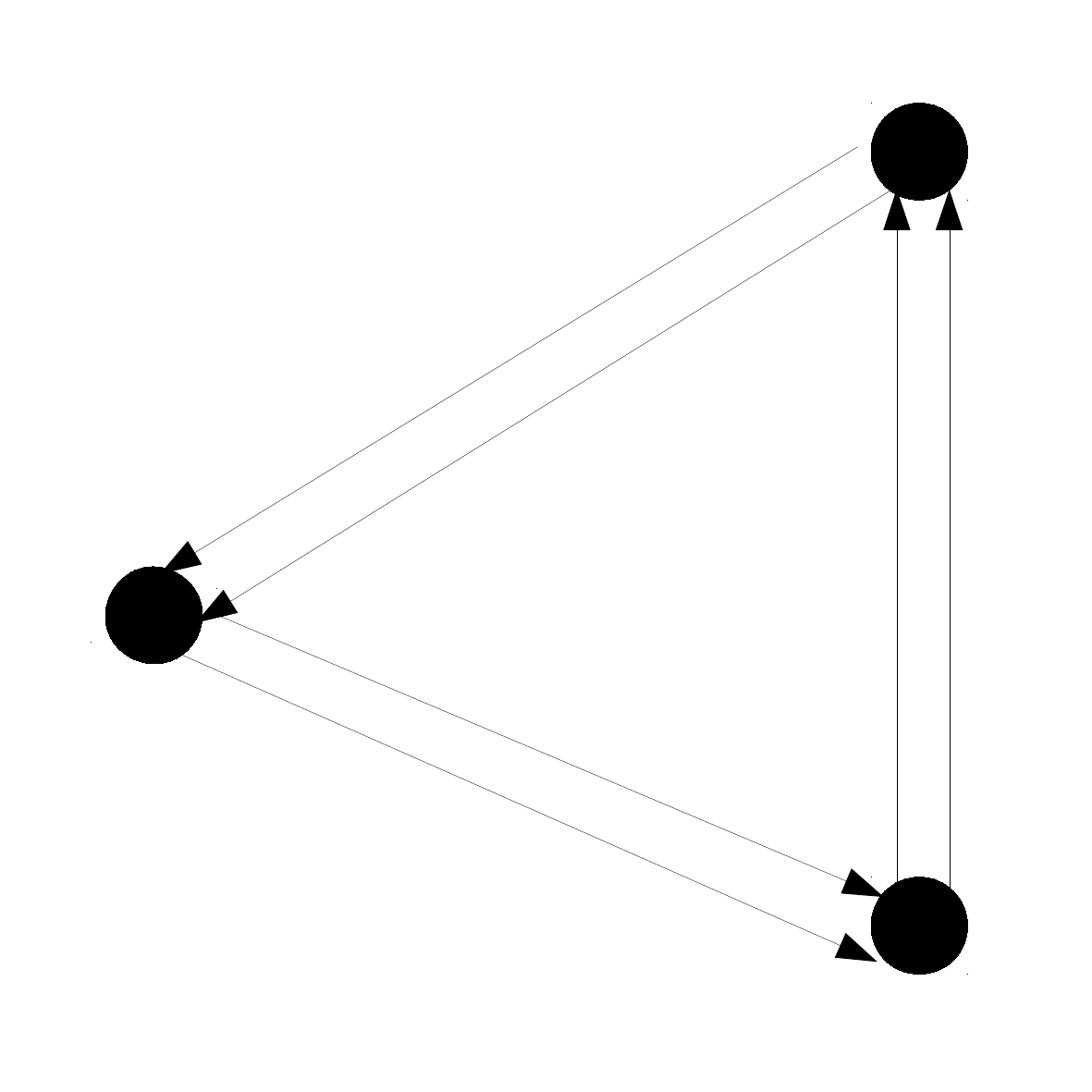}
\end{center}
\caption{The Quiver diagram corresponding to D5 branes probing $\mathbb{C}^2 / \mathbb{Z}_3$. The edges correspond to hyper-multiplets and the nodes to vector-multiplets.}
\label{f:triQuiv}
\end{figure}

Quiver diagrams forged a cast iron link between gauge theory and geometry. They are graphs that encode vector multiplets as nodes and hyper multiplets as edges. A hyper-multiplet corresponding to and edge connecting two nodes transforms in the fundamental representation of the gauge group associated to the first vector-multiplet and the anti-fundamental representation of the second. The idea was extended to cover non abelian orbifolds. The extended Dynkin diagram of the non abelian gauge group was found to correspond to the quiver describing the gauge theory matter content \cite{DouglasMoore96}. The field content of the quiver gauge theory living on a D5 brane probing $\mathbb{C}^2 / \mathbb{Z}_3$ is given in Figure \ref{f:triQuiv}.

The work of Maldacena in 1997 demonstrated a second intimate link between gauge theory and geometry \cite{Maldacena:1997re}. The near horizon limit of a system of N D3 branes in flat space can be viewed both as Type IIB string theory on $\mathrm{AdS}_5 \times S^5$ with $N$ units of the self dual 5-form RR flux and also as $\mathcal{N} = 4$ SYM with an $SU(N)$ gauge group. The duality between the two theories is known as the AdS / CFT correspondence.

Maldacena's original conjecture has been generalised to a duality between certain four dimensional conformal gauge theories and IIB string theory on $AdS_5 \times X_5$, where $X_5$ is a five dimensional Sasaki-Einstein (SE) manifold with 5 form flux. The conifold model is an example of this. In this case IIB string theory on $\mathrm{AdS}_5 \times (SU(2) \times SU(2)) / U(1)$ is thought to be a dual description of a special supersymmetric gauge theory \cite{Klebanov:1998hh}. One important feature of the duality is that the moduli space of the gauge theory is thought to correspond to the space transverse to the branes on the string theory side.

\begin{figure}[h]
\begin{center}
\begin{tabular}{cc}
\includegraphics[height=3.5cm]{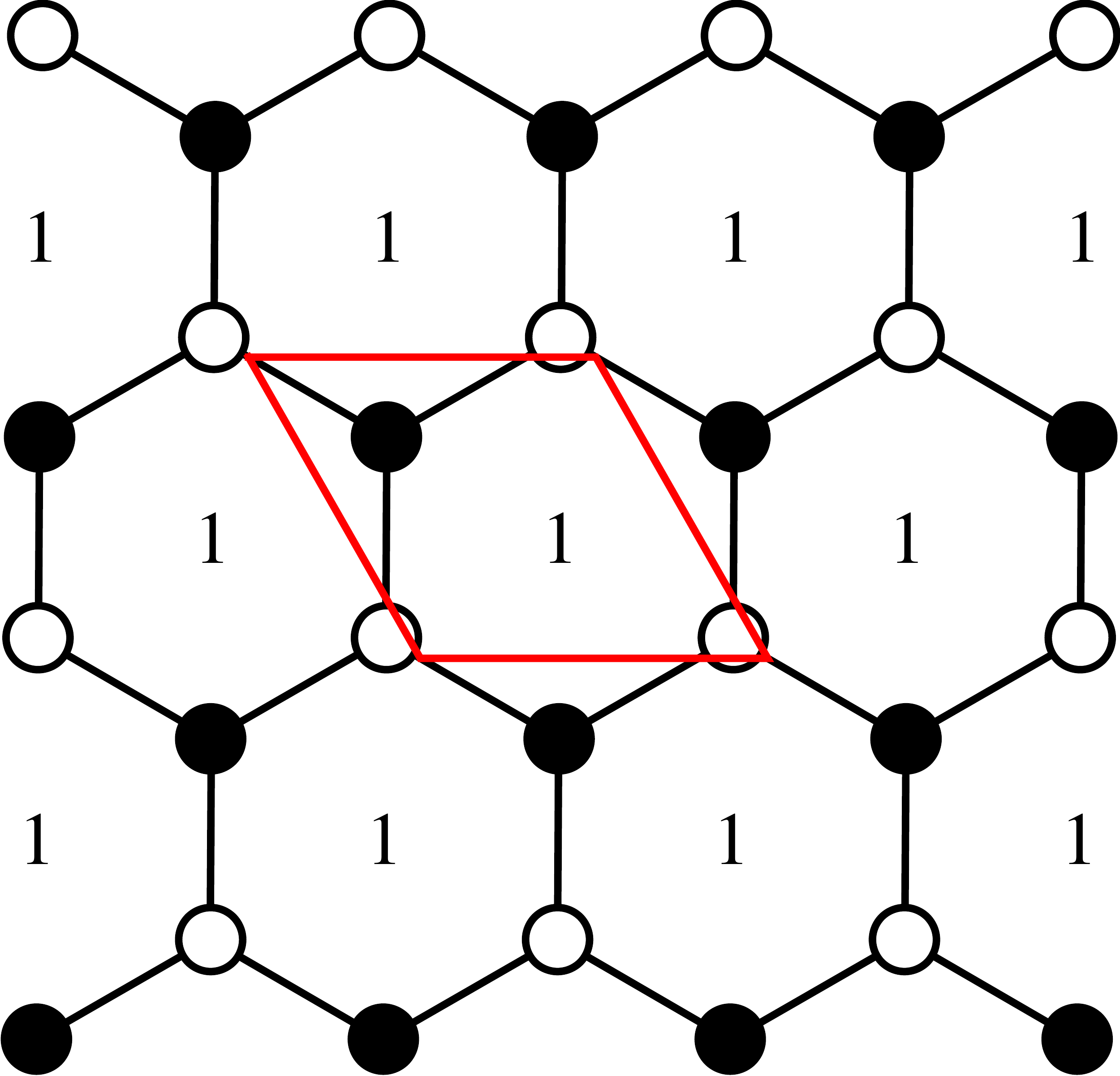} 
&
\includegraphics[height=3.5cm]{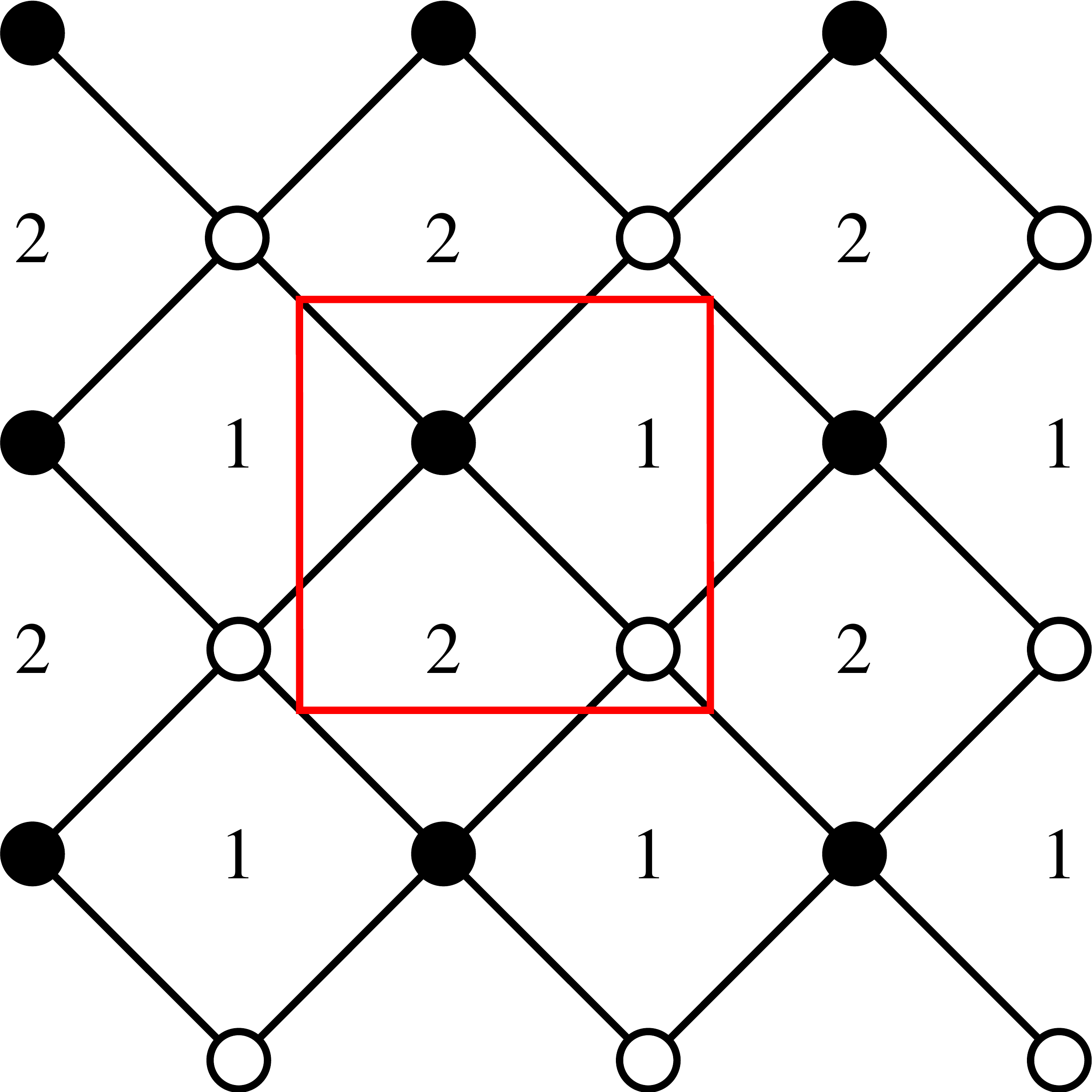}
\end{tabular}
\end{center}
\caption{Tilings that correspond to $\mathcal{N} = 4$ SYM (left) and the Conifold (right). The smallest repeating unit (fundamental domain) is shown in red.}
\label{fig:ConAndNIs4}
\end{figure}

The Brane Tiling has further strengthened the link between geometry and gauge theory. Brane Tilings describe gauge theories that are dual (in the sense of the AdS / CFT correspondence) to toric Calabi-Yau (CY) 3-fold singularities \cite{Kennaway2005,Hanany05,Franco:2005sm}. The CY 3-folds are formed by taking the (real) cone over a class of SE 5-folds. Many well known gauge theories are described by brane tilings. For instance, there is a brane tiling that corresponding to the famous $\mathcal{N} = 4$ SYM theory and another corresponding to the conifold model (Figure \ref{fig:ConAndNIs4}). There are also a plethora of theories that have a tiling description and have not yet been studied in detail in academic literature.

In Chapter \ref{ch:btandd3}, a brief overview of the ideas in toric geometry and quiver gauge theory that are most useful for understanding the rest of this thesis is given. We define a quiver gauge theory and then discuss some aspects of toric geometry including the concept of a toric diagram. We then turn our attention to the brane tiling and show how it is possible to quickly compute the toric data corresponding to the moduli space of a gauge theory that is described by a brane tiling.

An algorithm for generating brane tilings is discussed in Chapter \ref{ch:class}. The algorithm is based on the generation of quivers and then finding superpotentials that can be formed from these quivers. Tilings are then reconstructed from these quiver gauge theories. A catalogue of all brane tilings with at most 8 superpotential terms is given in Appendix \ref{a:tilings}. The chapter closely follows `On the Classification of Brane Tilings' \cite{Davey:2009bp}.

Toric CY singularities that are abelian orbifolds of $\BC^3$ are counted in Chapter \ref{ch:orbs}. Three equivalent methods of counting these orbifolds are explained. Firstly the counting is performed using tilings that can be constructed using only hexagonal faces. A method using 3-tuples is also demonstrated before a way of counting using the toric description of the orbifolds is shown. The chapter is an edited version of `An Introduction to Counting Orbifolds' \cite{OrbsIntro}, which is itself a review based on `Counting Orbifolds' \cite{OrbsFull}.

Supersymmetric Chern-Simons (CS) theories in 2+1 dimensions have attracted a lot of interest due to their proposed description of the M2-brane \cite{BaggerLambert07,Gustavsson07}.  A $U(N) \times U(N)$ CS theory at level $(k, -k)$ with bi-fundamental matter fields was subsequently proposed as a description of $N$ M2-branes on the $\BC^4/\BZ_k$ orbifold background \cite{ABJM08}. One recent and quite exciting development has been that we can use brane tilings (with a few modifications from the 3+1 dimensional case) to study 2+1 dimensional CS theories \cite{Hanany:2008cd, HananyZaff08}. These CS theories are conjectured to have an M-theory dual.

In Chapter \ref{ch:phases} we show how a CS theory can be defined using a brane tiling and how it is possible for several CS theories to have the same moduli space. The Higgs mechanism has been found to be useful for relating different CS theories and is investigated in Section \ref{Sec:Higgs}. The Chapter follows some parts of `Phases of M2-brane Theories' \cite{Davey:2009sr} and `Higgsing M2-brane Theories' \cite{Davey:2009qx}.

The concept of a Fano variety is discussed in Chapter \ref{Ch:Fanos}. Brane Tiling technology has been used to find Chern--Simons theories which correspond to 14 of the smooth toric fano 3-folds. The toric data corresponding to the moduli space of these 14 theories is calculated explicitly in Appendix \ref{fanoapp}. The chapter uses some of the results that are contained in `M2-Branes and Fano 3-folds' \cite{Davey:Fano} and also `Brane Tilings, M2-branes and Chern-Simons Theories' \cite{Davey:2009et}.

In Chapter \ref{Ch:Children} the concept of a parent and child tiling is introduced. Children of 4 different parent tilings are counted according to the number of fields added to the parent theory. Partition functions that count these children have been calculated using the discrete symmetry of the parent tiling together with a discrete Molien formula.

\chapter{Brane Tilings and D3-branes}
\label{ch:btandd3}
Brane tilings form an important link between quiver gauge theories and toric geometry \cite{Kennaway2005,Hanany05}. 
 Tilings represent a large class of superconformal theories with known gravity duals in 3+1 and also 2+1 dimensions and have proved useful for describing the physics of both D3-branes and also M2-branes probing Calabi-Yau singularities.

\begin{figure}[h!]
\begin{center}
\includegraphics[width=5cm]{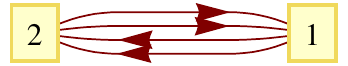}
\end{center}
\caption{A Typical Quiver Diagram}
\label{f:TypicalQuiv}
\end{figure}

In this chapter, we will first review some of the basics of quiver gauge theories and toric geometry and then go on to define exactly what a brane tiling is. Two excellent reviews which cover the idea of a brane tiling are \cite{Yamazaki:2008bt,Kennaway:2007tq}. For a more mathematical review on the subject, see \cite{Broomhead:2008an}.
\Section{Quiver Gauge Theories}
A quiver is simply an oriented graph: a collection of vertices together with a set of oriented edges. It is possible for an edge to start and end on the same node. It is also possible to have more than one edge connecting any two nodes. A typical quiver diagram is given in Figure \ref{f:TypicalQuiv}.

A quiver is much more than a graph to a physicist. It is possible to specify completely the Lagrangian of a large family of $\CN$=1 SUSY gauge theories from a quiver diagram together with information about the superpotential of the theory. Quiver gauge theories have been used to describe the world volume of D3-branes at Calabi--Yau singularities \cite{Hanany:2001py, Franco:2002ae, Feng:2004uq, Benvenuti:2004dw, Hanany:2005hq, ypq, AmaxZmin, 0608050, ZPQ,Hanany:2006uc,pprogram,Plog}.

There is a dictionary between a quiver and a gauge theory that it corresponds to. The vertices of a quiver correspond to the gauge groups of the theory. Here we shall concern ourselves with only $U(N)$ gauge groups, although quiver theories with $SO$ and $Sp$ gauge groups have been found to be be useful for understanding orientifolds in string theory \cite{Imamura:2008fd,orientifolds}. It is possible for different nodes in the quiver to correspond to gauge groups of different ranks, although here we shall only consider theories with a gauge symmetry of $\prod U(N)$. 

Quivers can be used to describe supersymmetric theories that have a matter content that consists of chiral superfields transforming under bi-fundamental representations of the gauge symmetry of the theory. Edges in the quiver correspond to these chiral superfields. A field that corresponds to an edge that links node $i$ to node $j$ transforms in the fundamental representation of gauge group $i$ and the anti-fundamental representation of gauge group $j$. Adjoint matter can be thought of as the case when $i=j$ and so the matter transforms in the adjoint representation of gauge group $i$. Such matter corresponds to an edge both starting and ending at the same node.

The data encoded in the quiver is not enough to specify a gauge theory completely. The superpotential of the theory is completely undetermined by the quiver. It is possible to find terms that could form part of a superpotential from observing that terms in the superpotential are gauge invariant and correspond to closed loops in the quiver.

To see this, recall the quiver in Figure \ref{f:TypicalQuiv}. Let us call the fields that transform in the $(\Box_1,\bar{\Box}_2)$ representation of the gauge symmetry $A$ and $B$ and let the fields that transform in the $(\Box_2,\bar{\Box}_1)$ representation of the gauge symmetry be called $C$ and $D$. Then it is possible to build at least 2 different superpotentials from this quiver. For instance both of the superpotentials that are given in \eqref{e:2DiffPots} are gauge invariant and both define gauge theories if paired with the quiver given in Figure \ref{f:TypicalQuiv}.

\bea
W_1 &=& {A}_{ij} {B}_{jk}  {C}_{kl} {D}_{li} -  {A}_{ij} {D}_{jk}  {C}_{kl} {B}_{li} \nn \\
W_2 &=& 0
\label{e:2DiffPots}
\eea

We can see that the gauge indices in \eqref{e:2DiffPots} are already becoming confusing. In the work that follows, the gauge indices in superpotential terms shall be suppressed. An overall trace will be implicit as we shall always contract the first and last indices so that the superpotential is gauge invariant.

From now on we shall only consider superpotentials that satisfy a `toric condition'. This condition is that each field in the quiver appears in the superpotential exactly twice -- once in a positive term and once in a negative term. We shall see that if a quiver gauge theory has a superpotential that satisfies this condition then the `mesonic' moduli space of the gauge theory can be described using the tools of toric geometry.

\subsection{Anomaly Cancellation}
\label{s:Acanc}
Quiver gauge theories are in general chiral and so we should expect there to be a condition for the theory to have vanishing gauge anomalies. It is known that for a $U(n)$ gauge theory, matter transforming under a representation $r$ of the gauge symmetry will have an anomaly coefficient $A(r)$ that satisfies $A(r) = -A(\bar{r})$ (see for example pg 676 of \cite{PeskinSc}). Here we restrict our attention to quiver gauge theories with a gauge symmetry of $\prod U(N)$. In this case the anomaly cancellation condition for each gauge group is
\beq
\sum_i A(r_i) = 0
\eeq
where the sum is taken over all matter transforming under a representation of the gauge group we have chosen.

As the matter in our quiver gauge theory transforms in bifundamental representations of the gauge symmetry, this means that for each node in our quiver there must be an equal number of incoming and outgoing arrows. In this thesis we will only consider quivers that satisfy this condition.
\Section{Toric Geometry}
In this section, some of the basics of Toric Geometry will be briefly sketched with a particular focus on the tools of toric geometry that are relevant to brane tilings. For a more detailed introduction to toric geometry, the reader is directed to \cite{Fulton, Greene:1996cy, ClayMath2003}. Some reviews that deserve attention are \cite{Leung:1997tw, ClossetReview}. This section closely follows \cite{Bouchard:2006ah}. 

\subsection{Homogeneous Coordinates}
One way of defining a toric variety is by using a homogeneous coordinate construction, which makes the geometries seem similar to complex projective spaces. This is sometimes known as the Cox representation of a toric variety.

Suppose we start with the complex space $\BC^m$ and let us let this space be acted upon by $(\BC^*)^p$. Let $U \subset \BC^m$ be those points which are left fixed by the action and for $p<m$ define
\beq
\cal{M} = ( {\BC}^{\mathnormal m} \backslash {\mathnormal U} ) / (\BC^*)^{\mathnormal p}
\label{e:DefToric}
\eeq
If a variety $\cal{M}$ can be written in the form above, it is said to be a toric variety.

A concrete example of a toric variety is $\mathbb{CP}^2$. We can embed this variety in $\BC^3$ by writing it as
\beq
\mathbb{CP}^2 = (\BC^3 \backslash \{ 0 \} ) / ( \BC^* )
\label{e:cp2define}
\eeq
where the action of $\BC^*$ is
\beq
(x,y,z) \sim \lambda (x,y,z) \; \; \mathrm{for} \; \; \lambda \in \BC^*
\eeq
To continue the discussion of toric varieties, it will be useful to make a few definitions.

Suppose $v_i$ are vectors in a lattice, which for the moment can be thought of as $\BZ^3$. Then a convex polyhedral cone (or cone for short) is a set
\beq
\sigma = \{ a_1 v_1 + a_2 v_2 + \ldots + a_k v_k | a_i \geq 0 \}
\eeq
if $\sigma \cap (-\sigma) = 0$. The vectors $v_i$ are said to generate the cone.

A collection of cones, $\Sigma$, is called a fan if
\begin{itemize}
 \item each face of a cone in $\Sigma$ is in $\Sigma$ and
\item  the intersection of two cones in $\Sigma$ is in $\Sigma$.
\end{itemize}
Let $\Sigma(1)$ be the set of all one dimensional cones in $\Sigma$ and let $v_i, i=1 \ldots k$ be vectors that generate $\Sigma(1)$. Then we can use this set of vectors to define a 3-dimensional toric variety. To each $v_i$ we assign a homogeneous coordinate $w_i \in \mathbb{C}$. We can then write the toric variety as a quotient of the form
\beq
\cal{M}_\mathnormal{\Sigma} = \mathnormal{ \left( \BC^k \backslash Z(\Sigma ) \right) / H }
\eeq
Where $H$ is the product of $(\mathbb{C}^*)^\mathnormal{k-3}$ and a finite abelian group. It is not hard to see how to extend this definition to cover 4-dimensional toric varieties. For now we shall only consider the case of a trivial finite group. $Z(\Sigma)$ is a set of points that must be removed from $\BC^k$ in order to make the quotient well defined \cite{Bouchard:2006ah}.

We can relate the action of $(\mathbb{C}^*)^\mathnormal{k-3}$ to the vectors $v_i$ in the following way. Suppose each $\mathbb{C}^*$ action can be written as an equivalence relation of the form
\beq
(w_1 , \ldots , w_k ) \sim (\lambda^{Q_{a}^1} w_1 , \ldots , \lambda^{Q_{a}^k} w_k )
\eeq
where $\lambda \in \BC^*$. Then the matrix $Q$ can be related to $v_i$ as:
\beq
\Sigma_{i=1}^k Q_{a}^i v_i = 0
\eeq
Generally, $Q^i_{a}$ are chosen to be integer valued and the greatest common divisor of $Q^i_{a}$ (for fixed a) is equal to 1
\subsubsection{An Example: $\mathbb{CP}^2$}
\begin{figure}[h!]
\begin{center}
\includegraphics[width=4cm]{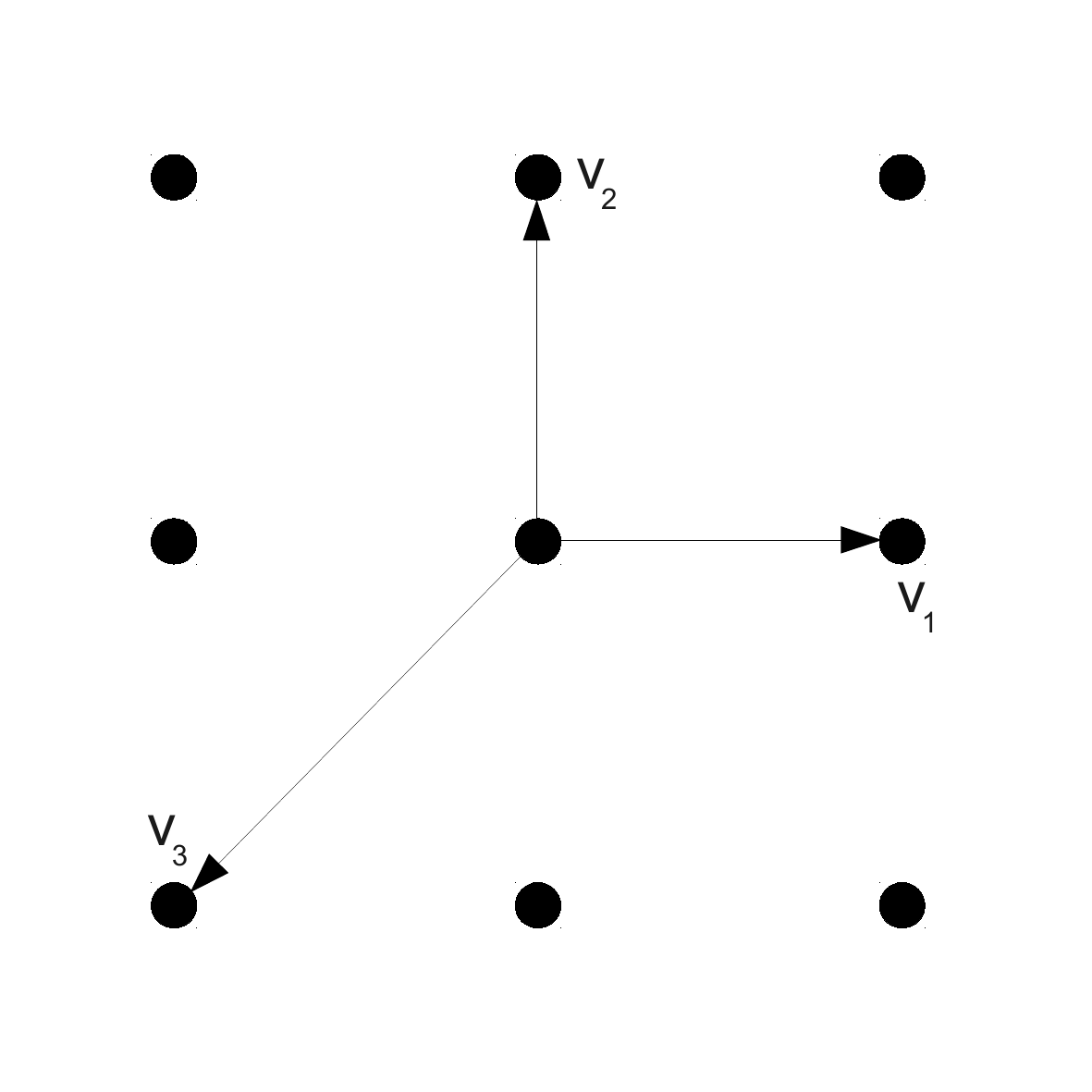}
\end{center}
\caption{The fan of $\mathbb{CP}^2$.}
\label{f:CP2Fan}
\end{figure}

We can illustrate these concepts using $\mathbb{CP}^2$ as an example. The fan of $\mathbb{CP}^2$ is given in \fref{f:CP2Fan}. There are three 1-dimensional cones that are generated by the vectors $v_1=(1,0), v_2=(0,1), v_3=(-1,-1)$.
To each $v_i$ we associate a homogeneous coordinate $w_i$, which shows that the variety can be written as a quotient of $\mathbb{C}^3$. We also have the relation that
\beq
1(1,0) + 1(0,1) + 1(-1,-1) = (0,0)
\eeq
Which means that $Q$ can be chosen to be equal to $(1,1,1)$. Therefore our space is
\beq
\cal{M}_\mathnormal{\Sigma} = \mathnormal{ \left( \BC^3 \backslash Z(\Sigma ) \right) / \BC^* }
\eeq
With the $\BC^*$ action being
\beq
(w_1,w_2,w_3) \sim \lambda (w_1,w_2,w_3) \; \; \mathrm{for} \; \; \lambda \in \BC^*
\eeq
The set $Z(\Sigma )$ is equal to $\{ 0 \}$ and so the space can be identified as $\mathbb{CP}^2$ as in \eqref{e:cp2define}.
\subsection{Toric Calabi-Yau 3-folds}
\label{s:ToricCY3}
There are several definitions of a Calabi-Yau that one can use. In this thesis, we concern ourselves only with toric Calabi-Yau (CY) manifolds and so make the two following (equivalent) definitions:
\begin{itemize}
 \item A toric manifold is Calabi-Yau if and only if the charges $Q^i_a$ satisfy the condition $\Sigma_{i=1}^k Q^i_a = 0$ for all $a$.
 \item A toric manifold defined by a fan $\Sigma$ is Calabi-Yau if and only if the generating vectors $v_i$ lie in some co-dimension 1 hyper-surface.
\end{itemize}
The first definition implies that the vectors $v_i$ can be chosen such that
\beq
v_i = {1 \choose \tilde{v}_i}
\eeq
We can then store the vectors $v_i$ as rows of a matrix $G$, i.e.
\beq
G = \{ v_1 , \ldots , v_k \}
\eeq
As every element of the first of row of $G$ is equal to 1, we can remove this row and call the resulting matrix $G_t$. The toric diagrams corresponding to Toric CY 3-folds that are given in the remainder of this thesis are formed from the columns of $G_t$. They are a set of lattice points in $\BZ^2$. It can be shown that all toric Calabi-Yau varieties are necessarily non-compact.

In the remainder of this thesis, the $Q$ matrix that defines a toric variety shall be called $Q_T$. This is to avoid confusion between this matrix and some other charge matrices that will be introduced. Later on, we shall discuss CY 4-folds which can be defined by a set of lattice points in $\BZ^3$. It is also worth while mentioning that the Cox representation of a toric variety is known as the linear sigma model description in physics literature.
\Section{Toric Geometry from Gauge Theory}

The vacuum moduli space is one of the most fundamental features of a supersymmetry gauge theory that one can investigate. The space can be thought of as an algebraic variety defined by the solutions to both F-terms and D-terms. Typically this space can be thought of as a union of various branches.

In this section, we will consider an algorithm which allows us to compute the mesonic moduli space of a supersymmetric quiver gauge theory with a superpotential that satisfies a toric condition. There is an algorithm that allows us to calculate the toric data associated to this moduli space \cite{FengToric, Franco:2006gc}. If the gauge theory lives in a stack of D3-branes, it is thought that its mesonic moduli space coincides with that space that the branes probe.

The first step in this moduli space computation is to find the moduli space when only F-terms are taken into account. This space is known as the Master Space (or $\cal{F}^\mathnormal{\flat}$) of the gauge theory, and can be studied in it own right \cite{Masterspace, MMasterspace}. 

Every superpotential in this thesis satisfies a `toric condition', that is each bi-fundamental (or adjoint) field occurs in the superpotential exactly twice: once in a positive term and once in a negative term. In order to analyse the Master Space of gauge theories that have a superpotential that satisfies this condition, it is useful to introduce the concept of a perfect matching.

A \emph{perfect matching} is a collection of fields in a quiver such that each of the $E$ field can be found exactly twice in the superpotential: once in a positive term and once in a negative term. The $c$ perfect matchings can be represented in a matrix $P_{E \times c}$ such that 
\beq
P_{ij}  = \left\{
\begin{array}{rl}
1 & \text{if field } i \text{ is in perfect matching } j\\
0 & \text{otherwise} 
\end{array} \right.
\eeq
Now let us define the a matrix $Q_F$:
\beq
Q_F = \text{Ker}(P)
\eeq
i.e. each row of $Q_F$ corresponds to a relation between perfect matchings. One can show that if the quiver gauge theory (with toric superpotential) has a gauge symmetry of $U(1)^g$, then are $c-g-2$ relations between perfect matchings when F-terms are taken into account. This means $Q_F$ can be written as a $(c-g-2) \times c$ matrix. The master space $(\firr{})$ can be thought of as the space of perfect matchings quotiented by the $\BC^*$ relations encoded in $Q_F$. From this point forwards we shall use the notation:
\beq
 \firr{} = \BC^\mathnormal{c} / / \mathnormal{Q_F}
\eeq
to mean the space formed by the $\BC^*$ quotient defined using $Q_F$.

Strictly speaking the variety above is not actually the full master space ($\f$), but the coherent component of the master space ($\firr{}$). This subtlety is not addressed here and the reader is directed to \cite{Masterspace} for further details. Henceforth we shall mildly abuse notation and it will be left implicit that we are dealing with the coherent component of the master space whenever we quotient by only F-terms.
\subsection{The Mesonic Moduli Space}
The mesonic moduli space of a quiver gauge theory can be thought of as the set of vacua of the gauge theory when both F-terms and D-terms are taken into account. The quiver gauge theories discussed in this thesis have a mesonic moduli space which is a toric Calabi-Yau 3-fold. It is possible to consider the case of a stack of D-branes probing a CY singularity. In this case the moduli space of the world-volume theory of the branes is thought to be a symmetrised product of the singularity corresponding to the one brane theory \cite{Berenstein:2002ge}. For now we shall concern ourselves with only the abelian, 1-brane theory.

For supersymmetric gauge theories, it can be shown that the mesonic moduli space can be written as a quotient of the master space by $g-1$ `baryonic' symmetries encoded in a charge matrix $Q_D$ \cite{Luty:1995sd}. We can write
\beq
\cal{M}^{\mathnormal{mes}} = \firr{} / / \mathnormal{Q_D}
\label{e:QDQuotient}
\eeq
These baryonic symmetries can be thought of as coming from the `independent' gauge symmetries of the theory, and there is a well known prescription for calculating $Q_D$. 

First of all, we can read off the way in which fields of the theory are charged under the $U(1)^g$ gauge symmetry. This information is encoded in the $g \times E$ quiver adjacency matrix $d$ which can be read off from the quiver diagram:
\beq
d_{ij}  = \left\{
\begin{array}{rl}
1 & \text{if arrow } j \text{ starts at node } i\\
-1 & \text{if arrow } j \text{ ends at node } i\\
0 & \text{otherwise} 
\end{array} \right.
\label{e:DijDefn}
\eeq
In order to compute $Q_D$, we then must convert the charges for fields that are encoded in $d$ to charges for perfect matchings. This can be done by using the perfect matching matrix $P$. We define $\tilde{Q}$ as follows:
\beq
\tilde{Q}_{g \times c} \; \cdot \; (P^T)_{c \times E} = d_{g \times E}
\label{e:QTildeDef}
\eeq
Not all of the charges in $\tilde{Q}$ are independent. We can see this from the defining equation of $d$ in \eqref{e:DijDefn}. As each edge in the quiver ends at one node and starts at one node, if we sum over all nodes we have
\beq
\Sigma_i d_{ij} = 0
\eeq
And so by \eqref{e:QTildeDef} we have
\beq
\Sigma_i \tilde{Q}_{ij} = 0
\label{e:Redund}
\eeq
We can get rid of this redundancy by defining
\beq
c_{1 \times g} = (1,1, \ldots , 1)
\eeq
and storing a basis of all vectors perpendicular to $c$ in the matrix $\text{Ker}(c)_{g-1 \times g}$, which we can choose to be
\beq
\text{Ker}(c) = \left(
\begin{array}{ccccc}
1&-1&0&0&\ldots \\
1&0&-1&0&\ldots \\
1&0&0&-1&\ldots \\
\vdots&&&&  \\
\end{array} \right)
\eeq
We then write the charge matrix for perfect matchings without the redundancy in \eqref{e:Redund}
\beq
Q_{D \; g-1 \times c} = \text{Ker}(c)_{g-1 \times g} \; \cdot \; \tilde{Q}_{g \times c}
\eeq
This charge matrix can be used to construct the quotient given in \eqref{e:QDQuotient}
\subsection{Relating the Charge Matrices to Toric Geometry}
\label{s:D3Charges}
It is possible to store both $Q_F$ and $Q_D$ in a larger matrix:
\beq
Q_{T \; c-3 \times c} = {Q_{F \; c-g-2 \times c} \choose Q_{D \; g-1 \times c}}
\eeq
And to write the mesonic moduli space in the language of toric geometry:
\beq
\cal{M} \mathnormal{= \BC^c / / Q_T}
\eeq
The quiver gauge theories in this thesis have mesonic moduli spaces which are Toric CY singularities. Therefore all of the $Q_T$ matrices computed satisfy the Calabi-Yau condition
\beq
\Sigma_j Q_{T \; ij} = 0
\eeq
The $Q_T$ matrix can be used to define the $G$ and $G_t$ matrices corresponding to the toric CY 3-fold. These concepts were discussed in Section \ref{s:ToricCY3}.
\Section{The Brane Tiling}
\label{s:TheBT}
The brane tiling has strengthened the link between ideas in toric geometry and quiver gauge theory\cite{Stienstra:2007dy}. Some fundamental aspects of brane tilings shall now be discussed.

\begin{figure}[h!]
\begin{center}
\begin{tabular}{ccc}
\includegraphics[height=5cm]{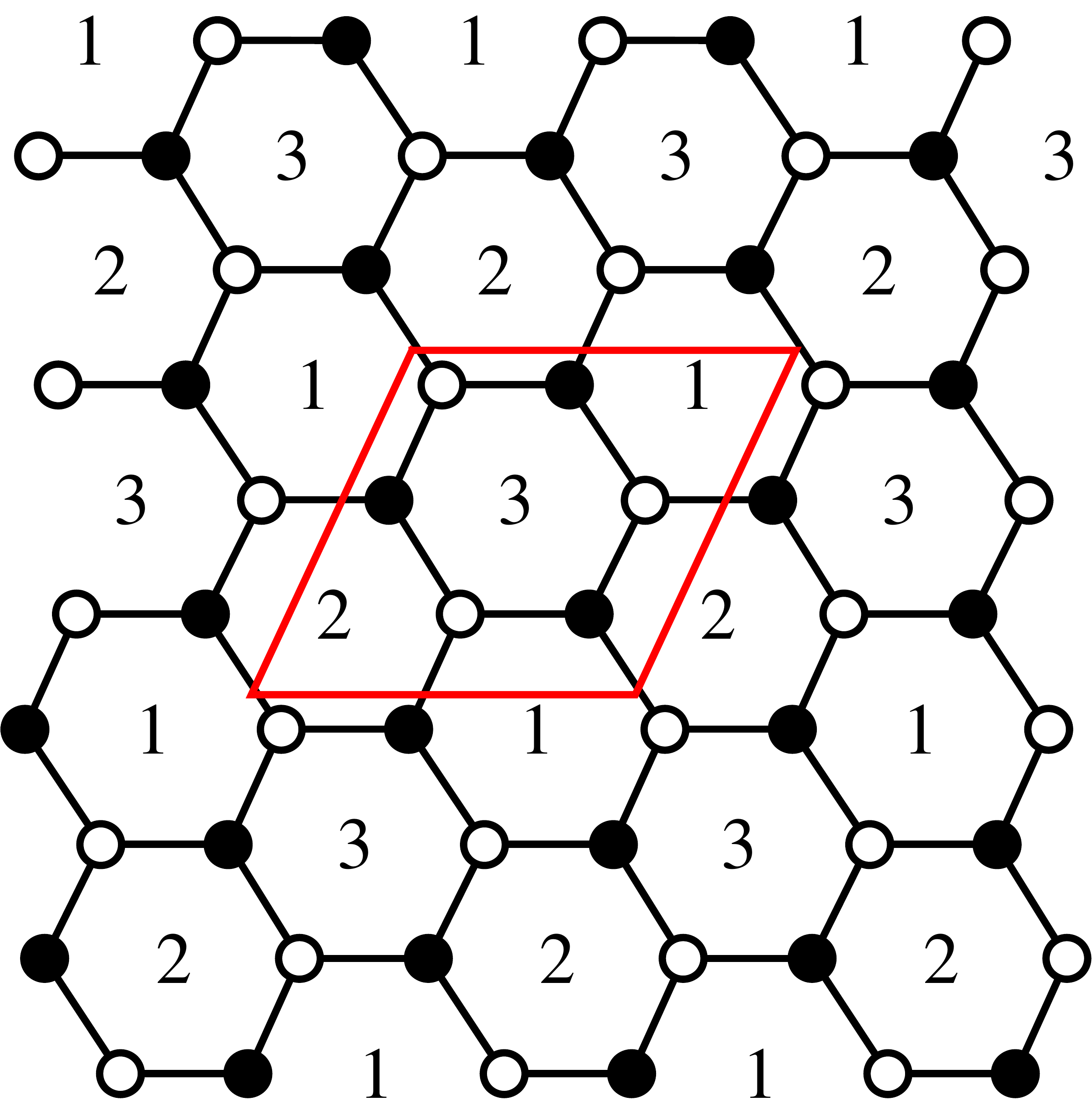}
\end{tabular}
\end{center}
\caption{A Typical Brane Tiling. The fundamental domain is drawn in red. A 2-torus can be formed by identifying opposite edges of this red parallelogram.}
\label{fig:btilingintro}
\end{figure}

A brane tiling (or dimer model) is a periodic bipartite graph on the plane. Alternatively, it is possible to draw a tiling on the surface of a 2-torus by taking the smallest repeating structure (known as the fundamental domain) and identifying opposite edges \cite{Kennaway2005}. The bipartite nature of the graph allows us to colour the nodes either white or black such that white nodes only connect to black nodes and vice versa. In this work we actually restrict attention further to all brane tilings that contain an equal number of black and white nodes. Such brane tilings are known as being `balanced'. A typical brane tiling is given in figure \ref{fig:btilingintro}. For this tiling, the smallest repeating unit consists of 6 nodes (3 black and 3 white) and 9 edges. Brane tilings can be used to represent certain quiver gauge theories which describe the world volume physics of D3-branes probing toric Calabi-Yau 3-fold singularities.

\subsection{Brane Tilings for D3-brane Theories}
Brane tilings were originally developed to describe certain (3+1)-dimensional superconformal field theories (SCFTs) that arise as worldvolume theories for certain branes in Type IIB string theory\cite{Kennaway2005,Hanany05,Franco:2005sm,Feng:2005gw}. Specifically, let us consider Type IIB string theory on $AdS_5 \times X_5$, where $X_5$ is a Sasaki-Einstein manifold. This string theory can be thought of as the gravity dual of a gauge theory living in a stack of D3-branes placed at the conical singularity of $Y_6$, \label{Y6}the cone over $X_5$ \cite{Klebanov:1998hh}. Brane tilings can be used to describe the gauge theory corresponding to (non-compact) toric Calabi-Yau 3-fold singularities.

There is a simple dictionary between a tiling and the (3+1)-dimensional gauge theory that it represents. Every face in the tiling corresponds to a $U(N)$ gauge group. Each edge in the tiling corresponds to a chiral field that transforms under a bi-fundamental representation of the two gauge groups that the edge sits next to in the tiling, with an orientation defined by the bipartite nature of the tiling. White (black) nodes in the tiling correspond to positive (negative) superpotential terms. Each term is a gauge invariant quantity formed by tracing over the fields that the node connects to. The relationship between a tiling, its graph dual - the periodic quiver and the gauge theory it represents is given (Table \ref{t:bt}). One can fully reconstruct a quiver gauge theory's Lagrangian with knowledge of the tiling.

\begin{table}[h!]
\begin{center}
\begin{tabular}{c|c|c}
Tiling & Periodic Quiver & Gauge Theory\\
\hline
Face & Node & $U(N)$ Gauge Group\\
Edge & Edge & Bi-fundamental Chiral Field \\
Node & Face & Superpotential Term
\end{tabular}
\end{center}
\caption{The relationship between a brane tiling, a periodic quiver and the field theory that they represent}
\label{t:bt}
\end{table}

Whereas a quiver diagram requires a superpotential to define a Lagrangian, a tiling fully specifies a quantum field theory. We can think of the tiling specifying a quiver and a superpotential and so specifying a Lagrangian. Also, due to the bipartite nature of the tiling, the anomaly cancellation condition discussed in Section \ref{s:Acanc} is automatically satisfied. These two features of a tiling make it a very appealing object.

\subsection{The Forward Process using the Kasteleyn Matrix}
\label{D3FFA}

It has been found that there is an alternative way of finding the toric description of the moduli space of such a gauge theory \cite{Hanany05}. This method does not rely on calculation of the charge matrices, but rather involves computing a weighted adjacency matrix of the tiling, which is known as the Kasteleyn matrix. The two methods of computing the toric description of the moduli space of a theory described by a brane tiling have been shown to be equivalent \cite{Franco:2006gc}. We shall now review the method which uses the Kasteleyn matrix.

\begin{figure}[h!]
\begin{center}
\includegraphics[totalheight=6cm]{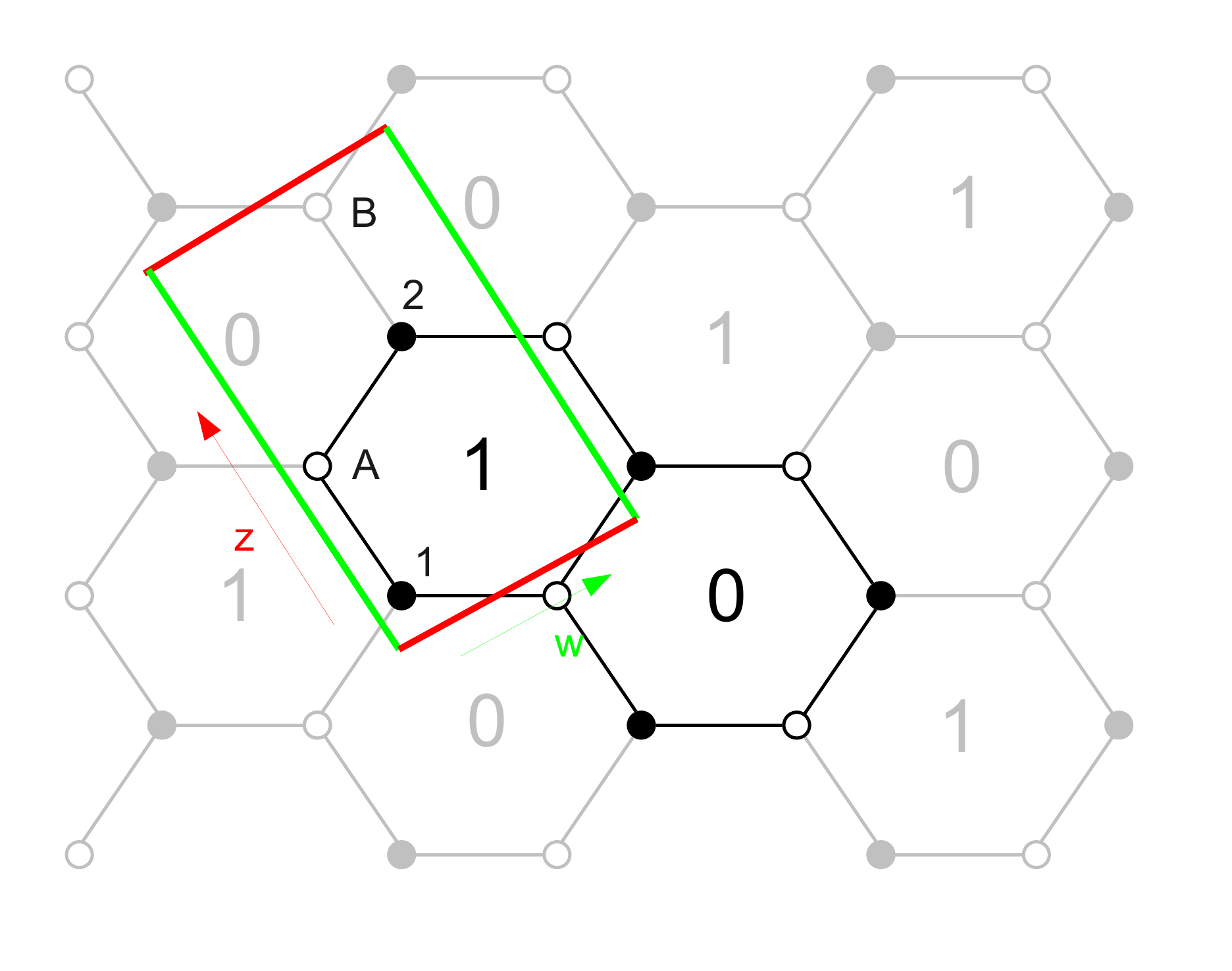} 
\end{center}
\caption{The 2 hexagon tiling. The rectangle represents the fundamental domain of the tiling. White nodes are labeled `A' and `B'. Black nodes are labeled `1' and `2'.}
  \label{f:Tile2Hex}
\end{figure}

The first step in this algorithm is to write down a brane tiling. Let us choose the tiling with 2 hexagons as our starting point (Figure \ref{f:Tile2Hex}).

\begin{figure}[h!]
\begin{center}
\includegraphics[totalheight=4cm]{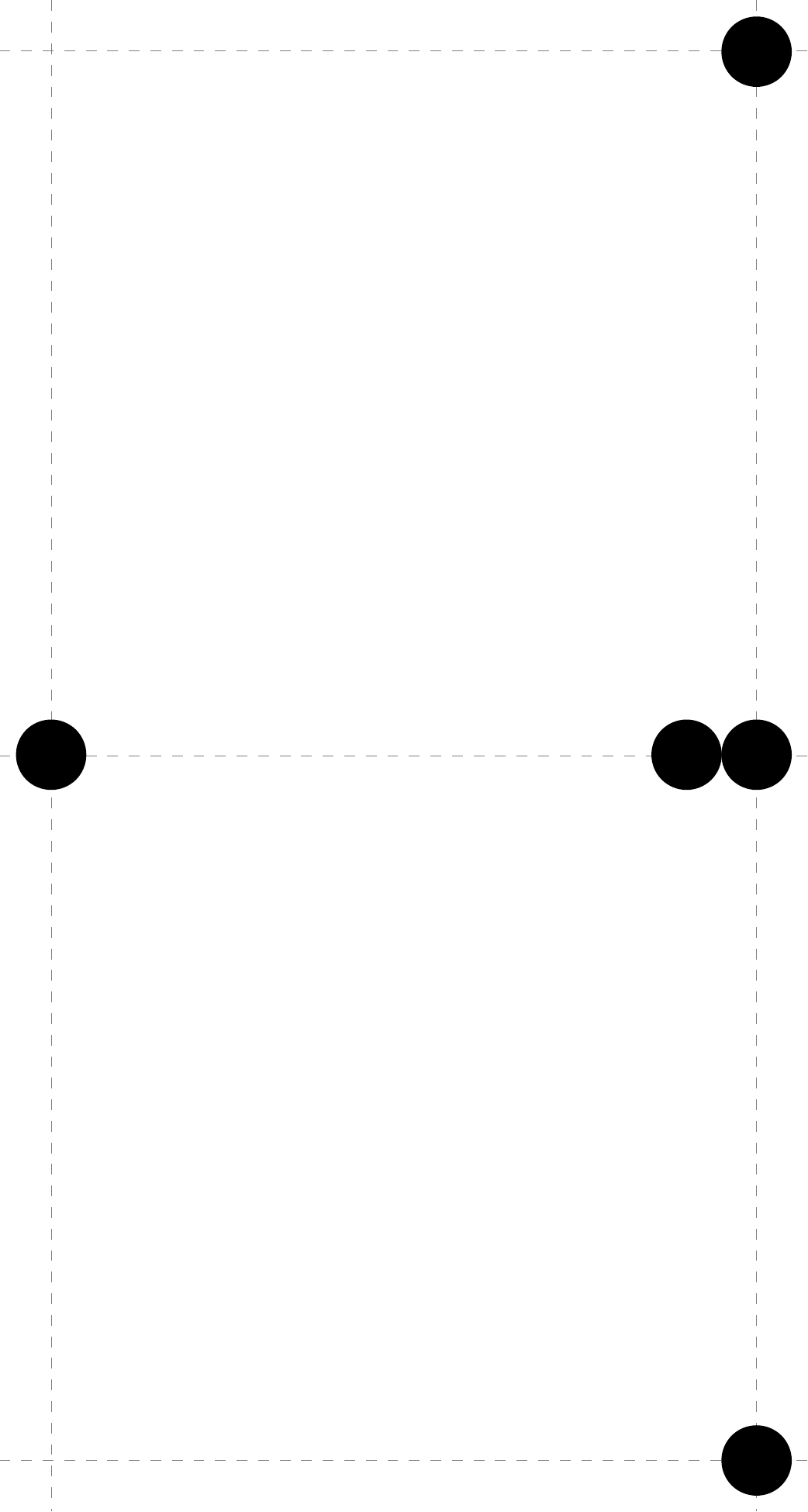} 
\end{center}
\caption{The toric diagram corresponding to the 2 hexagon tiling. The multiplicity of 2 at (1,0) is indicated on the diagram by a double point}
  \label{f:Toric2Hex}
\end{figure}

We must now write down a weighted adjacency matrix corresponding to the tiling. This matrix is known as the Kasteleyn matrix of the tiling. Our convention is that columns are indexed by white nodes and rows are indexed by black nodes. In order to construct the Kasteleyn matrix, the fundamental domain of the tiling is drawn. This fundamental domain is not unique, but this detail is not too important: any fundamental domain with no nodes on its edges is good enough for this algorithm to work. Two variables $w$ and $z$ are chosen and each edge is weighted according to how it crosses sides of the fundamental domain. If an edge crosses no sides it is given weight 1. Edges carry an orientation as they all connect a black node to a white node. If an edge crosses the $w$-boundary in a positive orientation, it is given a weight $w$.
Similarly if an edge crosses the $z$-boundary in a negative orientation, it is given a weight $1/z$. The Kasteleyn matrix for the two hexagon tiling (see Figure \ref{f:Tile2Hex}) is given in \ref{e:Kast2Hex}.
\bea
K =   \left(
\begin{array}{c|cc}
& A & B \\
\hline
1 & 1 &  z + z w  \\
2 & 1 + 1/w   & 1  \end{array}
\right) ~.
\label{e:Kast2Hex}
\eea
The next step is to compute the permanent of the Kasteleyn matrix. The permanent of a square matrix can be thought of as the determinant without signs of permutations taken into account. Each term in the permanent of an adjacency matrix of a bipartite graph corresponds to a perfect matching. The permanent of the Kasteleyn matrix corresponding to the 2 hexagon tiling is:
\beq
\mathrm{Perm}(K) = 1 + z( 1/w + 2 + w)
\eeq
It is possible to display this information on a $\BZ^2$ lattice. Each term in the permanent can be displayed as a point on this lattice. The exponent of $w$ in the term corresponds to one of the coordinates of the point, and the exponent of $z$ corresponds to the other coordinate. The shape formed is the toric data of the moduli space of the gauge theory that the brane tiling represents. The toric data corresponding to the 2 hexagon tiling is given in Figure \ref{f:Toric2Hex}.

\subsection{Inverse Process for D3-brane theories}
\label{d3inv}

It is interesting to ask whether it is possible to reverse the forward algorithm; that is if we start with a toric CY 3-fold, is it possible to construct a 3+1 dimensional quiver gauge theory that has this singularity as its mesonic moduli space. The string theory interpretation would be whether it is possible to construct a world-volume theory for D3-branes probing generic toric CY singularities. This question has been tackled and it is known how to construct at least one tiling -- and so at least one quiver gauge theory -- that corresponds to each toric CY 3-fold \cite{FengToric, Hanany:2005ss, Gulotta}. 

\subsection{A Brane Interpretation of the Tiling}

It is conjectured that a brane tiling can be interpreted as a brane construction in type IIB string theory \cite{Franco:2006gc}.

The brane tiling can be made from NS5-branes and D5-branes. An NS5-brane
is extended in the 0123 direction and wraps a holomorphic curve in the 4567 directions. The 4 and
6 directions are periodically identified giving rise to a 2-torus. It is this 2-torus that can be drawn as a tiling.
 D5-branes are extended in the 012346 directions and can be thought of as being
suspended within the `holes' of the NS5-brane in the 46 torus. Every stack of D5-branes gives
rise to a gauge group. Strings crossing every NS5-brane segment and connecting two D5-brane
stacks correspond to chiral multiplets transforming in the bi-fundamental representation of the
corresponding gauge groups. Gauge invariant superpotential terms are produced by the coupling of
massless string states at the nodes of the NS5-brane configuration.

\begin{figure}[h!]
\begin{center}
\includegraphics[totalheight=4cm]{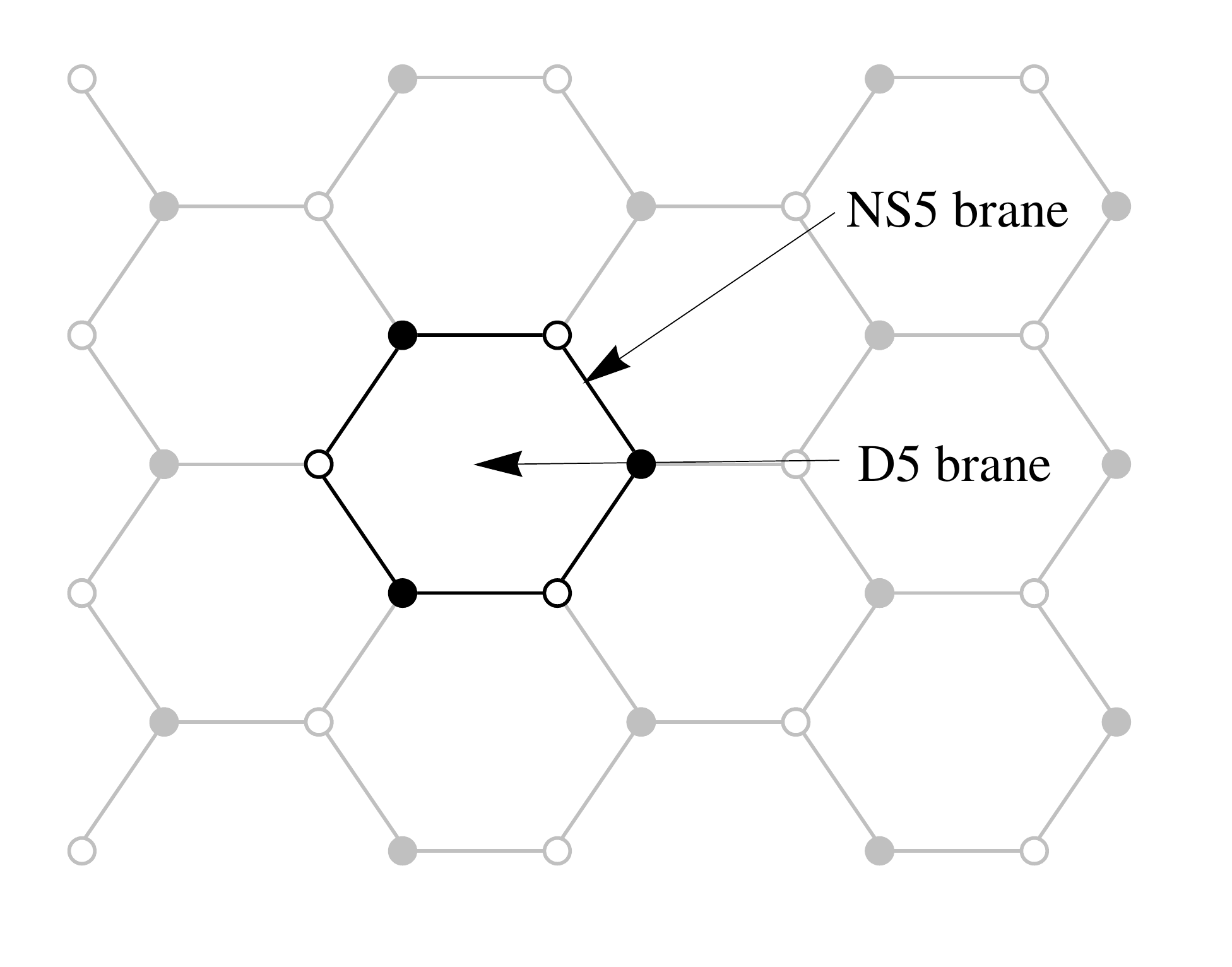} 
\end{center}
\caption{A string theory interpretation of the brane tiling}
  \label{f:NS5D5}
\end{figure}

This construction is conjectured
to be related to the D3-branes probing the singularity by two T-dualities. The suspended D5-branes
are dual to the probe D3-branes and the NS5-brane structure is dual to the singular geometry.

Regardless of whether this conjecture is true, the brane tiling is an incredibly easy way to visualize a large family of quiver gauge theories.

\subsection{Consistency of the Gauge Theory}
\label{ss:consist}

A quantum field theory can be thought of as an ultraviolet fixed point together with an infrared fixed point connected by a renormalization group flow \cite{AMax} \cite{Amax2}. Every quantum field theory (including those described by brane tilings) should flow to some conformal field theory at low energies. It is possible for the low energy theory to be trivial and only consist of non-interacting scalar fields, but a more interesting case is where one has an interacting fixed point. 

The IR limit of a large class of quantum field theories corresponding to brane tilings is known, although some `inconsistent' brane tilings exist which correspond to theories that have more complicated IR properties. These inconsistent tilings can correspond to gauge theories that are tachyonic \cite{Tachion}, while others are fractional Seiberg duals \cite{FractionalDuals} or mutations \cite{Mutations}. Luckily there is a simple and elegant consistency check we can perform on a tiling.

A tiling representing a (3+1)-dimensional gauge theory is thought to be `consistent' if and only if it has the same number of gauge groups as there are cycles for D-branes to wrap in the dual gravity theory \cite{Hanany:2005ss} \cite{Gulotta}. A glance at the tiling is sufficient to see the number of gauge groups of the quiver theory however the method we employ to count the number of gauge groups from the string theory side is a little more involved. One way of counting the relevant cycles is by computing the area enclosed by toric diagram produced by applying the fast forward algorithm to the tiling \cite{Hanany05}. Many of the tilings later shown in this chapter are labeled consistent or inconsistent based on this check.

\chapter{On the Classification of Brane Tilings}
\label{ch:class}
The complete classification of all brane tilings is still an open problem. Progress has recently been made by developing an algorithm that can -- at least in principle -- be used to generate all brane tilings with a given number of superpotential terms. Equivalently one could think of the task as generating all possible balanced bipartite graphs on a torus with a given number of nodes. This section follows the publication `On the Classification of Brane Tilings' \cite{Davey:2009bp}.

The total number of these tilings is, of course, infinite so it is important to figure out which parameters can be used to organize the classification of brane tilings. The natural parameters of a tiling are the number of nodes in the fundamental domain of the tiling $N_T$ and the number of tiles $G$. The number of edges in the fundamental domain $E$ is then fixed by the Euler condition:
\be
E=G+N_T.
\ee
We should remind ourselves that these numbers correspond to details of the quiver gauge theory that the tiling represents. The number of nodes in the quiver (or number of gauge groups) is equal to $G$, the number of bifundamental fields is $E$ and the number of terms in the superpotential is $N_T$.

Working directly with tilings is computationally quite difficult. As a tiling can be formed from a collection of highly irregular faces, it is not obvious how to set up a systematic calculation of the possible periodic tilings with some parameters $(N_T, G)$, especially without making any a priori assumptions about the shapes of the tiles. For that reason we choose quiver gauge theories as our main working objects, in a similar spirit to \cite{YangsTeam}. Our method of attack is to enumerate all possible quivers and superpotentials, and then check which ones admit a tiling description. As each brane tiling corresponds to a quiver gauge theory, we can be sure that every tiling will be generated.

The algorithm that has been developed to generate all possible tilings goes as follows:
\begin{enumerate}
\item 
Fix the order parameters $(N_T, G)$. 
\item 
Enumerate all distinct \emph{irreducible} quivers with $G$ nodes and $E=G+N_T$ fields.
\item 
For each quiver enumerate all possible superpotential terms satisfying the toric condition. This gives the full list of possible quiver gauge theories for $(N_T, G)$.
\item 
Try to reconstruct the tiling for each quiver gauge theory. If we succeed, we add it to the classification, otherwise we conclude that the gauge theory doesn't have a tiling description.
\end{enumerate}

Each step here requires further explanation. But let us postpone this and introduce the concept of doubling and explain exactly what is meant by the term irreducible quiver.

\subsection{The Doubling Process and Quadratic-Node Tilings}
\label{s:doubling}

\begin{figure}[h]
\begin{center}
\includegraphics[width=12cm]{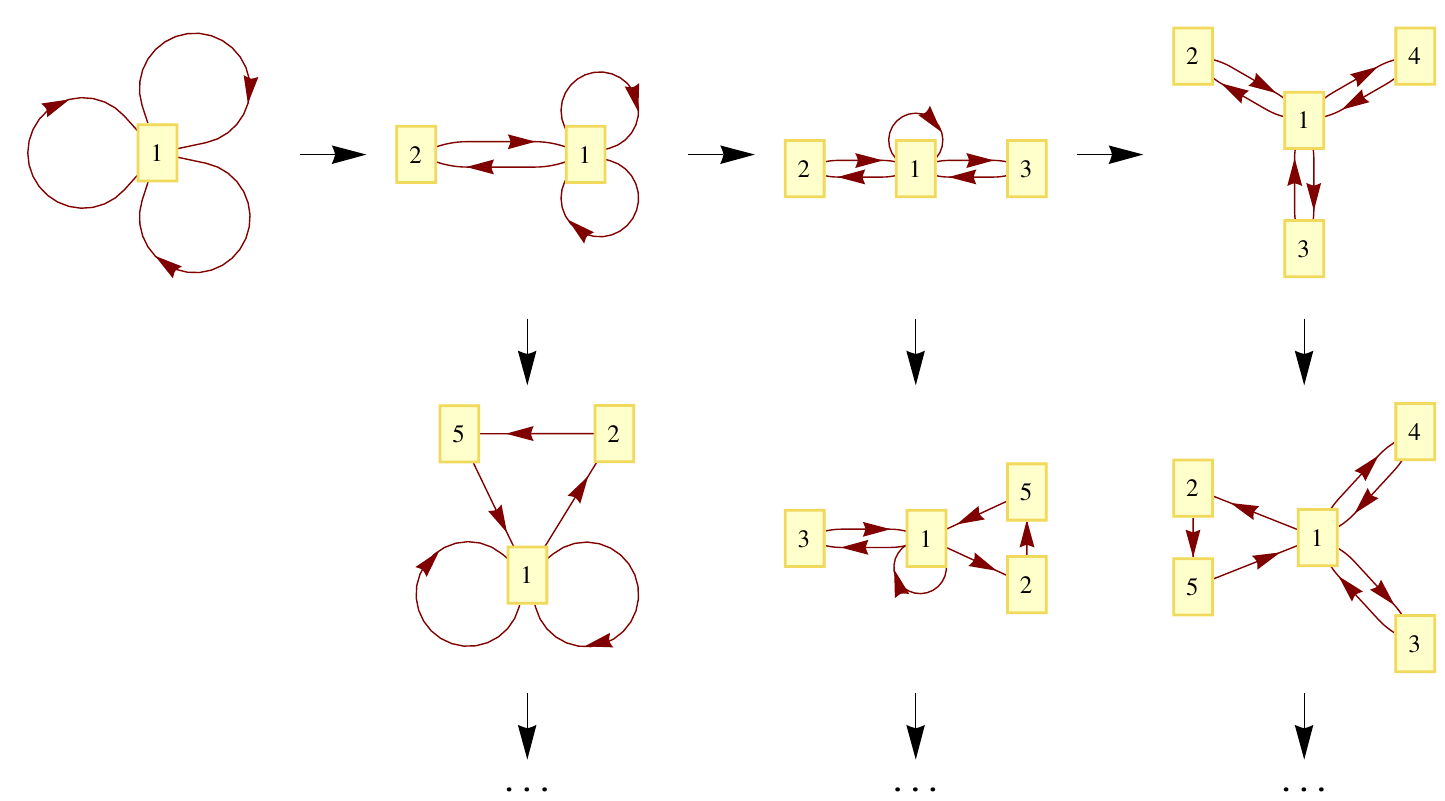}
\end{center}
\caption{Quivers generated by applying the doubling process to the $\BC^3$ quiver.}
\label{fig:doubling-C3}
\end{figure}

Let us consider an operation on a quiver diagram where we replace an edge with two edges, both connected to a node of valence 2. We shall call this process doubling. This process defines a new theory when applied to any of the fields in a quiver. For example, starting with the simple $\BC^3$ model we can construct an infinite number of models by repeatedly applying the doubling procedure (see Figure~\ref{fig:doubling-C3}). This process has a corresponding effect on the brane tiling. An edge in the tiling is replaced by two edges and a face surrounded by only these 2 edges. This is known as a double bond \cite{HananyZaff08,Davey:2009sr}.

\begin{figure}[h]
\begin{center}
\includegraphics{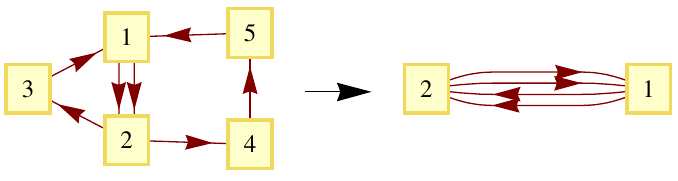}
\end{center}
\caption{Reduction of a quiver by removal of single-in, single-out nodes.}
\label{fig:reduction1}
\end{figure}

This doubling process is always reversible. If we are given a brane tiling with double-(or multi-)bonds, we can always remove them by the process of ``Higgsing". By Higgsing the right fields we can remove all nodes of valence 2 from the quiver (Figure~\ref{fig:reduction1}). Let us call quivers with at least one node of valence two ``reducible". If a quiver isn't reducible it is said to be ``irreducible'

For the moment, we will only consider irreducible quivers (or tilings with no double-(or multi-)bonds. All reducible quivers can easily be generated by applying the doubling process to the set of irreducible quivers. This is a crucial observation, because it lets us effectively ignore an infinite ``direction" in the space of tilings, thus allowing us to concentrate on the much smaller class of brane tilings, which are not related by this simple transformation.

\begin{figure}[h]
\begin{center}
\begin{tabular}{ccc}
\begin{minipage}{5cm}
\includegraphics[width=5cm]{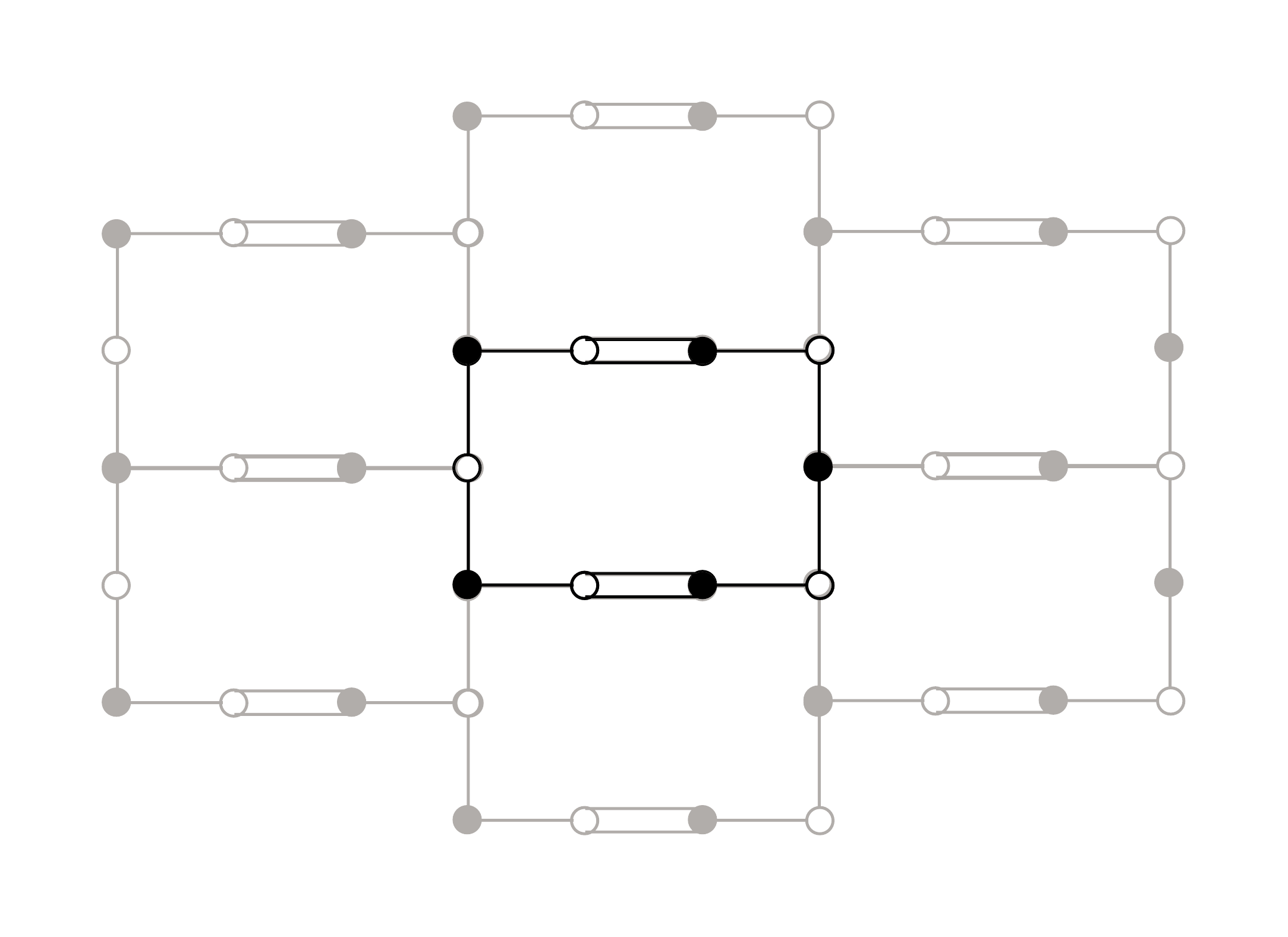}
\end{minipage} 
&  $\rightarrow$ &
\begin{minipage}{5cm}
\includegraphics[width=5cm]{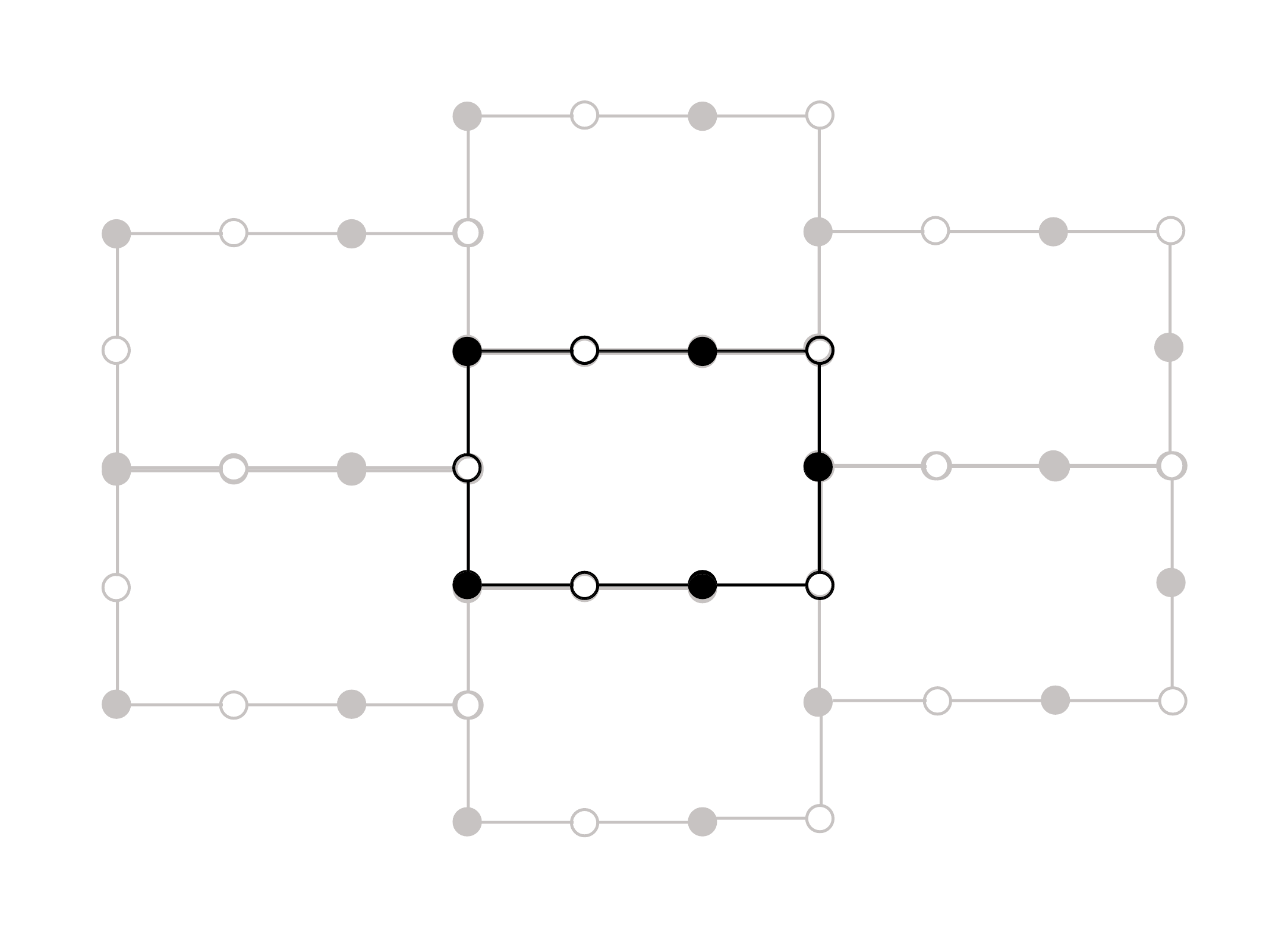}
\end{minipage} 
 \\
\begin{minipage}{5cm}
\begin{center} \includegraphics{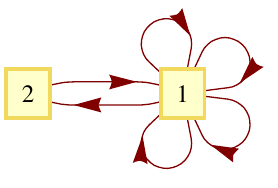} \end{center}
\end{minipage} 
& $\rightarrow$ &
\begin{minipage}{5cm}
\begin{center} \includegraphics{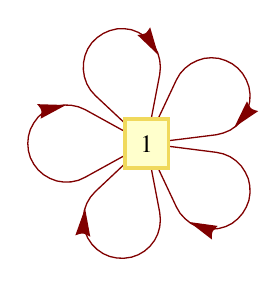} \end{center}
\end{minipage} 
\end{tabular}
\end{center}
\caption{Reduction of a quiver resulting in a quadratic-node tiling.}
\label{fig:reduction2}
\end{figure}

We have to note, however, that there is one caveat in the argument above. For some reducible quivers the Higgsing procedure results in a brane tiling, which has nodes connected only by two edges, as seen in Figure~\ref{fig:reduction2}. This means that the corresponding quiver gauge theory will have a superpotential with quadratic terms in it. We call such models quadratic-node tilings. 

\begin{figure}[h]
\begin{center}
\begin{tabular}{ccc}
\begin{minipage}{5cm}
\includegraphics[width=5cm]{class/N4-base.pdf}
\end{minipage} 
&  $\rightarrow$ &
\begin{minipage}{5cm}
\includegraphics[width=5cm]{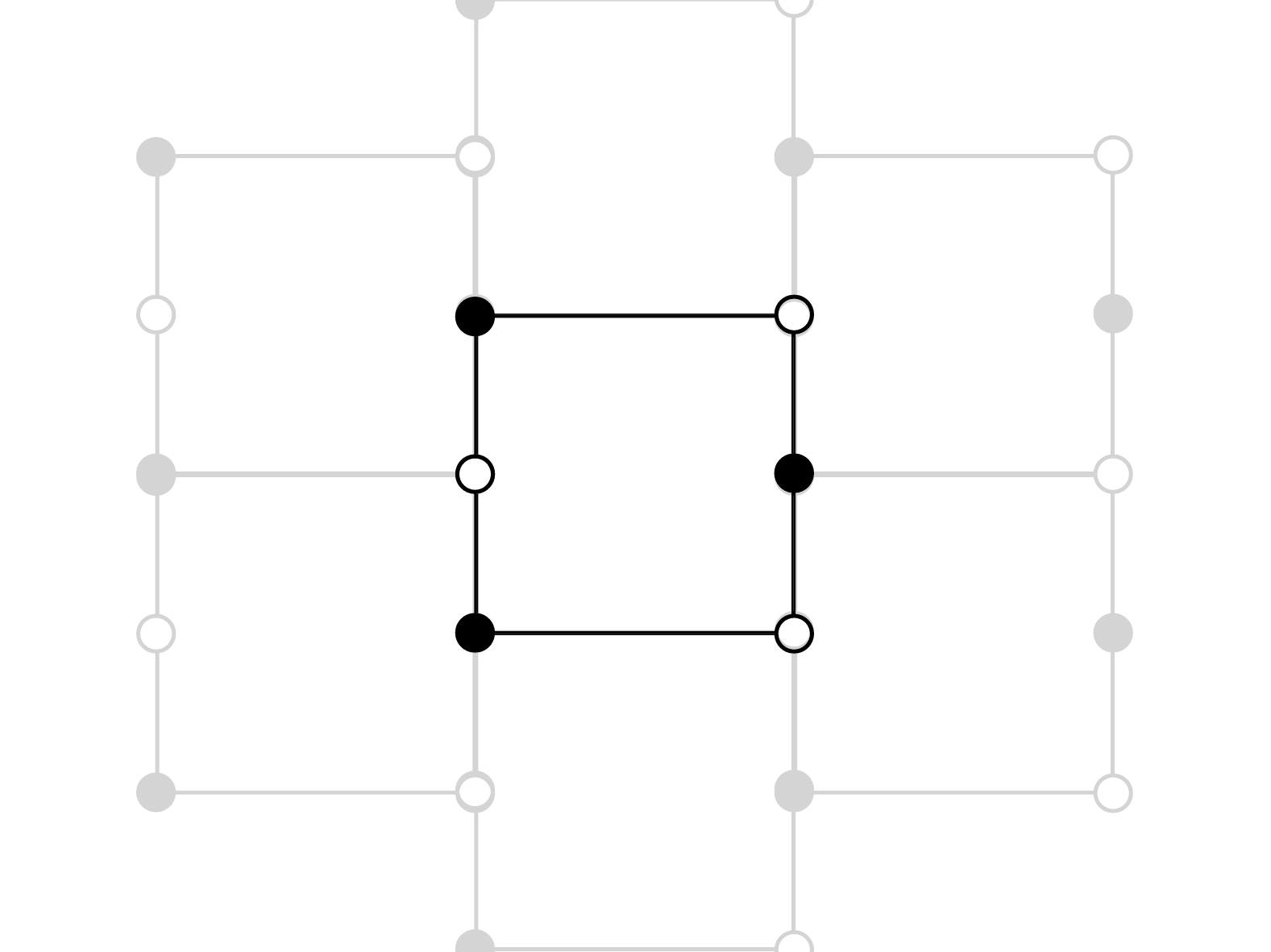}
\end{minipage} 
\end{tabular}
\end{center}
\caption{Reduction of a quadratic-node tiling.}
\label{fig:reduction3}
\end{figure}

The quadratic-node tilings are perfectly valid as bipartite tilings of a plane, however, they are not normally considered in the context of quiver gauge theories on D3-branes. This is because the quadratic superpotential terms indicate massive fields, which become non-dynamical in the infrared limit \cite{Hanany05}. Since we are interested in analyzing the IR limit of these gauge theories, the massive fields should be integrated out using their equations of motion. The corresponding effect on the tiling is that the quadratic node can be removed, gluing the two adjacent nodes together (see Figure~\ref{fig:reduction3}).

For this reason we exclude the tilings with quadratic nodes from our classification. However, this means that the models where quadratic nodes are only absent because of multi-edges (such as the one in Figure~\ref{fig:reduction2}) can not be recovered from the irreducible quivers simply by the doubling procedure. To get back such tilings from the classification in this paper we would have to combine the doubling procedure together with an insertion of two extra nodes.

For now we shall restrict our attention to the generation of brane tilings without multi-edges or quadratic nodes. Let us now describe our algorithm further.

\subsection{Order parameters}

The reader may recall that there are two parameters that we are going to use to order our classification - $N_T$ and $G$. There are a few simple arguments that have allowed us to put limits on the possible values of $G$ that need to be considered.

Firstly, let us consider the requirement that the quiver is irreducible. This is equivalent to saying that there should be no nodes in the quiver of valency 2. As the nodes in the quiver must have the same number of incoming and outgoing edges\footnote{This is a consequence of the bipartite nature of the tiling and also the aforementioned gauge anomaly cancellation condition.}, each node should be of valency 4 or higher. We also have the following relationship for any quiver:
\be
E = \frac{1}{2}\sum_{i=1}^{G}n_i,
\label{eq:adjacencies}
\ee
where $n_i$ is the order of node $i$ and the sum is taken over all nodes in the quiver. We therefore find the condition that
\be
E \geq 2 G
\ee
Using $E=G+N_T$ we have the condition
\be
N_T \geq G
\label{e:GCond}
\ee
Therefore if we want to build all irreducible brane tilings with a given number of superpotential terms, we know that $G$ must satisfy \eqref{e:GCond}.

A lower bound for $G$ for fixed $N_T$ can also be found. As the tilings are irreducible, this means the minimum order of all nodes is 3. Let us use \eref{eq:adjacencies} on the tiling, counting only edges and nodes in the fundamental domain. Now the edges are again fields and the nodes are the superpotential terms, giving us the bound:
\be
E \geq \frac{3}{2} N_T.
\ee
Using $E=G+N_T$ we get
\be
G \geq \frac{1}{2} N_T,
\ee
which is our lower bound on the parameter $G$ for given $N_T$, and so we have for fixed $N_T$
\be
\frac{1}{2} N_T \leq G \leq N_T
\label{e:GRange}
\ee
It is now clear how to organize the classification. We will consider each $N_T$ in an increasing order, exploring all possible values of $G$ satisfying \eqref{e:GRange} at each step. The number of possible superpotential terms $N_T$ is, of course, unbounded. A summary of the range of parameters that we must consider for low values of $N_T$ is given in Table \ref{tab:order}.

\begin{table}
\centering
\begin{tabular}{c|c|c|c|c}
$N_T$ & $G_{min}$ & $G_{max}$ & $E_{min}$ & $E_{max}$ \\
\hline
2 & 1 & 2 & 3 & 4 \\
4 & 2 & 4 & 6 & 8 \\
6 & 3 & 6 & 9 & 12
\end{tabular}
\caption{Values of the possible quiver parameters which must be explored.}
\label{tab:order}
\end{table}

\subsection{Finding Quivers}

Once we fix the parameters $(N_T, G)$, the next step is to enumerate all of the possible quiver graphs with a given number of nodes $G$ and edges $E$. The task is quite straightforward, but it has to be handled with a little care, to avoid the algorithm becoming too computationally expensive as $G$ and $E$ grow larger.

A na\"{i}ve approach would be to consider all possible ways of connecting $G$ nodes with $E$ edges. With $G(G-1)$ ways of drawing a directed edge, we would have the order of
\be
(G(G-1))^E
\ee
possible graphs to consider, which is clearly too large for, say, $G=6, E=12$. However, we are only interested in a very small fraction of these graphs. Nodes of quivers that correspond to brane tilings must have the same number of incoming as they do outgoing edges. The reader should recall that this corresponds to the anomaly cancellation condition (ACC) in 3+1 dimensions (Section \ref{s:Acanc}).

\begin{figure}[h]
\centering
\includegraphics{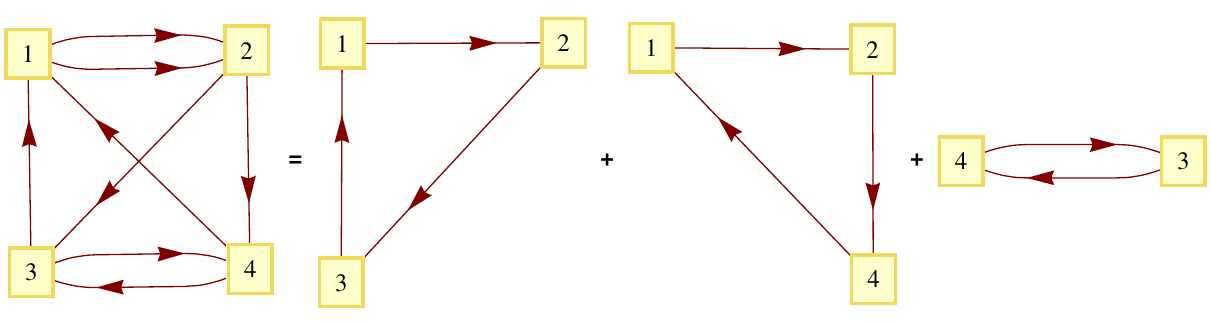}
\caption{Decomposition of a graph into cycles.}
\label{fig:decomposition}
\end{figure}

The key idea of this efficient algorithm for finding all possible quivers is to incorporate the ACC into the construction of the quiver. We achieve this by making the following observation: a graph has the same number of incoming and outgoing edges at each node if and only if it can be decomposed into a `sum' of cycles. By `sum' we mean that we take the union of nodes and the union of edges from the constituent cycles, while keeping the labels of the nodes intact (so that $1\ra 2\ra 3\ra 1$ is different from $1\ra 2\ra 4\ra 1$). An example of such a decomposition is shown in Figure~\ref{fig:decomposition}.

In order to build a complete list of quivers for a given $G$ and $E$, we must first consider all of the possible cycles over $G$ nodes. Then we take combinations of those cycles such that the total number of fields adds up to $E$. This way we have all of the quivers that satisfy the ACC.

\subsection{Finding Superpotentials}
\label{sec:superpotentials}
After finding the quivers, we must construct all possible quiver gauge theories. This is done by finding all of the superpotentials $W$ that could be associated to each quiver. By considering two important features these special quiver gauge theories must have, we can efficiently find all possible consistent superpotentials.

There are two useful constraints on the form of a quiver gauge theory's superpotential that we should consider. The first is that each term in $W$ has to be gauge-invariant. With the bi-fundamental (or adjoint) nature of the fields, this means that a field `ending' on a group factor $g$ has to be contracted with a field `starting' on $g$. If the field $X_{ij}$ transforms under the fundamental representation of gauge group $i$ and the anti-fundamental representation of gauge group $j$, a typical term in the superpotential will look like
\be
 Tr(X_{12}X_{23}X_{31}) \subset W
\ee
This condition has a nice interpretation in the quiver picture: gauge-invariant terms are just \emph{cycles} in the quiver. From this observation, we can see that the cycles generated in the quiver generation step of our algorithm will allow us to quickly generate all possible superpotentials.

The second constraint on the superpotential that we must consider is known as the `toric condition' \cite{ToricD}. It states that each field in the quiver gauge theory should appear in the superpotential exactly twice: once in a positive term and once in a negative term. The bipartite nature of the tiling is a manifestation of this toric condition. For every quiver, we take all ways in which cycles can make up the quiver and find all ways of combining these cycles into superpotentials that satisfy the toric condition. However only a small fraction of these models can actually admit a tiling description, and for that we need a final step in the algorithm.

\subsection{Reconstructing Tilings}

The final step in the algorithm is to check for whether a given quiver gauge theory can correspond to a brane tiling and then to find this tiling.

The way we proceed is by using an object called a periodic quiver. The periodic quiver is simply the graph dual of the tiling: nodes are gauge groups, fields are edges and faces are superpotential terms. Since the data generated so far comprises of a list of quivers and superpotentials, the task of finding the tilings reduces to whether we can `unfold' the quivers into bi-periodic graphs of the plane. If we can find a periodic quiver from an ordinary quiver and a superpotential, then we know that the model admits a tiling description, and we can easily find the tiling by taking the graph dual of the periodic quiver.

The algorithm used to produce the tilings goes as follows. We are given the quiver $Q$ and superpotential $W$ generated from previous steps of the algorithm. The idea is to try to build up the fundamental domain of the periodic quiver. To do that, firstly, we represent each term in $W$ by a polygon with edges around its perimeter representing fields. We choose the fields to have a clockwise orientation for positive terms and a counter-clockwise orientation for negative terms. These polygons (with directed edges) will be the faces of the periodic quiver. 

 Next, we fit these polygons together into one shape by gluing edges that represent the same field together. The process is always possible due to the toric condition on the superpotential. The shape generated is our candidate for the fundamental domain. The test this shape must pass is whether we can identify opposite edges in a way such that the resulting manifold is a 2-torus. If we can do this we have found a periodic quiver and so a brane tiling.

\begin{figure}[h]
\centering
\includegraphics[width=3.5cm]{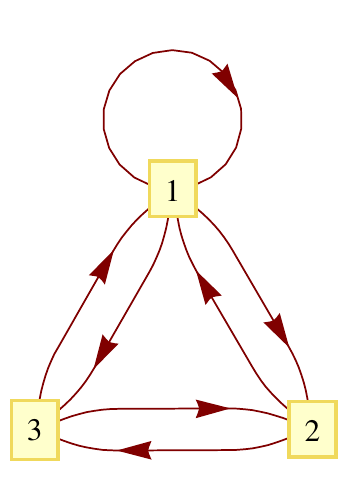} 
\caption{The $SPP$ quiver.}
\label{fig:spp-quiver}
\end{figure}

Let us illustrate this procedure with an example known as the suspended pinch point \cite{Hanany05}. The quiver is shown in Figure~\ref{fig:spp-quiver} and the superpotential is the following:
\be
W = \phi _1^{}.X_{12}^{}.X_{21}^{} -\phi _1^{}.X_{13}^{}.X_{31}^{} 
-X_{12}^{}.X_{23}^{}.X_{32}^{}.X_{21}^{} +X_{13}^{}.X_{32}^{}.X_{23}^{}.X_{31}^{}
\ee

\begin{figure}[h]
\centering
\begin{tabular}{c}
\includegraphics{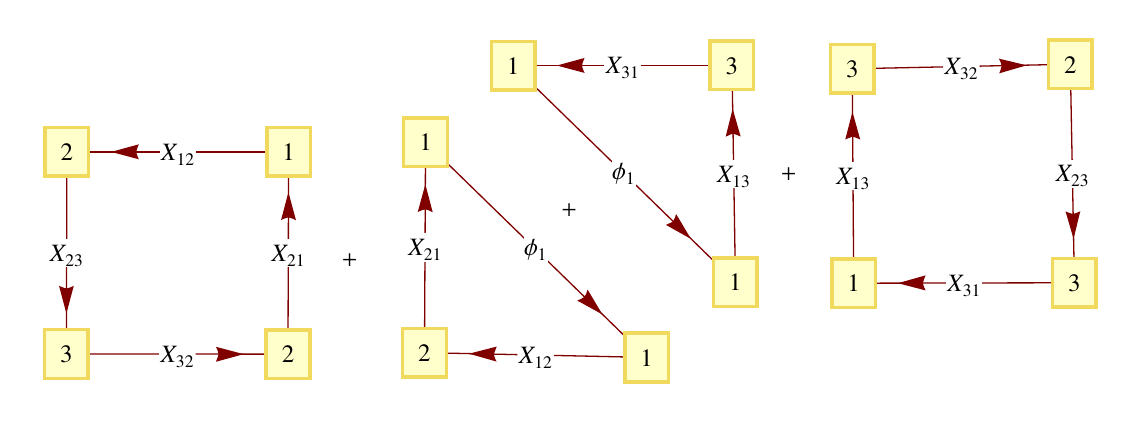} \\
$\downarrow$ \\
\includegraphics{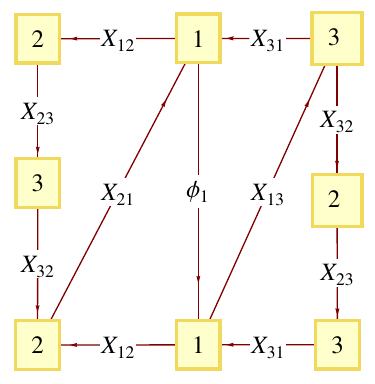}
\end{tabular}
\caption{Combining the superpotential terms into a fundamental domain of the periodic quiver.}
\label{fig:tiling1}
\end{figure}

There are four terms in the superpotential, which we represent by four polygons - two ``triangles" corresponding to the cubic terms and two ``squares" corresponding to the quartic terms (Figure~\ref{fig:tiling1}). Recall that the arrows around the faces go clockwise for positive and counter-clockwise for negative terms. We can now treat the problem just like a jigsaw puzzle: we have to put these pieces together allowing only edges corresponding to the same field to touch. If it is possible to fit these pieces together to form a 2-torus, we will have generated a graph that can be flattened out to form a periodic quiver.

Let us consider the SPP model and glue the four terms together into one shape, by identifying the three fields $X_{21}$, $\phi_1$ and $X_{13}$. This shape is our candidate for the fundamental domain. It is unimportant as to which three fields we pick to glue together; a different choice will just result in generating a different fundamental domain of the periodic quiver. Next we attempt to deform the shape into a rectangle that can be used to tile the plane. If this is possible we have found the model's periodic quiver\footnote{
In some more complicated cases, it is possible to generate a shape that has a pair of identical fields adjacent to each other. We simply glue together all of these repeating edges, until we have a shape with no such repeated edges. We then test whether this shape can be used to tile the plane.}.

We can see in Figure~\ref{fig:tiling1} that it is possible to find a periodic quiver for the SPP. By glancing at the rectangle, we can see that it is possible to use it to tile the plane with only edges corresponding to identical fields touching. We can equivalently see that the shape generated is really a 2-torus. The top and bottom sides of the rectangle can be identified directly along $(X_{12}, X_{31})$, effectively turning the rectangle into a cylinder. Then the ends of the cylinder each consist of $(X_{32}, X_{23})$, and even though they are not exactly the same on the rectangle, the cylinder can be ``twisted" so that the ends are correctly identified.

A key part of the algorithm is this important check for whether the resulting fundamental domain can be wrapped to make a torus. A given quiver gauge theory admits a tiling description if and only if this is possible. A simple shape that fails this check is one that has fields $(\phi_1, \phi_1, \phi_2, \phi_2)$ forming the perimeter of a rectangle.

\begin{figure}[h]
\begin{center}
\begin{tabular}{ccc}
\begin{minipage}{5cm}
\includegraphics[width=5cm]{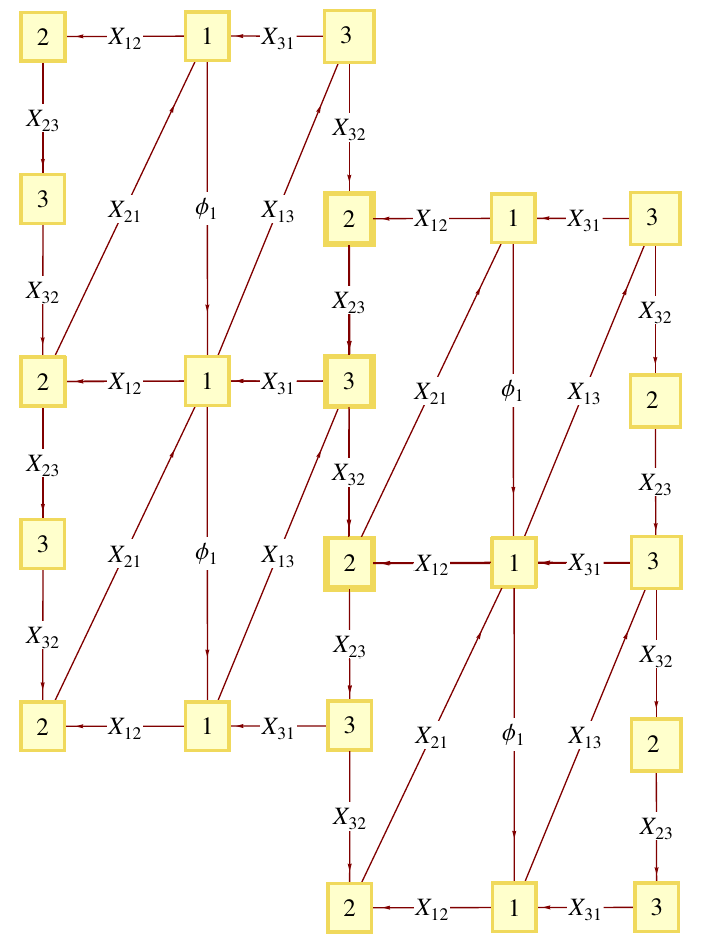}
\end{minipage} 
&  $\rightarrow$ &
\begin{minipage}{5cm}
\includegraphics[width=5cm]{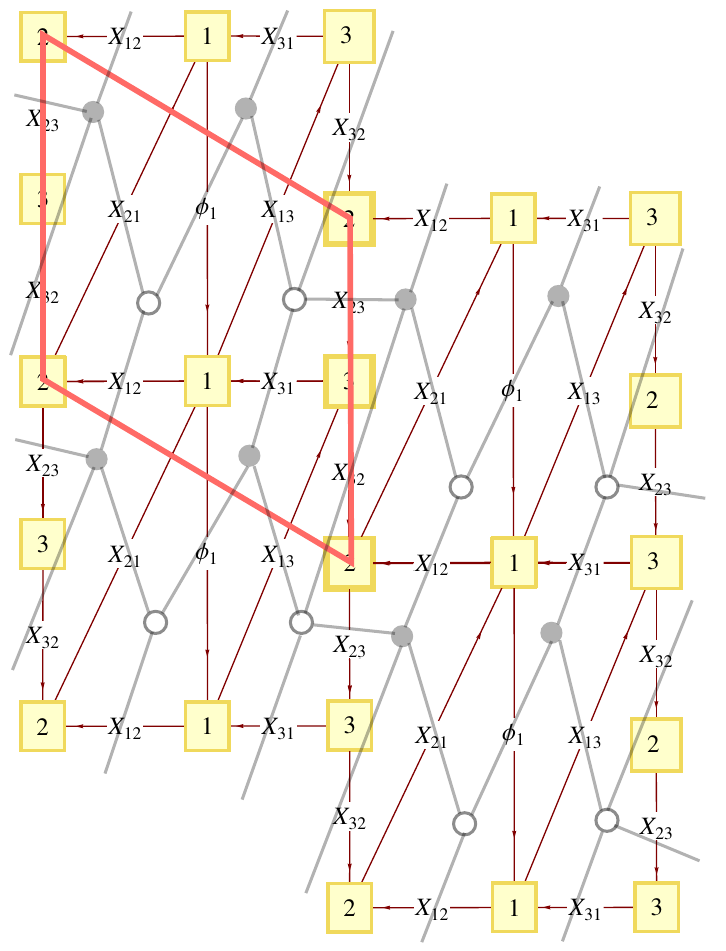}
\end{minipage} 
\end{tabular}
\end{center}
\caption{From the periodic quiver to the brane tiling for the $SPP$.}
\label{fig:tiling2SPP}
\end{figure}

If the construction of a periodic quiver works, we can easily extract the brane tiling from it by finding the dual graph. Firstly, we draw the periodic quiver with our `fundamental rectangle'. Then we insert a white or black node at the center of each face according to whether the arrows go clockwise or counter-clockwise around the perimeter of the face. By replacing edges as in Figure~\ref{fig:tiling2SPP} we build the dual graph (the brane tiling). In the case of the SPP, we see that the tiling consists of one hexagon and two quadrilaterals.

The reader should note that while the algorithm generates a complete list of tilings, it fails to produce aesthetically pleasing brane tilings. In order to display the tiling in terms of nice geometrical shapes we have had to rely on existing algorithms that are able to display large planar graphs neatly.

\subsection{A Model Overview}

An implementation of the algorithm described here has been used to generate all irreducible tilings that have at most 8 superpotential terms. An ordinary desktop computer was easily capable of generating these tilings. In this section we will briefly discuss some of the models found using this implementation. A list of the tilings generated that have at most 8 superpotential terms is given in Appendix \ref{a:tilings}.

\begin{figure}[h]
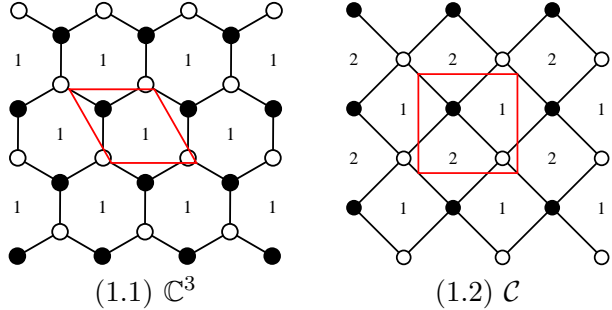

\begin{center}
\begin{tabular}{cc}
\begin{tabular}[b]{c}
\includegraphics[height=3.5cm]{class/N2-G1-1-tiling.pdf} \\
(1.1) $\BC^3$
\end{tabular}
&
\begin{tabular}[b]{c}
\includegraphics[height=3.5cm]{class/N2-G2-2-tiling.pdf} \\
(1.2) $\cal{C}$
\end{tabular}
\end{tabular}
\end{center}
\caption{Consistent tilings with two superpotential terms.}
\label{fig:tilingswith2}
\end{figure}

Let us start our discussion by considering the case of just two terms in the superpotential. In this case, we only need to consider the possibility of having either one or two gauge groups, and we find one possible tiling for each case. These are the most familiar models: the $\BC^3$ model corresponding to the one-hexagon tiling and the conifold $(\cal{C})$ model corresponding to the two-square tiling (see Figure~\ref{fig:tilingswith2}). Both of these tilings are consistent \cite{Kennaway2005}.

Let us now consider the 6 tilings generated that have four superpotential terms. With the minimal possibility of two gauge groups and six fields we find only the two-hexagon model corresponding to the geometry $\BC^2 / \BZ_2 \times \BC$ \cite{Kennaway2005}. Among the models with three and four gauge groups we have the $SPP$, Phase I of $\BF_0$ and Phase I of $L^{222}$ (Figure~\ref{fig:tilings4}). We also find two tilings which are inconsistent (see Appendix \ref{a:tilings}).

Another way of generating all of the tilings with four superpotential terms comes from considering the hexagon as the fundamental unit of a tiling. Let us start with the two-hexagon tiling. Adding new edges to a tiling keeps the number of superpotential terms the same but increases the number of gauge groups. We can find all tilings with 4 superpotential terms by adding edges across faces of the two-hexagon model. We find that there are two ways of adding one diagonal to one of the hexagons, which give the models with three gauge groups. If we add a 2nd diagonal, we can generate the remaining three tilings with four gauge groups. This procedure of finding the tilings by adding diagonals also works for the case of two superpotential terms. We start with the basic one-hexagon tiling and find the conifold model by adding one diagonal (see Figure \ref{fig:tilingswith2}).

\begin{figure}[h]
\begin{center}
\begin{tabular}{ccc}
\begin{tabular}[b]{c}
\includegraphics[width=3.5cm]{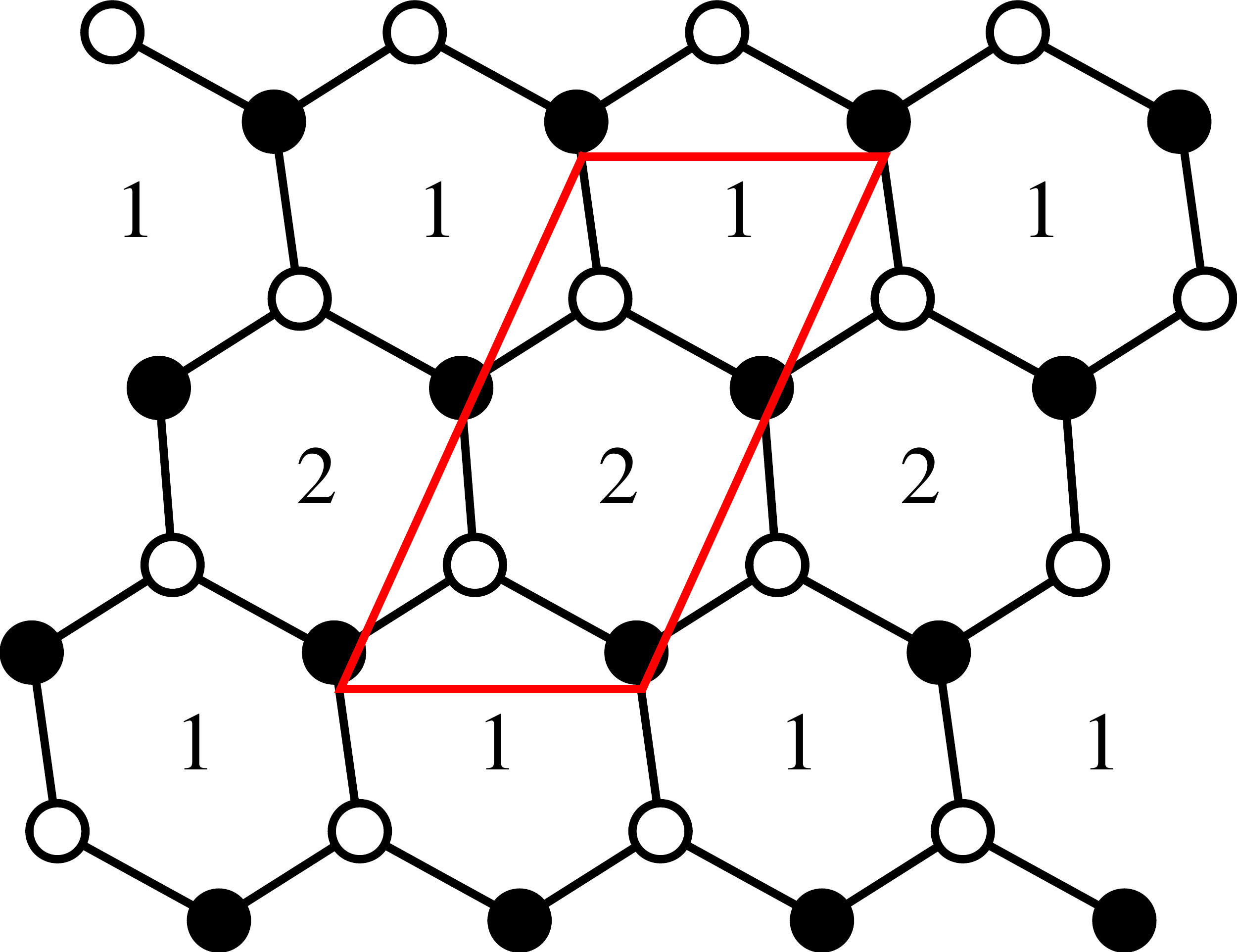} \\
(2.1) $\BC^2/\BZ_2 \times \BC$
\end{tabular}
&
\begin{tabular}[b]{c}
\includegraphics[width=3.5cm]{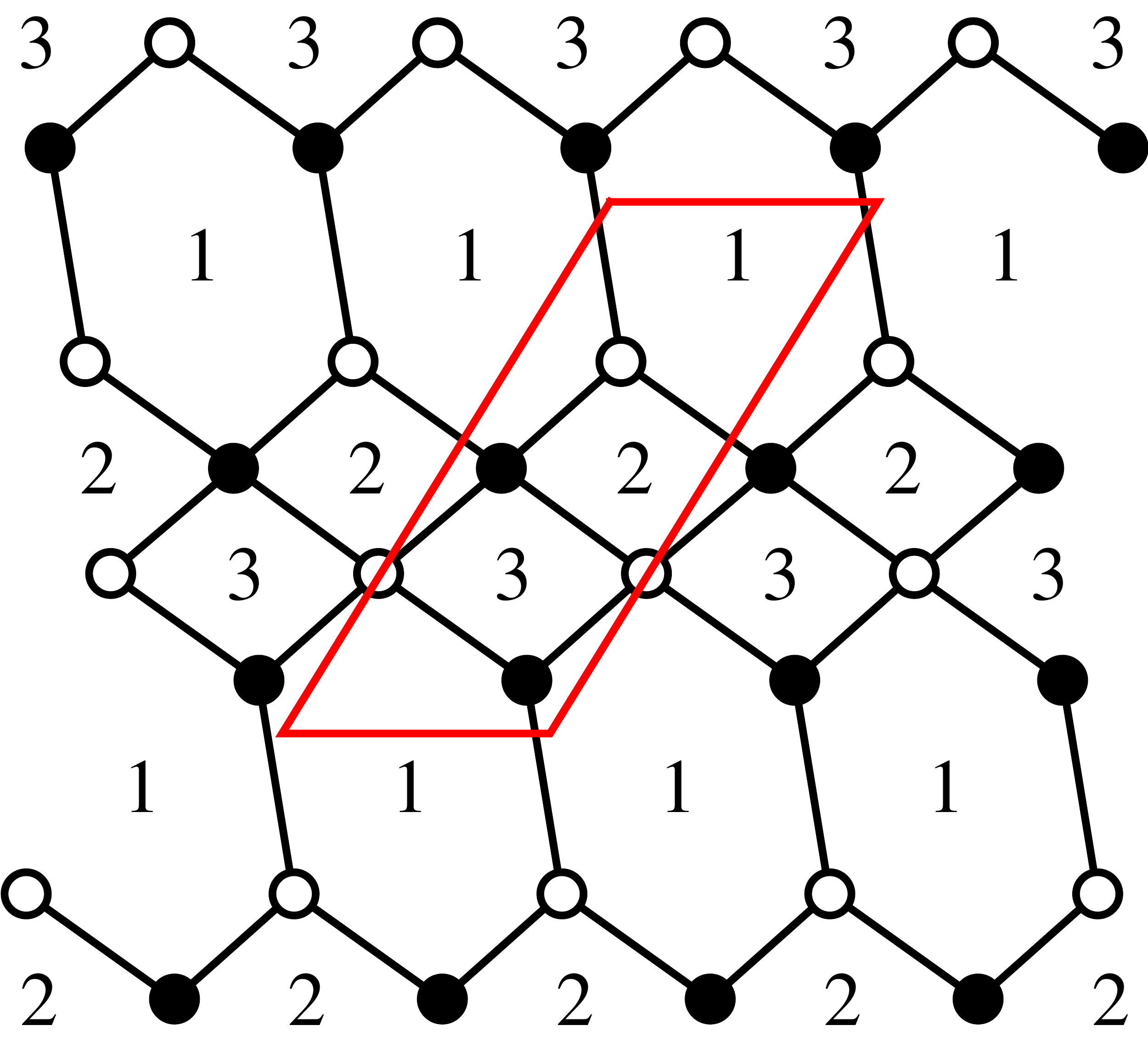} \\
(2.2) $SPP$
\end{tabular}
&
\begin{tabular}[b]{c}
\includegraphics[width=3.5cm]{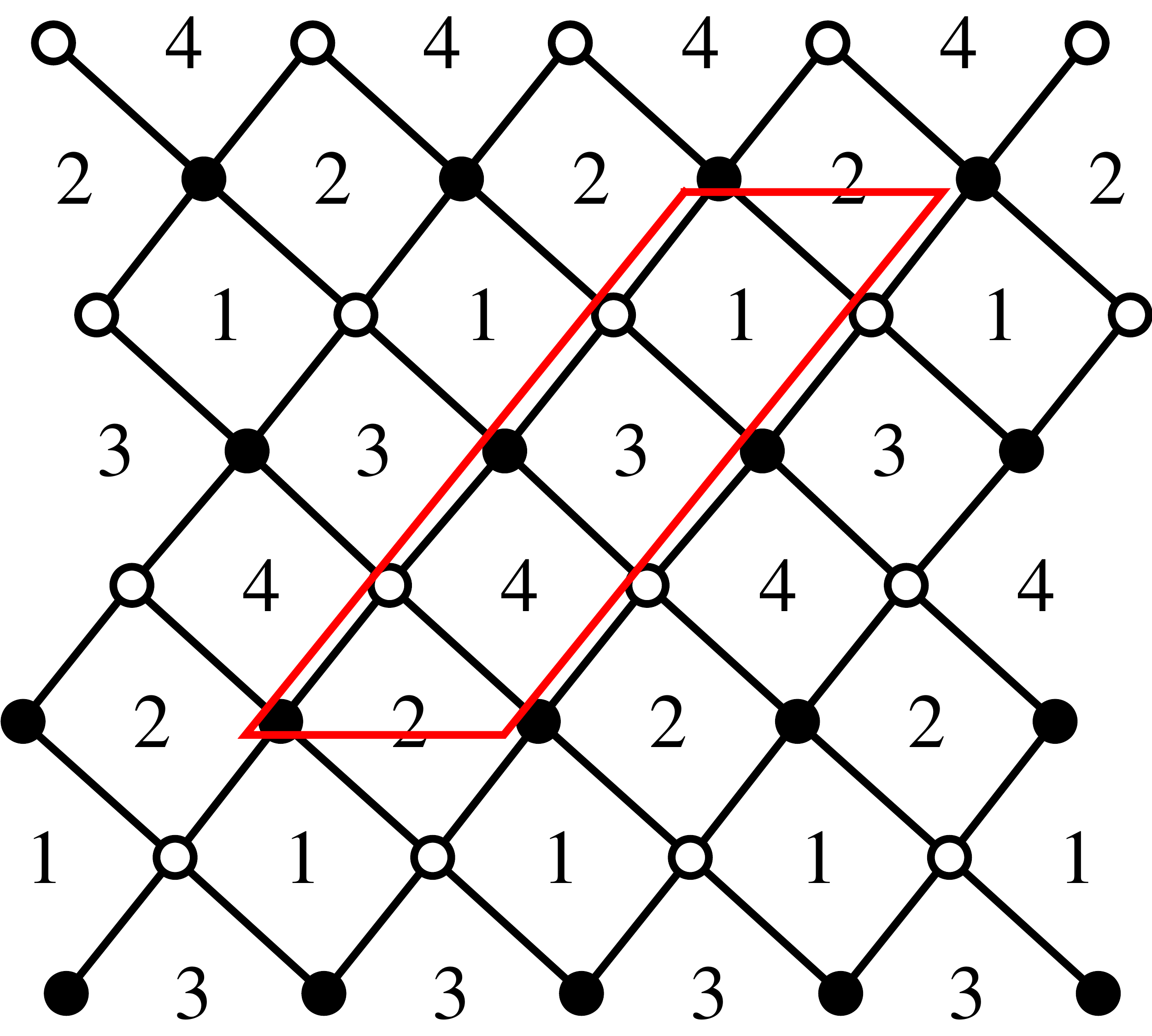} \\
(2.4) $L^{222}$ (I)
\end{tabular}
\\
&
\begin{tabular}[b]{c}
\includegraphics[width=3.5cm]{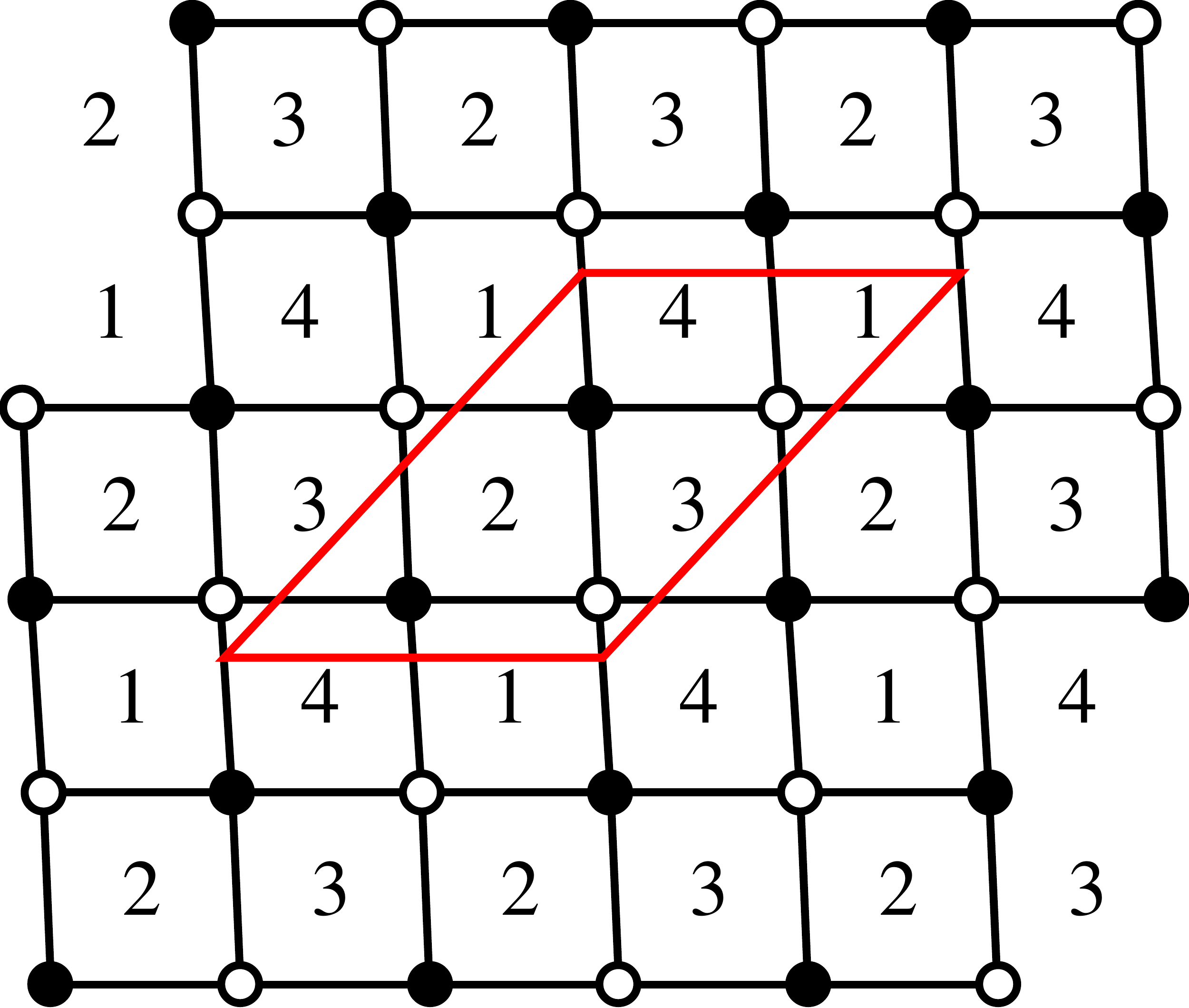} \\
(2.5) $\BF_0$ (I)
\end{tabular}
\end{tabular}
\end{center}
\caption{Consistent tilings with four superpotential terms.}
\label{fig:tilings4}
\end{figure} 

Let us now consider the models with six terms in the superpotential. Our algorithm generates a total of 37 different tilings, having from three to six gauge groups. Of these 37 tilings, 10 are consistent. We find that all of the consistent tilings are either phases of $L^{aba}$ or $Y^{p,q}$ families, or one of the del-Pezzo surfaces. Specifically, we find the models $dP_0$ (or $\BC^3 / \BZ_3$), $dP_1$, $dP_2$, $dP_3$, $L^{030}$ (or $\BC^2 / \BZ_3 \times \BC$), $L^{131}$, another phase of $L^{222}$, $L^{232}$, $L^{333}$ and $Y^{3,0}$. The other models are not as familiar, because they fail the usual tiling consistency condition.

We may wonder whether it is possible to quickly generate all tilings with 6 superpotential terms by adding diagonals to the 3 hexagon tilings, in a method similar to the 4 superpotential term case. Unfortunately this is not possible as there is a tiling containing an octagonal face. 

Our algorithm has been used to generate all tilings with at most 8 superpotential terms but it becomes computationally difficult to generate all tilings with 10 superpotential terms. One could ask whether it is possible to find a more efficient tiling generation algorithm using some more general base figure. One could start with some template and then add edges in all possible ways to generate tilings. To this date we have not found such a method that guarantees the generation of all brane tilings with 8 or more superpotential terms. This could be a direction of future research.

\chapter{Counting Orbifolds}
\label{ch:orbs}
Orbifolds have been studied intensively by both mathematicians and physicists. The understanding of how it is possible to compactify string theory on orbifolds \cite{preDixon,postDixon,Gaberdiel:1999ch} is seen as being a key advance. Orbifolds have also attracted interest in the study of conformal field theory \cite{Dixon:1986qv}, heterotic string theory \cite{Ibanez:1987pj} and cosmic strings \cite{Greene:1989ya}.

In this chapter we will see how brane tilings have proved to be useful in the study of certain orbifolds of $\BC^3$. This chapter will follow the recent works \cite{OrbsIntro,OrbsFull}.

\Section{What We Are Counting}

It has been found that D3-branes which probe non-compact abelian orbifolds of $\mathbb{C}^{3}$ \cite{DouglasMoore96,DouglasMoore97,DouglasGreeneMorrison97,Muto:1997pq,Lawrence:1998ja,HananyUranga98,Beasley:1999uz} have a world volume theory which is a $(3+1)$-dimensional quiver gauge theory \cite{Klebanov:1998hh,Acharya:1998db, Hanany:1998sd}. It is known that these world volume theories are very special in that they correspond to brane tilings that can be formed from only hexagonal faces.

In this chapter we are going to describe how it is possible to count these orbifolds of $\BC^3$. Perhaps surprisingly there have been relatively few systematic studies on enumerating orbifolds, although certain orbifold geometries have been studied in great detail. For instance, in the investigation of branes on orbifold singularities, it is widely known that there are two abelian orbifolds of the form $\BC^3/\Gamma$ at order $|\Gamma|=3$, which are $\BC^3 / \BZ_3$ -- sometimes known as the cone over $\mathrm{dP}_0$ -- and $\BC^2 / \BZ_2 \times \BC$. A question that has remained unanswered until quite recently is how many distinct abelian orbifolds of $\mathbb{C}^3$ are there for an arbitrary order of $\Gamma$.

Let us consider then the systematic study of abelian orbifolds of the form $\mathbb{C}^3 / \Gamma $ with $\Gamma$ being a finite abelian subgroup of $SU(3)$. We will count these orbifolds according to the order of the group $\Gamma$. These orbifolds are toric Calabi-Yau (CY) singularities.

Three methods of counting the aforementioned orbifolds shall be illustrated in this chapter. These are:
\begin{itemize}
\item
Counting all possible Brane Tilings that can be constructed using only hexagons. The details of this method can be found in Section \ref{s:TilingEnum}.
\item Using 3-tuples that specify actions of the generators of $\Gamma$ on $\BC^3$. There are some technical details which make this approach difficult. Full details of this method are given in Section \ref{s:OrbEnum}
\item
Exploiting the toric description of abelian orbifolds. Abelian orbifolds of $\mathbb{C}^{3}$ correspond to triangles on a $\BZ^2$ lattice. The counting of orbifolds using this method is covered in Section \ref{s:ToricEnum}.
\end{itemize}

All three of the methods above are found to give an identical counting of orbifolds of the form $\BC^3 / \Gamma$. The counting is given explicitly in Section \ref{s:Counting}. A formula for the partition function that counts these orbifolds is also given \cite{HananyOrlando10}. We will also discuss how it may be possible to generalise the methods to count higher dimensional orbifolds of the form $\BC^d/\Gamma$ for $d>3$.

\Section{Counting Orbifolds using Brane Tilings}
\label{s:TilingEnum}
As we have mentioned, one way in which it is possible to count abelian orbifolds of $\BC^3$ is by using brane tilings that have only hexagonal faces. As has been discussed in Section \ref{s:TheBT}, brane tilings are periodic bipartite graphs on the plane and can be used to describe quiver gauge theories which are world-volume theories of a D3-brane probing a toric CY singularity. Brane tilings formed from only hexagonal faces correspond to gauge theories whose moduli space is an abelian orbifold of $\BC^3$. The number of distinct faces or gauge groups in the corresponding quiver gauge theory is the order $|\Gamma|$ of the orbifold. Therefore, by counting all possible distinct hexagonal brane tilings formed by $|\Gamma|$ hexagons, one also counts abelian orbifolds of the form $\BC^3 / \Gamma$. The problem turns out to be equivalent to enumerating hexagonal lattices which has been studied the field of discrete mathematics \cite{Sloane97}.

\subsection{An Example - $\BC^3 / \BZ_3$}

Let us consider again the abelian orbifolds of $\BC^3$ at order $|\Gamma|=3$. Starting with $3$ distinct hexagons which we label from $1$ to $3$, we find that there are two different brane tiling constructions which are given in \ref{e:3HexTiles}.
\bea &~&
\underbrace{
\includegraphics[height=3cm, trim=1cm 0cm 3cm 0cm]{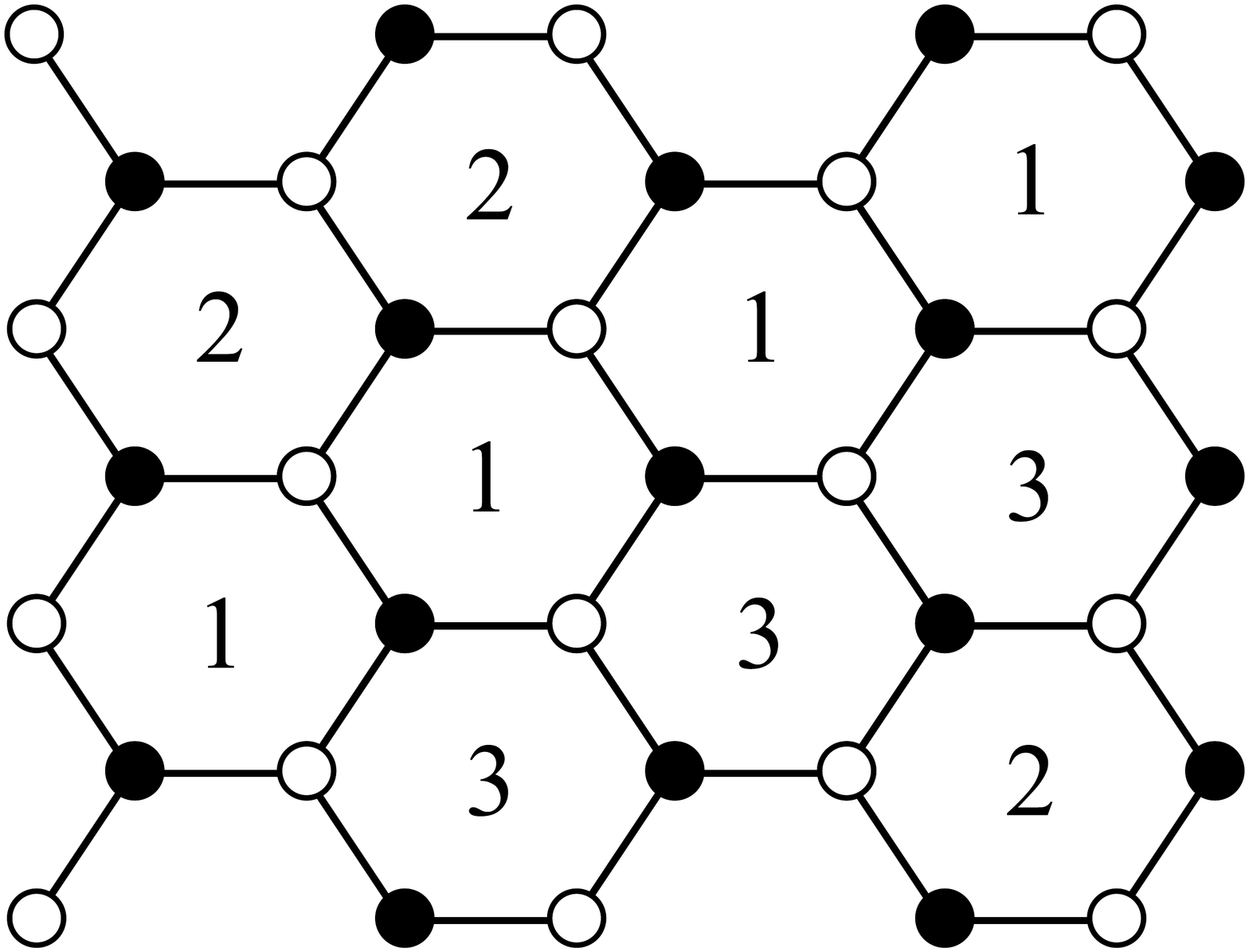}
\includegraphics[height=3cm, trim=1cm 0cm 3cm 0cm]{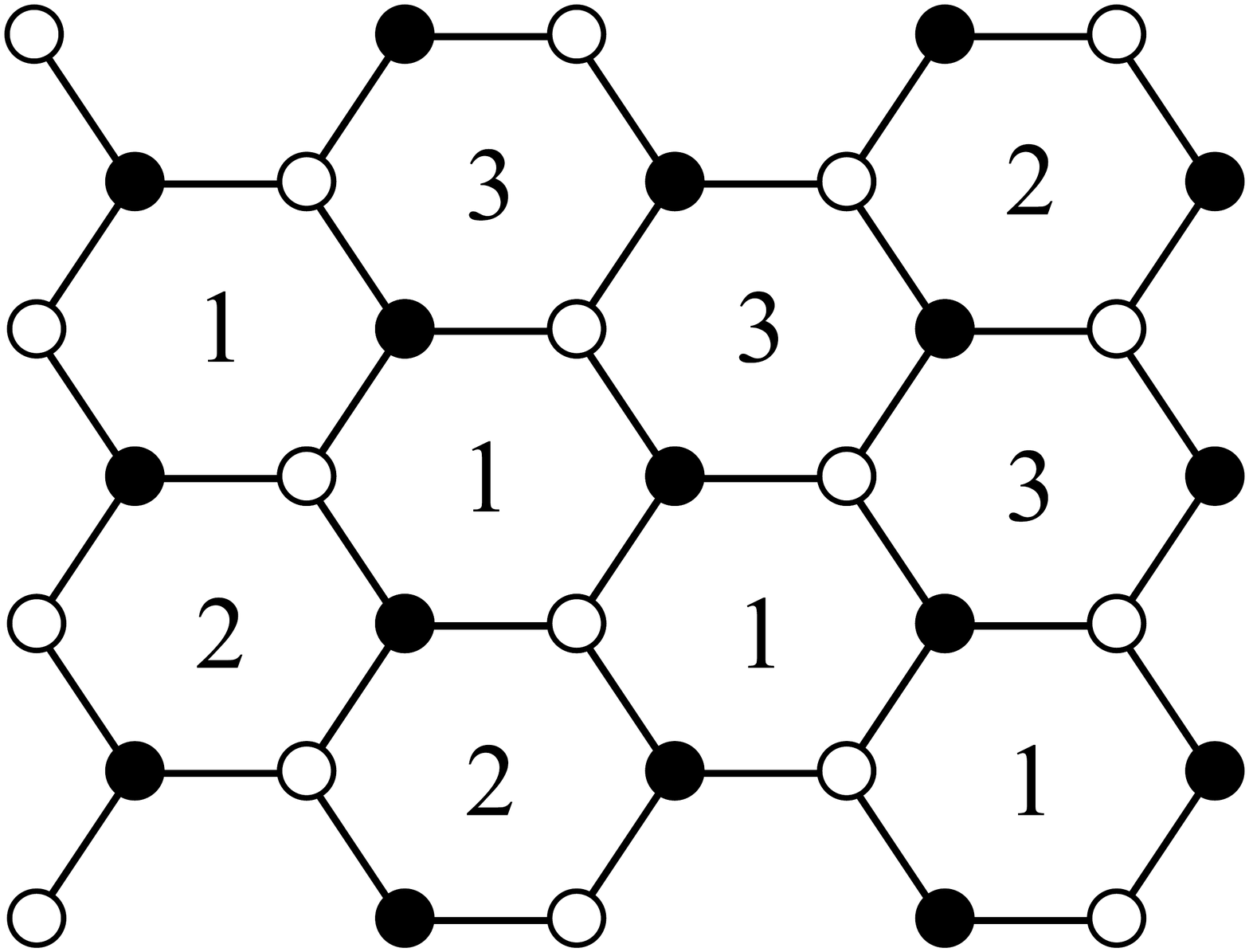}
\includegraphics[height=3cm, trim=1cm 0cm 3cm 0cm]{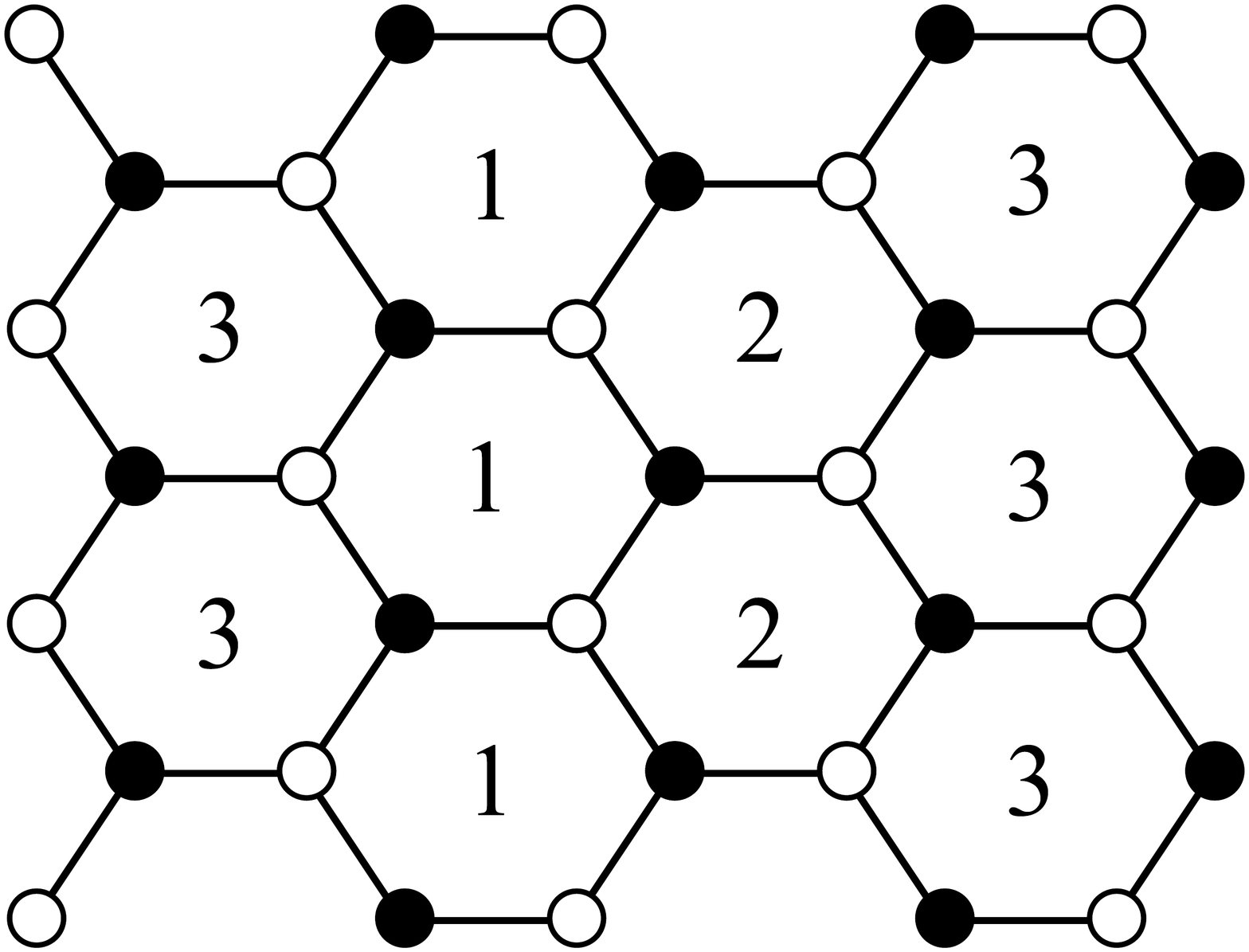}
}_{\mathbb{C}\times\mathbb{C}^2/\mathbb{Z}_3}
\nn \\ &~&
\underbrace{
\includegraphics[height=3cm, trim=1cm 0cm 3cm 0cm]{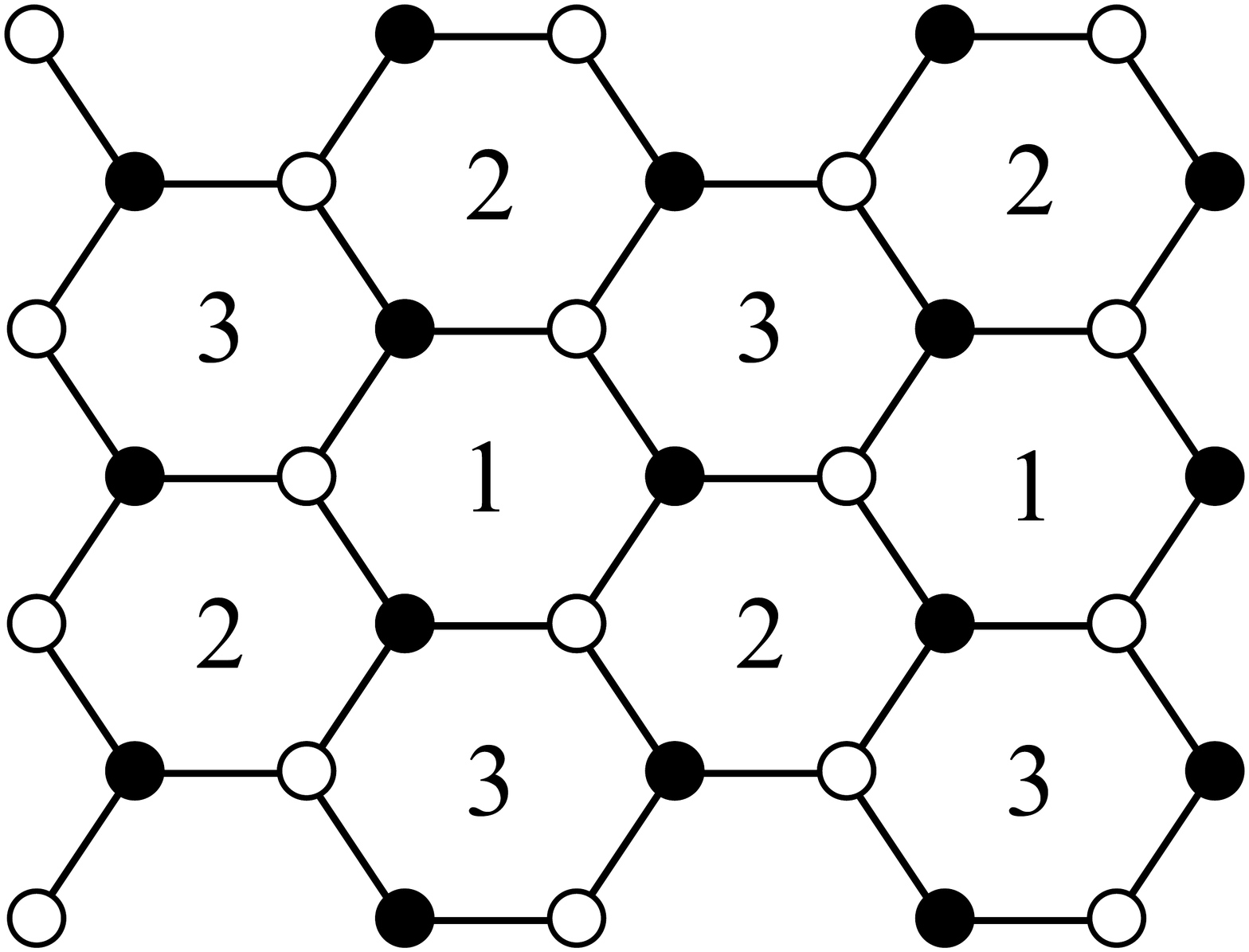}
}_{\mathbb{C}^3/\mathbb{Z}_3} \hfill
\label{e:3HexTiles}
\eea
The two distinct brane tilings that can be formed with 3 hexagons correspond to the orbifolds of the form $\mathbb{C}\times\mathbb{C}^2/\mathbb{Z}_3$ and $\mathbb{C}^3/\mathbb{Z}_3$.

\Section{Counting Orbifolds Using 3-tuples}
\label{s:OrbEnum}
A second way in which it has been possible to count orbifolds of $\mathbb{C}^{3}$ is by using a collection of 3-tuples. Let us consider the quotient formed when $\Gamma$, a finite abelian subgroup of $SU(3)$, acts on the space $\BC^3$. As we have mentioned, the resulting space is a toric non-compact Calabi-Yau (CY) singularity. 

As the group $\Gamma$ is abelian, it can be written as the product $\Gamma = \BZ_{n_1} \times \BZ_{n_2}$ with $|\Gamma| = n_1 n_2$. Let $g$ be a generator of one of the $\BZ_{n_i}$. Then as $g \in SU(3)$, it can be written as
\beq
g ~~=~~
\left( 
\begin{array}{ccc}
e^{\frac{i2\pi a_1}{n_i}} & 0 & 0\\
0&e^{\frac{i2\pi a_2}{n_i}} & 0 \\
0&0&e^{\frac{i2\pi a_3}{n_i}} \\
\end{array}
\right)
~~=~~\Diag \left(
e^{\frac{i2\pi a_1}{n_i}} ,
e^{\frac{i2\pi a_2}{n_i}} , 
e^{\frac{i2\pi a_3}{n_i}} 
\right)
\eeq
The action of the group $\BZ_{n_i}$ is therefore encoded by three integer parameters $a_i$ which satisfy $(a_1+a_2+a_3)=0 ~~(\bmod \; {n_i})$. We can keep track of this action in a 3-tuple $(a_1,a_2,-a_1-a_2)$. A list of these 3-tuples, each defining an action for a $\BZ_{n_i}$, can be used to define an orbifold action for the whole group $\Gamma$.

One way in which it is possible to count orbifolds is to simply consider all collections of 3-tuples that can form an action. One must then take into account that the same geometry could be defined by two different collections of 3-tuples.

\subsection{Over-counting Issues}
\label{s:OC}
There are different ways in which a set of 3-tuples that define an orbifold action can give rise to the same geometry:
\begin{itemize}
 \item There is a freedom of choosing the parameterization of $\BC^3$ by the coordinates $z_i$. One should consider two quotients equivalent if they are related to each other by a permutation of these coordinates.
\item The generators of each $\BZ_{n_i}$ are not necessarily unique. For instance, if one considers a generator $g \in \BZ_5$ then $g^2$, $g^3$ and $g^4$ are all generators of the group $\BZ_5$. Therefore if one has a 3-tuple $(a_1 , a_2 , a_3)$ that defines the action of some group $\BZ_n$ on $\BC^3$ then, for $\lambda$ co-prime to $n$, the 3-tuple $\lambda (a_1 , a_2 , a_3)$ defines an equivalent orbifold action. The convention used here is to only consider 3-tuples $(a_1,a_2,a_3)$ that satisfy $\mathrm{gcd}(a_1,a_2,a_3) =1$.
\item If $p$ and $q$ are co-prime, $\BZ_p \times \BZ_q = \BZ_{pq}$. Therefore orbifolds of composite order can be equivalent to orbifolds formed by a single $\mathbb{Z}_n$ acting on $\mathbb{C}^3$.
\end{itemize}
\subsection{An Example - $\BC^3 / \BZ_3$}
To explicitly illustrate some of the issues that are discussed above, let us consider the example of abelian orbifolds of the form $\BC^3 / \Gamma$ for $|\Gamma|=3$. The only abelian subgroup of $SU(3)$ of order $3$ is $\BZ_3$. By enumerating all 3-tuples that correspond to orbifolds actions of $\BZ_3$, one finds that there are 7 such 3-tuples. These are given in Table \ref{tab:3vecs}. After consideration of the over-counting issues given in Section \ref{s:OC}, it can be deduced that there are $2$ distinct abelian orbifolds of $\BC^3$ at order $3$. One orbifold has the orbifold action $(0,1,2)$ and is known in the literature as $\BC^2/\BZ_3 \times \BC$. The other orbifold has the action $(1,1,1)$ and is often referred to as $\BC^3 / \BZ_3$ or as the cone over the del Pezzo 0 $(\mathrm{dP}_0)$ surface.

\begin{table}
\begin{center}
\bmi{2.25in}
\begin{tabular}{|c|c|}
\hline
{\bf Orbifold Name} & {\bf Orbifold Action}\\
\hline
\multirow{6}{*}{$\BC^2/\BZ_3 \times \BC$} & $(0,1,2)$\\
 & $(0,2,1)$\\
 & $(1,0,2)$\\
 & $(2,0,1)$\\
 & $(1,2,0)$\\
 & $(2,1,0)$\\
\hline
$\BC^3 / \BZ_3$ & $(1,1,1)$\\
\hline
\end{tabular}
\emi
\end{center}
\caption{The two distinct orbifolds of the form $\mathbb{C}^3/\Gamma$ at order $|\Gamma|=3$. \label{tab:3vecs}}
\end{table}

\subsection{Consideration of $\BC^3 / (\BZ_n \times \BZ_m)$}
When considering orbifolds corresponding to groups of composite order, two 3-tuples must be used to keep track of the orbifold action. A detailed discussion for this case is given in \cite{OrbsFull}.
\Section{Counting Orbifolds using the Toric Description}
\label{s:ToricEnum}
A third way in which it is possible to count abelian orbifolds of $\BC^3$ is to use their toric description. As has been mentioned, a toric Calabi-Yau 3-fold can be represented by a convex polygon in a $\mathbb{Z}^{2}$ lattice. Two such polygons correspond to the same manifold if and only if they are related to each other by a $GL(2 , \BZ)$ transformation. Abelian orbifolds of $\mathbb{C}^{3}$ are toric and have lattice triangles as their toric diagrams. Therefore it is possible to count distinct abelian orbifolds of $\mathbb{C}^3$ by considering all triangles in a $\BZ^2$ lattice that are not related to each other by a $GL(2 , \BZ)$ transformation.

The area of a toric triangle in $\BZ^2$ equals the order of the group, $| \Gamma |$, in $\BC^3 /  \Gamma$. Therefore, to count orbifolds according to $|\Gamma|$, all toric triangles of area $|\Gamma|$ must be generated first. This can be done by multiplying each of the vectors that represent the vertices of a unit triangle by $2 \times 2$ integer valued matrices of determinant $|\Gamma|$.

As an example, it is possible to generate triangles of area $2$ by using integer valued $2 \times 2$ matrices of determinant $2$. One could multiply each of the vectors $\{ {0 \choose 0} , {1 \choose 0 }, {0 \choose 1} \}$ by the matrix  ${1 \; 0 \choose 0 \; 2} $ to get the vectors $\{ {0 \choose 0} , {1 \choose 0 }, {0 \choose 2} \}$ which corresponds to a triangle of area 2 in a $\BZ^2$ lattice. This procedure is shown diagrammatically in \eqref{eq:HNFZ2}
\beq
\includegraphics[height=0.7cm, trim=0cm 3cm 0cm 0cm ]{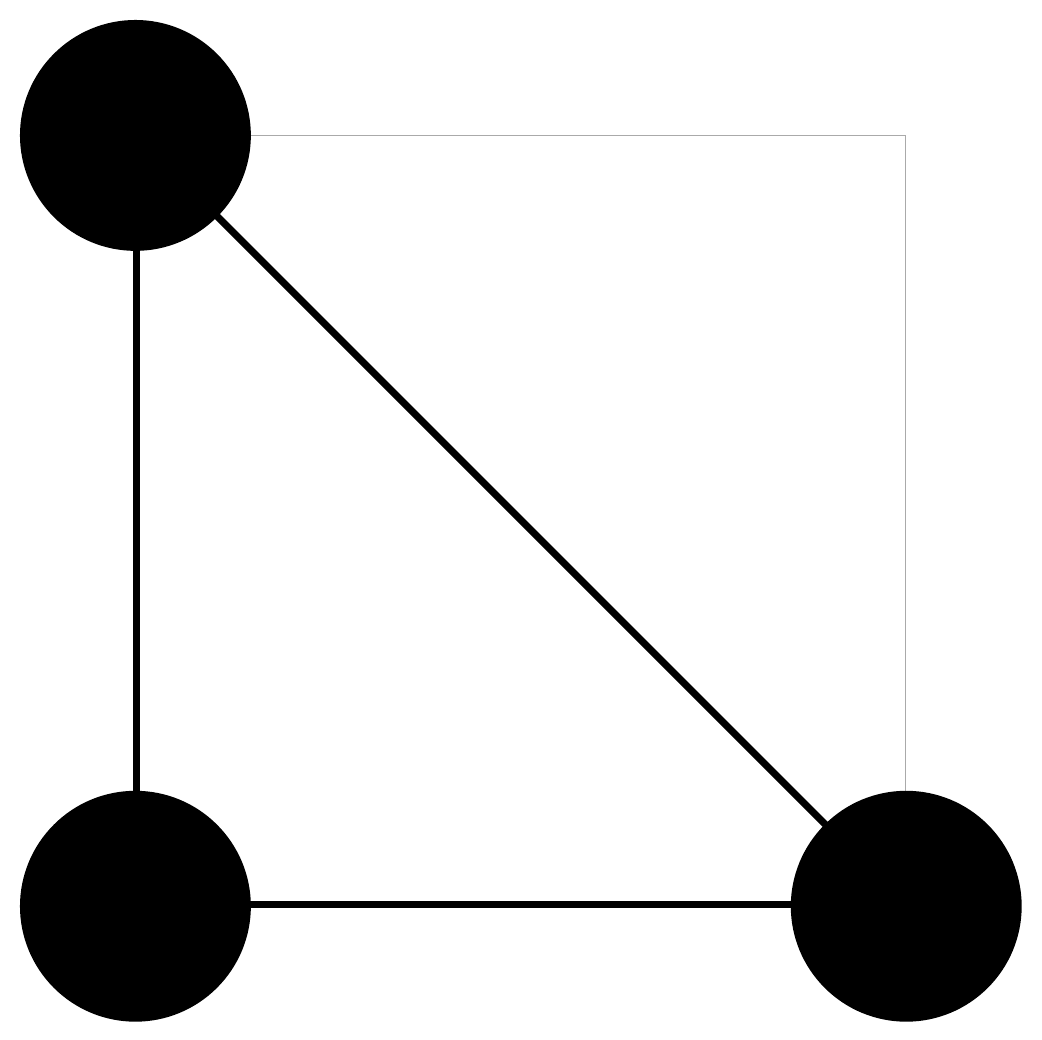} 
\times {1 \; 0 \choose 0 \; 2} = 
\includegraphics[height=1.4cm, trim=0cm 3cm 0cm 0cm ]{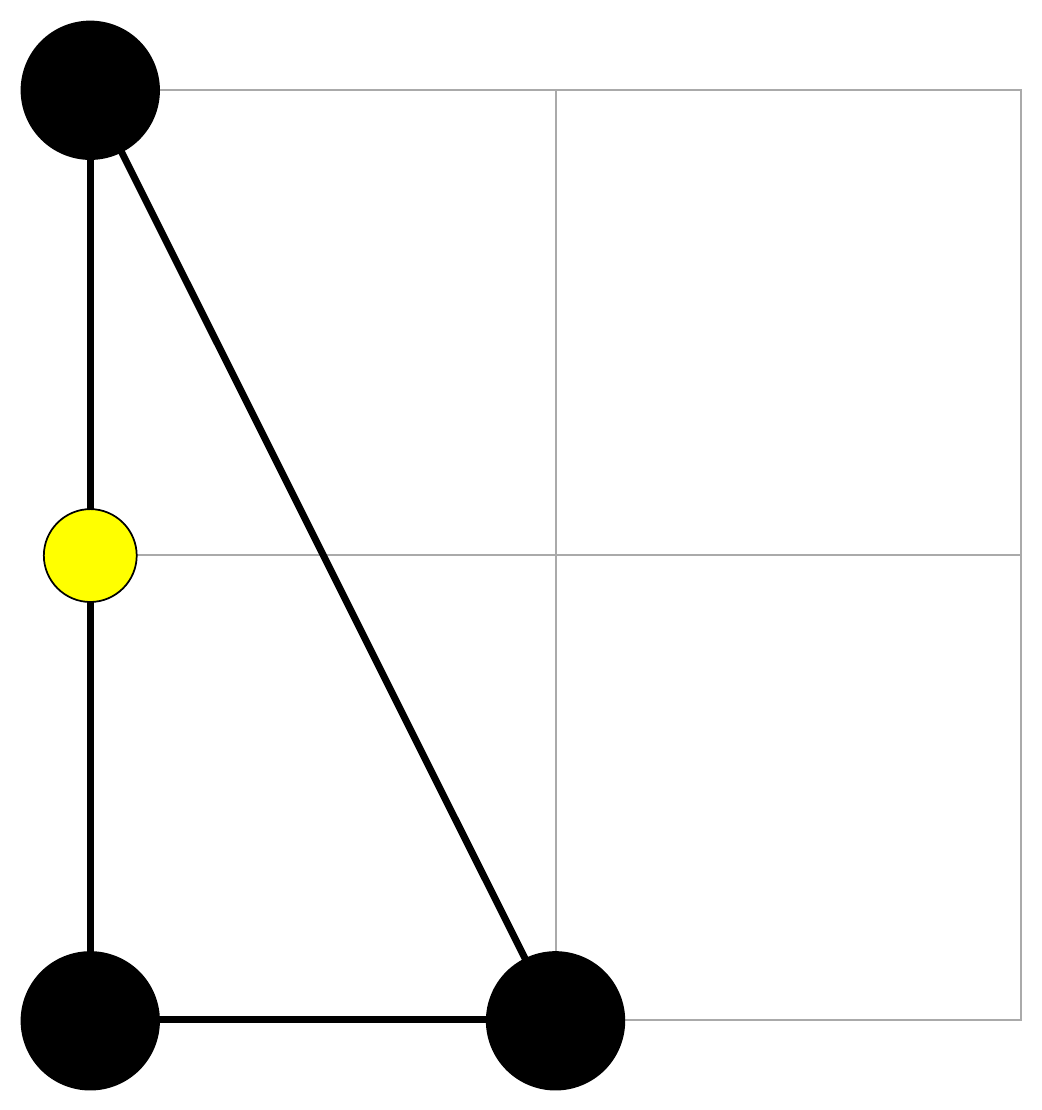} 
~~.
\label{eq:HNFZ2}
\eeq

The $2\times 2$ matrices one has to consider in order to cover all possible toric triangles of a given area are in Hermite Normal Form (HNF).

\subsection{Hermite Normal Form}
An upper triangular $2\times 2$ integer valued matrix of the form
\beq
M=
\left(
\begin{array}{cc}
a & b \\
0 & c
\end{array}
\right)~~,
\eeq
where $\det{M}=ac$ and $0\leq b < c$ is said to be in Hermite Normal Form (HNF). All $2 \times 2$ integer valued matrices can be written as the product of a matrix in HNF and a second matrix in $GL( 2 , \BZ)$. There are a finite number of integer valued matrices in HNF with any fixed determinant. Therefore, when generating triangles of a given area $|\Gamma|=\det{M}$, one only needs to consider this finite list of matrices in HNF in order to generate all distinct triangles.

\subsection{An Example - $\BC^3 / \BZ_3$}

Let us consider again the orbifolds of $\BC^3$ at order $|\Gamma|=3$. The HNF matrices of determinant $3$ and the corresponding toric triangles are
\beq
\underbrace{
\left( \begin{array}{cc}
1 & 0 \\
0 & 3
\end{array}\right)}_
{\includegraphics[width=1.7cm]{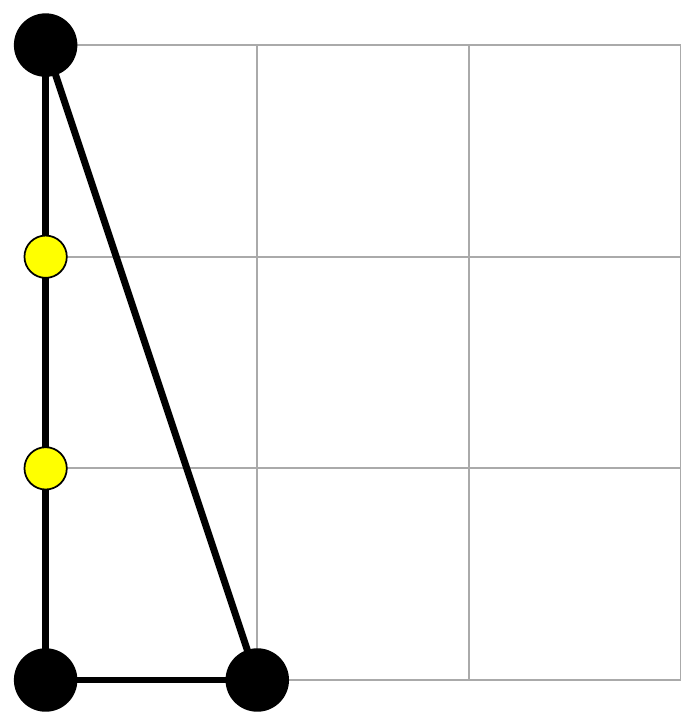}} ~~,~~
\underbrace{\left( \begin{array}{cc}
1 & 1 \\
0 & 3
\end{array}\right)}_
{\includegraphics[width=1.7cm]{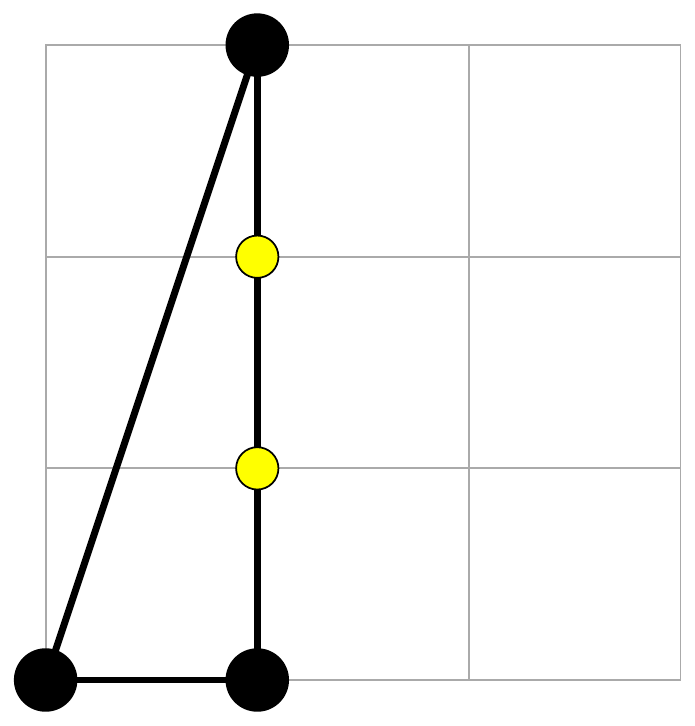}} ~~,~~
\underbrace{\left( \begin{array}{cc}
3 & 0 \\
0 & 1
\end{array}\right)}_
{\includegraphics[width=1.7cm]{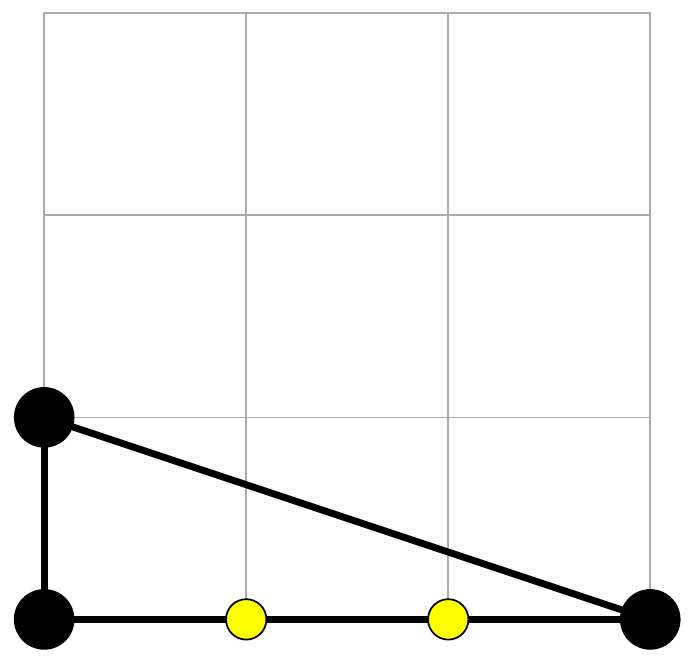}} ~~,~~
\underbrace{\left( \begin{array}{cc}
1 & 2 \\
0 & 3
\end{array}\right)}_
{\includegraphics[width=1.7cm]{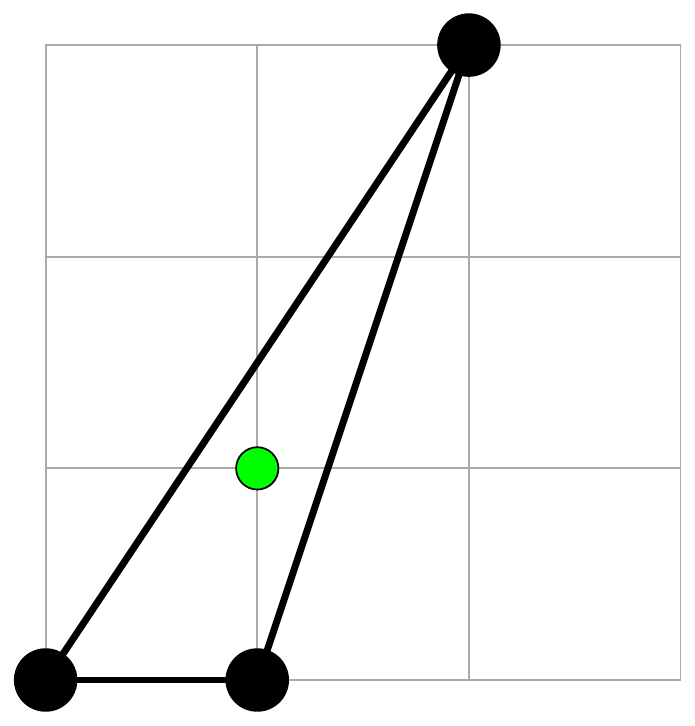}} ~~.
\label{HNFTriangles}
\eeq
Each of the triangles in \eqref{HNFTriangles} have an edge which is parallel to the x-axis because all $2 \times 2$ matrices in HNF have a lower left entry which is zero.
One observes that there are two distinct abelian orbifolds of $\mathbb{C}^3$ at order $|\Gamma|=3$, which correspond exactly to the two distinct orbifolds in Table \ref{tab:3vecs}.

\Section{Explicit Counting}
\label{s:Counting}
The three methods given above have been used to count abelian orbifolds of $\BC^3$. These three methods are equivalent and give the same counting. Let the number of orbifolds of the form $\BC^3 / \Gamma$ at order $|\Gamma | = n$ be $f(n)$. The first 50 values of $f(n)$ are given in Table \ref{t:fn}.\\
\begin{table}[h!]
  \begin{tabular}{c|cccccccccccccccccccccc}
    $n$ &
 1 		&
 2			&
 3			&
 4			&
 5			&
 6			&
 7			&
 8			&
 9			&
 10		&
 11		&
 12		&
 13		&
 14		&
 15	
 \\ \hline
    $f(n)$ &
 1			&
 1			&
 2			&
 3			&
 2			&
 3			&
 3			&
 5			&
 4			&
 4			&
 3			&
 8			&
 4			&
 5			&
 6			 \end{tabular}   \begin{tabular}{c|cccccccccccccccccccc}
    $n$ &
 16		&
 17		&
 18		&
 19		&
 20		&   
 21		&
 22		&
 23		&
 24		&
 25		&
 26		&
 27		&
 28		\\ \hline
    $f(n)$ 
    &
 5			&
 10	 &
 8			&
 7			&
 5			&
 15		&
 7			&
 8			&
 9			&
 13		&
 6			&
 14		&
 7		
  \end{tabular}
  \begin{tabular}{c|cccccccccccccccccccc}
    $n$ &
 29		&
 30		&
 31		&
 32		&
 33		&
 34		&
 35		&
 36		&
 37		&
 38		& 
 39		&
 40		&
 41	\\ \hline
    $f(n)$ 
    &
 10		&
 20		&
 8			&
 11 & 12		&
 20		&
 8			&
 18		&
 9			&
 17		&
 16		&
 13		&
 9			
  \end{tabular}
  \begin{tabular}{c|ccccccccc}
    $n$ &	
 42		&
 43		&
 44		&
 45		&
 46		&
 47		&
 48		&
 49		&
 50	\\ \hline
	  $f(n)$ 		&
 28		&
 12		&
 17				&
 15		&
 10		&
 10		&
 9			&
 4			&
 8		 
  \end{tabular}
\caption{The number of orbifolds of $\BC^3 / \Gamma $ for $n =1,\dots,50$}
\label{t:fn} 
\end{table}

By writing the sequence $f(n)$ in terms of a partition function $F(t) = \sum f(n) t^n$, one finds the formula \cite{HananyOrlando10}
\beq
F(t) = \sum_{m=1}^\infty \left[ \frac{1}{\left( 1 - t^m \right) \left( 1 + t^{2m} \right) \left( 1 - t^{3m} \right)} - 1 \right] 
~~.
\label{eq:GenFunction}
\eeq

\Section{Extensions to Higher Dimensional Orbifolds}

Two of the methods which have been used to count orbifolds of $\BC^3$ can be used to count orbifolds of higher dimensional spaces. In fact, the use of tuples and toric diagrams can be generalised to count any higher dimensional abelian orbifold of $\BC^d$ with $d > 3$ \cite{OrbsFull,RakNew}. 

It is possible to extend the idea of a 3-tuple that defines an action of a cyclic group on $\BC^3$ to a $d$-tuple that defines the action of a cyclic group on $\BC^d$. One can also use toric data to count orbifolds of $\BC^d$ for $d>3$. For instance, to count the abelian orbifolds of $\BC^4$, one must count distinct tetrahedra in a $\BZ^3$ lattice of a volume $|\Gamma|$. Higher dimensional simplices must be considered to count orbifolds of $\BC^d$ for $d>4$.

Currently, it is not well understood how to extend the idea of the brane tiling to describe and count all abelian orbifolds of $\BC^4$. It is possible that brane crystals \cite{Lee:2006hw, M2Crystal, Kim:2007ic} may offer a way of counting all distinct abelian orbifolds of $\BC^4$. This could be a direction for future research.\\

\chapter{Brane Tilings and M2-branes}
\label{ch:phases}
Supersymmetric Chern-Simons (CS) theories in 2+1 dimensions have attracted a lot of interest due to their proposed description of the M2-brane \cite{BaggerLambert07,BaggerLambert08a,BaggerLambert08b,Gustavsson07,Gustavsson08,VanRaamsdonk:2008ft,Lambert:2008et,Fuji:2008yj,Benna:2008zy}.  A $U(N) \times U(N)$ CS theory at level $(k, -k)$ with bi-fundamental matter fields was subsequently proposed as a description of $N$ M2-branes on the $\BC^4/\BZ_k$ orbifold background \cite{ABJM08}. At strong coupling ($N \gg k$), the ABJM theory is conjectured to be dual, in the sense of the AdS/CFT correspondence, to M-theory on $\mathrm{AdS}_4 \times S^7 / \mathbb{Z}_k$.
 After the proposal of this theory, a flurry of activity followed \cite{Hosomichi:2008jb,Terashima:2008ba,Kim:2008gn,Imamura:2008dt,Aharony:2008gk,Ooguri:2008dk,Jafferis:2008qz,Imamura:2008ji,Imamura:2009ur,Kim:2009wb} including the investigation of $\CN = 2$ CS theories with a more general quiver structure \cite{Martelli:2008si, Ueda:2008hx, Hanany:2008cd,Benishti:2009ky,Taki:2009wf}. A nice review on the subject has been written \cite{Klebanov:2009sg}.

\subsection{Strongly Coupled CS theories and AdS / CFT}

The ABJM theory at strong coupling is conjectured to be dual to M-theory on $\mathrm{AdS}_4 \times S^7 / \mathbb{Z}_k$. It is important to study both sides of the correspondence in more detail in order to better our understanding of this fascinating conjecture.

The AdS / CFT correspondence implies that gauge invariant scalar operators on the gauge theory side should be in a one-to-one correspondence with the Kaluza-Klein harmonics on $S^7$ \cite{ABJM08}. It is known that there are 35 harmonics that correspond to operators of dimension one. An analysis of these operators is a challenge, particularly because we must deal with the ABJM theory at strong coupling and so a perturbative study of the theory is difficult \cite{Klebanov:2009sg}.

Monopole operators are vital in order to understand both the supersymmetry enhancement from $\mathcal{N} = 6$ to $\mathcal{N}=8$ and also details of the spectrum of gauge invariants in the ABJM theory with $k = 1,2$ \cite{Benna:2009xd}. In particular, without these operators only 15 of the 35 operators of dimension 1 would be realised. Although a full study of monopole operators is beyond the scope of this thesis, it is interesting to note the 20 operators that involve these monopole operators can be written in the form:
\beq
Y_A^\dagger Y_B^\dagger \mathcal{M}^2, \qquad Y_A Y_B \mathcal{M}^{-2}
\eeq
where $\mathcal{M}^2$ and $\mathcal{M}^{-2}$ are monopole and anti-monopole operators respectively and $Y_A$ are fields with the same charges as the bi-fundamental chiral fields \cite{Klebanov:2009sg}.

It is vital to show that the monopole operators do not alter the `naive dimension' of the scalar bilinears and, as the gauge theory is strongly coupled, the monopole operators are difficult to analyse. A way of overcoming this problem has been to embed the ABJM theory into a $\mathcal{N}=3$ supersymmetric yang-mills theory and study the theory in the UV where it is weakly coupled \cite{Benna:2009xd}. It is possible to perform an analysis of monopole operators in the UV and then argue that their $SU(2)_R$ charges are not modified by the RG flow.

\subsection{Brane Tilings and Chern-Simons theories}

As we have mentioned in previous sections, brane tilings have proved to be useful tools in establishing a connection between 3+1 dimensional gauge theories and their moduli spaces. One recent and quite exciting development has been that we can use brane tilings (with a few modifications from the 3+1 dimensional case) to study 2+1 dimensional CS theories as well \cite{Hanany:2008cd, HananyZaff08}. All of the models we study here are brane tilings but the general class of quiver gauge theories is larger, since every brane tiling gives rise to a quiver but not every quiver gives rise to a brane tiling. It should be mentioned that all presently known M2-brane theories can be described by brane tilings.

In this chapter, we shall study supersymmetric CS theories which are known to describe M2-branes probing various toric Calabi--Yau 4 folds. In particular, we shall focus on the `forward algorithm' for M2-branes which allows us to obtain the toric data of the mesonic moduli space\footnote{Here we use the term mesonic moduli space to be the moduli space found after both F and D terms are taken into account. This space can be thought of as being the space perpendicular to the branes in M-theory}. We will also only concern ourselves with the 1-brane theory and so consider only theories with a moduli space which is a Calabi-Yau 4-fold. We will also sketch how it is possible to consider the `inverse algorithm' for M2-branes although this method has not yet been perfected \cite{Davey:2009sr}. 

\Section{Supersymmetric Chern--Simons Theory} \label{summary}
Let us consider 2+1 dimensional quiver Chern--Simons (CS) theories with $\CN=2$ supersymmetry (four supercharges). We will restrict our attention to CS theories that have a $U(N)^G$ gauge symmetry. These CS theories have no kinetic terms for the gauge fields but instead have CS terms. The theories also contain bi-fundamental and adjoint matter.  The Lagrangian of a Supersymmetric CS theory having a gauge symmetry of $U(N)^G$ and a total of $E$ fields is of the form:
\bea
\label{e:M2lagrange}
\mathcal{L} &=& -\int d^4 \theta\left( \sum\limits_{X_{ab}} X_{ab}^\dagger e^{-V_a} X_{ab} e^{V_b}
-i \sum\limits_{a=1}^G k_a \int\limits_0^1 dt V_a \bar{\mathcal{D}}^{\alpha}(e^{t V_a} \mathcal{D}_{\alpha} e^{-tV_a})
\right) \nn \\ &+& 
\int d^2 \theta W(X_{ab}) + \mathrm{c.c.}
\eea
In the equation above, $a$ indexes the factors in the gauge group, $X_{ab}$ are the superfields accordingly charged, $V_a$ are the vector multiplets, $\mathcal{D}$ is the superspace derivative, $W$ is the superpotential and $k_a$ are the CS levels which are integer valued. An overall trace is implicitly taken as all of the fields are matrix valued.

The first and third terms in \eref{e:M2lagrange} are the usual matter and superpotential terms respectively.
It can be useful to write the second term above, which includes the usual CS terms, explicitly in component notation.
The 2+1 dimensional $\CN=2$ vector multiplet $V_a$ consists of a gauge field $A_a$, a scalar field $\sigma_a$, a two-component Dirac spinor $\chi_a$, and an auxiliary scalar field $D_a$, all transforming in the adjoint representation of the gauge group $U(N_a)$.  
This can be viewed as a dimensional reduction of the 3+1 dimensional ${\CN}=1$ vector multiplet. 
In particular, $\sigma_a$ arise from the zero modes of the components of the vector fields in the direction along which we reduce.
In component notation, the CS terms, in Wess--Zumino (WZ) gauge, are given by
\bea \label{csterms}
S_{\mathrm{CS}}\, = \, \sum_{a=1}^G \frac{k_a}{4\pi}\int  \mathrm{Tr} \,\left( A_a \wedge \ud A_a + \frac{2}{3} A_a\wedge A_a\wedge A_a - \bar\chi_a \chi_a +
2D_a \sigma_a \right)~.
\eea

\subsection{The vacuum equations.} From \eref{e:M2lagrange}, it is possible to obtain the following vacuum equations\cite{Hanany:2008cd}:
\begin{eqnarray}
\nn \partial_{X_{ab}} W &=& 0~, \\
\nn \mu_a(X) := \sum\limits_{b=1}^G X_{ab} X_{ab}^\dagger - 
\sum\limits_{c=1}^G  X_{ca}^\dagger X_{ca} + [X_{aa}, X_{aa}^\dagger] &=&  4k_a\sigma_a~, \\
\label{DF} \sigma_a X_{ab} - X_{ab} \sigma_b &=& 0 \ .
\label{e:CSModSpace}
\end{eqnarray}
The first set of equations above are referred to as the F-term equations.  The second set of equations seem to be similar to the D-term equations of $\CN=1$ gauge theories in 3+1 dimensions whereas the third set don't seem to have a 3+1 dimensional analogue. We call the space of all solutions to \eref{DF} the `mesonic moduli space' ($\CMm$). This space has the interpretation of being the geometry that the M2-branes probe.

\subsection{Connection to M2-branes.} In the rest of this thesis, it will be assumed that:
\begin{itemize}
\item All gauge groups are abelian or $U(1)$. This has the physical interpretation that we are only considering a single M2-brane probe.
\item The superpotential $W$ satisfies a toric condition. Each chiral multiplet appears precisely twice in $W$; once with a positive sign and once with a negative sign. 
\item $\CMm$ shall be a toric Calabi-Yau 4-fold. This is a strong restriction on the CS theories that we shall consider.
\end{itemize}

\subsection{The Classical Moduli Space of Abelian Theories}  

From the second equation of \eref{DF}, we can see that as the theory is abelian,
\bea
\sum_a k_a \sigma_a = 0~. \label{ks}
\eea 
The third equation of \eref{DF} sets all $\sigma_a$ to a single field, say $\sigma$.  From \eref{ks}, we see that for $\sigma \neq 0$, we must impose the following constraints on the CS levels:
\bea
\left(k_1, \ldots, k_G \right)  \neq 0~, \qquad \quad \sum_{a=1}^{G} k_a = 0~.  \label{k-con}
\eea
Note that if the last equality is not satisfied, then $\sigma$ is identically zero and \eref{DF} reduces to the usual vacuum equations for 3+1 dimensional gauge theories.  In this case the mesonic moduli space is 3 dimensional.  Thus, \eref{k-con} are indeed necessary conditions for the mesonic moduli space to be 4 dimensional, as we require from our brane picture.
For simplicity, we also take
\bea
 \gcd(\{k_a\}) = 1
\eea
so that we do not have to consider orbifold actions on the moduli space.  However, it is easy to generalise to the case of higher $\gcd(\{k_a\})$, and several explicit examples are given in \cite{Hanany:2008cd, Hanany:2008qc}.

\subsection{A Note on Quantum Corrections}

Let us briefly discuss possible quantum corrections to $\cal{N}$ $= 2$ CS theories.

Firstly it is known that the Chern-Simons levels $k_a$ are not renormalized beyond 1-loop \cite{Kapustin:1994mt}. One can argue that $k_a$ must be integer valued in order for a path integral of the theory the be invariant under large gauge transformations. It is also known that quantum corrections at two-loop or higher must be suppressed by a factor of $1/k_a$ which is not in general integer valued.

There is also an argument which forbids a dynamically generated superpotential \cite{Gaiotto:2007qi} although it is known that coefficients of the superpotential are in general renormalized \cite{Jafferis:2009th}. It is also known that the K\"{a}hler potential of the theory will in general receive corrections and that these will be either irrelevant or can be absorbed through a rescaling of $X_{ab}$.

It is interesting to consider possible corrections to the moduli space of the CS theory. In general $\cal{N}$ ${= 2}$ CS theories receive quantum corrections to the metric on the moduli space. Explicitly it is known that there can be at least two-loop corrections that can result in a cone-shaped metric \cite{Gaiotto:2007qi}. 

In the remainder of this work, we shall focus on the classical moduli space of CS theories.

\Section{Brane Tilings}

Just like the (3+1)--dimensional case, it is possible to describe certain (2+1)--dimensional quiver Chern--Simons theories using bipartite graphs on $T^2$. The dictionary between a tiling and a Chern-Simons theory is summarised in Table \ref{t:btm2}. The major difference is that tilings that correspond to Chern--Simons theories must come equipped with Chern--Simons levels $k_a$ if they are to completely specify our (2+1)--dimensional theory. This data can either be written on the tiling (an integer can be written in each of the faces of the tiling) or can be supplied in the form of a vector.

\begin{table}[h!]
\begin{center}
\begin{tabular}{c|c}
Tiling &  CS Theory\\
\hline
Face &  $U(N)$ Gauge Group\\
Edge &  Bi-fundamental Field \\
Node &  Superpotential Term \\
$k_a$ & CS levels \\
\end{tabular}
\end{center}
\caption{The relationship between a brane tiling and the CS theory that it represents}
\label{t:btm2}
\end{table}

\subsubsection{Brane Realisation} It is possible to think of the brane tiling corresponding to a (2+1)--dimensional theory as a system of D4 and NS5-branes in Type IIA string theory on $\BR^{1,7}\times T^2$. This idea is discussed in detail in \cite{Imamura:2008qs, Imamura:2008nn}.
It is also known that there is a relation between M-theory on a Calabi-Yau fourfold singularity with type IIA string theory on a Calabi-Yau threefold fibered over a real line, with RR 2-form fluxes turned on \cite{Aganagic:2009zk}.

\subsubsection{Chern--Simons levels for fields}

It is possible to encode $k_a$ -- the CS levels for gauge groups -- into CS levels for fields. A link can be made by using the incidence matrix of the quiver CS theory. This incidence matrix $d$ encodes a quiver diagram and is defined in \eqref{M2incidence}.

\bea \label{M2incidence}
d_{ai} = \left \{ 
\begin{array}{rl}
+1 & \quad \text{if edge $i$ is outgoing from the node $a$}~, \\
-1 & \quad \text{if edge $i$ is incoming to node $a$}~, \\
0 &  \quad \text{if edge $i$ is not connected to node $a$}~.  
\end{array} \right .
\eea

It is always possible to assign integers $n_i$ to the edges $i$ such that the CS levels of the gauge groups are given by\footnote{This way of representing $k_a$ is introduced in \cite{Hanany:2008cd} and is also used in \cite{Imamura:2008qs}.}
\bea
k_a = \sum_i d_{ai} n_{i}~. \label{kn}
\eea
Due to the bipartite nature of the tiling, the relation $\sum_{a} k_a = 0$ is always satisfied if the CS levels are written in this way. These variables $n_i$ will be useful in developing and understanding the forward process for M2-branes.

\subsection{The Forward Process for M2-branes}

It is possible to quickly compute the toric data associated to the moduli space of a CS theory from knowledge of a tiling equipped with a set of CS levels $n_i$. The data corresponding to a toric CY 4-fold singularity can be written as a convex set of lattice points in $\BZ^3$. There are two equivalent ways of performing this calculation and they are given below.

\subsection{The Forward Process using the Kasteleyn Matrix}
\label{s:FFKast}

\begin{figure}[ht]
\begin{center}
  \includegraphics[totalheight=5.5cm]{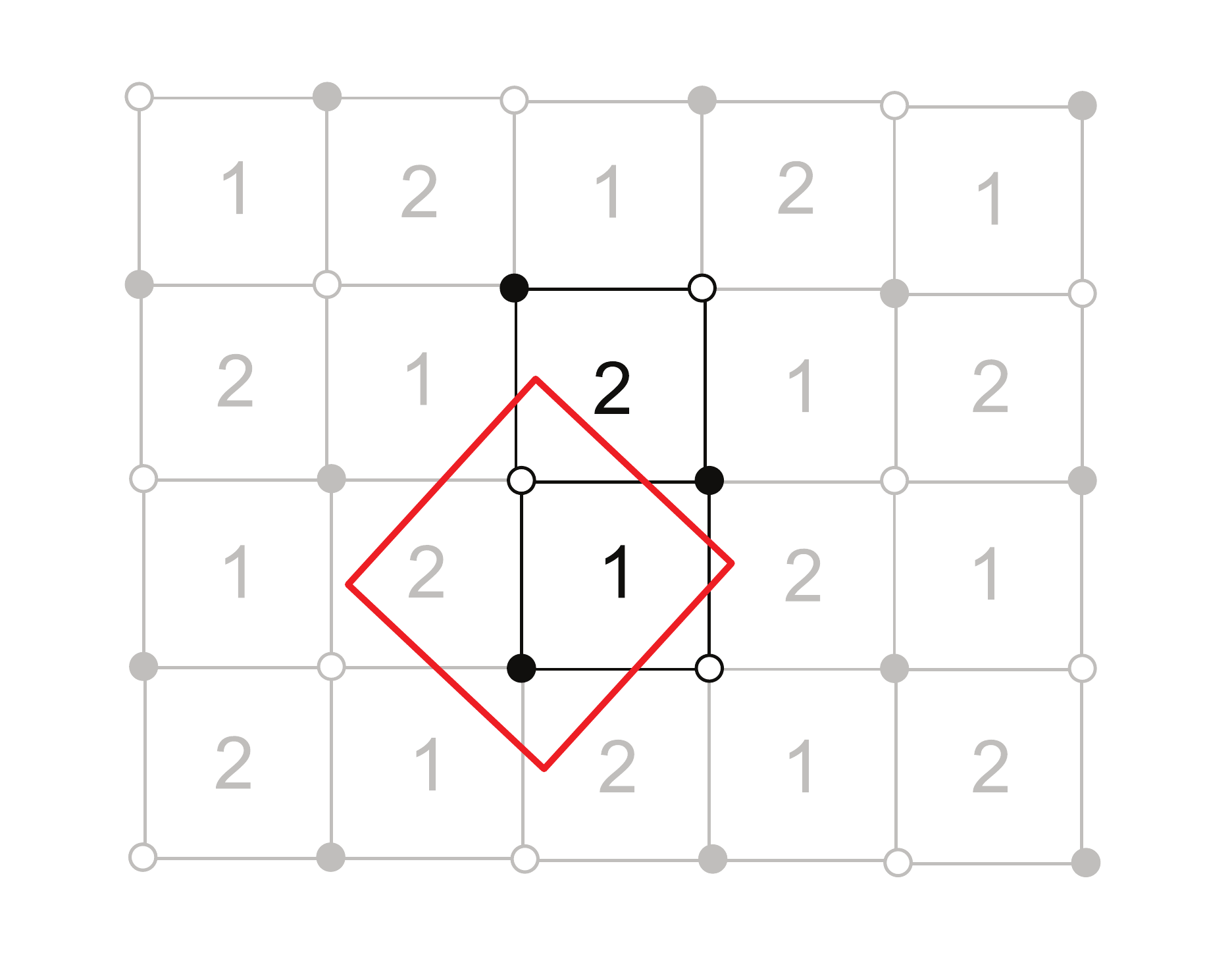}
 \caption{The two square tiling}
  \label{f:2SqTiling}
\end{center}
\end{figure}

Let us now consider the M2-brane analogue of the D3-brane algorithm that was discussed in Section \ref{D3FFA}. Suppose that we start with a tiling and a set of Chern--Simons terms. For example, we can start with the 2 square tiling given in \fref{f:2SqTiling}. This tiling, together with CS levels $k_a = (1,-1)$ corresponds to the M2-brane theory known as the ABJM model \cite{ABJM08}.

To proceed with the algorithm, we must write down a weighted adjacency matrix corresponding to the tiling. This is the Kasteleyn matrix of the tiling and is similar to the matrix discussed in section \ref{D3FFA}. Columns of the matrix are still indexed by white nodes and rows are indexed by black nodes. In order to construct the Kasteleyn matrix, the fundamental domain of the tiling is drawn and weights are given to edges. This time, three variables $x$, $y$ and $z$ are used to give weights to edges. Two of the variables ($x$ and $y$) are used to weight edges according to how they cross the sides of the fundamental domain. This is done in an identical way to the D3-brane case.

\begin{figure}[ht]
\begin{center}
   \includegraphics[totalheight=7cm]{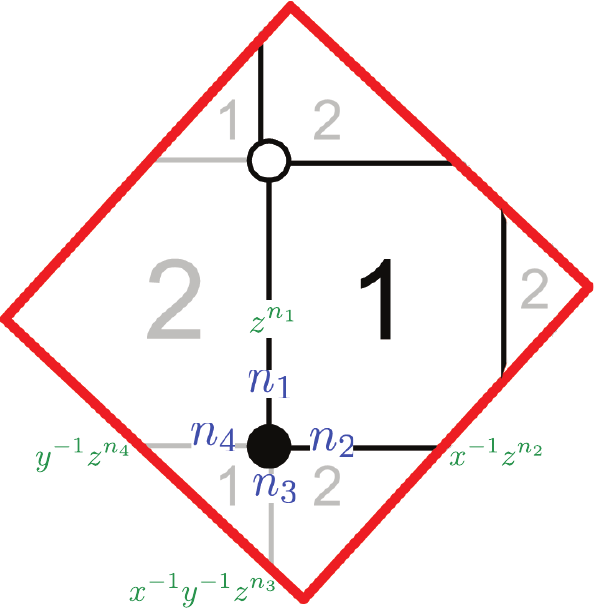}
 \caption{The 2 square tiling with assignments of Chern--Simons levels. Assignments of the integers $n_i$ to the edges are shown in blue. Weights of the edges are shown in green.}
  \label{f:2SqWithCS}
\end{center}
\end{figure}

The $z$ variable is used to encode the CS levels for each field. Each edge is given a weight $z^{n_i}$ where $n_i$ is the CS level of field $i$. The fundamental domain of the 2 square tiling with edges given appropriate weights is given in \fref{f:2SqWithCS}.

The next step is to compute the permanent of the Kasteleyn matrix. The permanent of the Kasteleyn matrix corresponding to the 2 square tiling is given in \eqref{Kabjm}. In this case it is trivial to compute the permanent as the Kasteleyn matrix is a $1 \times 1$ matrix.

\beq \label{Kabjm}
K = \mathrm{Perm} (K) =  z^{n_1} +  x^{-1} z^{n_2} +  x^{-1} y^{-1} z^{n_3} + y^{-1} z^{n_4}
\eeq

The next step is to pick Chern-Simons levels for the fields of our tiling. As we have mentioned, these levels must be integer valued. In our 2 square example we may pick these integers to be
\beq
n_3= 1 \; \; n_1=n_2=n_4=0
\eeq
Which gives
\beq
\mathrm{Perm} (K) = 1 +  x^{-1}  + x^{-1} y^{-1} z +   y^{-1} 
\label{e:permk2sq}
\eeq

\begin{figure}[ht]
\begin{center}
  \includegraphics[totalheight=3.0cm]{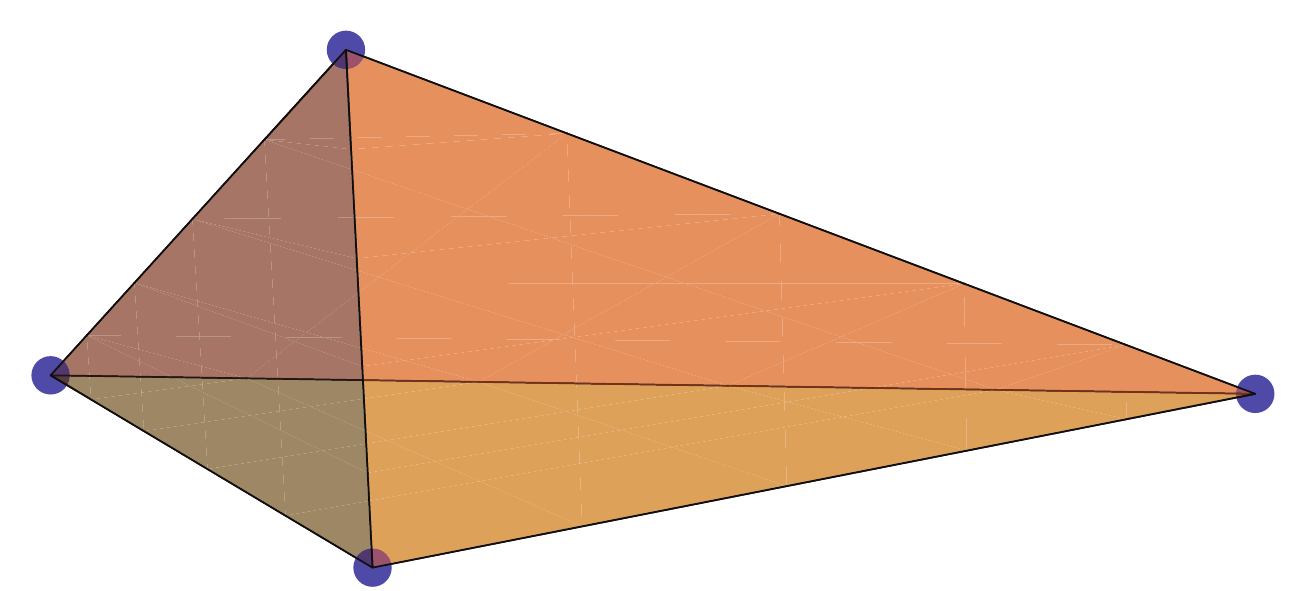}
 \caption{The toric diagram of $\BC^4$. Vertices of the tetrahedron correspond to the lattice points.}
  \label{f:torphase1c4}
\end{center}
\end{figure}

It is possible to display this information on a $\BZ^3$ lattice. Each term in the permanent can be displayed as a point on this lattice. A term of the form $x^a y^b z^c$ corresponds to point in the $\BZ^3$ lattice with coordinates $(a,b,c)$. The permanent of a matrix therefore corresponds to a collection of lattice points in $\BZ^3$. In this work we only consider tilings with CS levels that form convex shapes in $\BZ^3$ after this forward algorithm has been applied. The information that we have computed is the toric data of the Calabi-Yau singularity that is the moduli space of the CS theory defined by the tiling. The toric data corresponding to the permanent given in \eqref{e:permk2sq} is displayed in \fref{f:torphase1c4}

We shall now go on to describe an equivalent method of computing the toric data corresponding to the moduli space of a CS theory that can be described by the tiling. The equivalence of the two methods is proved in `Phases of M2-brane Theories' \cite{Davey:2009sr}

\subsection{The Forward Process using Charge Matrices}

In this section we will discuss a second method of computing the toric data corresponding to the moduli space of a CS theory described by a brane tiling. This method is very similar to the method of finding the moduli space of a 3+1 dimensional quiver theory that was discussed in Section \ref{s:D3Charges} and is discussed in detail in \cite{HananyZaff08}.

The first step in this method is to compute the perfect matching matrix of a brane tiling. This is done in an identical way to the 3+1 dimensional case. Columns of the perfect matching matrix correspond to perfect matchings and rows to fields. The matrix is filled entirely with wither 1s or 0. $P_{ij} =  1$ if field $i$ is in perfect matching $j$ and 0 otherwise. For the 2 square tiling (see \fref{f:2SqTiling}), the perfect matching matrix is the $4 \times 4$ identity matrix as the tiling has only 1 white node and 1 black node.  

We can compute the null-space of the perfect matching matrix $P$. This matrix can be thought of as encoding relations between perfect matchings and is known as $Q_F$. Just like the 3+1 dimensional case, a tiling with $c$ perfect matchings and $g$ gauge groups will have $Q_F$ being a $(c-g-2) \times c$ matrix.

The moduli space of vacua with only the F-terms taken into account is known as the Master space \cite{Masterspace, MMasterspace} of the gauge theory. This space can be thought of as the space of perfect matchings, $\BC^c$, modded out by the relations encoded in $Q_F$, i.e.
\bea
\firr{} = \BC^c//Q_F~. \label{sympq}
\eea
In this way, the matrix $Q_F$ can be regarded as a charge matrix associated with the F-terms.

In order to find a description for the full moduli space, we must now take into account the final two sets of equations in \eqref{DF}.

\subsubsection{From the Master Space to the Mesonic Moduli Space}

The D-term constraints for 2+1 dimensional quiver CS theories are similar, but not identical to the 3+1 dimensional case. The D-terms can be found in \eqref{e:CSModSpace} and can be summarized by
\beq
\mu_a (X) = 4 k_a \sigma_a
\label{e:MMap}
\eeq
As $\Sigma_a k_a = 0$ we automatically have that
\beq
\Sigma_a \mu_a (X) =0
\eeq
This redundancy in the D-terms mirrors the 3+1 dimensional case. There is another redundancy in \eqref{e:MMap} which is that the combination of $\mu_a$ that fall parallel to $k_a$ is equal to the field $\sigma$. Therefore there are actually $g-2$ constraints that come from D-terms. We can define the following $2 \times G$ matrix that we shall call $C$:
\begin{equation}\label{C}
C =\left(\begin{matrix}
1 & 1 & 1 & \ldots & 1 \\ k_1 & k_2 & k_3 & \ldots & k_g
\end{matrix}\right) \ .
\end{equation}
 We can now compute $\ker(C)$ whose rows are basis vectors of the null space of $C$. $\ker(C)$ is a $(G-2) \times G$ matrix and can be thought of as a tool that can be used to avoid the two aforementioned redundancy issues.

Just like the 3+1 dimensional case, we can define $\widetilde{Q}$ to be a $G \times c$ matrix as follows:
\bea
d_{G \times E} = \widetilde{Q}_{G \times c} \cdot (P^t)_{c \times E}~, \label{qtilde}
\eea
The reader is reminded that $d_{G \times E}$ is the previously defined incidence matrix of the quiver.

  We now compute
\bea
(Q_D)_{(G-2) \times c} = \ker{(C)}_{(G-2) \times G} \cdot \widetilde{Q}_{G \times c}~. \label{QD}
\eea 
$Q_D$ therefore stores the ways in which the perfect matchings are charged according to the gauge symmetry of the theory. It is also built to circumvent the issues with the two redundant D-terms.
 
We can now write the mesonic moduli space of the CS theory as
\bea
\CMm = \firr{} // Q_D = \left( \BC^c//Q_F \right) // Q_D~. \label{quoteFD}
\eea 
It is now possible to find the toric description of $\CMm$ using the charge matrices $Q_F$ and $Q_D$. To do this, we construct a $(c-4) \times c$ matrix $Q_T$ as follows:
\bea \label{Qtdef}
(Q_T)_{(c-4) \times c} =
\left( \begin{array}{c}
(Q_D)_{(G-2)\times c} \\
(Q_F)_{(c-G-2) \times c} 
\end{array} \right)~.
\eea
Then we can define a $4 \times c$ matrix 
\bea
G = \ker(Q_T) \label{Gt}
\eea
whose rows are basis vectors of the null space of $Q_T$.

The matrix $G$ stores the toric data of the Calabi-Yau 4-fold which is the $\CMm$ of the CS theory described by the tiling. It is always possible to choose the first row of $G$ to be $(1, \ldots, 1)$. This is because this vector lives in the null spaces of both $Q_F$ and $Q_D$.

We can think of ${(G)}_{(4 \times c)}$ as a collection of $c$ 4-vectors that lie in a 3 dimensional hyperplane. By removing the first row of $G$ we can obtain a $3 \times c$ matrix $G_t$. The columns of $G_t$ give the coordinates of points in the 3-dimensional toric diagram. In this work, we only consider CS theories that correspond to a convex collection of points in the $\BZ^3$ lattice after the forward algorithm has been applied.

\subsubsection{A Summary of the Forward Algorithm Using Charge Matrices.} 
We summarise the forward algorithm that was discussed above here:
\begin{itemize}
 \item From the tiling and CS levels, read off $ d_{G \times E} $, $C_{2 \times G}$ and $ P_{E \times c}$
\item Let $(Q_F)_{(c-G-2)\times c} = \ker (P)$.
\item Find $\widetilde{Q}_{G \times c}$ using $d_{G \times E} = \widetilde{Q}_{G \times c} \cdot (P^t)_{c \times E}$
\item Let $(Q_D)_{(G-2) \times c} = 	 \ker{(C)}_{(G-2) \times G} \cdot \widetilde{Q}_{G \times c}$
\item Write $(Q_T)_{(c-4) \times c} =
\left( \begin{array}{c}
(Q_D)_{(G-2)\times c} \\
(Q_F)_{(c-G-2) \times c} 
\end{array} \right)$
\item Find the toric data $(G)_{4 \times c} = {\ker}(Q_T) $
\end{itemize}

\subsection{Uniqueness of Toric Data}

The reader might be worried about the ambiguities that arise when one carries out the forward algorithm on a tiling. For instance, when computing toric data using the Kasteleyn method, our choice of fundamental domain was not unique. Our choices of variables $x$ and $y$ were also arbitrary. Another example of this ambiguity is that in \eqref{Gt}, we attempt to find the null-space of a matrix and represent this in terms of vectors that are stored in the matrix $G$. This collection of vectors is by no means unique and so different implementations of the forward algorithm may give rise to different matrices $G$.

These ambiguities are no cause for concern. 3 dimensional toric data is unique up to $GL(3,\BZ)$ transformations. Therefore it is possible that different implementations of the forward algorithm may give rise to different sets of points in $\BZ^3$, but these sets of points should be related to each other by one of these transformations. In particular if both versions of the forward algorithm are applied to the same tiling the resulting toric diagrams should be related to each other by one of the aforementioned transformations.

\Section{Examples of Brane Tilings for M2-branes}

In the previous section, we applied the version of the forward algorithm that uses the Kasteleyn matrix to a theory described by the 2 square tiling. Let us now consider applying the other version of the forward algorithm to this theory to see explicitly the similarities and differences between the two methods.

\subsection{The 2 Square Tiling}

In Section \ref{s:FFKast}, the forward algorithm was applied to the 2 square tiling given in \fref{f:2SqTiling} with some particular choice of CS levels. Let us now consider 
the second way of computing the moduli space of the theory.

From the tiling in \fref{f:2SqTiling}, we can read off the following matrices:

\bea
 d_{G \times E} & = & \left( \begin{array}{cccc}
          1&1&-1&-1\\
-1&-1&1&1
         \end{array} \right)
 \; \;
 C_{2 \times G}  =  \left( \begin{array}{cc}
          1&1\\
1&-1
         \end{array} \right)
\nn \\
  P_{E \times c}& = & \left( \begin{array}{cccc}
  1&0&0&0\\
  0&1&0&0\\
  0&0&1&0\\
  0&0&0&1
         \end{array} \right)
\label{e:2SqdCP}
\eea

As $P$ is the $4 \times 4$ identity matrix we find that
\beq
\widetilde{Q}_{G \times c} = d_{G \times E}
\eeq
and
\beq
 \mathrm{Ker} (C) = \{ \} \; \; \implies Q_D = \{ \}
\eeq
Therefore the total charge matrix $Q_T = \{ \}$. This allows us to find the toric data:
\beq
G = \left( \begin{array}{cccc} 1&1&1&1\\1&0&0&0\\0&1&0&0\\0&0&1&0 \end{array} \right)
\eeq
After removing the first row, the columns give the coordinates of points in the toric diagram:
\beq
G_t = \left( \begin{array}{cccc} 1&0&0&0\\0&1&0&0\\0&0&1&0 \end{array} \right)~. \label{gtabjm}
\eeq
This toric data corresponds to the 4 corners of the tetrahedron given in Figure \ref{f:torphase1c4}. The toric data encoded in \eqref{gtabjm} is equivalent (up to $GL(3,\BZ)$ transformations) to the toric data computed earlier using the Kasteleyn method.

\Section{Toric Duality}

It is possible for more than one tiling (with CS levels) to correspond to the same Calabi--Yau singularity. Such theories are known as different toric phases of a model and the phenomena is known as toric duality.  This duality has been studied in detail in the D3-brane case and has been discussed in section \ref{d3inv}.  Recently the M2-brane analogue of this effect has been studied \cite{HananyZaff08, Franco:2008um, Amariti:2009rb} and a number of models have been classified and systematically studied \cite{Hanany:2008gx}. A second phase of $\BC^4$ shall now be investigated.

\subsection{A Toric Dual of the ABJM theory: The One Hexagon Model with a `Double Bond'}
\begin{figure}[ht]
\begin{center}
  \includegraphics[totalheight=2cm]{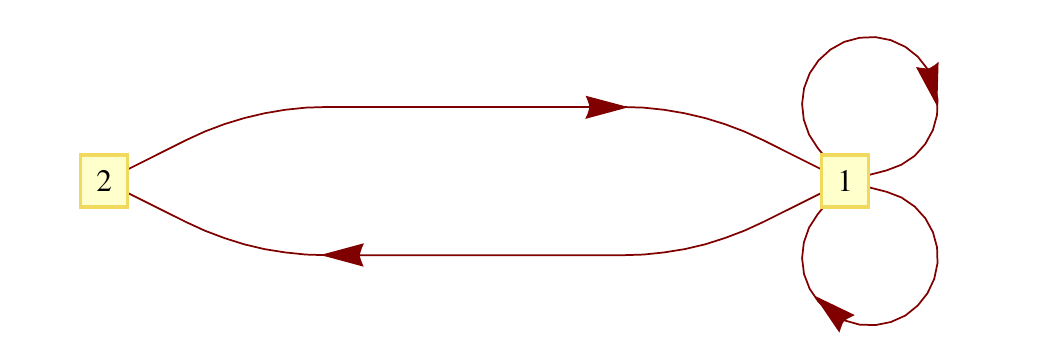}
  \hskip 0.5cm
  \includegraphics[totalheight=4.5cm]{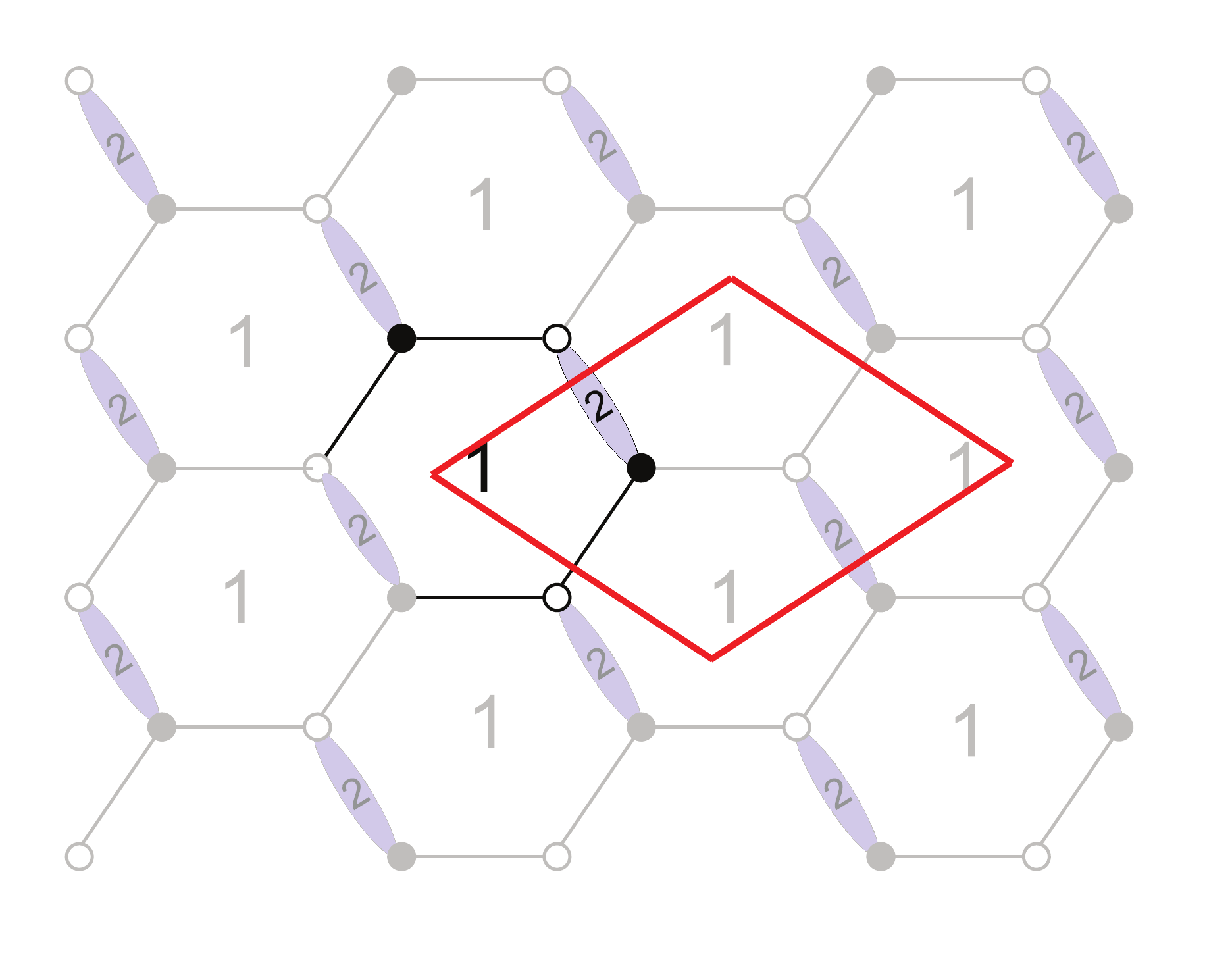}
 \caption{The quiver diagram and tiling corresponding to the second phase of $\BC^4$.}
  \label{f:1hex1db}
\end{center}
\end{figure}

The 1 hexagon tiling with 1 `double bond' is drawn alongside its quiver in Figure \ref{f:1hex1db} \cite{HananyZaff08}. The `double bond' is simply a face in the tiling with only two edges.

This theory has a gauge symmetry which is a product of two gauge groups.  There are 2 bi-fundamental fields $X_{12}$ and $X_{21}$ as well as 2 adjoint fields which we will call $\phi_1^1$ and $\phi_1^2$.   The superpotential is given by
\bea
W = \tr(X_{21}[\phi_1^1,\phi_1^2]X_{12})~.
\eea
We will take the Chern--Simons levels to be $k_1=-k_2=1$.

\begin{figure}[ht]
\begin{center}
   \includegraphics[totalheight=6.0cm]{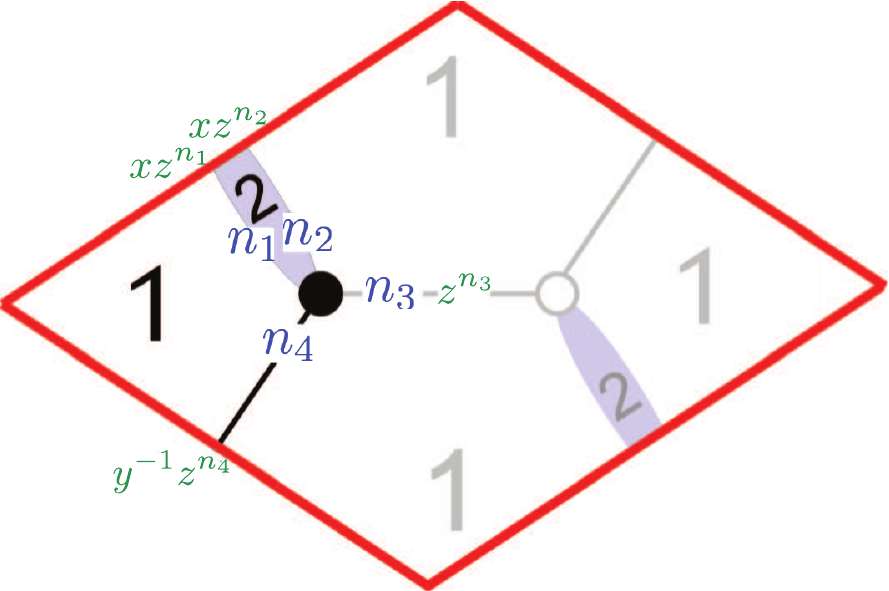}
 \caption{The fundamental domain of the tiling corresponding to the second phase of $\BC^4$. Assignments of the integers $n_i$ to the edges are shown in blue and the weights for these edges are shown in green.}
  \label{f:fdphase2c4}
\end{center}
\end{figure}

We will demonstrate the two methods of constructing the toric diagram that were mentioned earlier.

\subsubsection{Toric Data via the Kasteleyn Matrix}

First of all we pick the CS levels for edges, integers $n_i$. The edges are weighted according to these integers in Figure \ref{f:fdphase2c4}.  From the relationship between CS levels for edges and gauge groups, we find that :
\bea
\text{Gauge group 1~:} \qquad k_1 &=& 1 = -n_1 + n_2 ~, \nn \\
\text{Gauge group 2~:} \qquad k_2 &=& -1 = n_1 - n_2 ~.
\eea  
Because of this, we choose
\bea
n_2= 1,\quad n_1=n_3=n_4=0~.
\eea
We can now construct the Kasteleyn matrix for this model. Since the fundamental domain contains only one black node and one white node, the Kasteleyn matrix is a $1\times 1$ matrix and is therefore equal to its permanent:
\bea \label{permKph2c4}
K &=&  z^{n_3} +  y^{-1} z^{n_4} + x z^{n_1} + x z^{n_2} \nn \\
&=& 1 +  y^{-1} +  x  +  x z \qquad \text{(for $n_2=1,~n_1=n_3=n_4=0$)}~. 
\eea
The powers of $x, y$ and $z$ in each term of $K$ give the 3 coordinates of each point in the toric diagram.  
We can represent each 3-vector as a column in the following matrix, which we shall call $G_K$:
\bea
G_K = \left(
\begin{array}{cccc}
 1 & 0 & 1 & 0 \\
 0 & -1 & 0 & 0 \\
 0 & 0 & 1 & 0
\end{array}
\right)~.
\label{GK1Hex1DB}
\eea
In the work that follows, all $G_K$ matrices shall store toric data and shall be constructed via the Kasteleyn method.
\subsubsection{Toric Data via the Charge Matrices}
It is also possible to compute the toric data by using the charge matrices. We first read off the following matrices from the tiling:
\bea
 d_{G \times E} & = & \left( \begin{array}{cccc}
          -1&1&0&0\\
1&-1&0&0
         \end{array} \right)
 \; \;
 C_{2 \times G}  =  \left( \begin{array}{cc}
          1&1\\
1&-1
         \end{array} \right)
\nn \\
  P_{E \times c}& = & \left( \begin{array}{cccc}
  1&0&0&0\\
  0&1&0&0\\
  0&0&1&0\\
  0&0&0&1
         \end{array} \right)
\label{e:1Hex1DBdCP}
\eea

As $P$ is the $4 \times 4$ identity matrix, we find that $Q_F = \{ \}$ and also that
\beq
\widetilde{Q}_{G \times c} = d_{G \times E}
\eeq
and
\beq
 \mathrm{Ker} (C) = \{ \} \; \; \implies Q_D = \{ \}
\eeq
Therefore the total charge matrix $Q_T = \{ \}$. This allows us to find the toric data:
\beq
G = \left( \begin{array}{cccc} 1&1&1&1\\1&0&0&0\\0&1&0&0\\0&0&1&0 \end{array} \right)
\eeq
After removing the first row, the columns of the following matrix give the coordinates of points in the toric diagram:
\beq
G_t = \left( \begin{array}{cccc} 1&0&0&0\\0&1&0&0\\0&0&1&0 \end{array} \right)~. \label{gt1Hex1DB}
\eeq
Which is identical toric data to that found for the other phase of $\BC^4$. Also, it is possible to relate \eqref{gt1Hex1DB} to \eqref{GK1Hex1DB} by a $GL(3,\BZ)$ transformation.

\Section{Finding phases of $\cal{C} \times \BC$ using an inverse method}
\label{s:M2Inverse}
It has been possible to find different phases of an M2-brane model by using a method involving the projection of 3-dimensional toric data. Let us describe this method using an example which is known as the $\cal{C} \times \BC$ model.

\begin{figure}[h]
\begin{center}
  \includegraphics[totalheight=4.0cm]{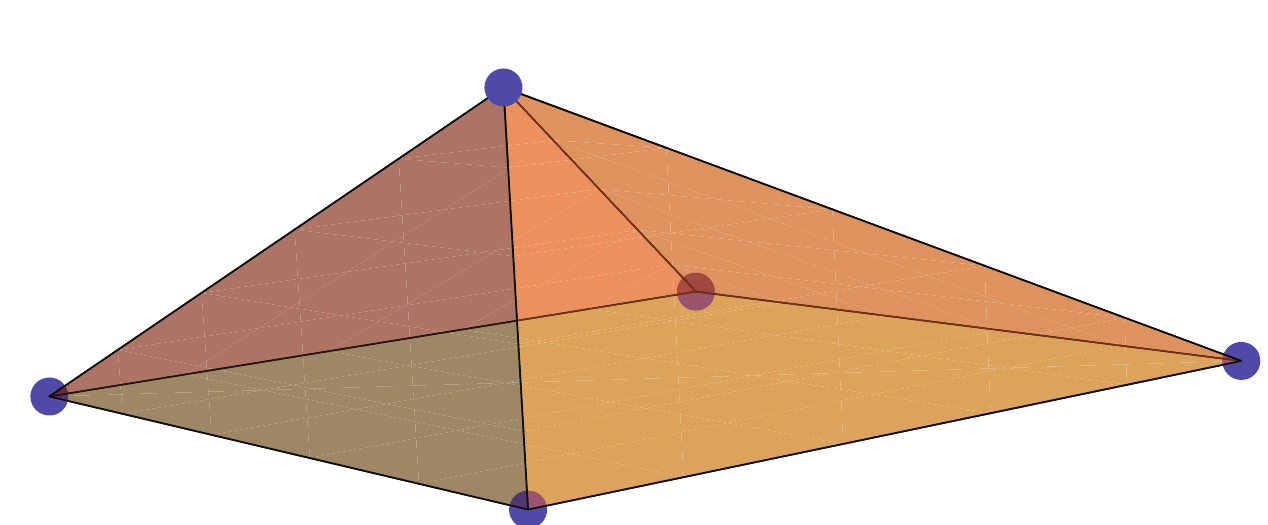}
 \caption{The toric diagram of the $\cal{C} \times \BC$ theory.}
  \label{f:torconxc}
\end{center}
\end{figure}

The toric diagram of the geometry known as $\cal{C} \times \BC$ is given in \fref{f:torconxc}. The coordinates of the vertices of the toric diagram are given as columns of $G_t$ in \eqref{e:conxcgt}

\beq
G_t = \left(
\begin{array}{ccccc}
 1 & 0 & 1 & 0 & 0 \\
 1 & 0 & 0 & 1 & 0 \\
 0 & 0 & 0 & 0 & 1
\end{array} \right)
\label{e:conxcgt}
\eeq

As we have mentioned previously the matrix $G_t$ that we chose to define our geometry is not unique. Any matrix that can be transformed into $G_t$ using a $GL(3,\BZ)$ transformation would have done the job just as well. Examples of such alternative matrices are $G'_t$ or $G''_t$ which are displayed in \eqref{e:conxcgt2}.

\bea
G'_t &=& \left(
\begin{array}{ccccc}
 0 & 0 & -1 & 1 & 0 \\
 0 & 0 & 0 & 0 & 1 \\
 1 & 0 & 1 & 0 & 0
\end{array}
\right)
\nn \\
G''_t &=& \left(
\begin{array}{ccccc}
 1&0 & 1&  0 & 0   \\
 0 & 0 & 0 &  0 &1   \\
 1 & 0 & 0  & 1 & 0 
\end{array}
\right)~.
 \label{e:conxcgt2}
\eea

The next step in the inverse process is to remove the third row of the $G_t$ matrix and see whether the columns of the resulting matrix form a convex shape in a $\BZ^2$ lattice. If such a convex shape is formed, we find a (3+1)-dimensional theory that has a moduli space defined by the toric data corresponding to this 2-dimensional shape.
To make this clearer, let us consider the following projection of the $G_t$ from \eqref{e:conxcgt}:

\beq
G_t \rightarrow \left(
\begin{array}{ccccc}
 1 & 0 & 1 & 0 & 0 \\
 1 & 0 & 0 & 1 & 0 
\end{array} \right)
\label{e:conxcredGt}
\eeq

\begin{figure}[h]
\begin{center}
  \includegraphics[totalheight=2.5cm]{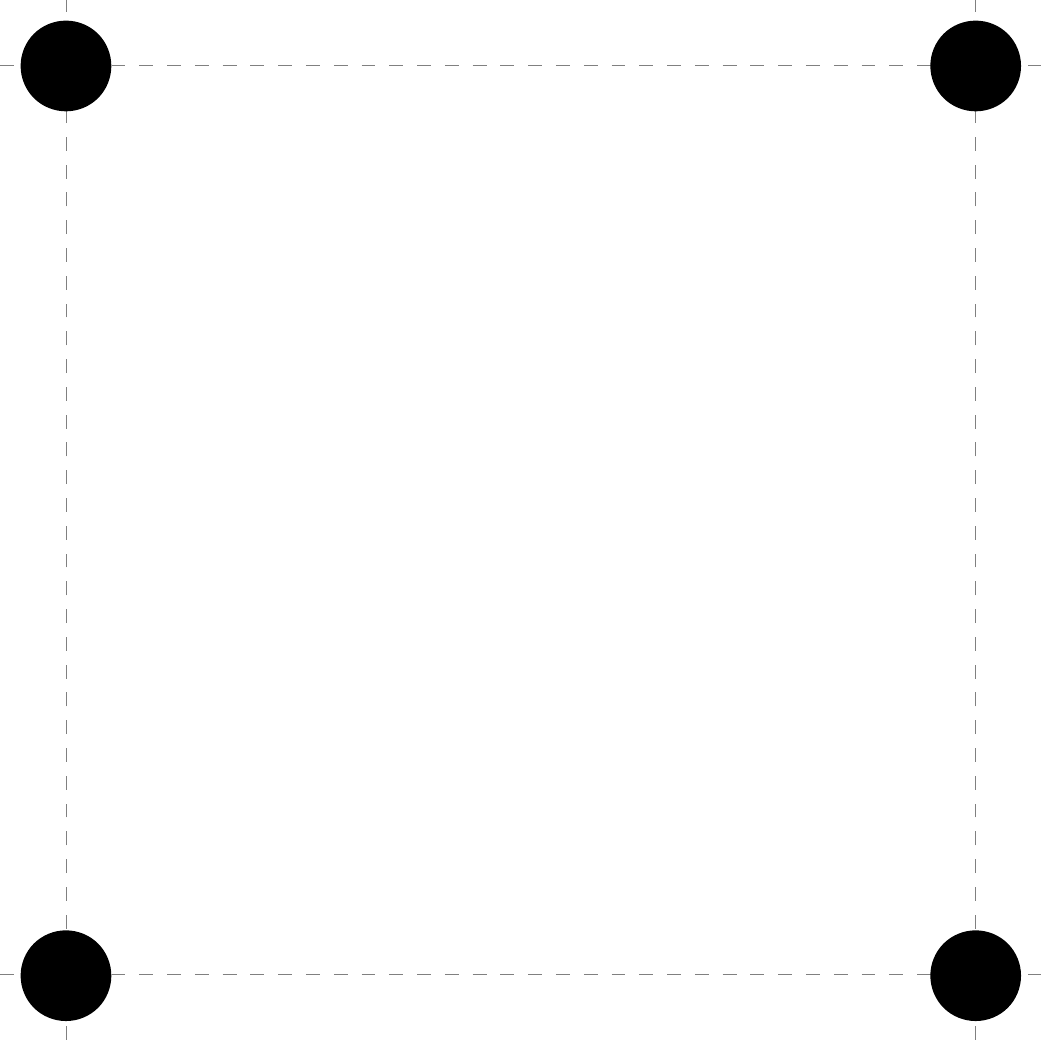}
 \caption{The toric diagram corresponding to the projected $G_t$ matrix given in \eqref{e:conxcredGt}.}
  \label{f:conxcgtprojtoric}
\end{center}
\end{figure}

The resulting 2-dimensional shape given by the projection in \eqref{e:conxcredGt} is given in \fref{f:conxcgtprojtoric}. The tiling corresponding to this 2-dimensional shape is the 2-square tiling. We can now ask whether this 2-square tiling can give rise to a CS theory that has $\cal{C} \times \BC$ as its moduli space.

\begin{figure}[h]
\begin{center}
   \includegraphics[totalheight=6cm]{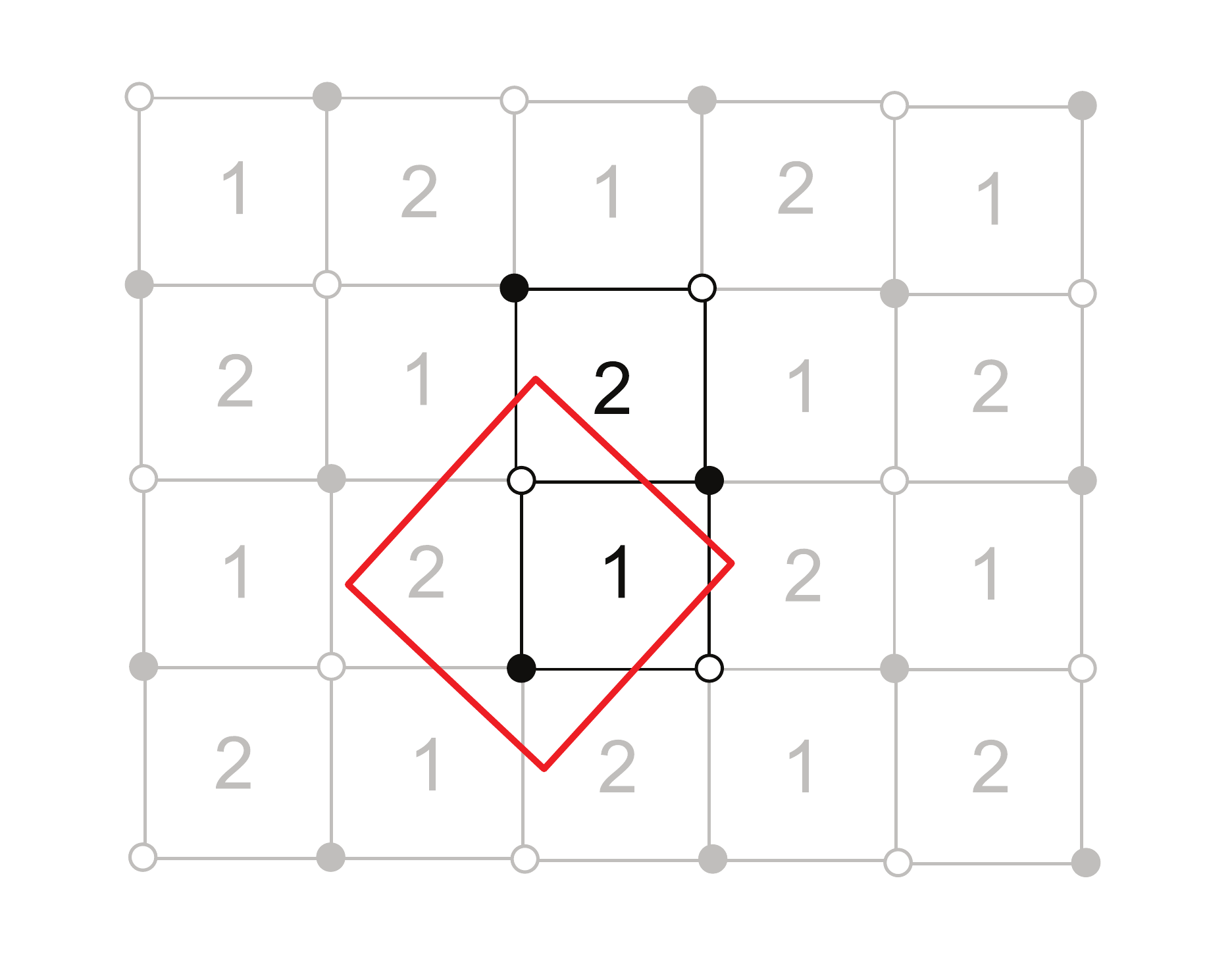}
 \caption{The 2 square tiling with assignments of Chern--Simons levels.}
  \label{f:2SqWithCS2}
\end{center}
\end{figure}

The 2 square tiling is given in \fref{f:2SqWithCS2}. The Kasteleyn matrix corresponding to this tiling is given in \eqref{Kabjm2}.

\beq \label{Kabjm2}
K = \mathrm{Perm} (K) =  z^{n_1} +  x z^{n_2} +   y z^{n_3} + x y z^{n_4}
\eeq

We can ask whether it is possible to allocate a set of CS levels to the 2 square tiling such that the toric data that we can extract from the tiling matches that displayed in \eqref{e:conxcgt}. Sadly it is not possible to do this, however we can use the idea of `double-bonds' to find a suitable theory.

\subsection{Phase I: The Two Square Tiling a `Double Bond'}
\label{s:Ph1Conxc}

\begin{figure}[ht]
\begin{center}
   \includegraphics[totalheight=4cm]{phases/tilphase1c4.pdf}
\hspace{1cm}
 \includegraphics[totalheight=4cm]{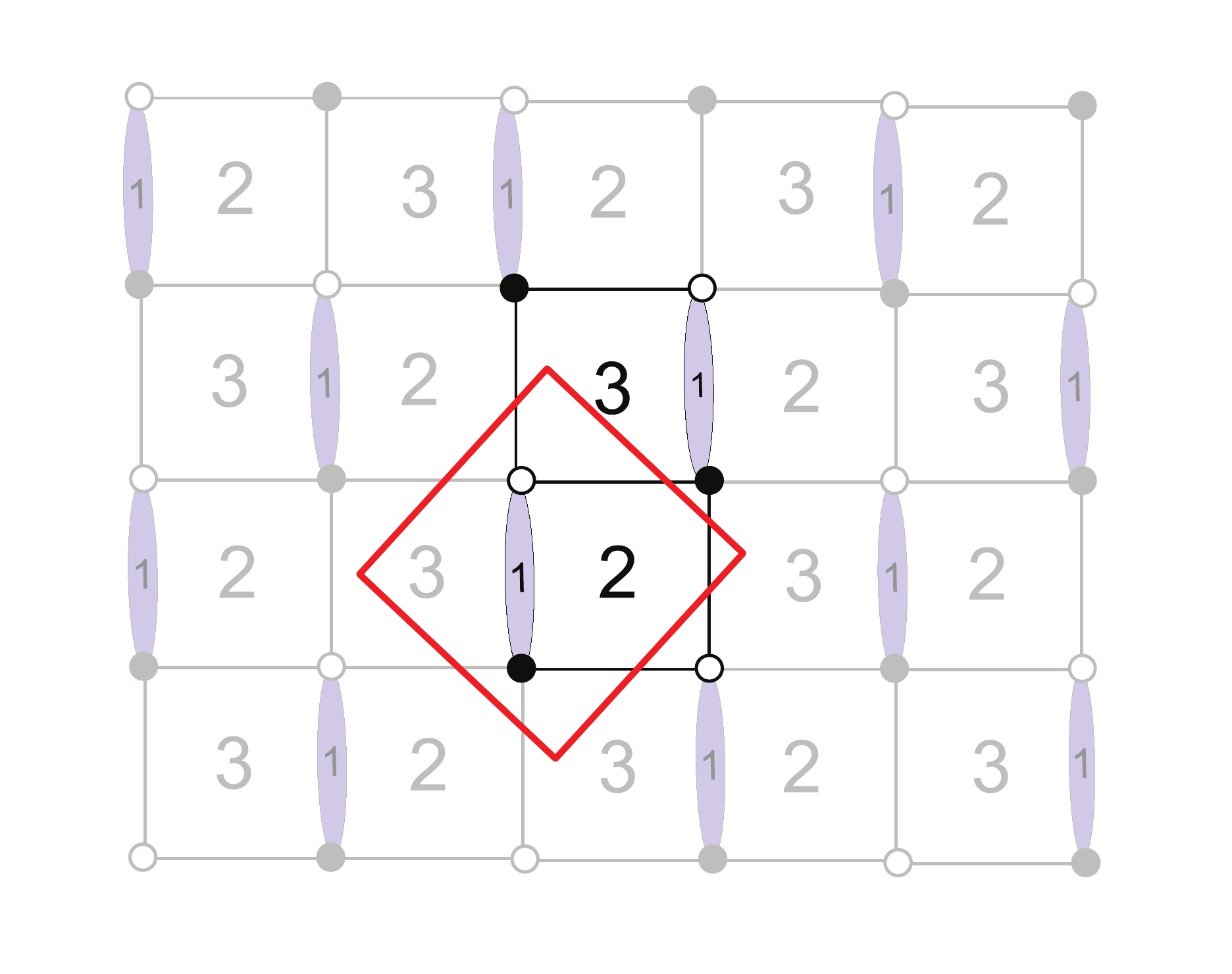}
 \caption{The 2 square tiling (left) and the 2 square tiling with 1 double bond (right).}
  \label{f:2SqandChild}
\end{center}
\end{figure}

The reason that the 2 square tiling can't be used to find a CS theory corresponding to the $\cal{C} \times \BC$ geometry has to do with the number of toric points that we have attempted to `grow' from the 2 dimensional toric diagram. The 2 square tiling has only 4 terms in its Kasteleyn, whereas the geometry we are trying to fit the theory to has 5 toric coordinates that should be filled. The way that we can find a suitable model is to add a `double bond' to the 2 square tiling as in \fref{f:2SqandChild}

\begin{figure}[ht]
\begin{center}
   \includegraphics[totalheight=6.0cm]{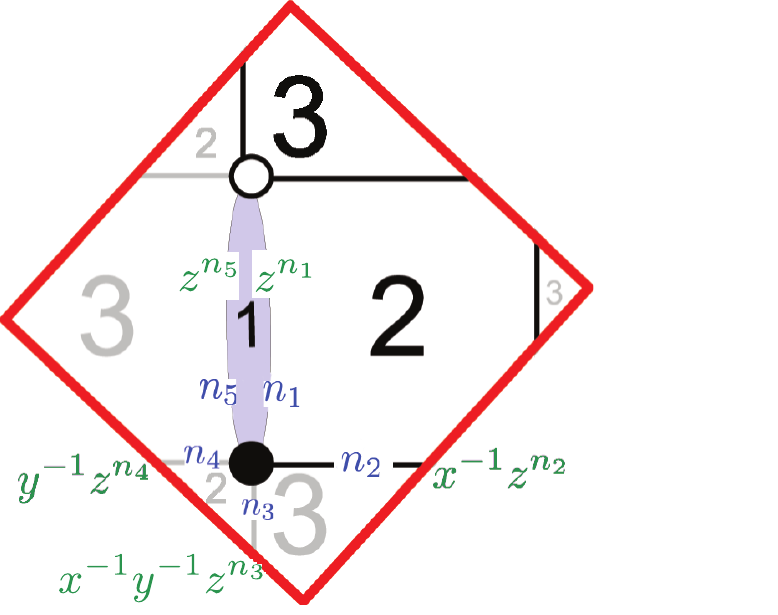}
 \caption{The fundamental domain of 2 square tiling with 1 double bond. Assignments of the integers $n_i$ to the edges are shown in blue and the weights for the edges are shown in green.}
  \label{f:2Sq1DBFundDom}
\end{center}
\end{figure}

The tiling comprising of 2 squares and 1 double bond is given with labeled fields in \fref{f:2Sq1DBFundDom}. The `double bond' adds a useful additional term to the Kasteleyn which can be found in \eqref{e:2sq1dbKast}. 

\bea \label{e:2sq1dbKast}
K &= &\mathrm{Perm} (K) =  z^{n_1} +  x z^{n_2} +   y z^{n_3} + x y z^{n_4} + z^{n_5} \nn \\
& =&  1 +  x  +   y  + x y + z \; \; \mathrm{for} \; \; n_5 =1, \; \; \mathrm{all} \; \; \mathrm{others} \; \; 0
\eea
By picking the CS levels to be $n_5=1$ and all others 0, we find a theory that corresponds exactly to the toric data given in \eqref{e:conxcgt}. In doing this, we have found one of the toric phases of $\cal{C} \times \BC$.

\subsection{The charge matrices}  
We will now examine the moduli space of this phase of the $\cal{C} \times \BC$ using charge matrices. From the tiling given in \fref{f:2Sq1DBFundDom} we can see that the perfect matchings are in 1 to 1 correspondence with the fields of the gauge theory. Therefore we can write
\beq
P= \left(
\begin{array}{ccccc}
 1 & 0 & 0 & 0 & 0 \\
 0 & 1 & 0 & 0 & 0 \\
 0 & 0 & 1 & 0 & 0 \\
 0 & 0 & 0 & 1 & 0 \\
 0 & 0 & 0 & 0 & 1
\end{array}
\right)
\eeq
Since there is a one-to-one correspondence between the quiver fields and the perfect matchings, it follows that 
\bea
Q_F=0~. \label{QFD1C}
\eea
Therefore $\firr{} = \BC^5$. We also have
\beq
C= \left(
\begin{array}{ccc}
 1 & 1 & 1  \\
 -1 &0 & 1
\end{array}
\right) \implies
\mathrm{Ker}(C) = \frac{1}{3}(1,-2,1)
\eeq
As $P$ is the $5 \times 5$ identity matrix, we have
\beq
\tilde{Q} = d = \left(
\begin{array}{ccccc}
1& 0& 0& 0& -1\\
-1&1& -1& 1& 0\\
0&-1& 1& -1& 1
\end{array}
\right)
\eeq
and so
\beq
Q_D = \mathrm{Ker}(C) \cdot \tilde{Q} = (1, -1, 1, -1, 0)  \label{QDD1C}
\eeq  
The total charge matrix is given by
\bea
Q_T = Q_D = (1,-1,1,-1,0)~.  \label{QtD1C}
\eea
From this, we can obtain the toric data of the singularity. This is encoded in the following matrix:
\beq
 \left(
\begin{array}{ccccc}
1& 1& 0& 0& 0\\
1& 0& 0& 1& 0\\
0& 0& 0& 0& 1 \end{array} \right)
\eeq
Which corresponds to the toric data of the $\cal{C} \times \BC$ and can be related to $G_t$ by a permutation of perfect matchings (or toric coordinates).

We can find the other phases of the $\cal{C} \times \BC$ by considering the toric data given in either of the matrices in \eqref{e:conxcgt2}. Before we go on to explore these other phases, let us summarise the inverse method. 

\subsection{A Summary of the Inverse Method}

Our inverse method for toric Calabi-Yau 4-fold singularities is as follows.

\begin{itemize}
 \item Construct the matrix $G_t$ encoding the toric data of the singularity. Columns of $G_t$ correspond to the three coordinates of a point of the toric diagram.
\item Apply elements of $GL(3,\BZ)$ to $G_t$ to create a list, $\cal{L}$, of equivalent toric data.\footnote{There are an infinite number of elements of $GL(3,\BZ)$. Any implementation of this algorithm should pick a `reasonably' large set of such matrices. A suitable set could be matrices composed of elements that have an absolute value that is less than a given number.}
\item For each element of $\cal{L}$, remove the third row and test whether its columns form a convex set of points in a $\BZ^2$ lattice.
\item If a matrix passes the last test find all brane tilings that, as (3+1)-dimensional theories, correspond to this set of 2-dimensional lattice points.
\item Add Chern-Simons levels to the brane tilings found in the last step as well as these tilings with `double-bonds'\footnote{There is no reason why we should not also consider tilings with `triple-bonds' or indeed `n-bonds'}.
\item Compute the permanent of the Kasteleyn matrix for these theories and see whether the CS levels can be chosen to match the toric data encoded in $G_t$.
\item Each CS theory found is a toric phase of the model described by the toric data in $G_t$.
\end{itemize}

We should mention that this inverse process for M2-branes may give 0,1,2 or more phases of a toric CY 4-fold singularity. This is different to the D3-brane case, in which we are guaranteed at least one tiling for a CY 3-fold singularity.

Now we have described the inverse method, let us discuss the two other toric phases of the $\cal{C} \times \BC$.

\subsection{Phase II: The Two-Hexagon Tiling}
\label{s:Ph2Conxc}
\label{C2Z2}
\begin{figure}[ht]
\begin{center}
   \includegraphics[totalheight=3.5cm]{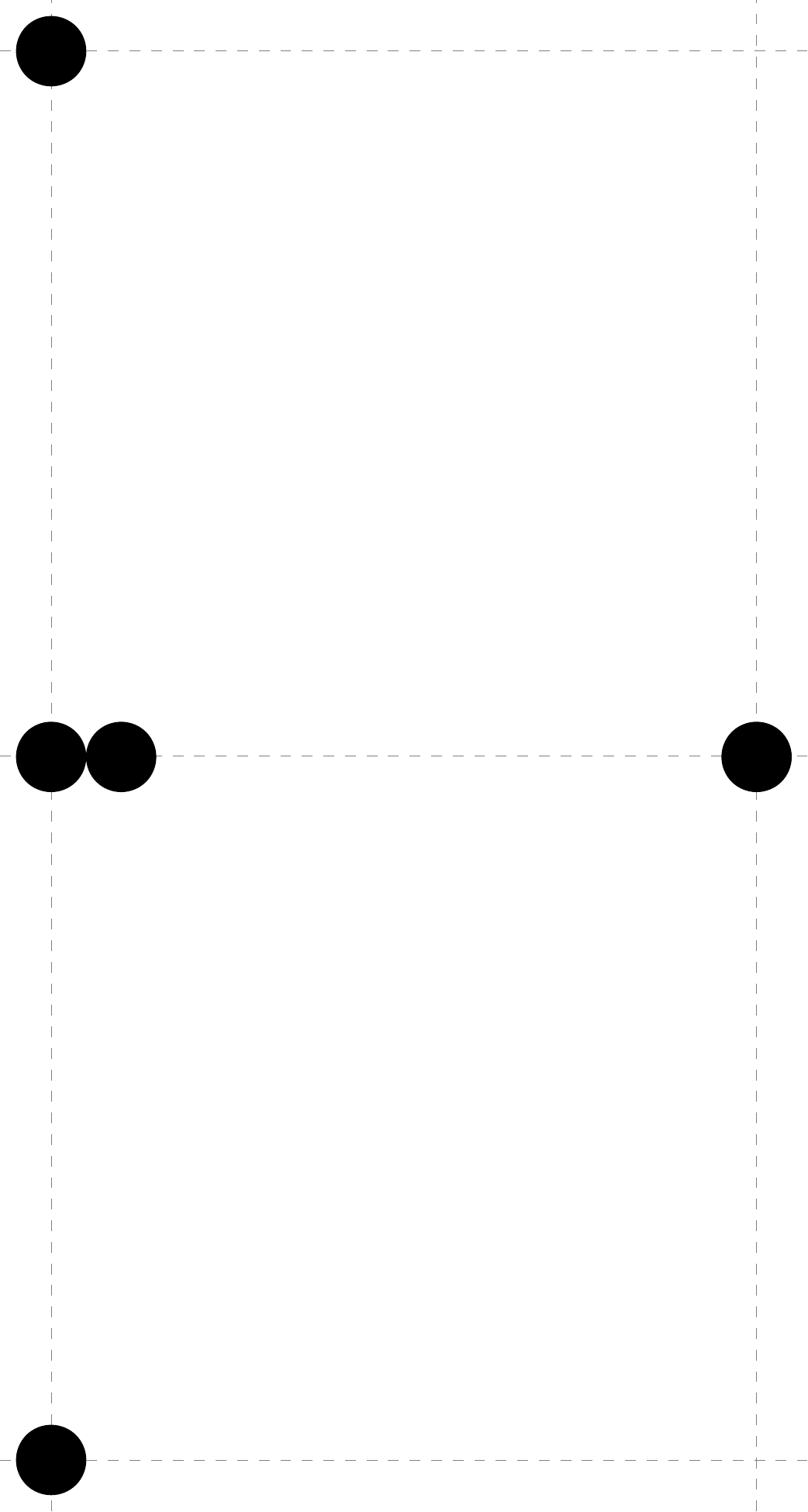}
\hspace{1cm}
  \includegraphics[totalheight=4cm]{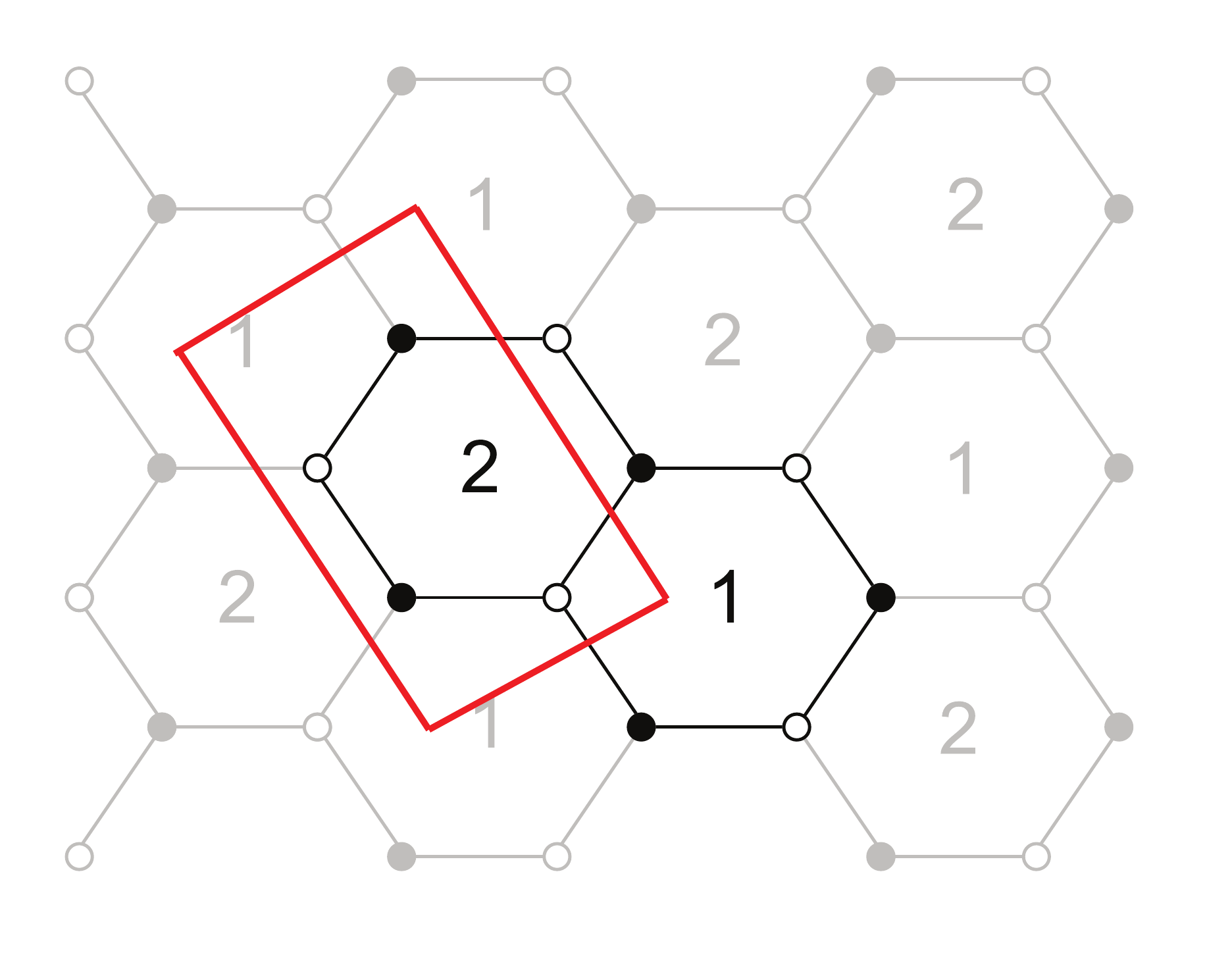}
 \caption{The projected toric data for phase II of the $\cal{C} \times \BC$ displayed as a collection of lattice points (left) and the the brane tiling that it corresponds to (right)}
  \label{f:phaseiiconxcproj}
\end{center}
\end{figure}

Let us try to build a phase of $\cal{C} \times \BC$ starting from the $G'_t$ matrix given in \eqref{e:conxcgt2}. We can delete the third row of this matrix to find 2-dimensional toric data. The $G'_t$ matrix is given again in \eqref{e:gtprimeconxc} along with the matrix formed when its third row is deleted.

\beq
G'_t = \left(
\begin{array}{ccccc}
 0 & 0 & -1 & 1 & 0 \\
 0 & 0 & 0 & 0 & 1 \\
 1 & 0 & 1 & 0 & 0
\end{array}
\right) \rightarrow 
\left(
\begin{array}{ccccc}
 0 & 0 & -1 & 1 & 0 \\
 0 & 0 & 0 & 0 & 1 
\end{array}
\right)
\label{e:gtprimeconxc}
\eeq
Once the third row of $G'_t$ has been deleted, we can relate the 2-dimensional toric data to a brane tiling that corresponds to a 3+1 dimensional theory. The projected toric data, along with the tiling it corresponds to (the 2 hexagon model) are given in \fref{f:phaseiiconxcproj}.

This computation shows that the two hexagon model is a candidate for being a phase of $\cal{C} \times \BC$, although we must test whether it is possible to assign Chern-Simons levels so that the geometry produced by the forward algorithm is exactly $\cal{C} \times \BC$. It is known that it it possible to assign Chern-Simons in such a way \cite{Hanany:2008cd, HananyZaff08, Davey:2009sr}.

The tiling given in \fref{f:phaseiiconxcproj} has two gauge groups and six chiral multiplets denoted as $\phi_1, \phi_2, X_{12}^1, X_{12}^2, X_{21}^1, X_{21}^2$. In 3+1 dimensions this tiling corresponds to the $\BC^2/\BZ_2 \times \BC$ theory. The superpotential is given by
\begin{equation} \label{spph1conc}
W = \tr \left( \phi_1 (X_{12}^2 X_{21}^1 - X_{12}^1 X_{21}^2 ) + {\phi}_2 (X_{21}^2 X_{12}^1 - X_{21}^1 X_{12}^2) \right) \ .
\end{equation}
\subsubsection{The Kasteleyn matrix}
\begin{figure}[ht]
\begin{center}
   \includegraphics[totalheight=6.0cm]{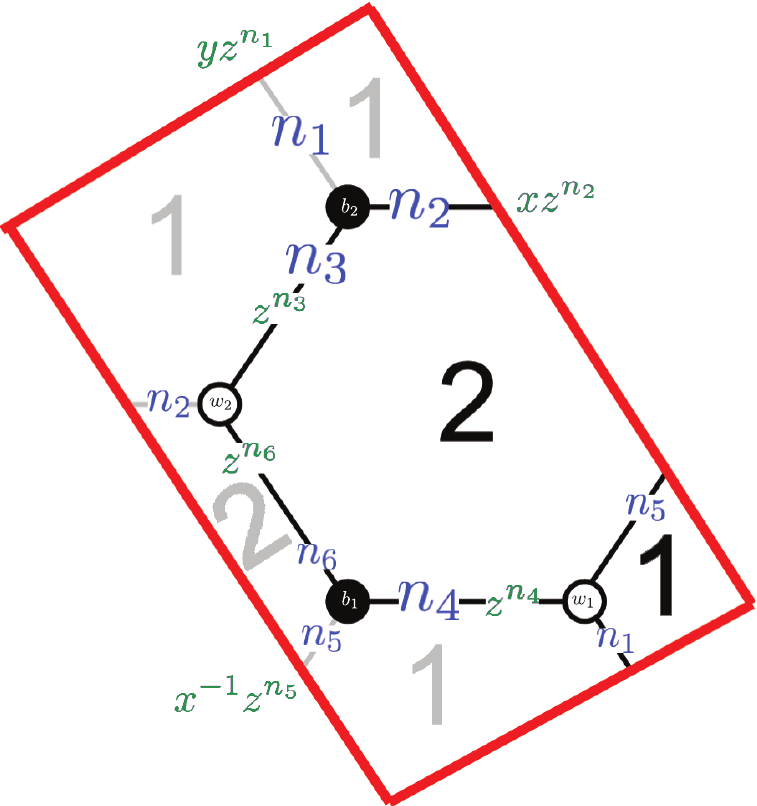}
 \caption{The fundamental domain of the 2 hexagon tiling. Assignments of the integers $n_i$ to the edges are shown in blue and the weights for these edges are shown in green.}
  \label{f:fdphase2conxc}
\end{center}
\end{figure}

We assign the CS levels to the edges $(n_i)$ according to Figure \ref{f:fdphase2conxc}.  Using the rule given in \eref{kn}, we can find out how the CS levels for gauge groups relate to the CS levels for fields. This dictionary is given in \eqref{e:cslevelsconixcph2}.
\bea
\text{Gauge group 1~:} \qquad k_1 &=& 1  = -n_2+n_3+n_4 - n_5 ~, \nn \\
\text{Gauge group 2~:} \qquad k_2 &=& -1 = n_2 - n_3 -n_4 + n_5 ~.
\label{e:cslevelsconixcph2}
\eea  
We choose $n_3= 1, n_i=0$ for $i\neq3$. This corresponds to $k_1 = -k_2 = 1$

It is possible to use \fref{f:fdphase2conxc} to construct the Kasteleyn matrix. This is given below in \eqref{e:kastph2conxc}.
\beq
K =   \left(
\begin{array}{c|cc}
\; & w_1 & w_2 \\
\hline
b_1 &  x^{-1} z^{n_5}+ z^{n_4}  & \  z^{n_6}  \\
 b_2 &  y z^{n_1}      & \  x z^{n_2}+  z^{n_3}
\end{array}
\right) ~. \label{e:kastph2conxc}
\eeq
The permanent of the Kasteleyn matrix is
\bea
\perm~K &=&   z^{n_2+n_5}  + x z^{n_2+n_4}  + x^{-1} z^{n_3+n_5} +  z^{n_3+n_4} +  y z^{n_1+n_6} \nn \\
&=& 1  +  x   +  x^{-1} z +  z +  y  \nn \\  
&&\text{(for $n_3 = 1$ and $n_i =0$ otherwise)} ~ .
\label{permph2conxc}
\eea

The toric data corresponding to the moduli space of this theory can be extracted from the permanent of the Kasteleyn matrix that was given in above in \eqref{permph2conxc}. Each term in this permanent corresponds to a column of the $G'_t$ matrix given in \eqref{e:gtprimeconxc}.

\subsubsection{The charge matrices}
Let us now use charge matrices to investigate this phase of the $\cal{C} \times \BC$. First, let us write down the perfect matchings corresponding to the 2 hexagon tiling given in \fref{f:fdphase2conxc}. We write each perfect matching as a collection of fields as follows:
\bea \label{pmph2conxc}
p_1 = \{ X^{1}_{12}, X^{2}_{12} \},\;\; p_2 = \{X^{2}_{21}, X^{2}_{12} \} , \;\; p_3 = \{X^{1}_{12}, X^{1}_{21} \}, \;\; p_4 = \{ X^{1}_{21}, X^2_{21} \} \nn \\  p_5=  \{ \phi_{1}, \phi_{2} \} \ . \qquad
\eea
This correspondence can be summarised in the perfect matching matrix:
\beq
P=\left(\begin{array} {c|ccccc}
 \;& p_1&p_2&p_3&p_4&p_5\\
  \hline 
  X^{1}_{12}&1&0 &1&0&0\\
   X^{2}_{12}&1&0 &0&1&0\\
  X^{1}_{21}&0&1&1&0&0\\
  X^{2}_{21}&0&1 &0&1&0\\
  \phi_{1}&0&0&0&0&1\\
  \phi_{2}&0&0&0&0&1
\end{array}
\right).
\eeq
From this matrix, we can calculate the null space of $P$ which we call $Q_F$:
\bea
Q_F = (1,1,-1,-1,0)~. \label{q1}
\eea
Since the number of gauge groups is $G=2$, it follows that $Q_D$ is trivial. One could interpret this as the lack of baryonic charges that come from the D-terms. $Q_T = Q_F$ and so the mesonic moduli space is equal to the Master space and is given by the quotient:
\bea
\CMm = \firr{}= \BC^5//(1,1,-1,-1,0)~. \label{quot2}
\eea
The toric data corresponding to this space can be found by computing the Kernel of $Q_T$. This data is encoded in $G'_t$ and can be found in \eqref{e:gtprimeconxc}.

\subsection{Phase III: The 2 Double-Bonded One-Hexagon Model}
\label{s:Ph3Conxc}
\label{1hex2db}
\begin{figure}[ht]
\begin{center}
   \includegraphics[totalheight=3.5cm]{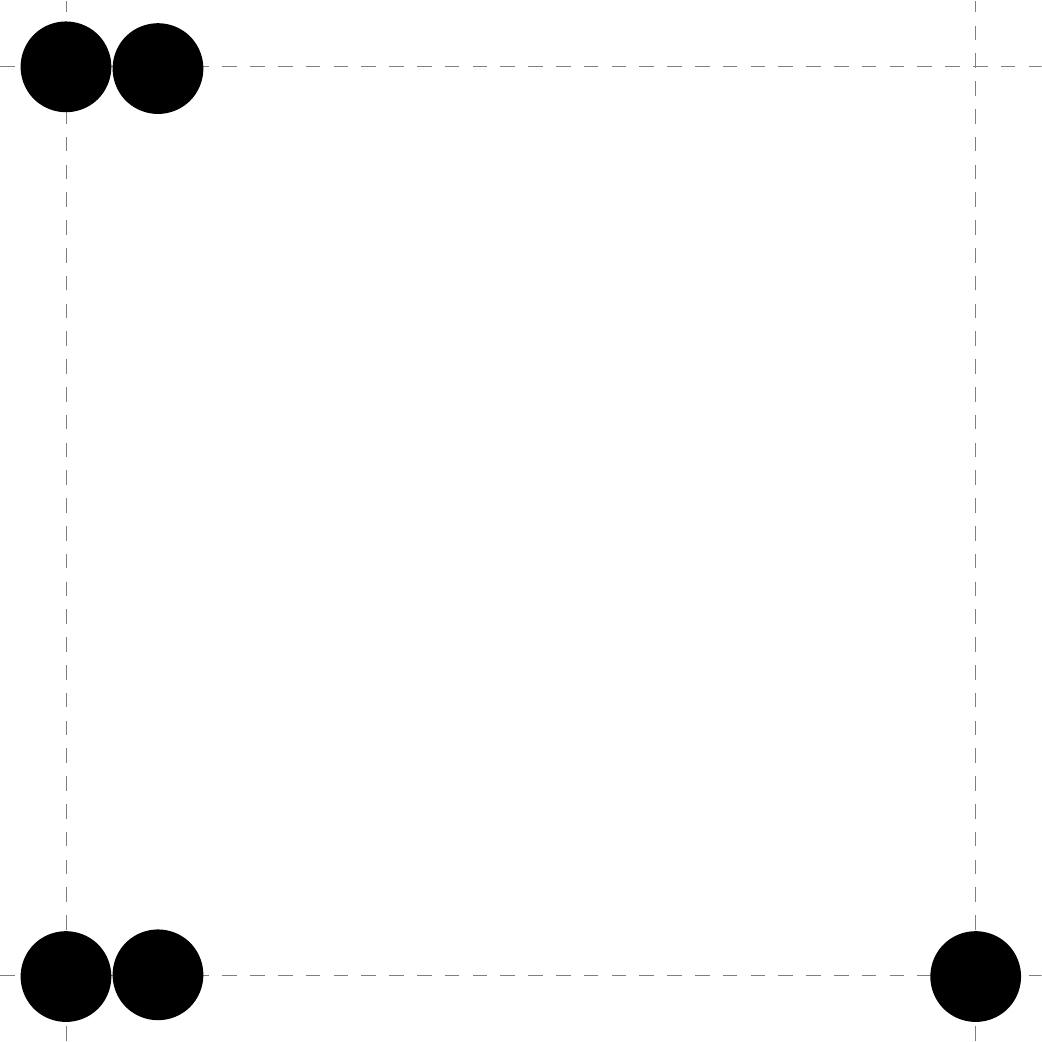}
\hspace{1cm}
  \includegraphics[totalheight=4.0cm]{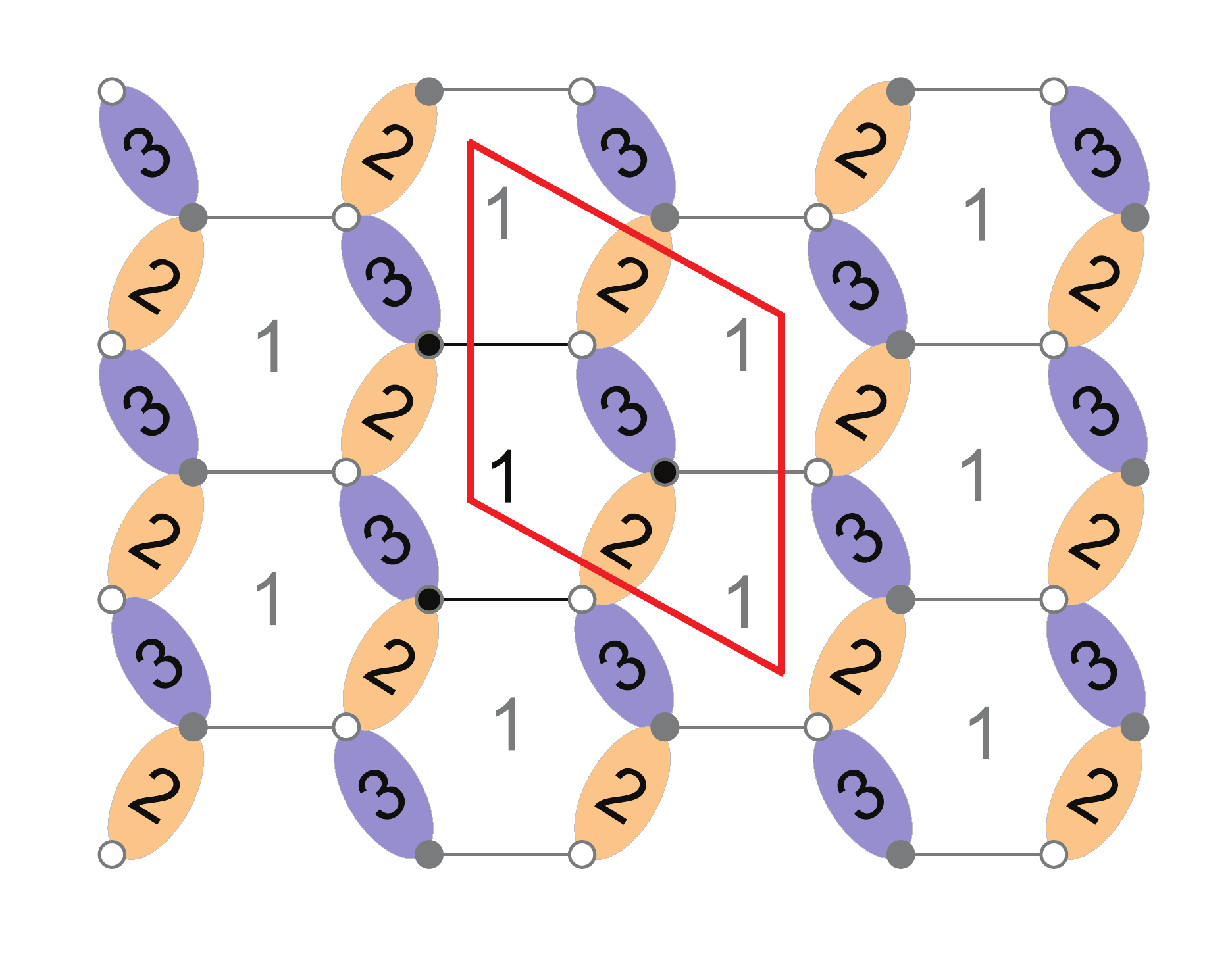}
 \caption{The projected toric data for phase III of the $\cal{C} \times \BC$ displayed as a collection of lattice points (left) and the the brane tiling that it corresponds to (right)}
  \label{f:phase3conxcproj}
\end{center}
\end{figure}

We will now attempt to build a third phase of $\cal{C} \times \BC$ starting from the $G''_t$ matrix given in \eqref{e:conxcgt2}. We can delete the third row of this matrix to find 2-dimensional toric data. The $G''_t$ matrix is given again in \eqref{e:gtprime2conxc} along with the matrix formed when its third row is deleted.

\beq
G''_t = \left(
\begin{array}{ccccc}
 1&0 & 1&  0 & 0   \\
 0 & 0 & 0 &  0 &1   \\
 1 & 0 & 0  & 1 & 0 \end{array}
\right) \rightarrow 
\left(
\begin{array}{ccccc}
 1&0 & 1&  0 & 0   \\
 0 & 0 & 0 &  0 &1  
\end{array}
\right)
\label{e:gtprime2conxc}
\eeq
Just as for the second phase, we can delete the third row of $G''_t$ and relate the 2-dimensional toric data to a brane tiling that corresponds to a 3+1 dimensional theory. The projected toric data, along with the tiling it corresponds to (the 1 hexagon model) are given in \fref{f:phase3conxcproj}. Two double bonds are added so that the moduli space of the theory can be fitted to be the $\cal{C} \times \BC$

The theory corresponding to the tiling shown in \fref{f:phase3conxcproj} was introduced in \cite{Hanany:2008gx} as part of a classification procedure for all models that have 2 terms in the superpotential.  The theory has 3 gauge groups and five chiral multiplets which we will denote as $X_{12}, X_{21}, X_{13}, X_{31}, \phi_1$, with a superpotential:
\bea
W= \tr \left( \phi_1 X_{12} X_{21} X_{13} X_{31} - \phi_1 X_{13} X_{31} X_{12} X_{21} \right)~. 
\eea
We will now demonstrate two methods of constructing the toric diagram
\subsubsection{The Kasteleyn matrix}
\begin{figure}[ht]
\begin{center}
   \includegraphics[totalheight=6.0cm]{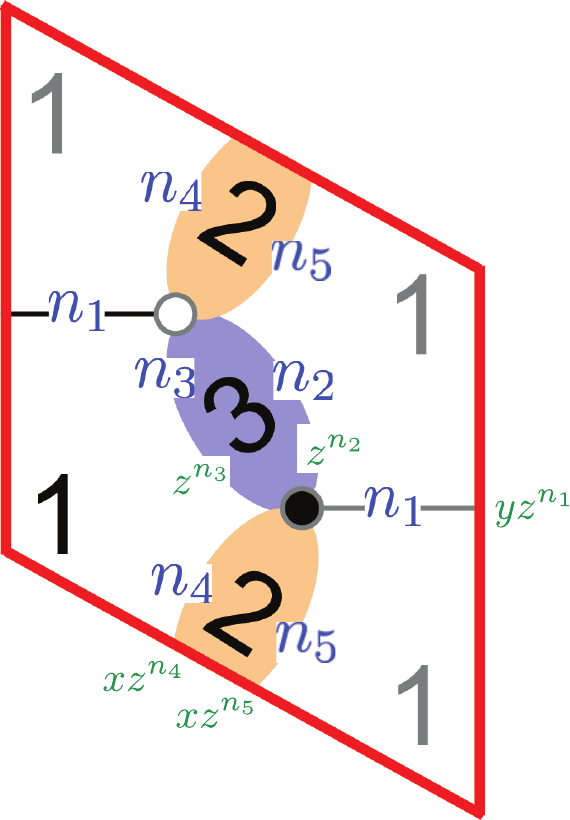}
 \caption{The fundamental domain of the tiling corresponding to phase III of $\cal{C} \times \BC$. Assignments of the integers $n_i$ to the edges are shown in blue and the weights for these edges are shown in green.}
  \label{f:fdphase3conxc}
\end{center}
\end{figure}

 We assign the integers $n_i$ to the edges according to Figure \ref{f:fdphase3conxc}. 

Chern Simons levels for fields relate to levels for gauge groups via the following dictionary: 
\bea
\text{Gauge group 1~:} \qquad k_1 &=& 0  = n_2 - n_3 + n_4 -n_5 ~, \nn \\
\text{Gauge group 2~:} \qquad k_2 &=& 1 = -n_4 +n_5 ~,  \nn \\
\text{Gauge group 3~:} \qquad k_3 &=& -1 = -n_2 + n_3~.
\eea  
We choose $n_2 = n_5 = 1$ and $n_i=0$ otherwise. This corresponds to the choice $k_1 = 0,~k_2 = 1,~ k_3 =-1$.

We can construct the Kasteleyn matrix, which in this case, is just a $1\times 1$ matrix and so coincides with its permanent:
\bea \label{Kph3conxc}
K &=&  y z^{n_1} + z^{n_2} +  z^{n_3} + x z^{n_4} + x z^{n_5}  \nn \\
&=&   y +  z  + 1 +  x + x z \quad \text{(for $n_2 = n_5 = 1$ and $n_i =0$ otherwise)} ~. \nn\\
\eea
The powers of $x, y$ and $z$ in each term of $K$ give the coordinates of each point in the toric diagram.
We collect these points in the columns of the following matrix, which we find is equal to $G''_t$:
\bea
 \left(
\begin{array}{ccccc}
 1&0 & 1&  0 & 0   \\
 0 & 0 & 0 &  0 &1   \\
 1 & 0 & 0  & 1 & 0 
\end{array}
\right) = G''_t
\eea

 \subsection{The charge matrices}  
It is also possible to construct the toric data of the moduli space of this theory by using charge matrices.
The perfect matching matrix of this phase of $\cal{C} \times \BC$ is the $5 \times 5$ identity matrix
\beq
 P = \left(
\begin{array}{ccccc}
 1&0 & 0&  0 & 0   \\
 0 & 1 & 0 &  0 &0   \\
 0 & 0 & 1  & 0 & 0  \\
 0 & 0 & 0  & 1 & 0  \\
 0 & 0 & 0  & 0 & 1  \\
\end{array}
\right)
\eeq
As the perfect matchings are in one-to-one correspondence with the quiver fields, it follows that
\bea
Q_F=0~.
\eea  
Therefore we have $\firr{} = \BC^5$. We also have
\beq
d =  \left(
\begin{array}{ccccc}
 0 & 1  & -1 & 1 & -1 \\
 0 & 0   & 0 & -1 & 1 \\
 0 & -1  & 1 & 0 & 0
\end{array}
\right) = \tilde{Q}
\eeq
and
\beq
C =  \left(
\begin{array}{ccc}
 1 & 1  & 1 \\
 0 & 1   & -1
\end{array}
\right) \implies 
\mathrm{Ker}(C) = \frac{1}{3}( -2 , 1 ,1 )
\eeq
And so we have
\bea
Q_D = \mathrm{Ker}(C) \cdot \tilde{Q} = (0,-1,1,-1,1) \label{QDph3conxc}
\eea  
The total charge matrix is then given by
\bea
Q_T = Q_D = (0,-1,1,-1,1)
\eea
Hence, the toric data is given by columns of
\bea
\left(
\begin{array}{ccccc}
 0 & 1 & 0 & 0 & 1 \\
 1 & 0 & 0 & 0 & 0 \\
 0 & 1 & 1 & 0 & 0 
\end{array}
\right)
\eea
This is the toric data for $\cal{C} \times \BC$. Although the matrix above is not exactly equal to $G''_t$, we can permute columns of the matrices to make the two match. This is not a problem, just a sign that the order of terms we wrote down in the Kasteleyn matrix in \eqref{Kph3conxc} does not match the labeling of fields in \fref{f:fdphase3conxc}.

\subsection{A Comparison between Phases of the $\cal{C} \times \BC$ Theory}
Let us make a comparison between phases of the $\cal{C} \times \BC$ theory:
\begin{itemize}
\item There are exactly 5 perfect matchings in each of the different phases of the model. The dimensionality of $G_t, G'_t$ and $G''_t$ are the same.
\item The quiver fields of Phases I and III are the perfect matchings, whereas the there are two quiver fields in some of the perfect matchings in Phase II. 
\item The Master spaces of Phases I and III and the space of perfect matchings in Phase II are identical; they are $\BC^5$.  For Phase II, the Master space is the mesonic moduli space.      
\item The mesonic moduli space of each of the three phases is $\cal{C} \times \BC$.
\end{itemize}
The master space and quiver fields are not the same in the different toric phases. This makes toric duality quite an interesting and rich phenomenon to study.

It is possible to analyse toric duality for other Chern--Simons theories including those corresponding to three phases of $D_3$ and also two phases of $Q^{1,1,1} / \BZ_2$. In `Phases of M2-brane Theories' \cite{Davey:2009sr} we consider these theories, together with the models already discussed in this chapter. The moduli space of all of the theories is discussed in a greater level of detail than here. Hilbert series of the mesonic moduli space as well as the Master Space are calculated for all models and there is a discussion of the generators of the mesonic moduli space.
\Section{Higgsing M2-brane Theories}
\label{Sec:Higgs}

In this section we will illustrate how it is possible to use brane tilings to show how different M2-brane theories are related via the Higgs mechanism. In particular, we will focus on how the $\cal{C} \times \BC$ model can be `Higgsed' to phases of the $\BC^4$ theory. This section will follow some sections of `Higgsing M2-brane Theories' \cite{Davey:2009qx}.

\begin{figure}[ht]
\begin{center}
   \includegraphics[totalheight=3.5cm]{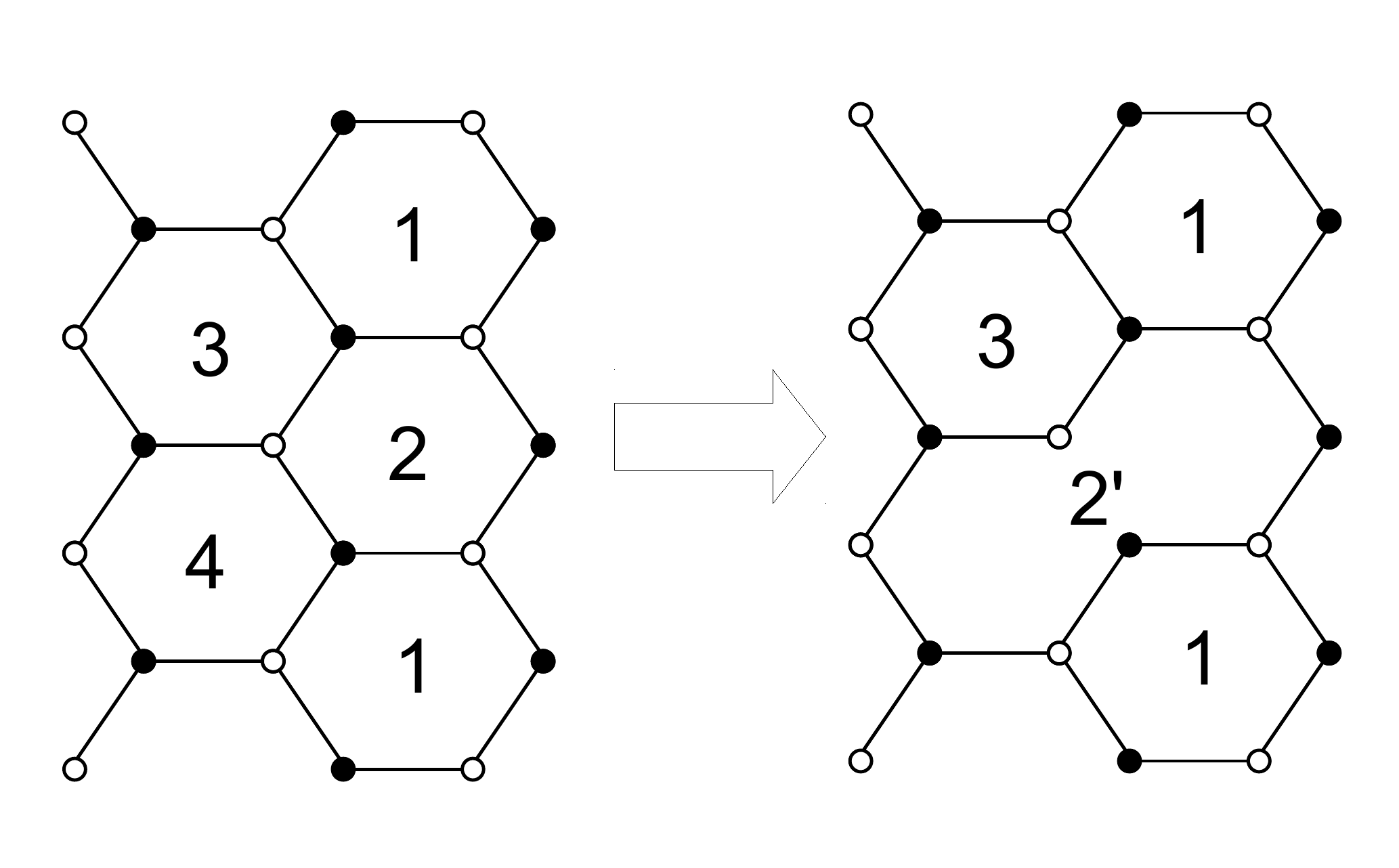}
\hspace{0.5cm}
 \includegraphics[totalheight=3.5cm]{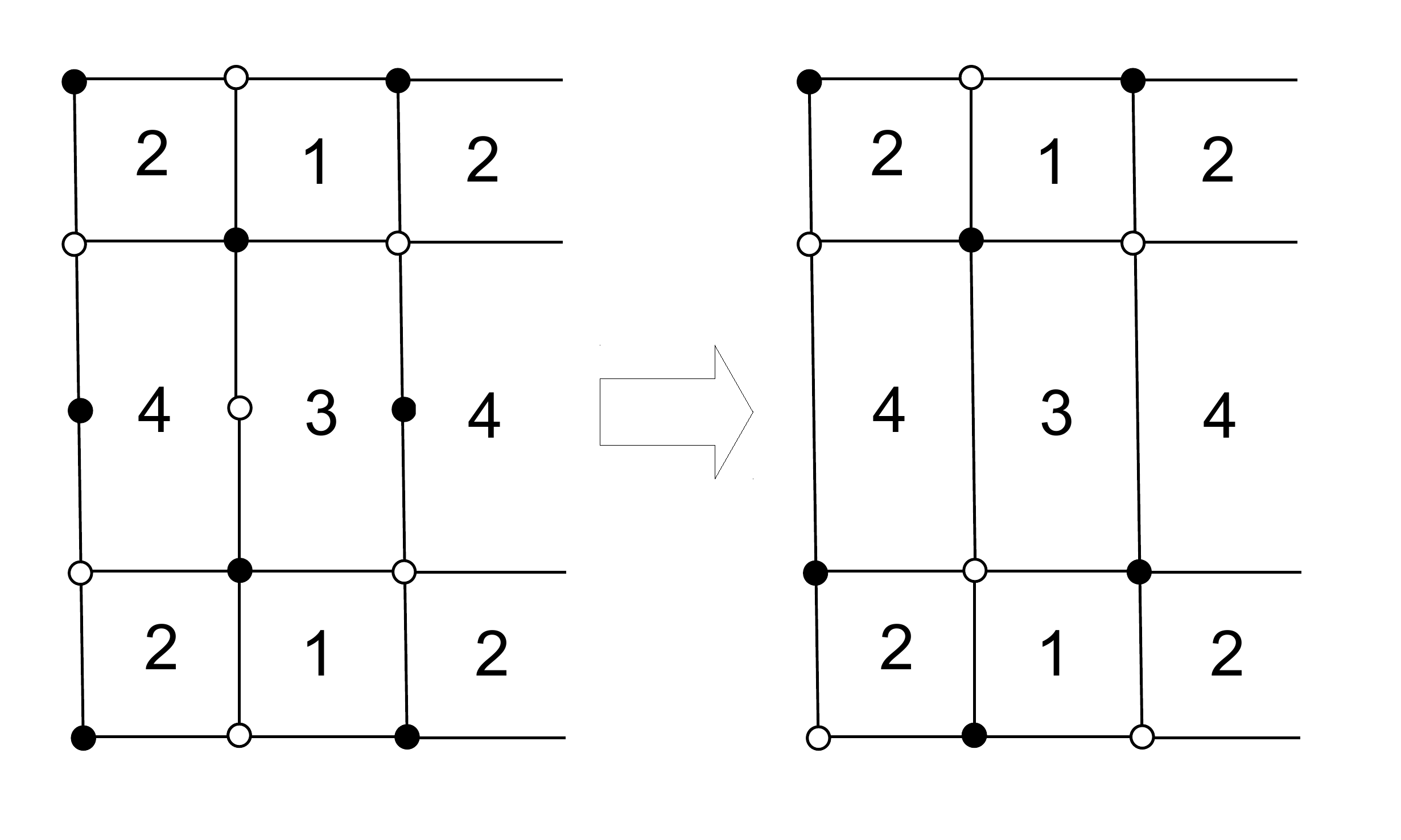}
 \caption{The effect of the two types of Higgsing on the brane tiling. The removal of an edge resulting in the reduction of the number of faces by one (left) and the removal of a node of valence two (left).}
  \label{f:2HiggsingEffects}
\end{center}
\end{figure}

Let us consider the effect of giving a vacuum expectation value (VEV) to a gauge field of a known M2-brane model. By flowing to an energy scale much lower than the scale set by the VEV, we can obtain a new field theory by `integrating out' the massive field. For a theory described by a brane tiling this effect corresponds to the removal of an edge in the tiling. This could be done by either removing an edge that separates two faces in the tiling, which would decrease the total number of faces in the tiling by one, or collapsing two vertices adjacent to a bivalent vertex into a single vertex of higher valence \cite{Kennaway2005,Hanany05}. The effect of these two types of Higgsing on the tiling are illustrated in \fref{f:2HiggsingEffects}.

\begin{figure}[ht]
\begin{center}
   \includegraphics[totalheight=2.5cm]{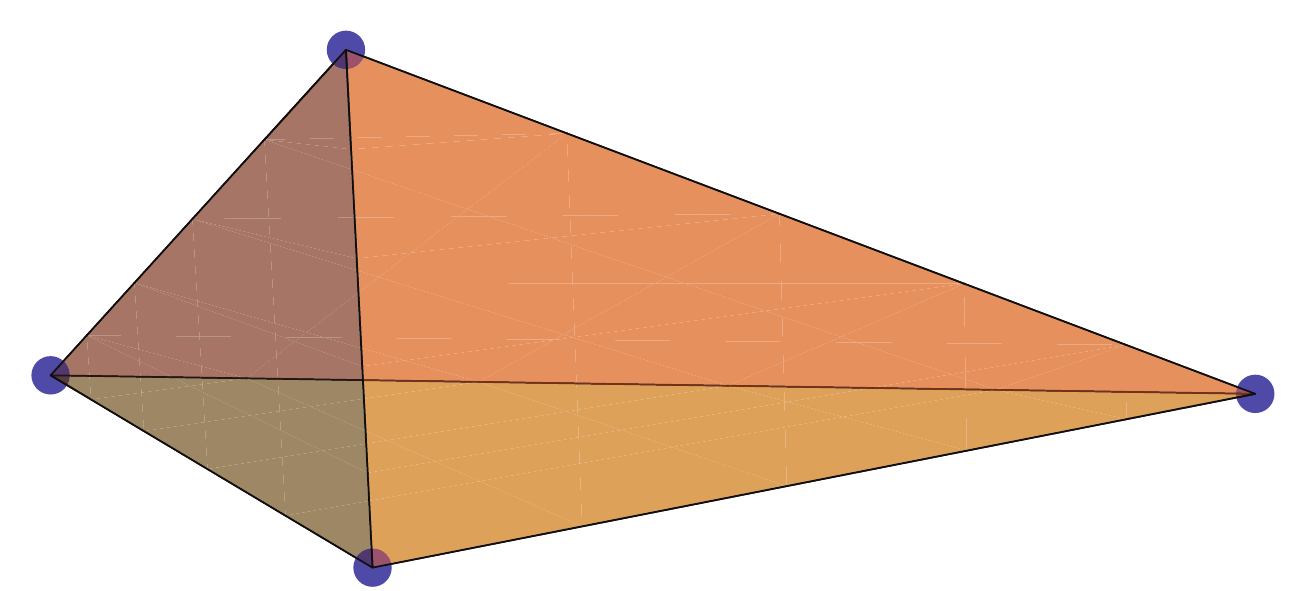}
\hspace{0.1cm}
 \includegraphics[totalheight=2.5cm]{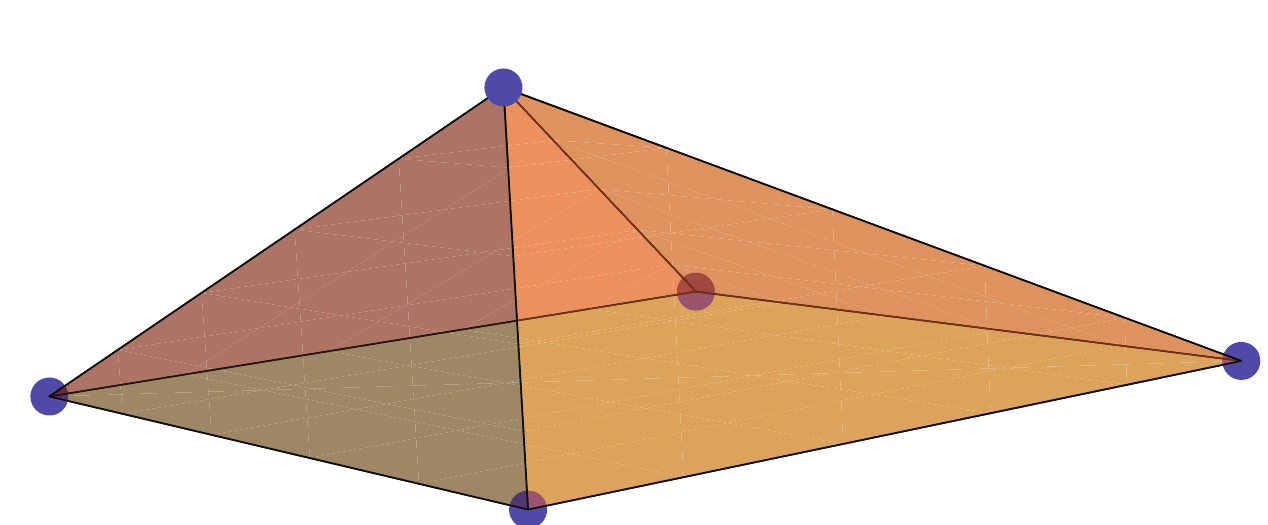}
 \caption{The removal of a point in the toric diagram of $\cal{C} \times \BC$ to obtain the toric diagram of $\BC^4$. Such an effect is known as partial resolution.}
  \label{f:ParResC4Conxc}
\end{center}
\end{figure}

One can think about how the toric data corresponding to the Higgsed theory is related to the toric data of the original theory. It is possible that one or more points of the original toric diagram could be removed when the theory is Higgsed. Such an effect is known as a partial resolution. An example of this is that $\BC^4$ can be though of as a partial resolution of the $\cal{C} \times \BC$ (see \fref{f:ParResC4Conxc}).  The methods of partial resolutions have been studied in detail for $(3 + 1)$-dimensional theories \cite{Kennaway2005, Hanany05, FengToric, Feng:2001xr, ToricD, Beasley:1999uz, Park:1999ep}, and recently have been discussed in the context of M2-brane theories \cite{Franco:2008um, Franco:2009sp}. In this section we will take the standpoint that the partial resolution is an effect of the Higgsing of a field in the tiling and that we can see this effect by applying the forward algorithm to the tiling.

We will now analyse how phases of $\cal{C} \times \BC$ can be Higgsed to phases of $\BC^4$.

\subsection{Higgsing Phase I of $\cal{C} \times \BC$}
\begin{figure}[ht]
\begin{center}
 \includegraphics[totalheight=4cm]{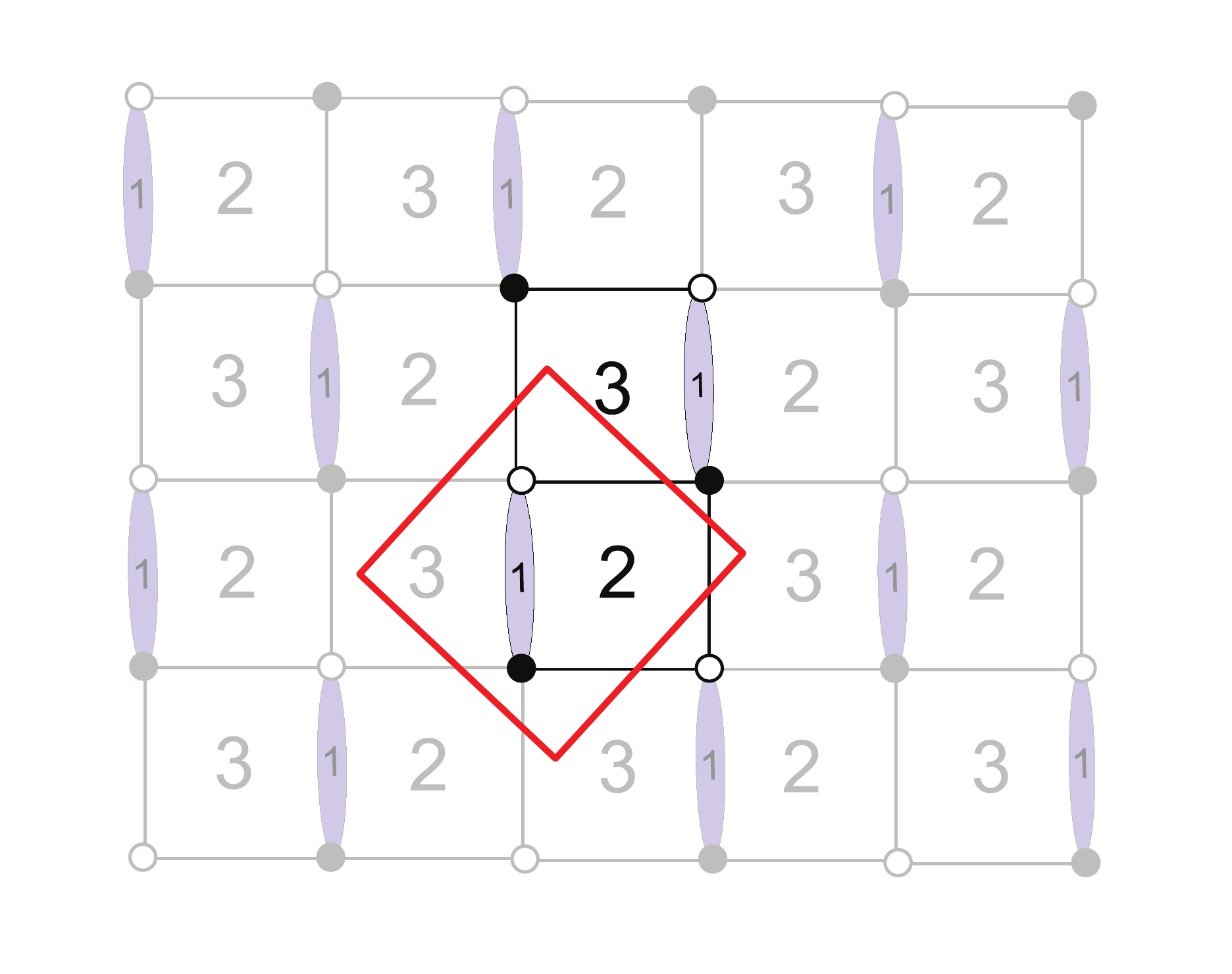}
 \caption{The 2 square tiling with 1 double bond.}
  \label{f:2Sq1DBhiggs}
\end{center}
\end{figure}

The details of this phase of the \conxc were given in Section \ref{s:Ph1Conxc}. For convenience, we shall give a brief summary here. The tiling corresponding to this phase of the \conxc is given in \fref{f:2Sq1DBhiggs}

The theory has 3 gauge groups and 5 chiral multiplets which we will call $X_{13}, X_{23}, X_{21}, X_{32}^1, X_{32}^2$. The superpotential is:
\bea
W = \tr \left(\epsilon_{ij} X_{21} X_{13} X^i_{32} X_{23} X^j_{32}\right)~. 
\eea
We choose the CS levels to be 
\bea
k_1 = 1,~k_2 = -1,~ k_3 =0~.
\eea
This corresponds to a choosing a CS level for field $X_{21} = 1$ with all the others 0.

\subsubsection{Giving a VEV to $X_{13}$ resulting in Phase I of $\BC^4$}
\begin{figure}[ht]
\begin{center}
 \includegraphics[totalheight=4cm]{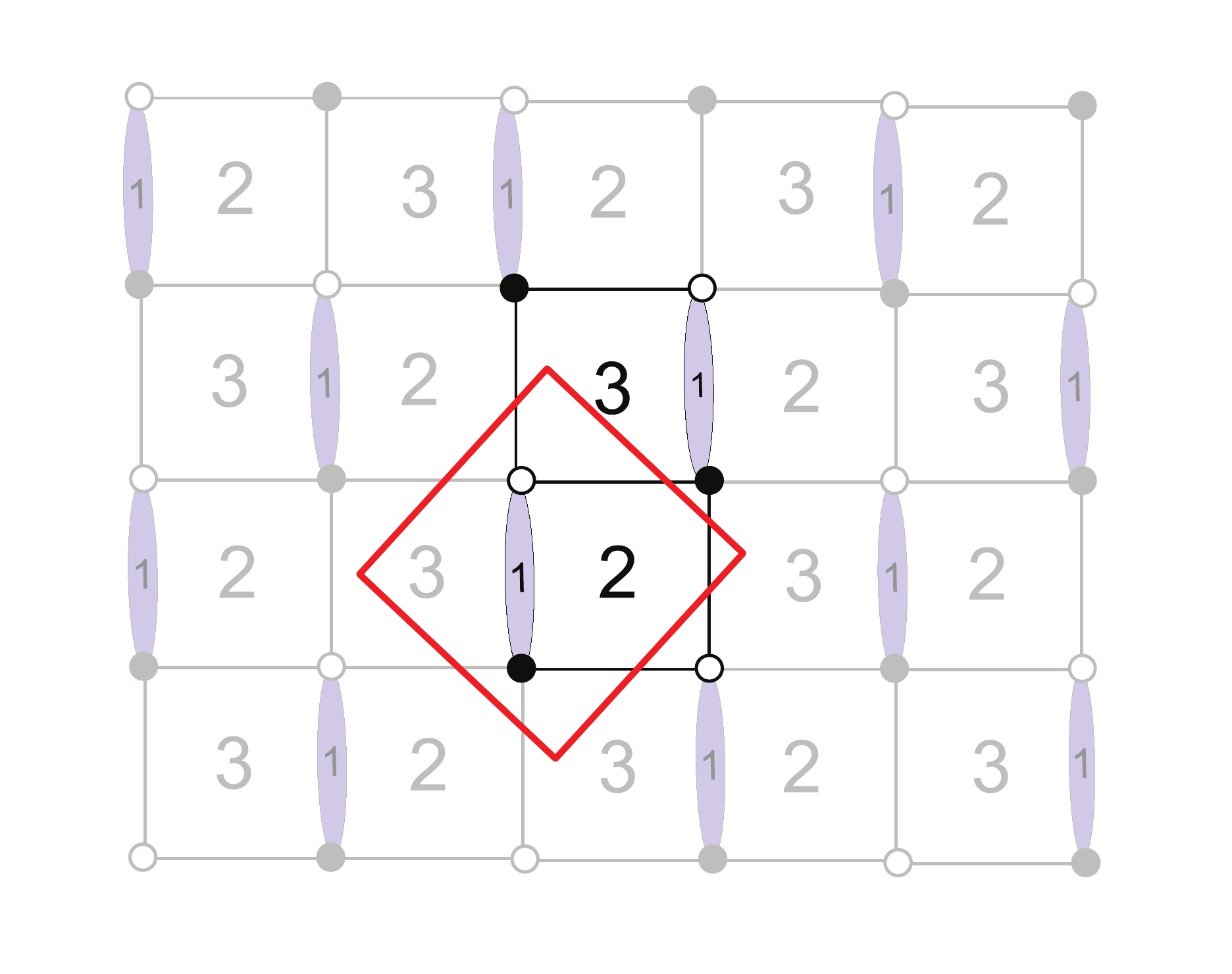}
\hspace{1cm}
 \includegraphics[totalheight=4cm]{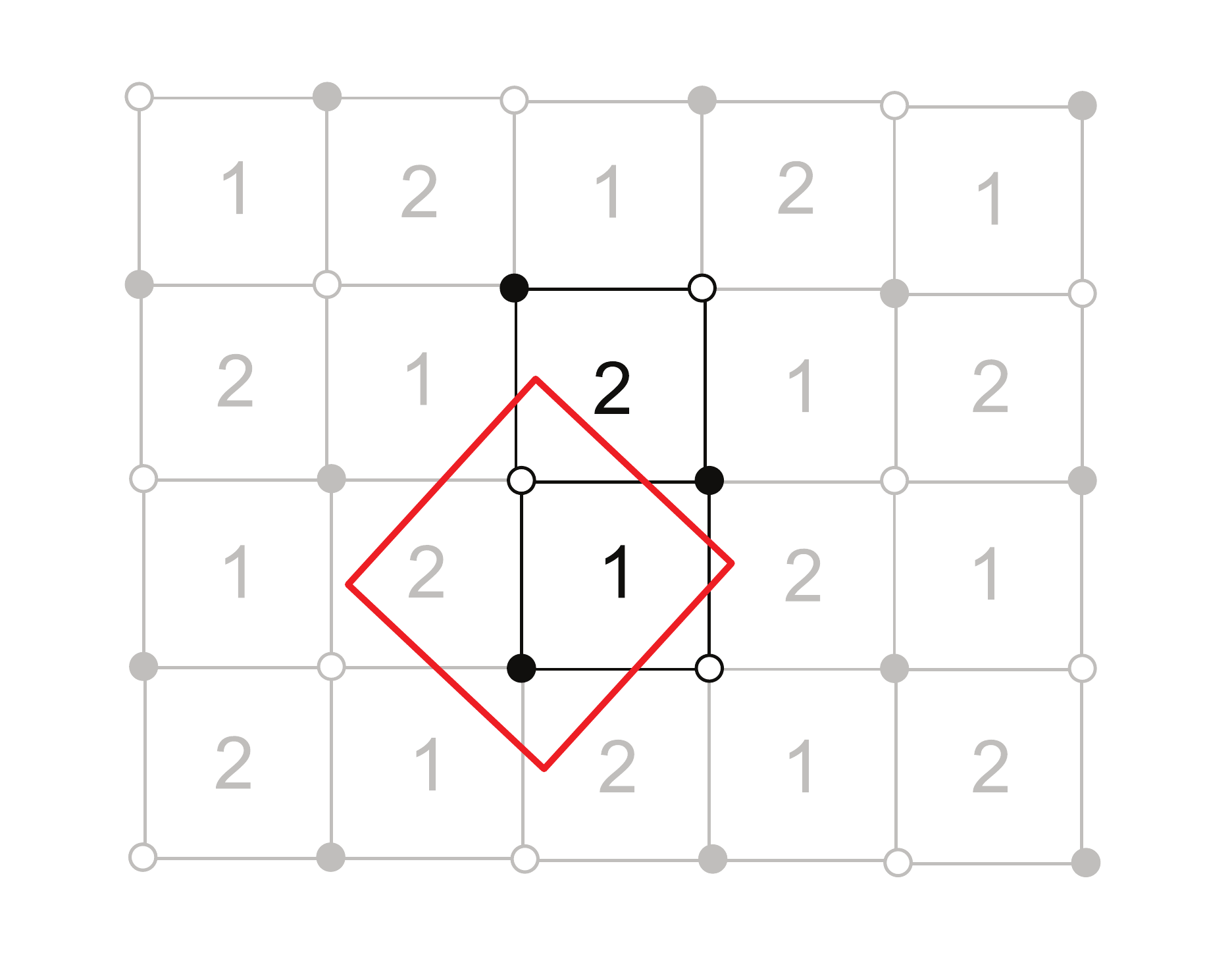}
 \caption{The 2 square tiling with 1 double bond and the tiling resulting in the removal of field $X_{13}$.}
  \label{f:HiggsingConxcPh1-1}
\end{center}
\end{figure}

Let us turn on a VEV to $X_{13}$. Flowing to an energy scale much lower than the scale set by the VEV, we obtain a new field theory resulting in the removal of this field in the tiling \fref{f:HiggsingConxcPh1-1}.

The new superpotential is 
\bea
W = \tr \left( \epsilon_{ij} X^2_{12} X^i_{21} X^1_{12} X^j_{21} \right)~. 
\eea
The CS levels associated with the Higgsed gauge groups (gauge groups 2 and 3 in the old tiling) are added, and so the new CS levels are
\bea
k_1 = 1, \quad k_2 = -1~.
\eea
The resulting theory is therefore Phase I of $\BC^4$ (the ABJM theory).

\subsubsection{Giving a VEV to $X_{23}$ resulting in Phase II of $\BC^4$}
\begin{figure}[ht]
\begin{center}
 \includegraphics[totalheight=4cm]{phases/tilphase2conxc.pdf}
\hspace{1cm}
 \includegraphics[totalheight=3.2cm]{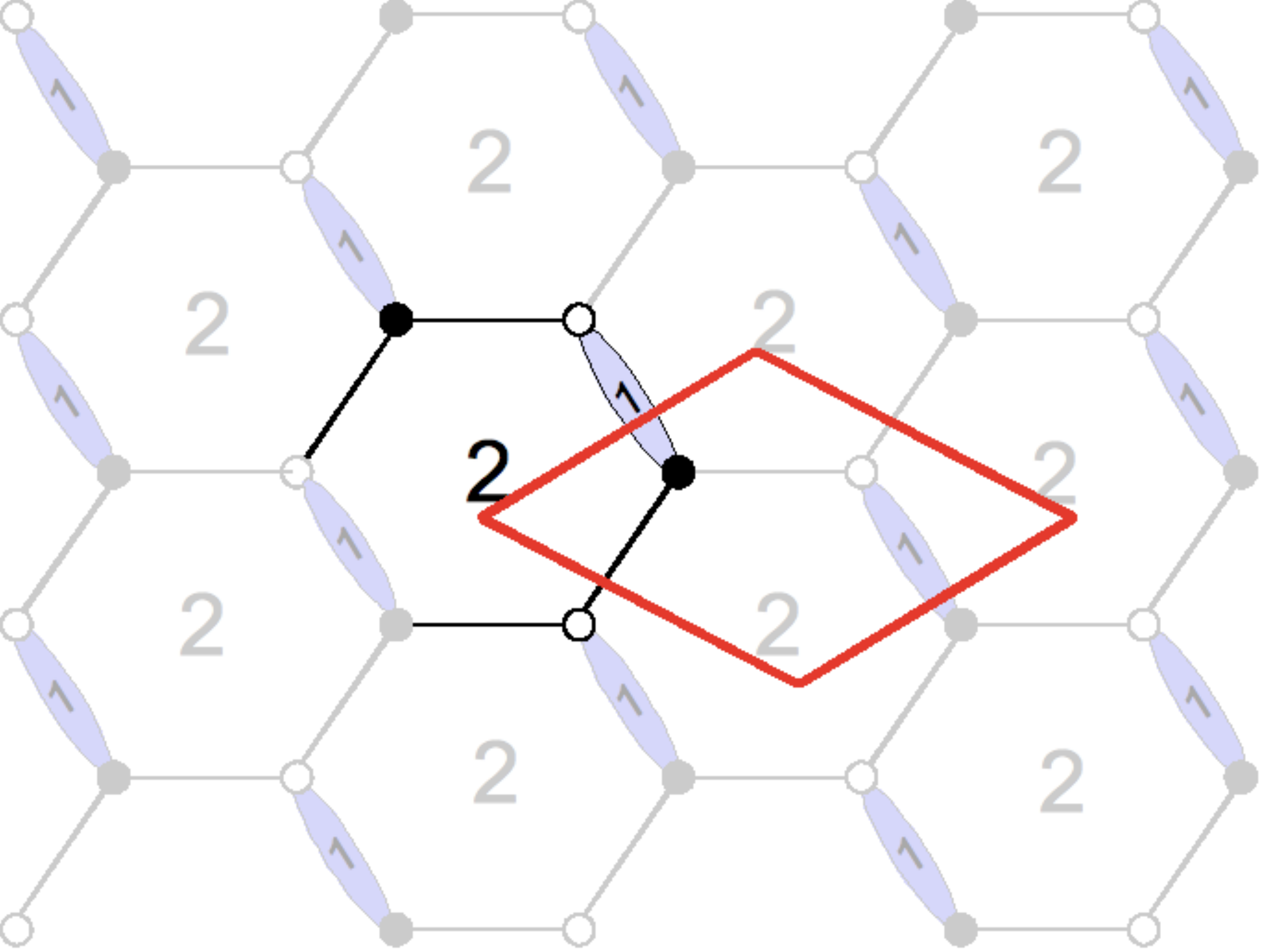}
 \caption{The 2 square tiling with 1 double bond and the tiling resulting in the removal of field $X_{23}$.}
  \label{f:HiggsingConxcPh1-2}
\end{center}
\end{figure}

Let us turn on a VEV to $X_{23}$. Faces 2 and 3 are merged into one larger face and the resulting tiling is given in \fref{f:HiggsingConxcPh1-2}.  
The new superpotential is given by
\bea
W = \tr(X_{21}X_{12}[\phi_2^1,\phi_2^2])~.
\eea

The new CS levels are $k_1 = 1$ and $k_2 = -1$. The resulting theory is therefore identified as being Phase II of the $\BC^4$ theory.

\subsection{Higgsing Phase II of $\CC\times \BC$}
\begin{figure}[ht]
\begin{center}
 \includegraphics[totalheight=4cm]{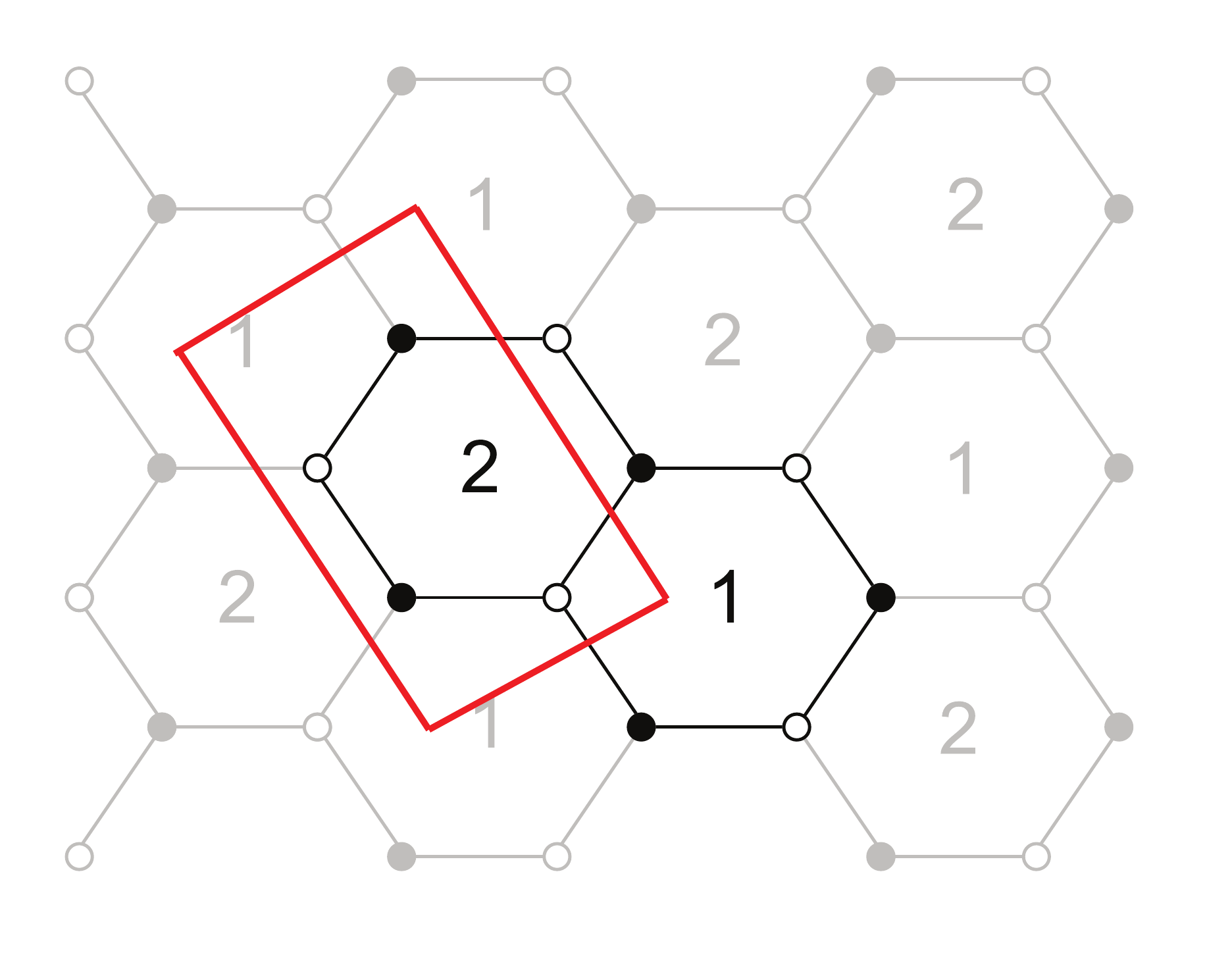}
 \caption{The 2 hexagon tiling that corresponds to phase II of \conxc.}
  \label{f:2Sq1DB}
\end{center}
\end{figure}

The details of this phase of \conxc were given in Section \ref{s:Ph2Conxc}. For convenience, we shall give a brief summary here. The tiling corresponding to this phase of the \conxc is given in \fref{f:2Sq1DB}

This theory has 2 gauge groups and 6 chiral multiplets which we shall call:
\beq
\phi_1, \phi_2, X_{12}^1, X_{12}^2, X_{21}^1, X_{21}^2
\eeq
The superpotential of the theory is equal to:
\begin{equation} 
W =  \phi_1 (X_{12}^2 X_{21}^1 - X_{12}^1 X_{21}^2 ) + {\phi}_2 (X_{21}^2 X_{12}^1 - X_{21}^1 X_{12}^2) 
\end{equation}
We will take the Chern--Simons levels to be $k_1=-k_2=1$.

\subsubsection{Giving VEV to any of $X^i_{12}$ or $X^i_{21}$}
\begin{figure}[ht]
\begin{center}
 \includegraphics[totalheight=7cm]{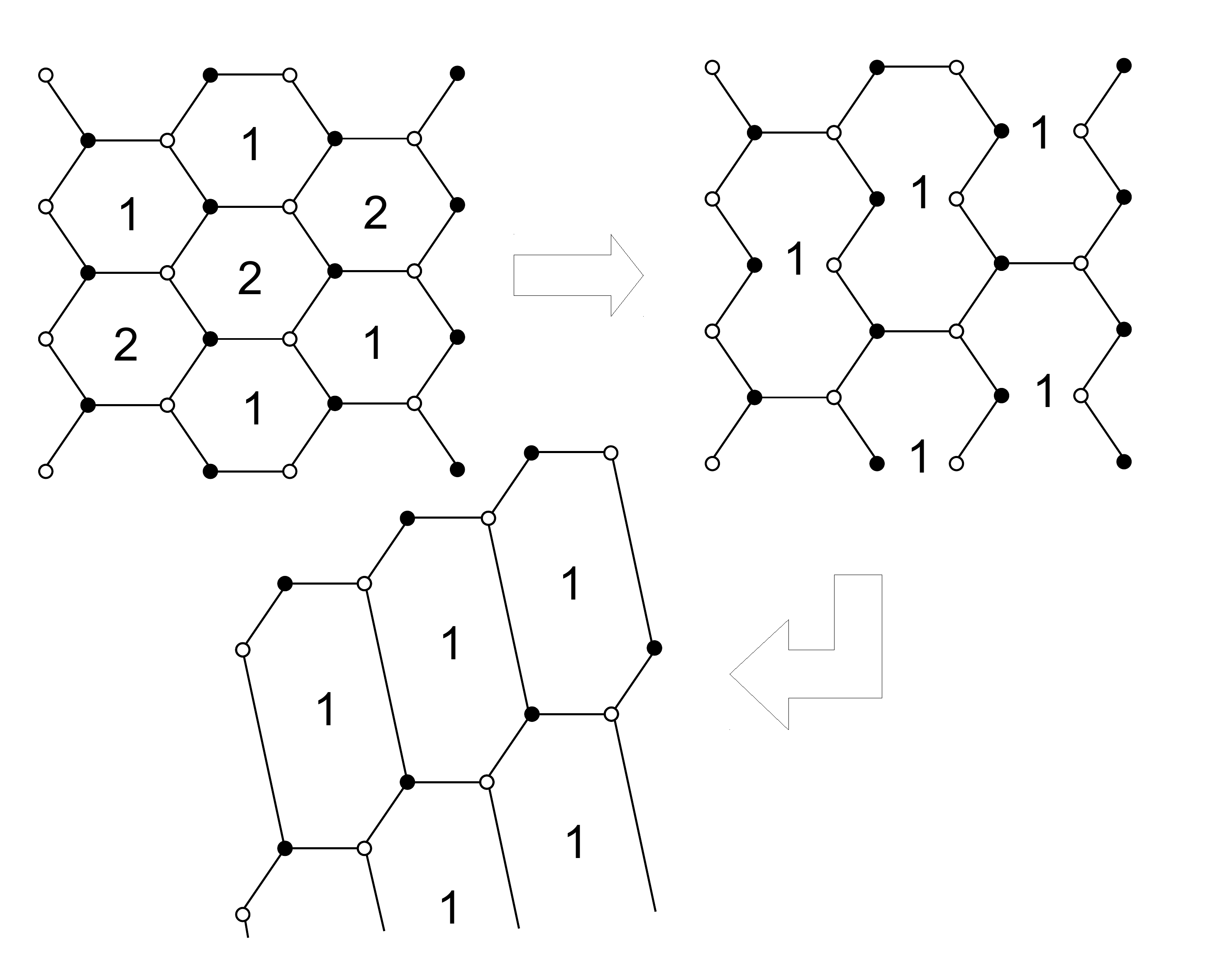}
 \caption{The Higgsing of the 2 hexagon model. Bivalent vertices are fully collapsed in the second step.}
  \label{f:ConxcHiggsPh2}
\end{center}
\end{figure}

Using symmetry arguments, we can see that giving a VEV to any of $X^1_{12}$, $X^2_{12}$, $X^1_{21}$ or $X^2_{21}$ should give the same moduli space. Without loss of generality let us give a VEV to $X^1_{12}$.  We should remove one of the edges that separate the faces that correspond to gauge groups 1 and 2, and collapse the two vertices adjacent to a bivalent vertex into a single vertex of higher valence \cite{Hanany05}. This is shown in \fref{f:ConxcHiggsPh2}

The theory that is a result of this Higgsing has only 1 gauge group and 3 adjoint fields. It can be represented by the one-hexagon tiling. As there is only one gauge group, the CS level must be $k = 0$. The usual forward algorithm for M2-brane tilings fails with this theory. If we apply it in a n\"aive way we find the moduli space is $\BC^3$. We expect this to be only a branch of the moduli space and that there is an additional complex degree of freedom due to a gauge kinetic term making the full (mesonic) moduli space $\BC^4$.

\subsection{Higgsing Phase III of $\CC \times \BC$}

\begin{figure}[ht]
\begin{center}
  \includegraphics[totalheight=4.0cm]{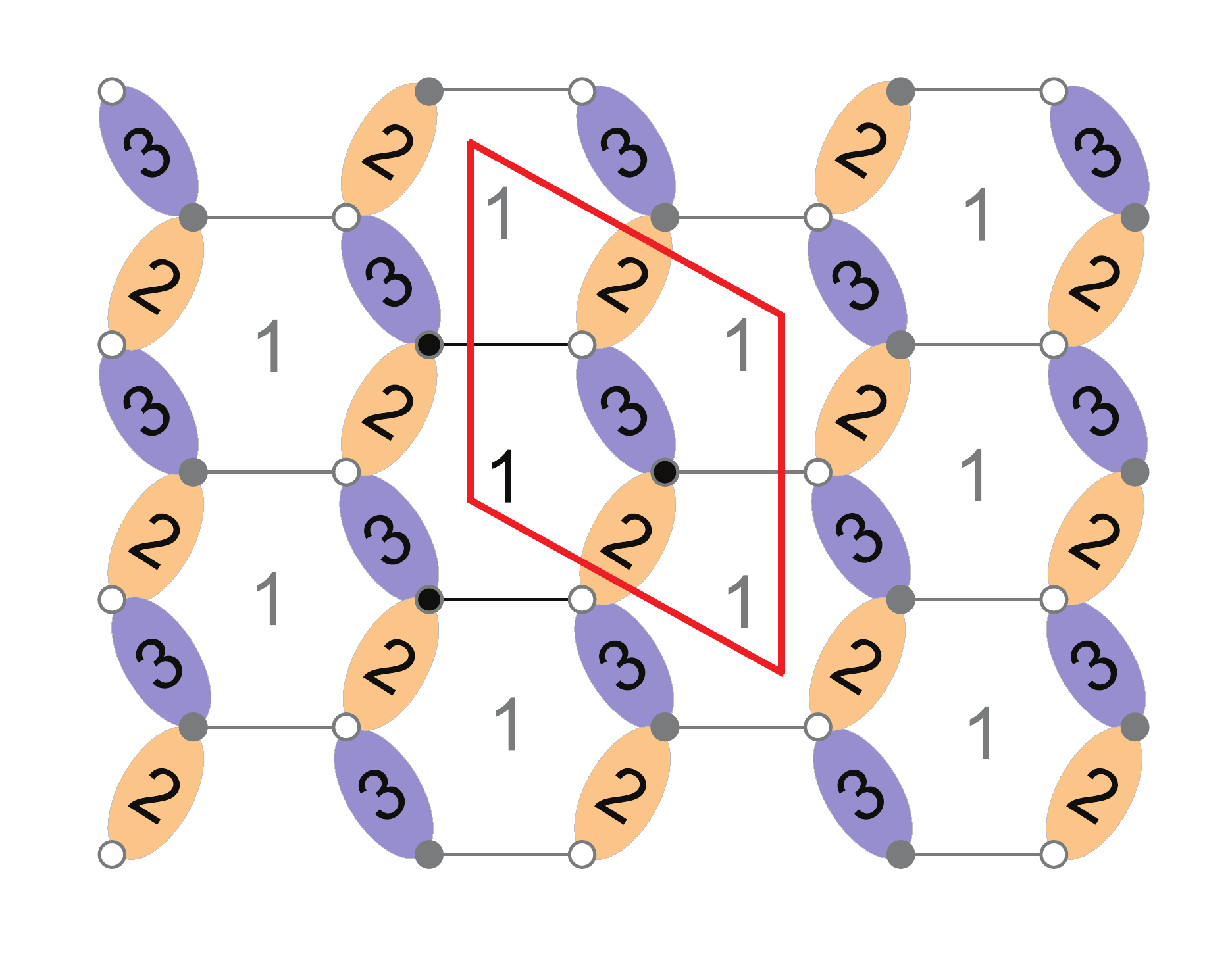}
 \caption{The tiling of phase III of $\CC \times \BC$}
  \label{f:phase3conxctilhiggs}
\end{center}
\end{figure}

The details of this phase of the \conxc were given in Section \ref{s:Ph3Conxc}. For convenience, we shall give a brief summary here. The tiling corresponding to this phase of the \conxc is given in \fref{f:phase3conxctilhiggs}

The theory has 3 gauge groups and 5 chiral multiplets which we shall call $X_{12}$, $X_{21}$, $X_{13}$, $X_{31}$ and $\phi_1$. The superpotential is given by
\beq
W=  \phi_1 \left[X_{12} X_{21}, X_{13} X_{31}\right] 
\eeq
We pick CS levels to be
\beq
k_1 = 0,~k_2 = 1,~ k_3 =-1
\eeq

\subsubsection{Giving a VEV to any of $X_{12}$, $X_{21}$, $X_{13}$, $X_{31}$}
\begin{figure}[h]
\begin{center}
  \includegraphics[totalheight=4.0cm]{phases/tilph3conxc.pdf}
\hspace{0.5cm}
  \includegraphics[totalheight=4.0cm]{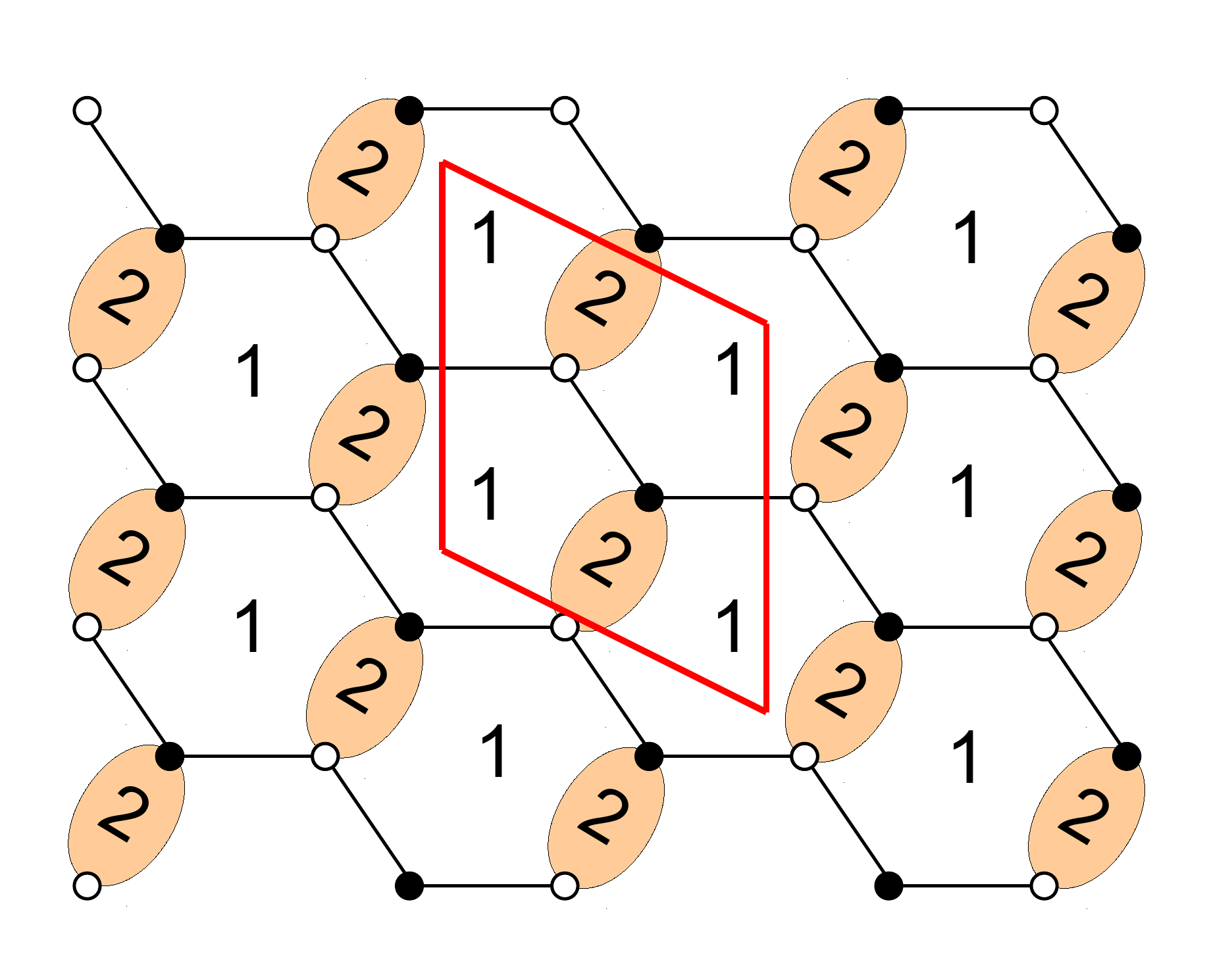}
 \caption{The effect of higgsing $X_{13}$ on the tiling of phase III of $\CC \times \BC$}
  \label{f:phase3conxchiggs2tiles}
\end{center}
\end{figure}

By symmetry we can argue that giving a VEV to any of the bi-fundamental fields leads to the same field theory, up to relabeling gauge groups and fields.
Without loss of generality, let us examine the case in which $X_{13}$ acquires a VEV. From the tiling shown in Figure \ref{f:phase3conxchiggs2tiles}, we see that removing the edge corresponding to $X_{13}$ amounts to combining gauge group 1 and 3, so that the double bond corresponding to the gauge group 3 disappears. The resulting tiling is therefore a single hexagon model with one double-bond.

Higgsing the theory corresponds to a choice of CS levels equal to $k_1 = 1$, $k_2=-1$. The resulting theory is Phase II of $\BC^4$.

\subsection{The Higgs mechanism and other M2-brane models}
\begin{figure}[h]
\begin{center}
  \includegraphics[totalheight=2.0cm]{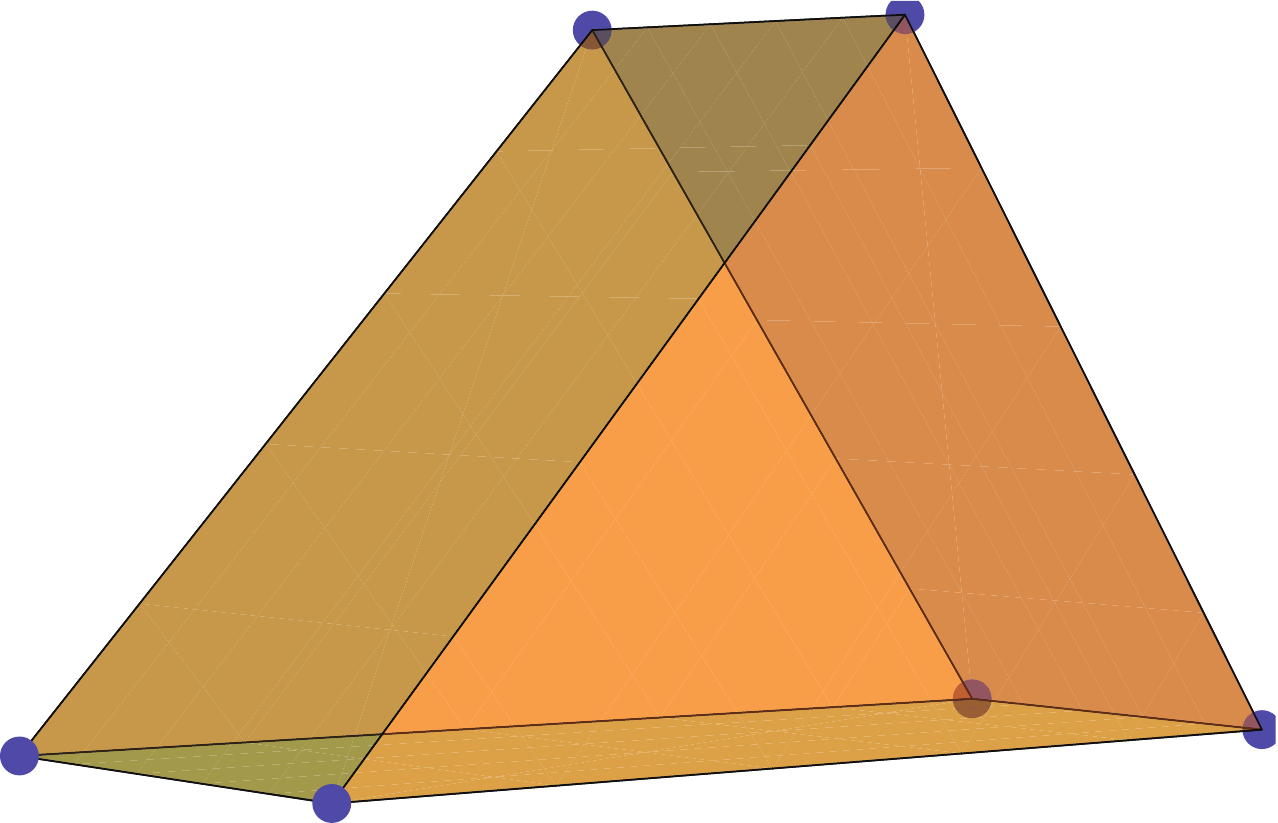}
  \includegraphics[totalheight=2.0cm]{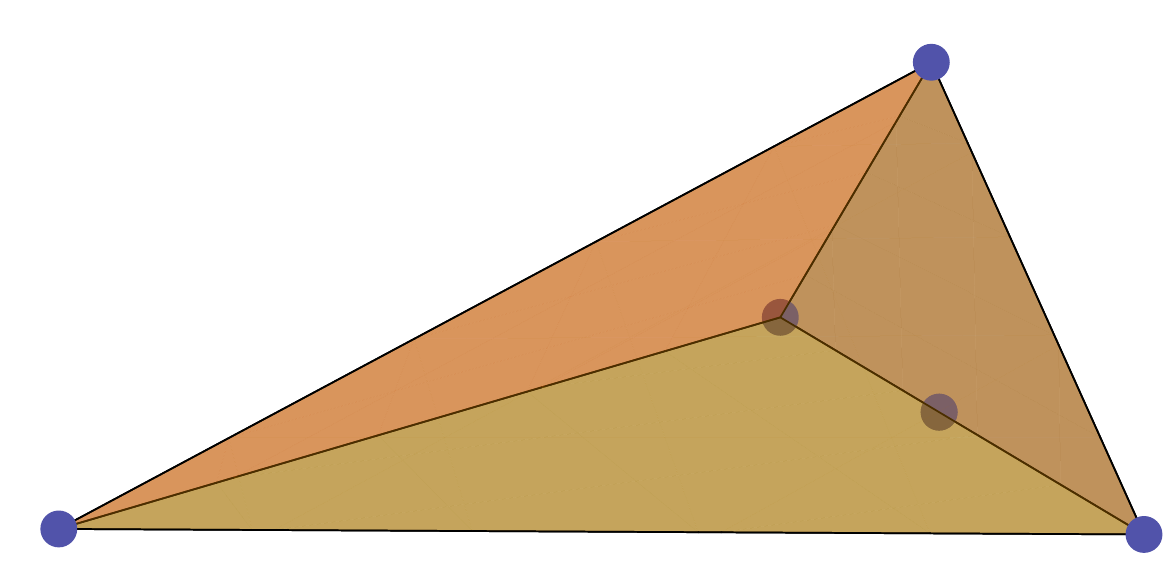}
  \includegraphics[totalheight=2.0cm]{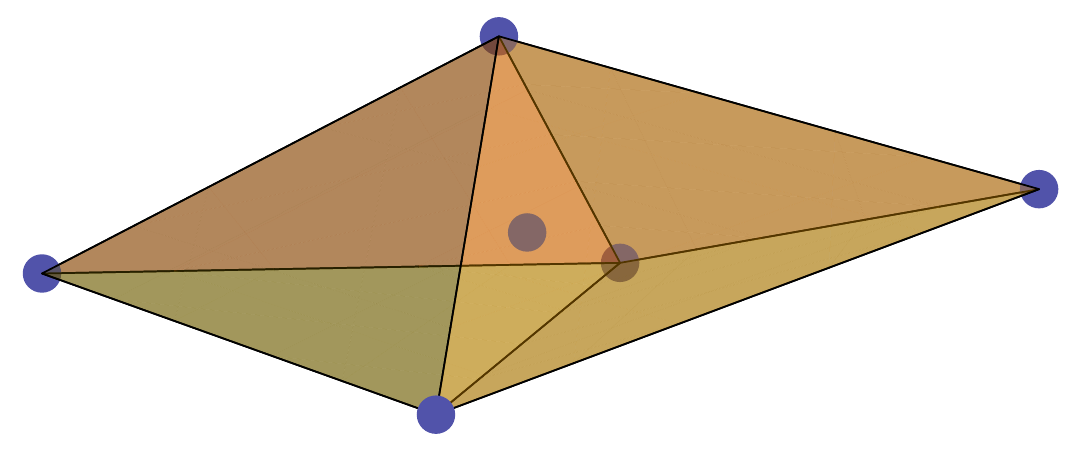}\\
$D_3$ \hspace{2.5cm}
$\left(\BC^2/\BZ_2\right) \times \BC^2$ \hspace{2.5cm}
 $M^{111}$ \\
\vspace{0.5cm}
  \includegraphics[totalheight=2.5cm]{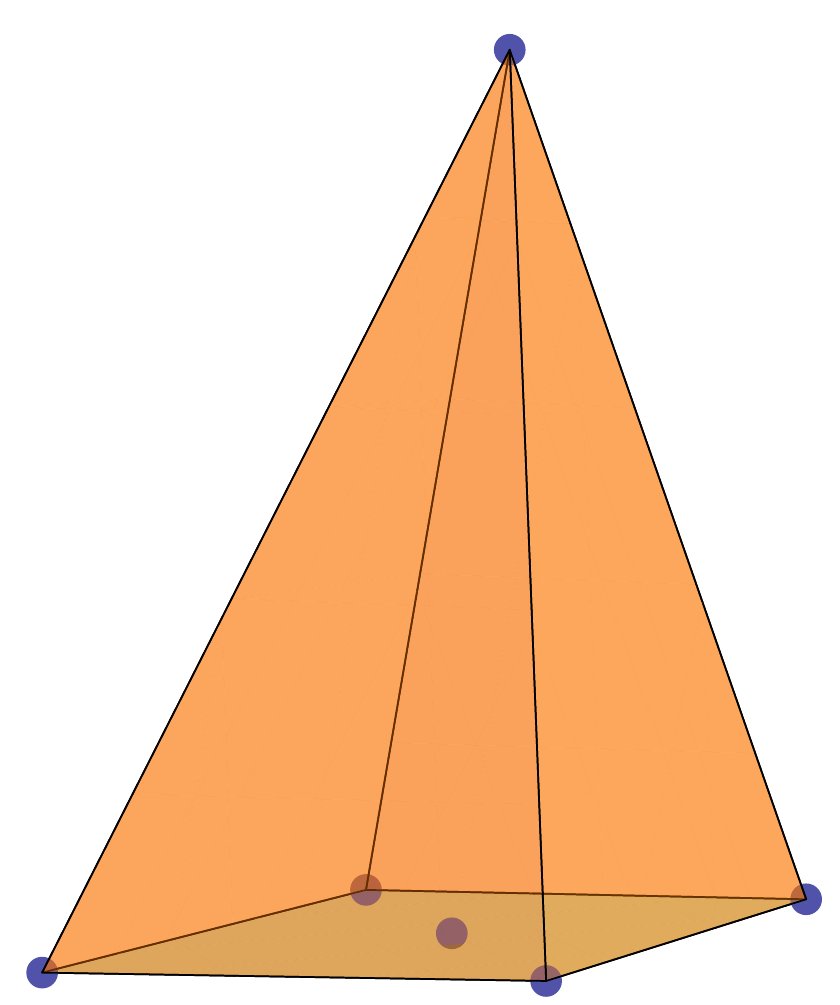}
\hspace{2cm}
  \includegraphics[totalheight=2.5cm]{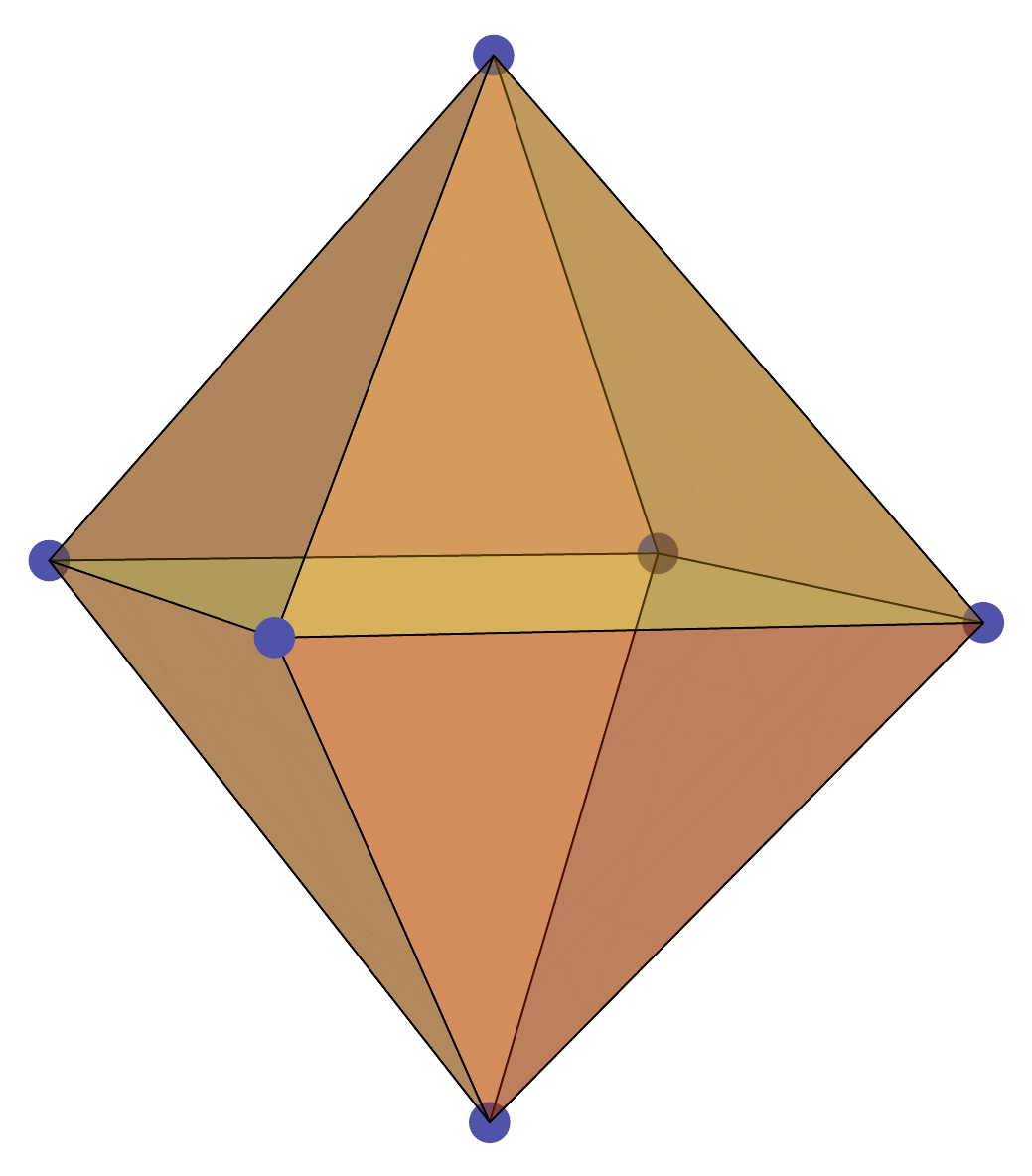}
\hspace{2cm}
  \includegraphics[totalheight=2.5cm]{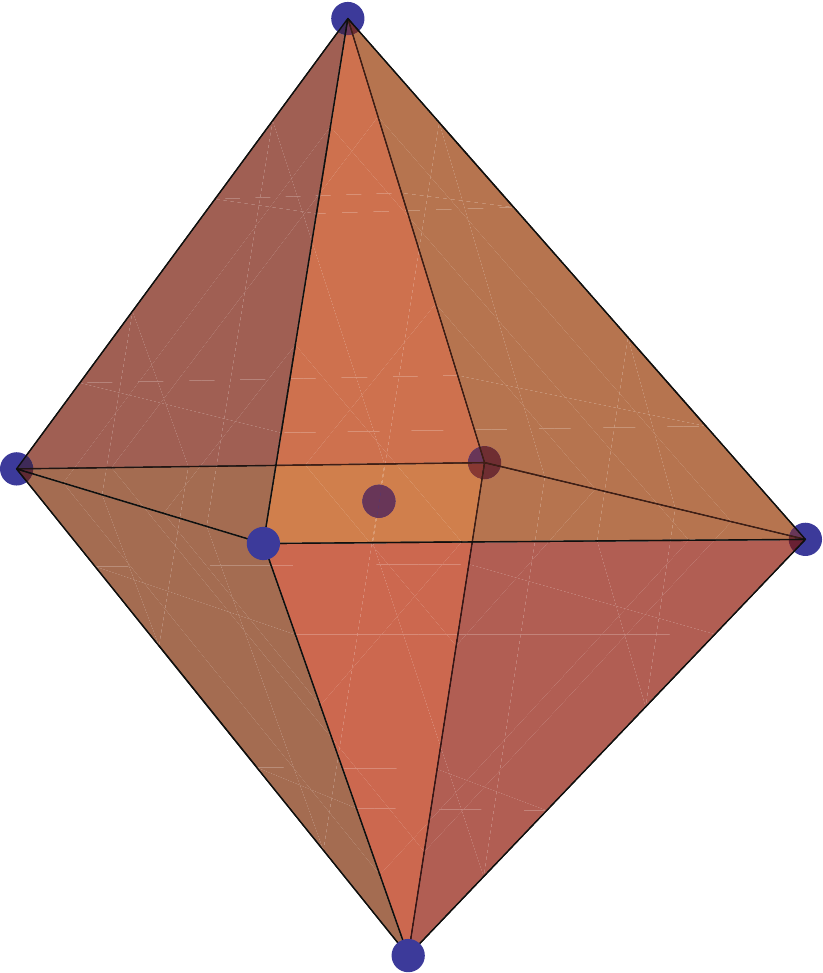}\\
$\BF_0 \times \BC$\hspace{3cm}
$Q^{1,1,1}$\hspace{3cm}
$Q^{1,1,1}/\BZ_2$
 \caption{The toric diagrams of (top left to bottom right) $D_3$, $\BC^2 / \BZ_2 \times \BC^2$, $M^{1,1,1}$, $\BF_0 \times \BC$, $Q^{1,1,1}$ and $Q^{1,1,1}/\BZ_2$.}
  \label{f:morehiggsing}
\end{center}
\end{figure}

In this chapter we have outlined how it is possible to relate phases of $\BC^4$ to phases of \conxc via the Higgs mechanism. There are many other M2-brane theories that can be related in similar ways. In `Higgsing M2-brane Theories' \cite{Davey:2009qx}, relationships between $\BC^4$ and \conxc and other M2-brane models are explored using the Higgs mechanism. The theories discussed correspond to the geometries known as $D_3$, $\BC^2 / \BZ_2 \times \BC^2$, $M^{1,1,1}$, $\BF_0 \times \BC$, $Q^{1,1,1}$ and $Q^{1,1,1}/\BZ_2$. The toric data of these models is given in \fref{f:morehiggsing} and the interested reader is directed to \cite{Davey:2009qx} for further discussion.

\chapter{Brane Tilings and Fano 3-Folds}
\label{Ch:Fanos}
This chapter focuses on supersymmetric CS theories on M2-branes probing a special class of CY 4-folds which can be formed by taking the complex cone over the smooth toric 
Fano 3-folds \cite{Davey:Fano,Davey:2009et}.

The Fano 2-folds are well known in the string theory literature. For instance the smooth toric Fano 2-folds have played an important role in the study of supersymmetric gauge theories that live on D3-branes probing a CY 3-fold given by the complex cone over a smooth toric Fano 2-fold. There are 5 fano 2-folds and they are more commonly known as the zeroth Hirzebruch surface $\BF_0$ or the del Pezzo surfaces $dP_{n=0,1,2,3}$\footnote{The other del Pezzo surfaces are Fano varieties but are not toric.}. The study of supersymmetric gauge theories corresponding to these CY 3-folds led to the discovery of the first examples of toric duality for $(3+1)$-dimensional gauge theories \cite{FengToric, Feng:2001xr, ToricD, Feng:2002fv, Feng:2001bn, Franco:2003ea, Franco:2003ja, FrancoToric, master3}. The del Pezzo surfaces have also been studied in a more phenomenological context \cite{Verlinde:2005jr, Malyshev:2007yb, Conlon:2008wa, Blumenhagen:2008zz, Blumenhagen:2009gk, Blumenhagen:2009yv, Balasubramanian:2009tv}.

One of the features of the toric Fano varieties is that for every complex dimension there always exists a finite number of smooth toric fanos \cite{KMM1, KMM2}. It is known that there are 18 smooth toric fano 3-folds \cite{database, toricfano4}, each of which can be used to construct a toric CY 4-fold. In this chapter some CS gauge theories that can be described by brane tilings are investigated that have these CY 4-folds as their moduli space.

\Section{The Fano Varieties}

A mathematician would probably define a fano variety by saying that it admits an ample anti-canonical sheaf. In this work we shall consider such varieties and we make the further restriction that the variety should be smooth and admit a toric description, even though many examples of non-smooth cases are well-known.

In one complex dimension the only Fano variety is $\BP^1$, which can also be thought of as the real 2-sphere. It is a classical result that in 2 complex dimensions there are exactly 10 Fano varieties, up to deformations: the zeroth Hirzebruch surface, $\BF_0 = \BP^1\times \BP^1$, and the 9 del Pezzo surfaces $dP_{n=0,\ldots,8}$. Of these 10 Fano varieties, only $\BF_0$ and $dP_{n=0,1,2,3}$ are toric and so can be investigated using brane tiling technology.

The first important results towards a classification of Fano 3-folds were obtained by Iskovskih \cite{isk1, isk2}, and a complete classification was given by Mori and Mukai \cite{MM3} (see also \cite{murre, cutkosky}). They found 88 varieties up to deformations of which 18 are toric \cite{database, toricfano3, toricfano4}. A complete classification of higher dimensional smooth Fano varieties is still an open problem \cite{toricfano4, Kreuzer:2007zy, Oebro07b}.

\subsection{The smooth toric Fano three-folds}

Before we enter into a discussion about the construction of the world-volume theory of an M2-brane probing a CY 4-fold formed from a smooth toric fano 3-fold, let us discuss this interesting class of geometries a little more. 

As we have mentioned there have been many previous studies of fano varieties and there are at least two naming systems that have been developed for them. In this thesis we shall be unbiased and use them both.

The first and perhaps more informative naming system exploits the toric description of the fano varieties. We give each variety a `Name'
$\cB_i$, $\cC_i$, $\cD_i$, $\cE_i$ or $\cF_i$ according to the number of external points the toric diagram corresponding to the variety has\footnote{The reason why both $\cC_i$ and $\cD_i$ are used to denote varieties having 6 external points has to do with the structure of the toric diagram \cite{toricfano4}.}. The exception is the $\BP^3$, which is just called $\BP^3$. The names are summarised in Table \ref{t:fanonames}

\begin{table}[h!]
\[
\hspace{0cm}
\begin{array}{|c||c|c|c|c|c|c|}
\hline
\mbox{Number of external points} &  \;\;4\;\;  &  \;\;5\;\;  &  \;\;6\;\;  &  \;\;6\;\;  &  \;\;7\;\;  &  \;\;8\;\;\\
\hline
\mbox{Number of varieties}       &  \;\;1\;\;  &  \;\;4\;\;  &  \;\;5\;\;  &  \;\;2\;\;  &  \;\;4\;\;  &  \;\;2\;\; \\
\hline
\mbox{Name}              &  \;\;\BP^3\;\;  &  \;\;\cB_i\;\;  &  \;\;\cC_i\;\;  &  \;\;\cD_i\;\;  &  \;\;\cE_i\;\;  &  \;\;\cF_i\;\; \\
\hline
\end{array}
\]
\caption{The smooth toric Fano three-folds are counted according to the number of external points in their toric diagram.}
\label{t:fanonames}
\end{table}

\begin{table}[h!]
\begin{center}
\[
\hspace{0cm}
\begin{array}{
|c|c|c|c|}
\hline
\mbox{Name}
& \mbox{ID \cite{database}}
  & \mbox{Toric Data} & \mbox{Symmetry}

  \\
\hline \hline
\BP^3 & 4 & 
\tmat{1 &-1 & 0 & 0 & 0 \\
      0 & 1 &-1 & 0 & 0 \\
      0 & 0 & 1 &-1 & 0}
& U(4)
	\\
	\hline 
\cB_4 & 24 &
\tmat{ 1 &-1 & 0 & 0 & 0 & 0 \\
       0 & 1 &-1 & 0 & 0 & 0 \\
       0 & 0 & 0 & 1 & -1 & 0 } 
 &  SU(3) \times SU(2) \times U(1)
	\\
	\hline
\cB_1 & 35 &
\tmat{
 1 &-1 & 0 & 0 & 0 & 0 \\
 0 & 1 &-1 & 0 & 0 & 0 \\
 0 & 0 & 2 &-1 & 1 & 0
}
&  SU(3) \times U(1)^2
	\\
	\hline
\cB_2 &  36&
\tmat{
 1 &-1 & 0 & 0 & 0 & 0 \\
 0 & 1 &-1 & 0 & 0 & 0 \\
 0 & 0 & 1 &-1 & 1 & 0
} 
 & SU(3) \times U(1)^2  
	\\
	\hline
\cC_3 & 62 &
\tmat{
 1 &-1 & 0 & 0 & 0 & 0 & 0 \\
 0 & 0 & 1 &-1 & 0 & 0 & 0 \\
 0 & 0 & 0 & 0 & 1 &-1 & 0 }
 & SU(2)^3 \times U(1)
	\\
	\hline
\cC_4 & 123  &
\tmat{
 1 &-1 & 0 & 0 & 0 & 0 & 0 \\
 0 & 0 & 1 &-1 & 0 & 0 & 0 \\
 0 & 0 & 1 & 1 &-1 & 0 & 0 }
 & SU(2)^2 \times U(1)^2 
	\\
	\hline
\cC_5 & 68 &
\tmat{ 1 &-1 & 0 &  0 & 0 & 0 & 0 \\
       0 & 0 & 1 & -1 & 0 & 0 & 0 \\
       0 & 1 & 0 & -1 & 1 &-1 & 0} & 
 SU(2)^2 \times U(1)^2
	\\
	\hline
\cB_3 & 37  &
\tmat{
 1 &-1 & 0 & 0 & 0 & 0 \\
 0 & 0 & 1 &-1 & 0 & 0 \\
 0 & 1 & 0 &-1 &-1 & 0
} 
 &SU(2)^2 \times U(1)^2  
	\\
	\hline
\cC_1 & 105& 
\tmat{
 1 &-1 & 0 &  0 & 0 & 0 & 0 \\
 0 & 0 & 1 & -1 & 0 & 0 & 0 \\
 0 & 1 & 0 &  1 &-1 & 1 &  0 } & 
 SU(2)^2 \times U(1)^2 
	\\
	\hline
\cC_2 & 136  &
\tmat{
 1 &-1 & 0 & 0 & 0 & 0 & 0 \\
 0 & 1 &-1 &-1 & 0 & 0 & 0 \\
 0 & 0 & 1 & 2 &-1 & 1 & 0} & SU(2) \times U(1)^3
	\\
	\hline
\cD_1 & 131 &
\tmat{
 1 &-1 & 0 & 0 & 0 & 0 & 0 \\
 0 & 1 & 1 &-1 & 0 & 0 & 0 \\
 0 & 0 & 0 & 1 &-1 & 1 & 0} 
 & SU(2) \times U(1)^3 
	\\
	\hline
\cD_2 & 139 &
\tmat{
 1 &-1 & 0 & 0 & 0 & 0 & 0 \\
 0 & 1 &-1 &-1 & 0 & 0 & 0 \\
 0 & 0 & 0 & 1 & 1 &-1 & 0} 
 & SU(2) \times U(1)^3 
	\\
	\hline
\cE_1 & 218  & 
\tmat{1 &-1 & 0 & 0 & 0 & 0 & 0 & 0 \\
      0 & 1 & 1 & 1 &-1 & 0 & 0 & 0 \\
      0 & 0 & 0 & 1 &-1 &-1 & 1 & 0} &
SU(2) \times U(1)^3
	\\
	\hline
\cE_2 & 275  &
\tmat{
 1 &-1 & 0 & 0 & 0 & 0 & 0 & 0 \\
 0 & 0 & 1 &-1 &-1 & 0 & 0 & 0 \\
 0 & 1 & 0 & 0 & 1 &-1 & 1 & 0} &
SU(2) \times U(1)^3
	\\
	\hline
\cE_3 & 266 &
\tmat{
 1 &-1 & 0 & 0 & 0 & 0 & 0 & 0 \\
 0 & 0 & 1 &-1 & 1 & 0 & 0 & 0 \\
 0 & 0 & 0 & 0 & 1 & 1 &-1 & 0} & 
SU(2) \times U(1)^3
	\\
	\hline
\cE_4 & 271 &
\tmat{
 1 &-1 & 0 & 0 & 0 & 0 & 0 & 0 \\
 0 & 1 & 1 &-1 &-1 & 0 & 0 & 0 \\
 0 & 0 & 0 & 0 & 1 & 1 &-1 & 0} &
SU(2) \times U(1)^3
	\\
	\hline
\cF_2 & 369 & 
\tmat{
 1 &-1 & 0 & 0 & 0 & 0 & 0 & 0 & 0 \\
 0 & 1 & 1 &-1 & 1 &-1 & 0 & 0 & 0 \\
 0 & 0 & 0 & 0 & 1 &-1 &-1 & 1 & 0} & 
SU(2) \times U(1)^3
\\
\hline
\cF_1 &  324 &
\tmat{
  1 & -1 & 0 &  0 & 0 & 0 &  0 &  0 & 0  \\
  0 &  0 & 1 & -1 & 1 &-1 &  0 &  0 & 0  \\
  0 &  0 & 0 &  0 & 1 &-1 & -1 &  1 & 0 
} & 
SU(2) \times U(1)^3
	\\
	\hline
\end{array}
\]
\end{center}
\caption{The 18 smooth toric Fano 3-folds and some important geometric data \cite{Hanany:2009vx}.}
\label{t:fanotable}
\end{table}

There is a second naming system used for these geometries which is used in an online database of fano varieties \cite{database}. Each variety is given a `Fano no.' which we refer to as the `ID' of the variety. The ID of all of the fano varieties that we shall deal with in this thesis are given in \tref{t:fanotable}.

The starting point for the M2-brane inverse method (discussed in Section \ref{s:M2Inverse}) is the toric data of the singularity that the M2-brane is to probe. As has been mentioned in the previous chapter the toric description of a geometry can be encoded in a matrix which we call $G_t$. Each column of this matrix corresponds to a point in the toric diagram of the singularity. The $G_t$ matrices of the smooth toric fanos are given in \tref{t:fanotable}.

The point $\left(0,0,0\right)$ is a column of each of the matrices in \tref{t:fanotable}. This is no coincidence as each of the toric diagrams corresponding to the fano 3-folds and fano 2-folds have a single internal point.

It is interesting to consider the symmetries of the CY 4-folds that are constructed by taking a complex cone over the smooth toric fanos. The fourth column of \tref{t:fanotable} encodes this information. The symmetry of the smooth toric fanos (apart for $\CP^3$) is of the form:
\bea
SU(3)^{a}\times SU(2)^{b}\times U(1)^{c},
\eea
Since the symmetry group of the CY must be of rank 4 there is the following restriction:
\bea
2a + b + c = 4. \; \; \mathrm{with} \; \; a,b,c \geq 0
\eea
which is consistent with the symmetries listed in \tref{t:fanotable}. The order of the rows in this table are determined by the amount of symmetry of the corresponding CY. The manifolds with the greatest number of non-abelian factors of highest rank come closest to the top. 

\subsection{Symmetry of a fano from $G_t$}

It is possible to find the symmetry of a fano geometry from analysis of the $G_t$ matrix associated to it. It turns out that it is always possible to put $G_t$ in a form such that the simple roots of the non-abelian symmetries of the mesonic moduli space are explicit. Let us take Fano 24 (sometimes known as the cone over $M^{1,1,1}$) as a concrete example. The symmetry of the mesonic moduli space of the theory is $SU(3)\times SU(2)\times U(1)$. The $G_t$ matrix of this theory can be written as:
\bea
G_t = \left(
\begin{array}{cccccc}
 1 & -1 &  0 &  0 &  0 & 0 \\ 
 0 &  1 & -1 &  0 &  0 & 0 \\
 0 &  0 &  0 &  1 & -1 & 0 
\end{array}
\right)~.
\eea
The first two rows of this matrix contain the simple roots of $SU(3)$ and the third row contains the simple root of $SU(2)$.

Let us see how this holds for a general model with a moduli space containing an $SU(2)$ global symmetry. It is known that perfect matching matrices parameterise the moduli space and we can think of the moduli space as the quotient
\beq
\CMm =  \BC^c//Q_T 
\eeq 
If there is an $SU(2)$ global symmetry, two of the perfect matchings must be equally charged. Therefore we can write $Q_T$ with two identical columns, i.e.
\beq
Q_T = \left(
\begin{array}{cccc}
 a_1 & a_1 &   & \cdots   \\ 
 a_2 & a_2 &  & \cdots \\
 a_3 & a_3 &   & \cdots \\
\vdots & \vdots & & 
\end{array}
\right)~.
\eeq
and so, as $G_t = \mathrm{Ker}(Q_T)$, we can write 
\beq
G_t \supset  ( 1 , -1 , 0  , 0 , \ldots ) 
\eeq
It is not hard to see how this argument can be extended to the case where the mesonic moduli space has a global symmetry of $SU(3)$ or $SU(4)$.

\subsection{Constructing theories corresponding to the fano 3-folds}

\begin{table}[ht!]
\begin{center}
\begin{tabular}{cc|cc}
\begin{tabular}[b]{c} Tiling and \\ CS levels \end{tabular} & \begin{tabular}[b]{c} Toric data \\ and fano ID \end{tabular} &
\begin{tabular}[b]{c} Tiling and \\ CS levels \end{tabular} & \begin{tabular}[b]{c} Toric data \\ and fano ID \end{tabular}  \\ \hline \hline
\begin{tabular}[b]{c}
\includegraphics[width=1.8cm]{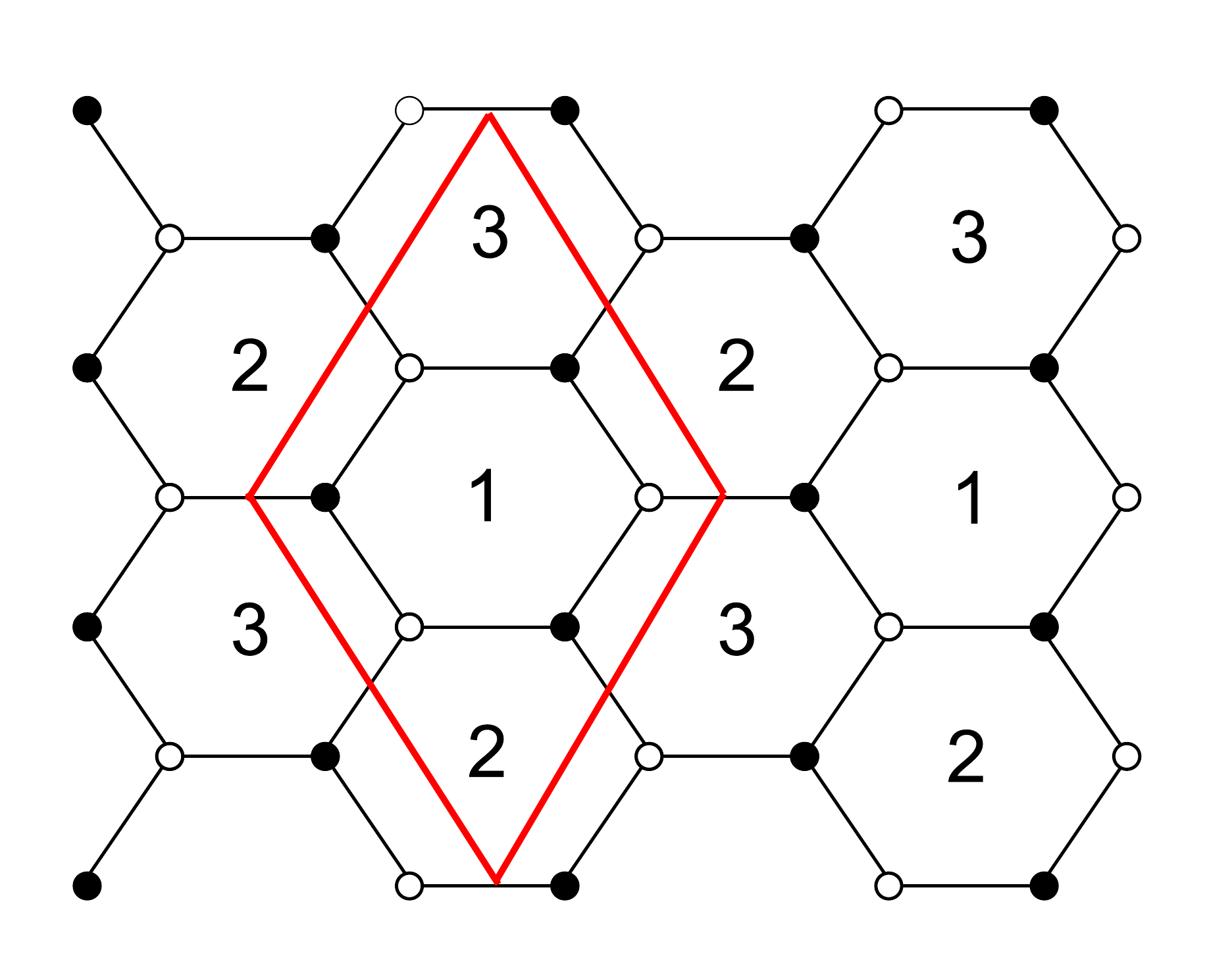} \\
(1, -2, 1)
\end{tabular}
&
\begin{tabular}[b]{c}
\includegraphics[width=1.8cm]{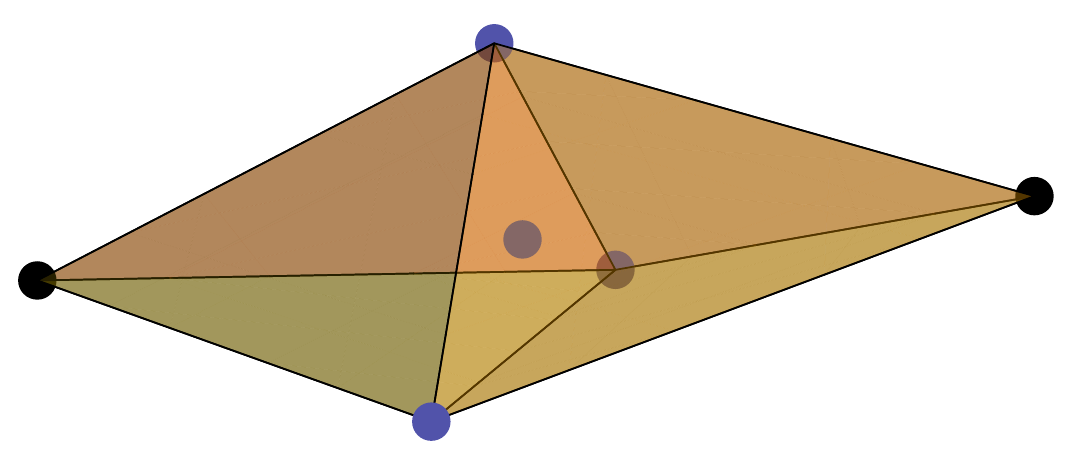} \\
\# 24
\end{tabular}
&
\begin{tabular}[b]{c}
\includegraphics[width=1.8cm]{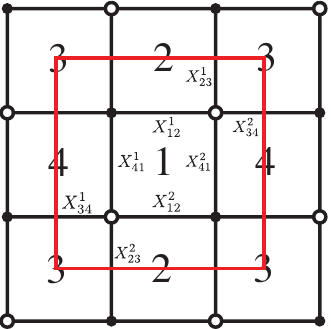} \\
(1,-1,-1,1)
\end{tabular}
&
\begin{tabular}[b]{c}
\includegraphics[width=1.8cm]{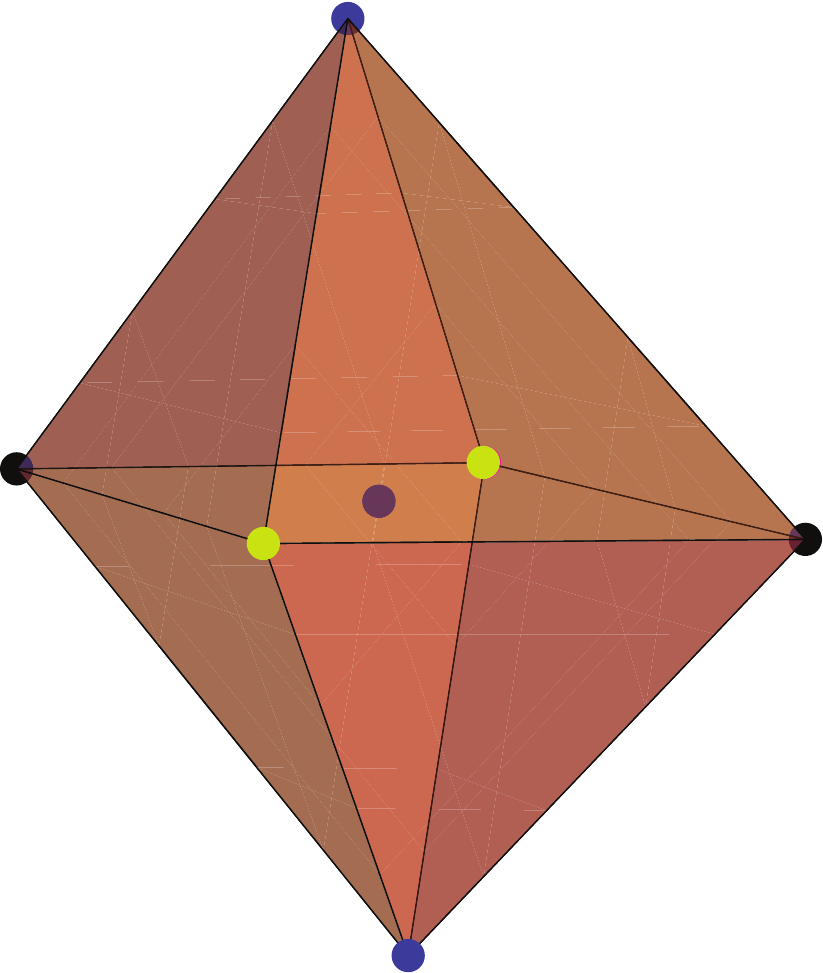} \\
\# 62
\end{tabular}
\\ \hline 
\begin{tabular}[b]{c}
\includegraphics[width=1.8cm]{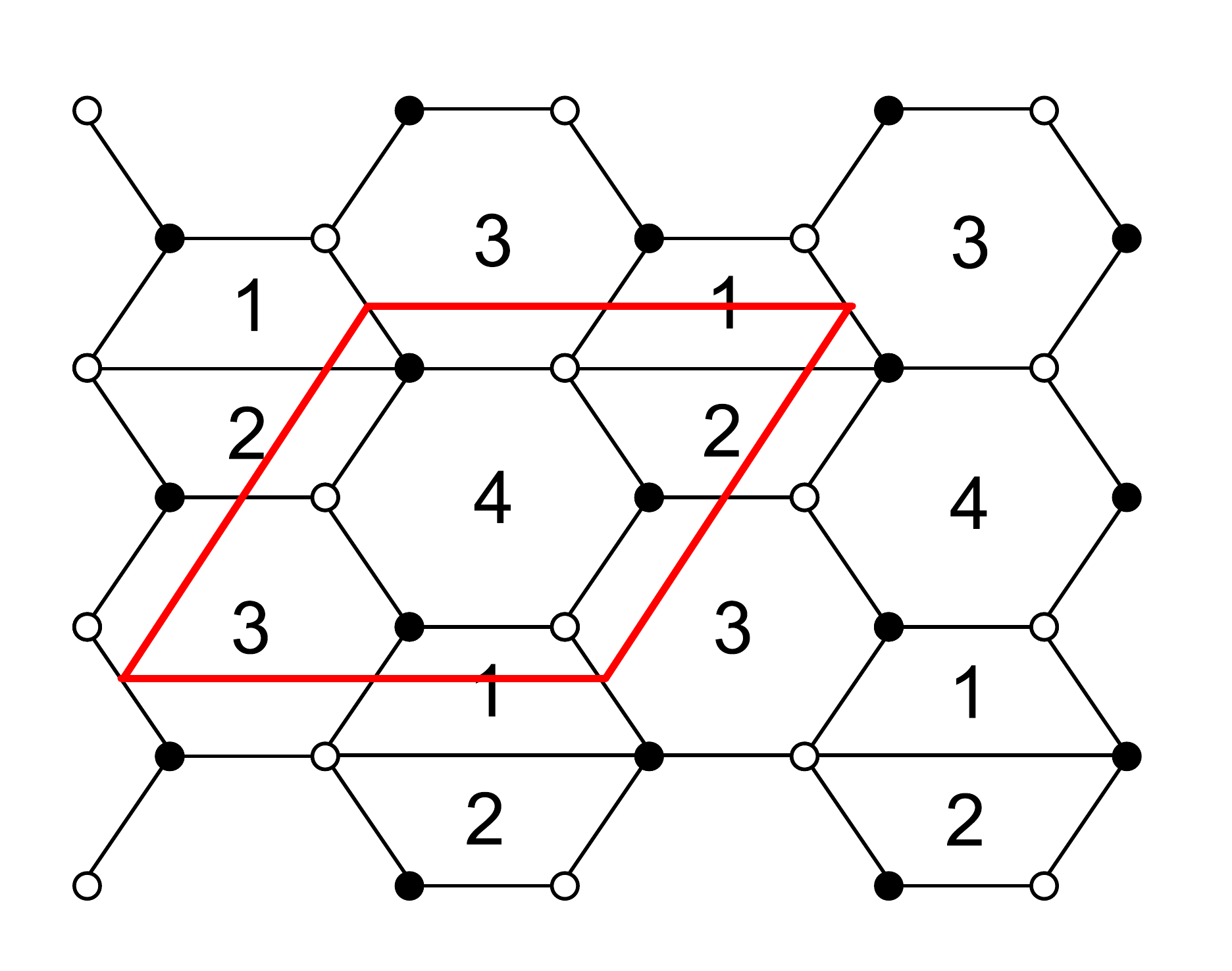} \\
(1,1,-1,-1)
\end{tabular}
&
\begin{tabular}[b]{c}
\includegraphics[width=1.8cm]{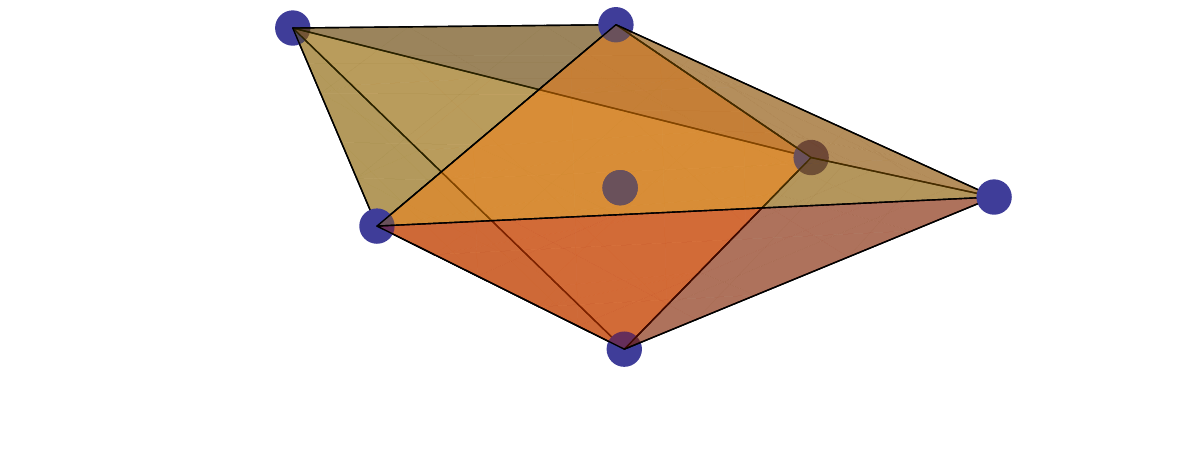} \\
\# 123
\end{tabular}
&
\begin{tabular}[b]{c}
\includegraphics[width=1.8cm]{FinalFano/tiling62I.pdf} \\
(1,-2,1,0)
\end{tabular}
&
\begin{tabular}[b]{c}
\includegraphics[width=1.8cm]{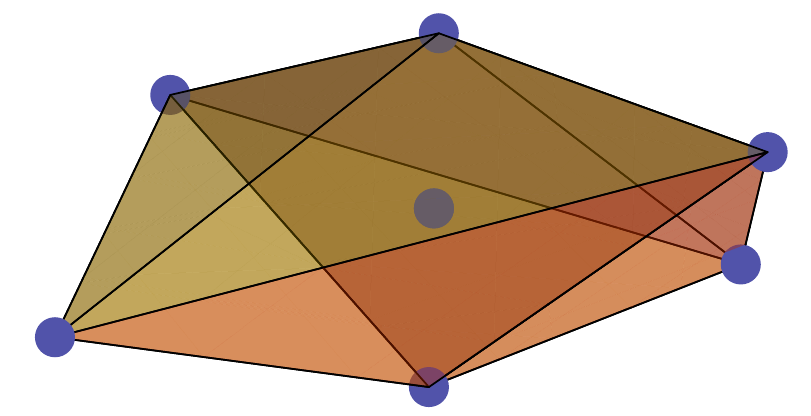} \\
\# 68
\end{tabular}
\\ \hline 
\begin{tabular}[b]{c}
\includegraphics[width=1.8cm]{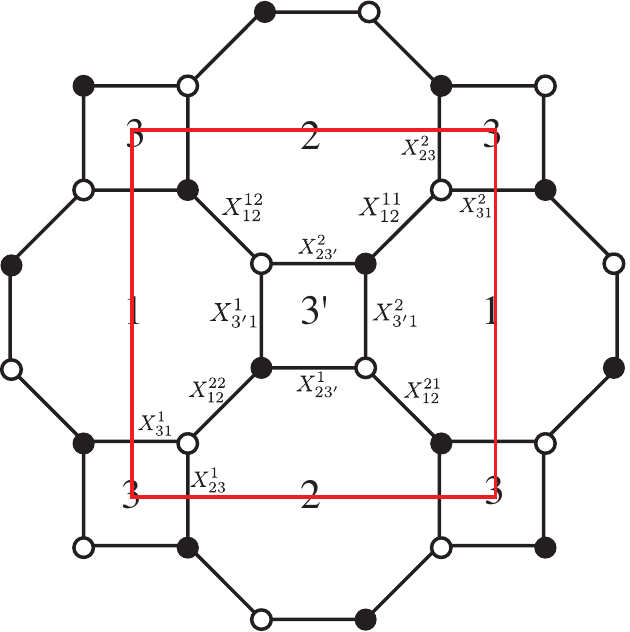} \\
(2,0,-1,-1)
\end{tabular}
&
\begin{tabular}[b]{c}
\includegraphics[width=1.8cm]{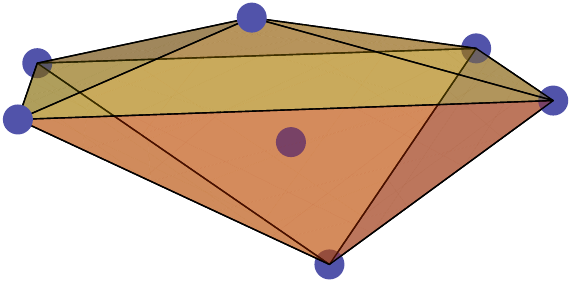} \\
\# 105
\end{tabular}
&
\begin{tabular}[b]{c}
\includegraphics[width=1.8cm]{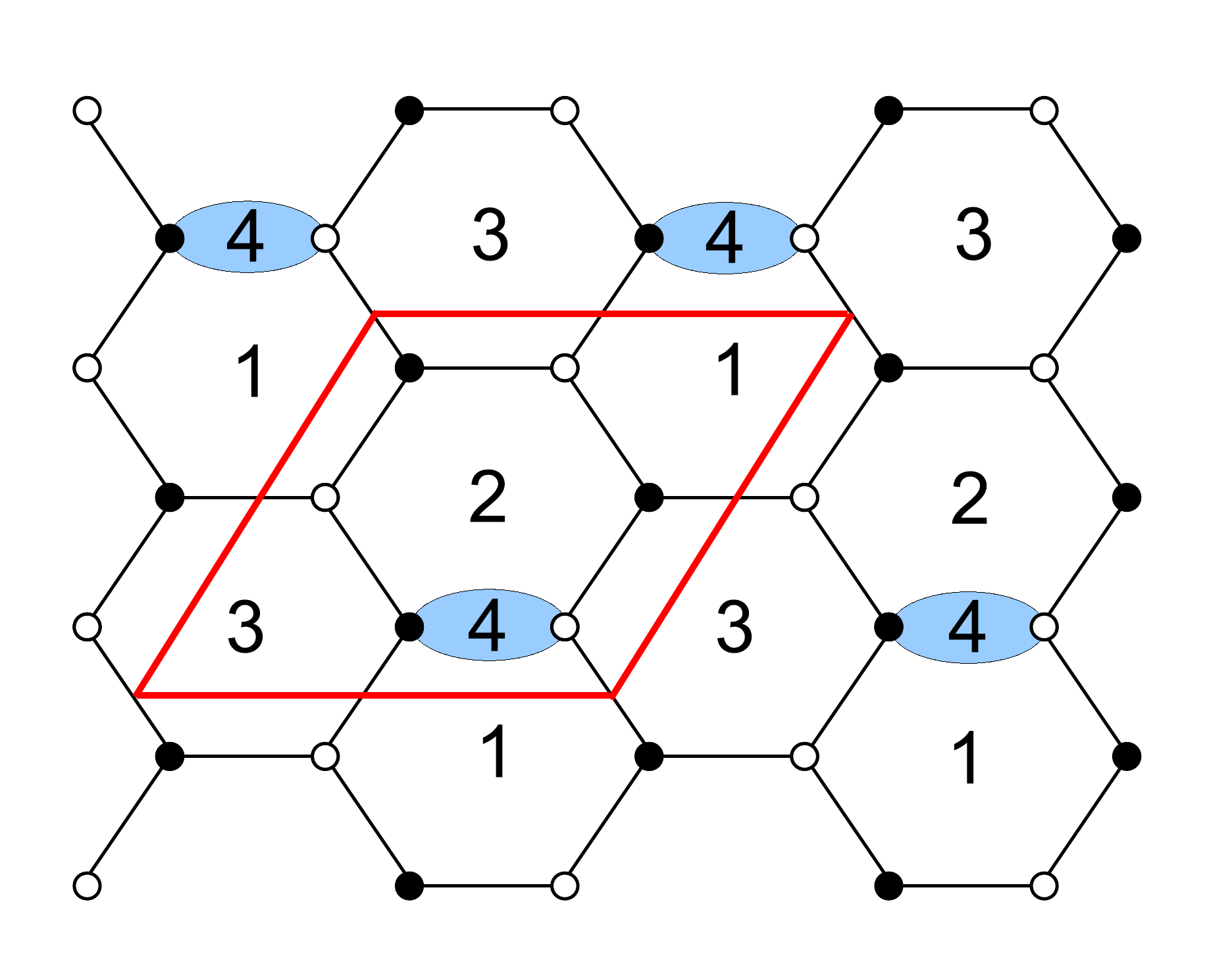} \\
(-1,2,0,-1)
\end{tabular}
&
\begin{tabular}[b]{c}
\includegraphics[width=1.8cm]{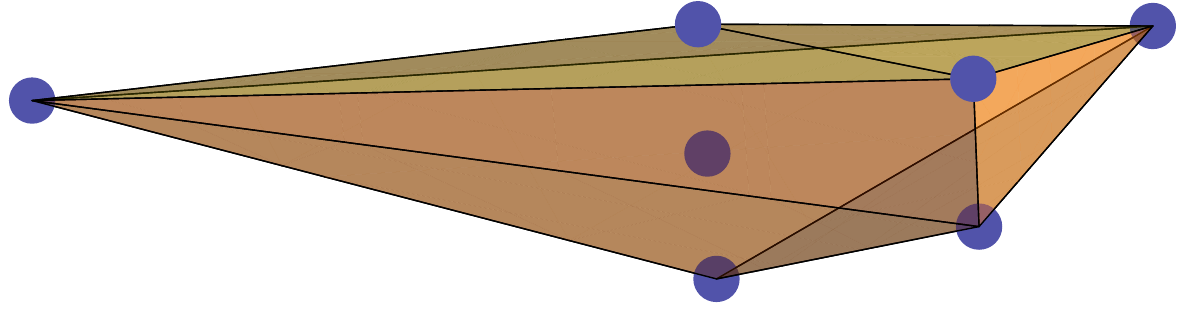} \\
\# 136
\end{tabular}
\\ \hline 
\begin{tabular}[b]{c}
\includegraphics[width=1.8cm]{FinalFano/tiling123.pdf} \\
(-1,-1,0,2)
\end{tabular}
&
\begin{tabular}[b]{c}
\includegraphics[width=1.8cm]{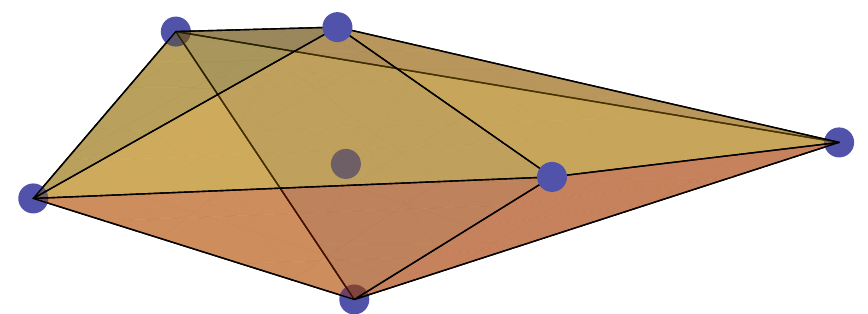} \\
\# 131
\end{tabular}
&
\begin{tabular}[b]{c}
\includegraphics[width=1.8cm]{FinalFano/tiling136.pdf} \\
(-1,1,1,-1)
\end{tabular}
&
\begin{tabular}[b]{c}
\includegraphics[width=1.8cm]{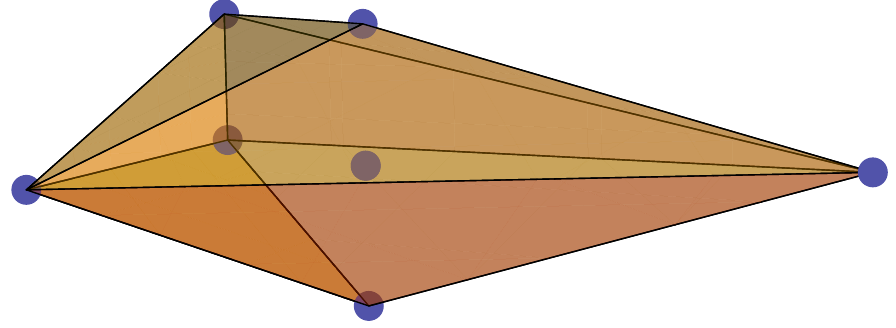} \\
\# 139
\end{tabular}
\\ \hline
\begin{tabular}[b]{c}
\includegraphics[width=1.8cm]{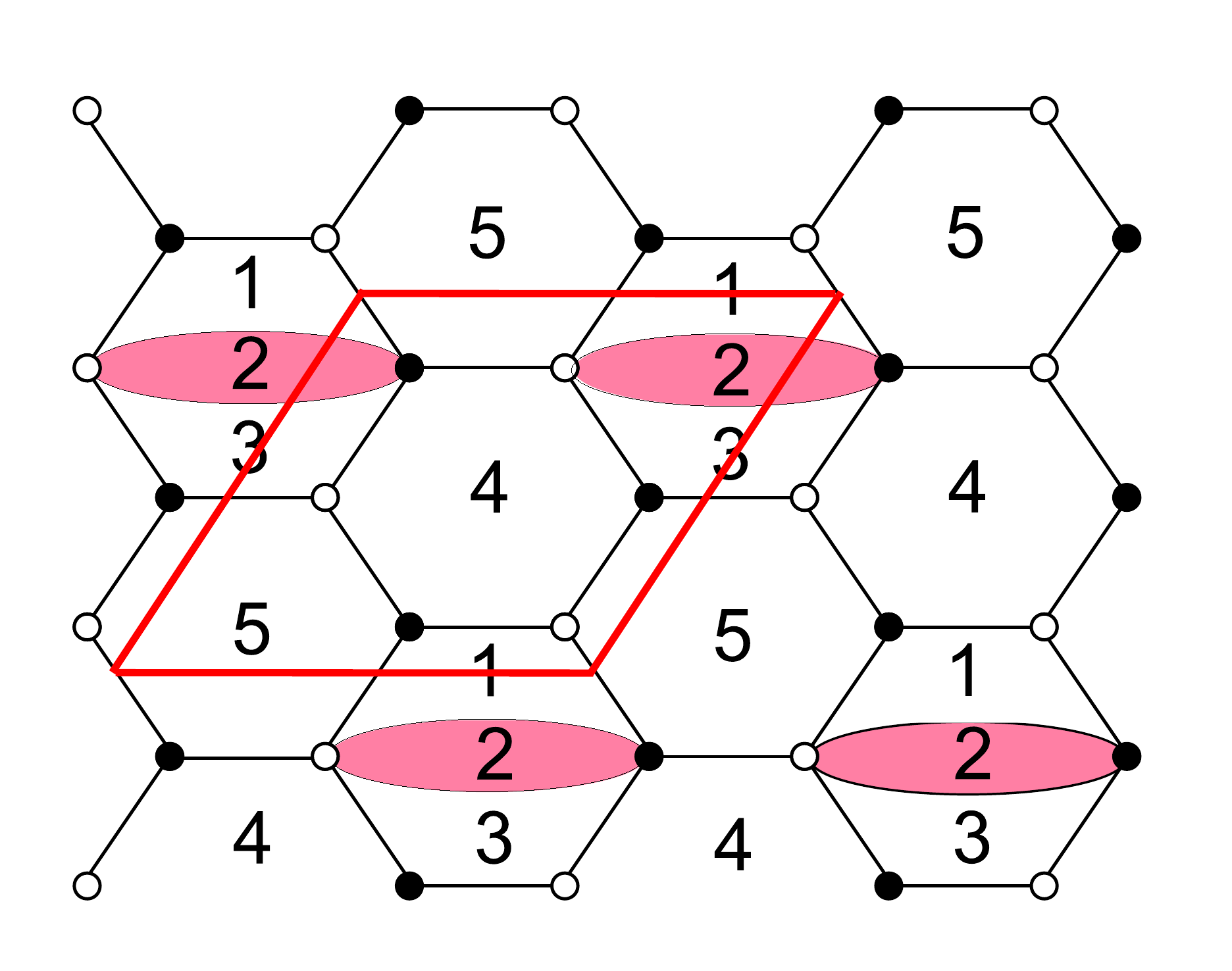} \\
(1,-1,0,-1,1)
\end{tabular}
&
\begin{tabular}[b]{c}
\includegraphics[width=1.8cm]{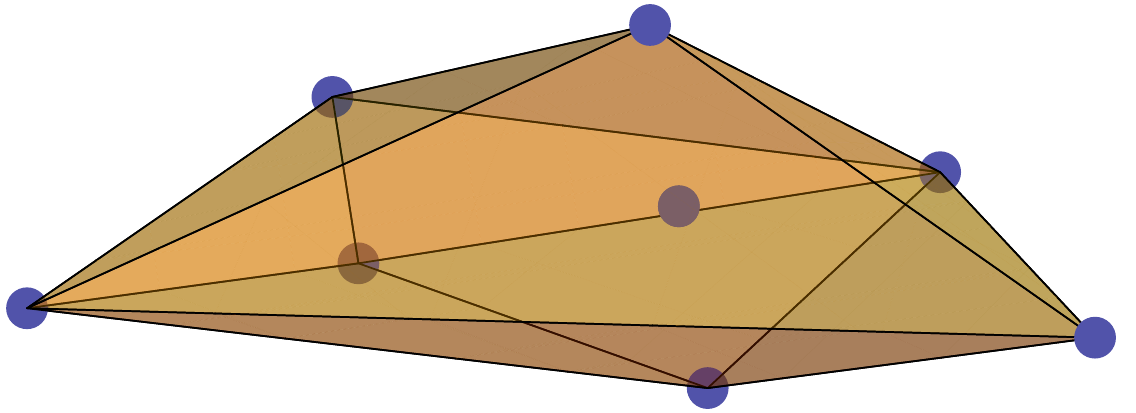} \\
\# 218
\end{tabular}
&
\begin{tabular}[b]{c}
\includegraphics[width=1.8cm]{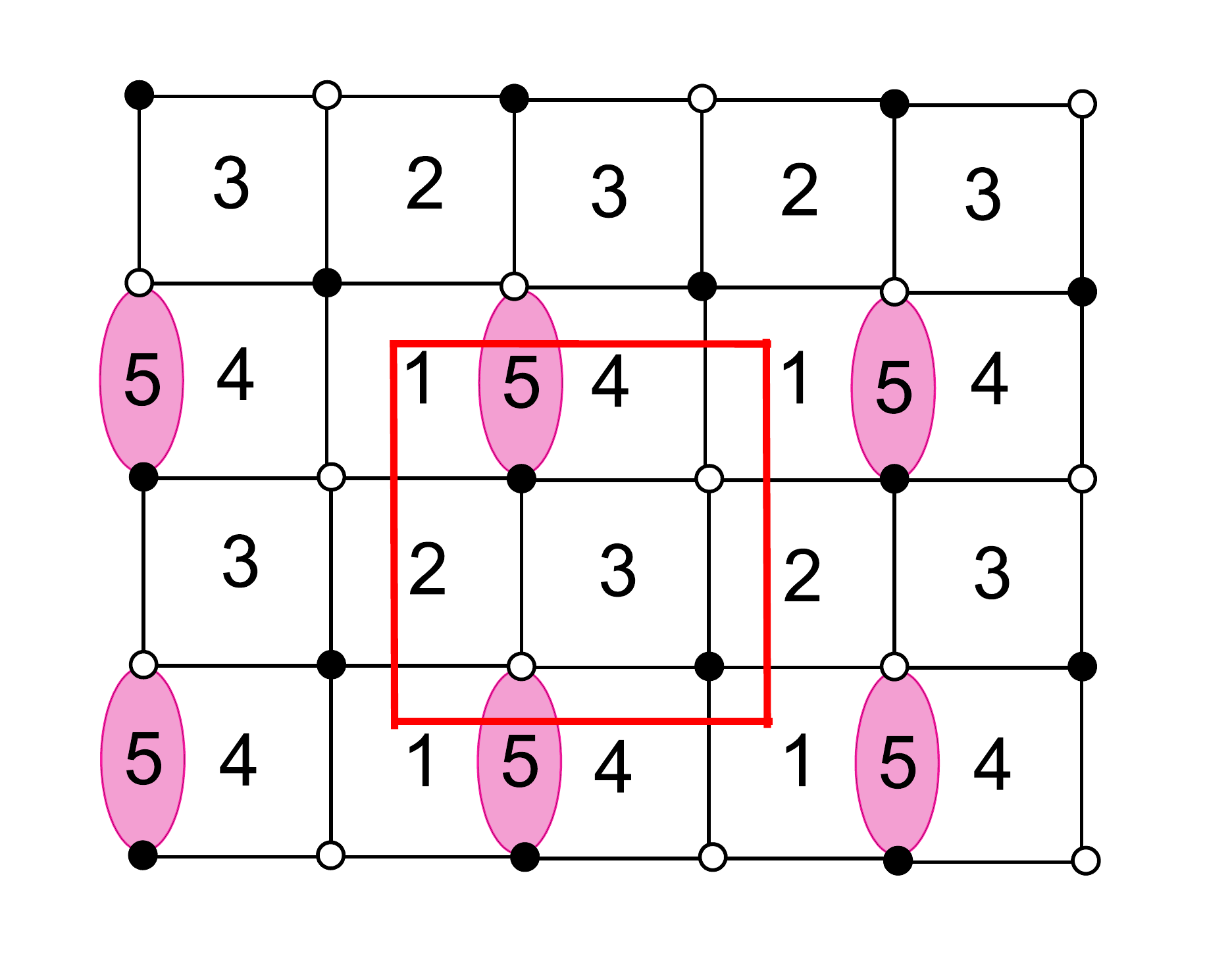} \\
(1,0,-1,-1,1)
\end{tabular}
&
\begin{tabular}[b]{c}
\includegraphics[width=1.8cm]{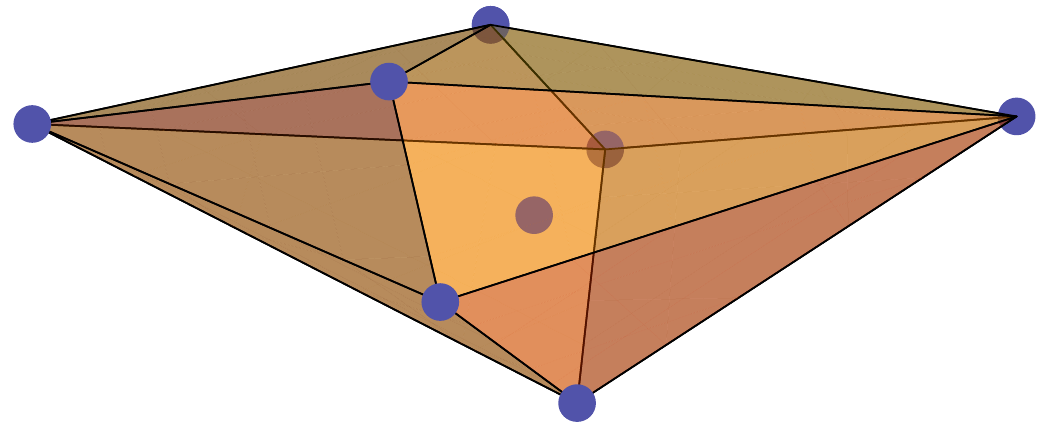} \\
\# 275
\end{tabular}
\\ \hline 
\begin{tabular}[b]{c}
\includegraphics[width=1.8cm]{FinalFano/tiling275.pdf} \\
(1,1,-1,0,-1)
\end{tabular}
&
\begin{tabular}[b]{c}
\includegraphics[width=1.8cm]{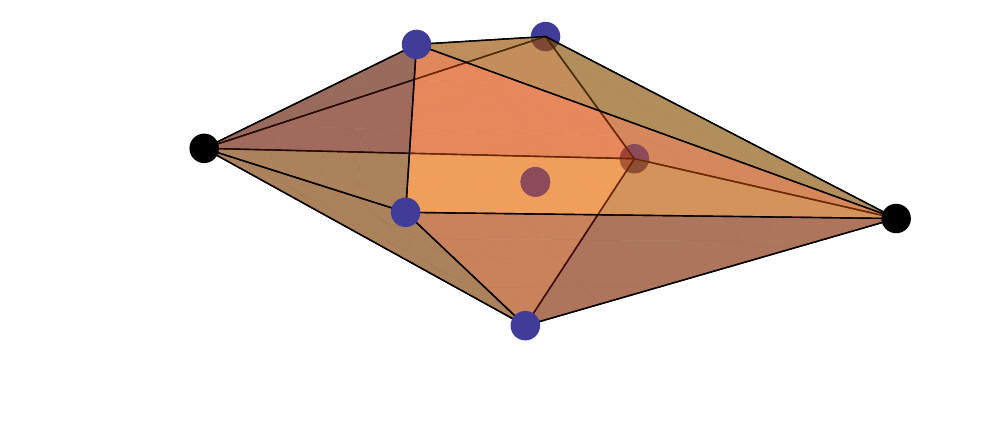} \\
\# 266
\end{tabular}
&
\begin{tabular}[b]{c}
\includegraphics[width=1.8cm]{FinalFano/tiling275.pdf} \\
(1,-1,0,-1,1)
\end{tabular}
&
\begin{tabular}[b]{c}
\includegraphics[width=1.8cm]{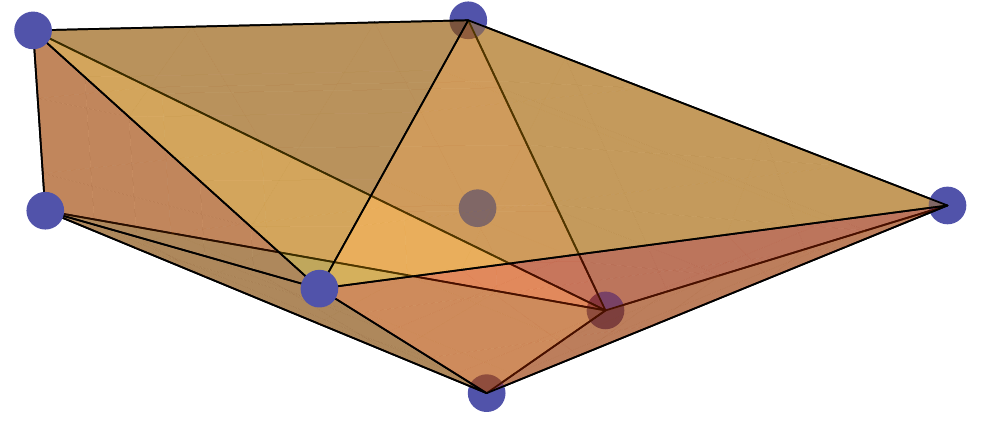} \\
\# 271
\end{tabular}
\\ \hline 
\begin{tabular}[b]{c}
\includegraphics[width=1.8cm]{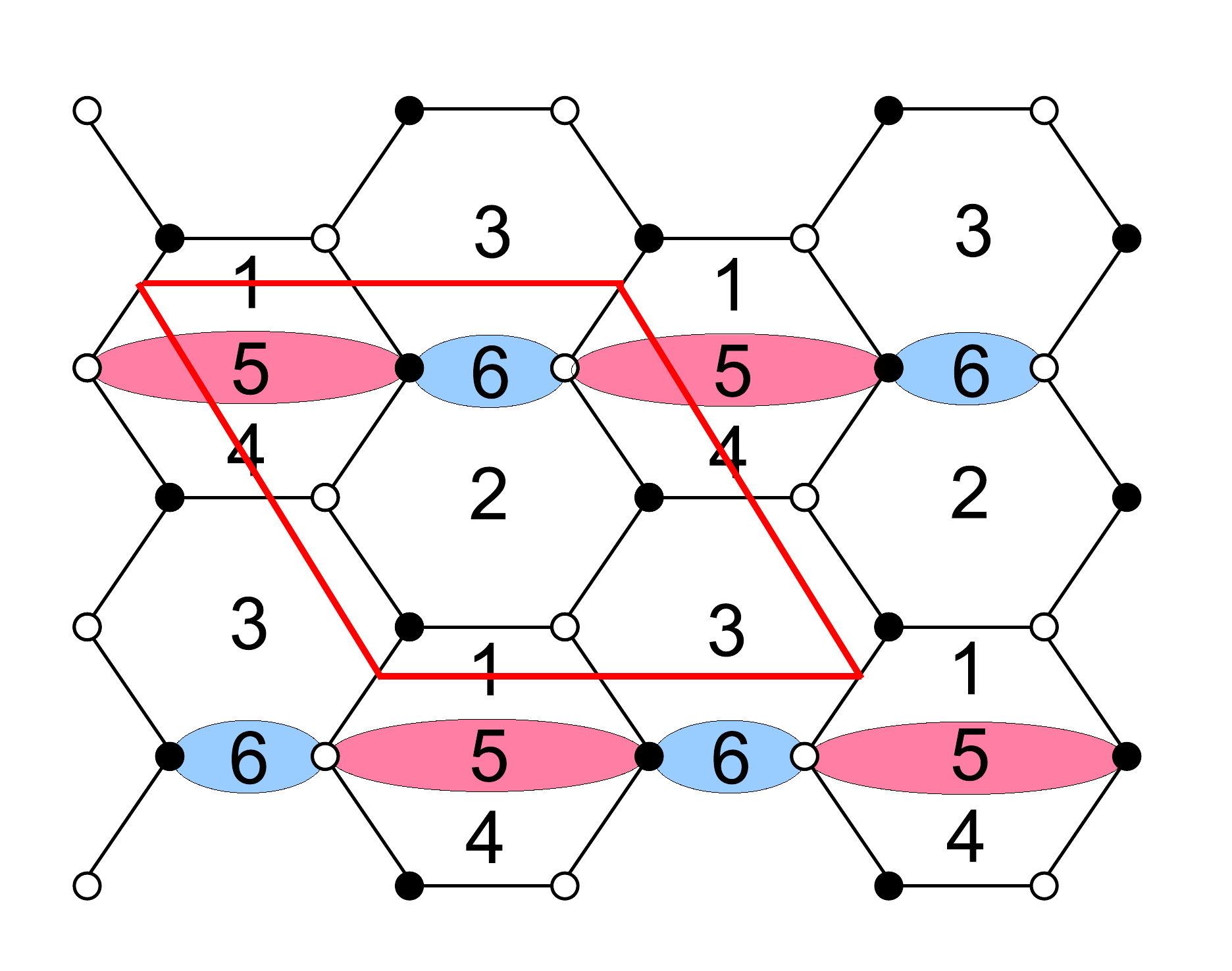} \\
(0,-1,0,-1,1,1)
\end{tabular}
&
\begin{tabular}[b]{c}
\includegraphics[width=1.8cm]{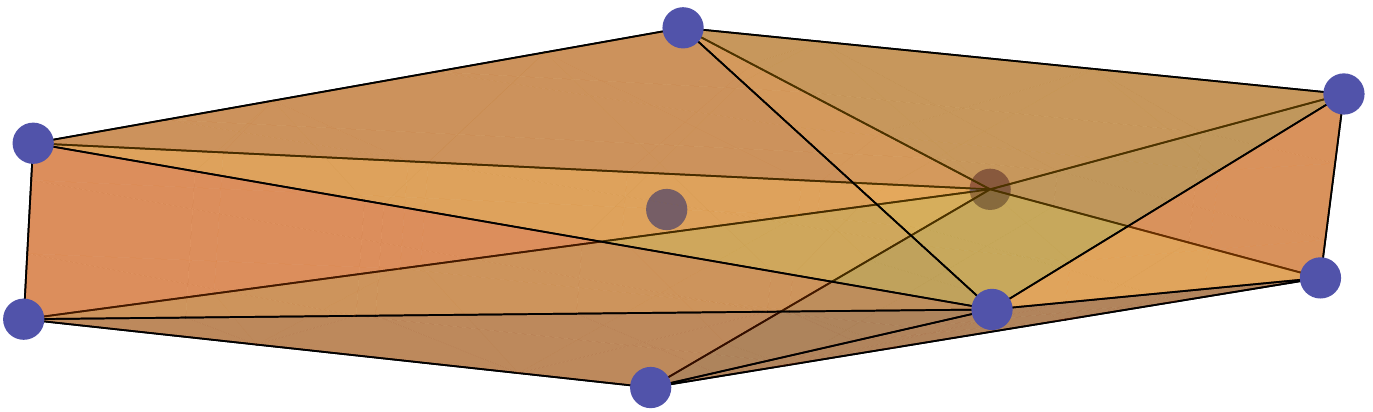} \\
\# 369
\end{tabular}
&
\begin{tabular}[b]{c}
\includegraphics[width=1.8cm]{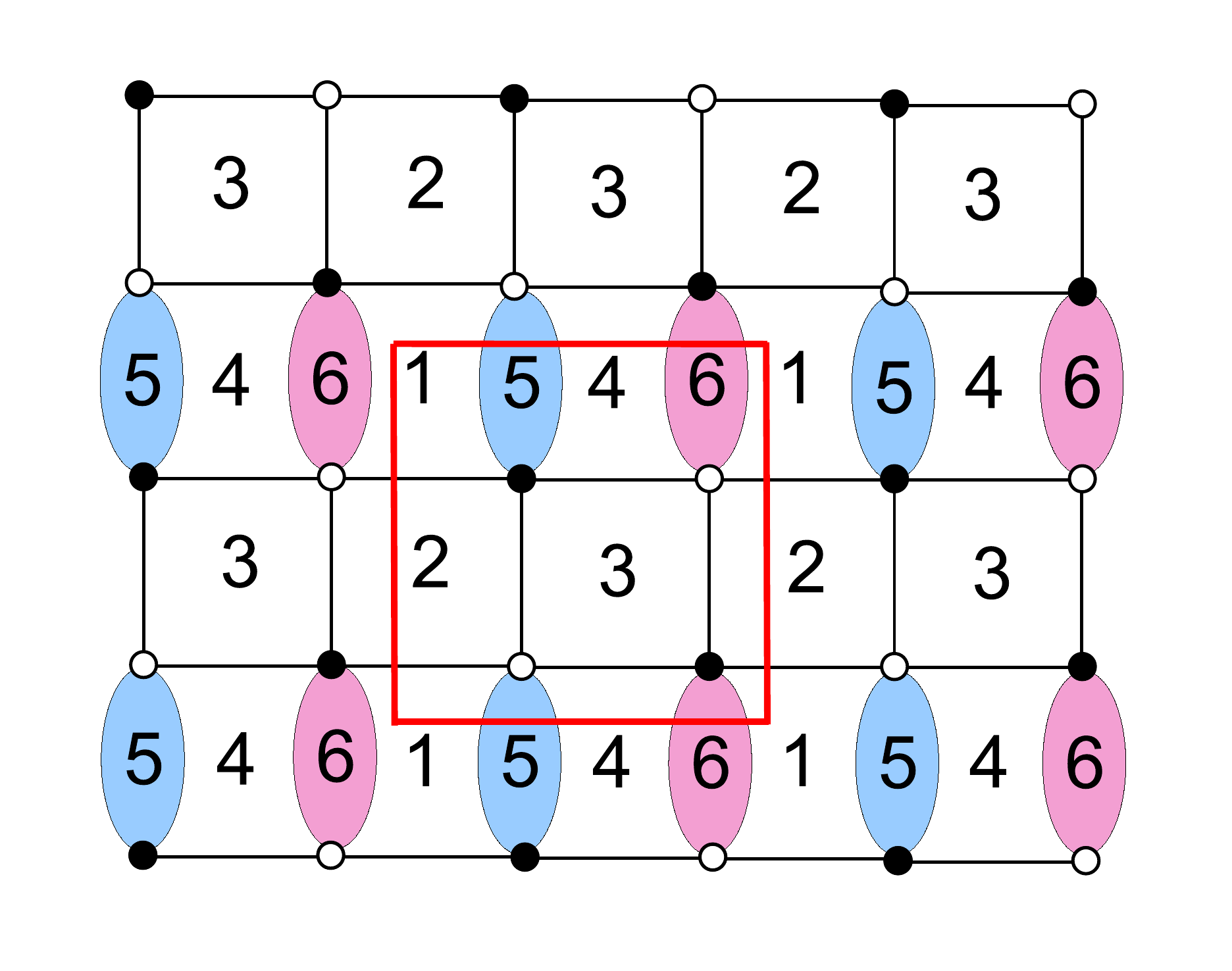} \\
(0,0,0,0,-1,1)
\end{tabular}
&
\begin{tabular}[b]{c}
\includegraphics[width=1.8cm]{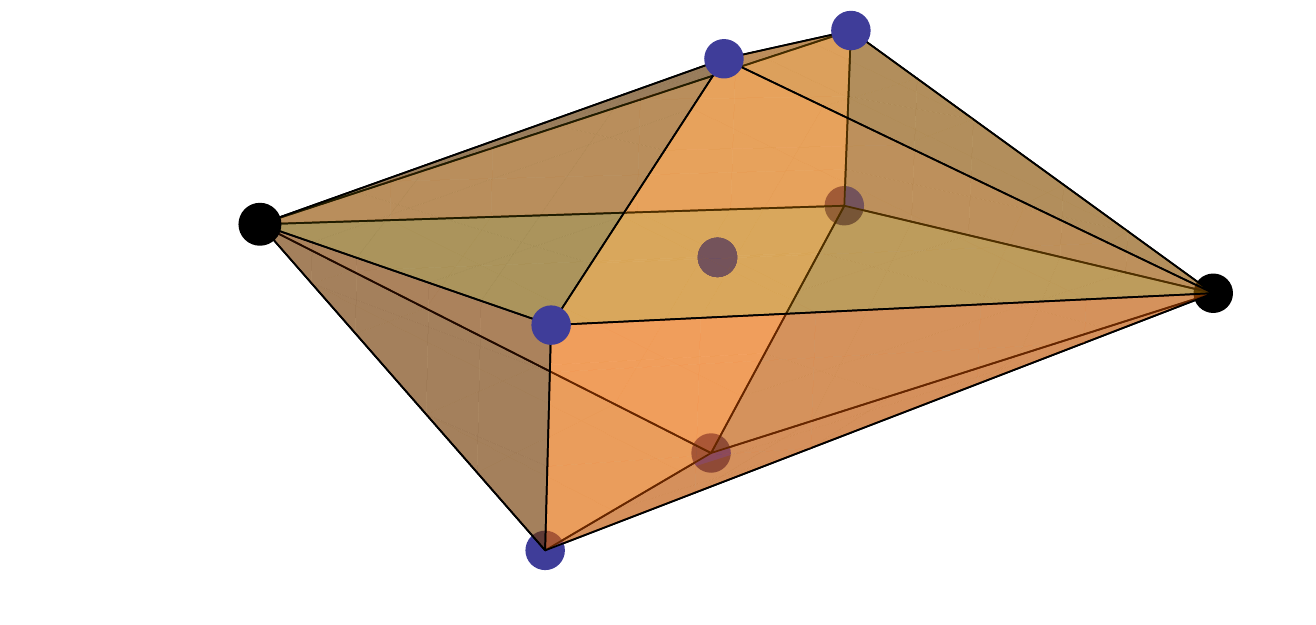} \\
\# 324
\end{tabular}
\end{tabular}
\end{center}
\caption{Tilings and CS levels that correspond to 14 of the 18 smooth toric Fano 3-folds.}
\label{t:14tilings}
\end{table}

The inverse method for M2-branes (which was discussed in Section \ref{s:M2Inverse}) has been used to find tilings that correspond to 14 of the 18 fano 3-folds. A summary of the tilings and CS levels found that correspond to these 14 fanos can be found in \tref{t:14tilings}. A more involved discussion of these 14 fanos and their corresponding CS theories can be be found in Appendix \ref{fanoapp}. The forward algorithm is applied to each of the theories and the non abelian global symmetry of the moduli space is verified.

Further calculations involving these 14 theories that correspond to fano 3-folds are presented in the work `M2-Branes and Fano 3-folds' \cite{Davey:Fano}. Starting from tilings the forward algorithm has been used to determine the Hilbert series, the generators of the mesonic moduli space and the spectrum of scaling dimensions of the chiral fields of each of the theories. The work demonstrates the strength of the forward algorithm - a detailed analysis of the structure of a CS gauge theory can be carried out by a small number of relatively simple computations.

\Section{$\BP^3, \cB_1, \cB_2$ and $\cB_3$ (Toric Fanos 4, 35, 36 and 37)}

\begin{figure}[h]
\begin{center}
  \includegraphics[totalheight=5.0cm]{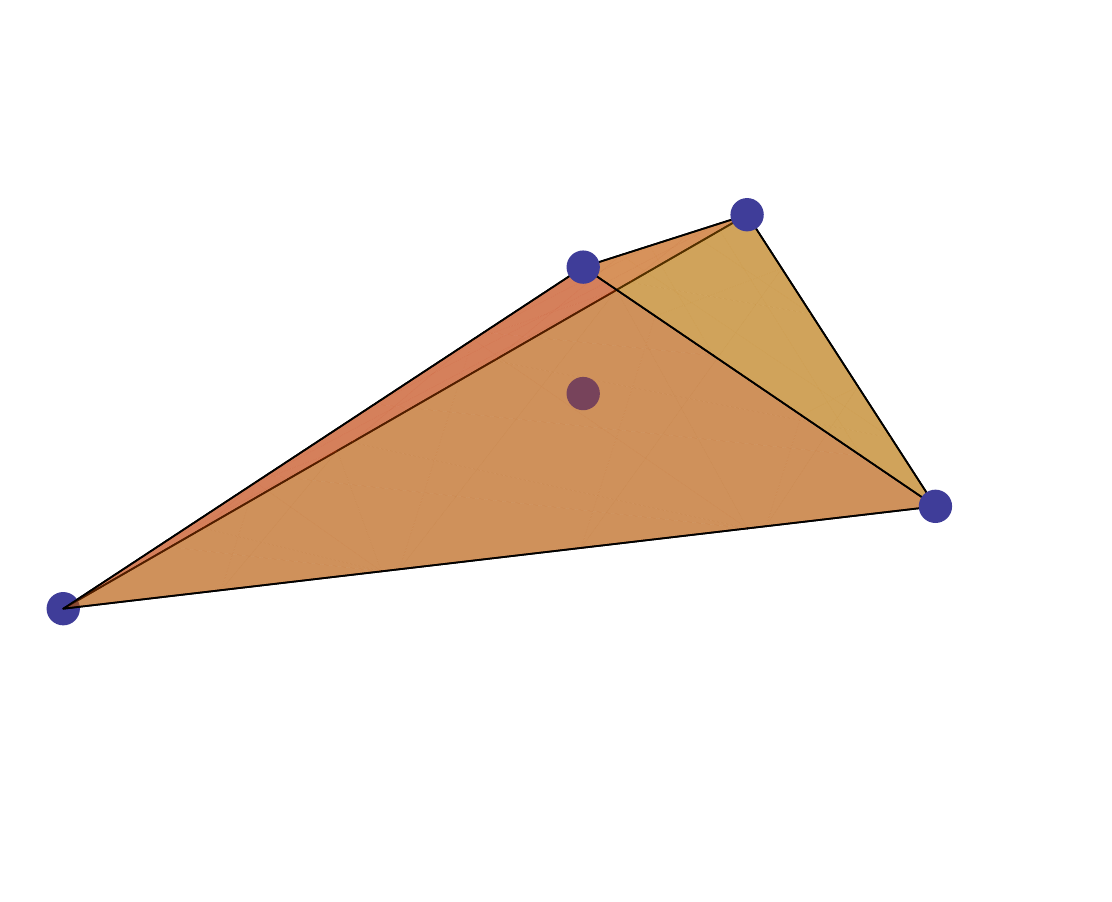}
\hfill
  \includegraphics[totalheight=5.0cm]{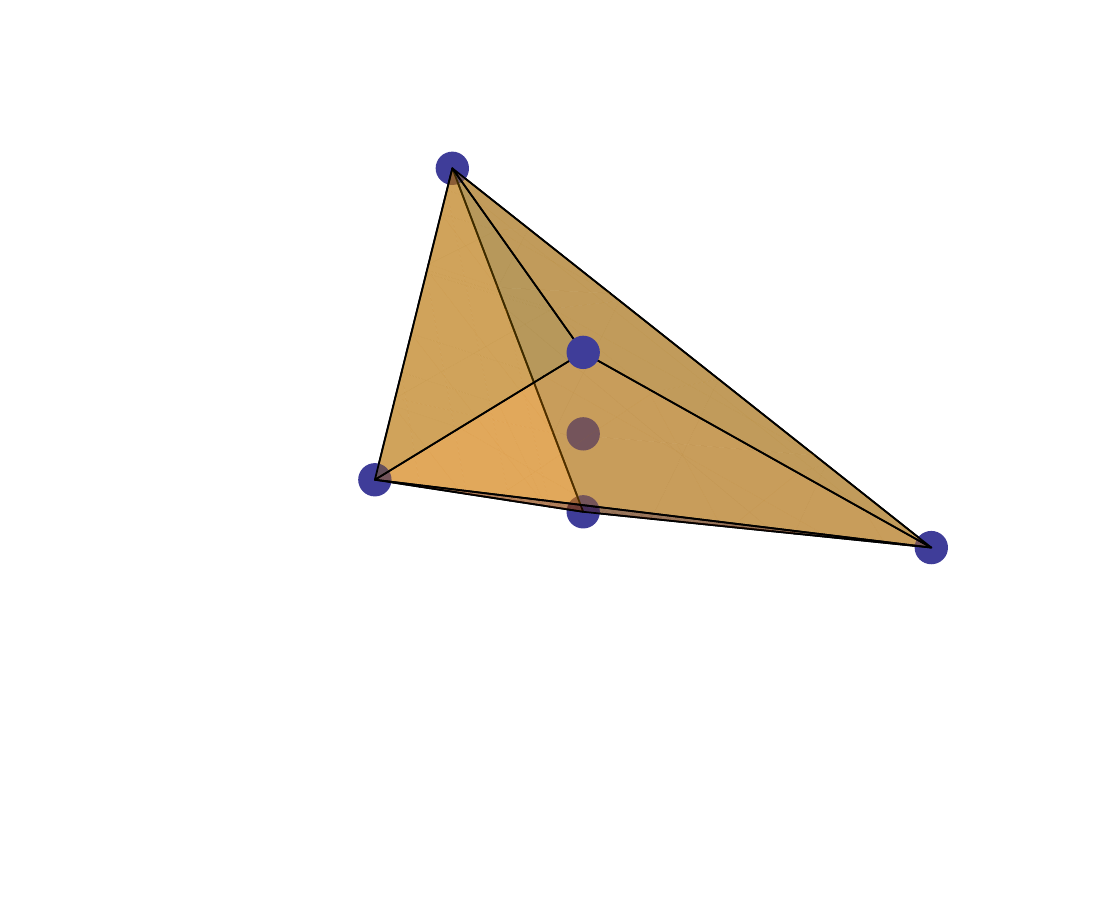}\\
$\BP^3$ \hspace{5cm}
 $\cB_1$ \\
\vspace{0.5cm}
  \includegraphics[totalheight=5cm]{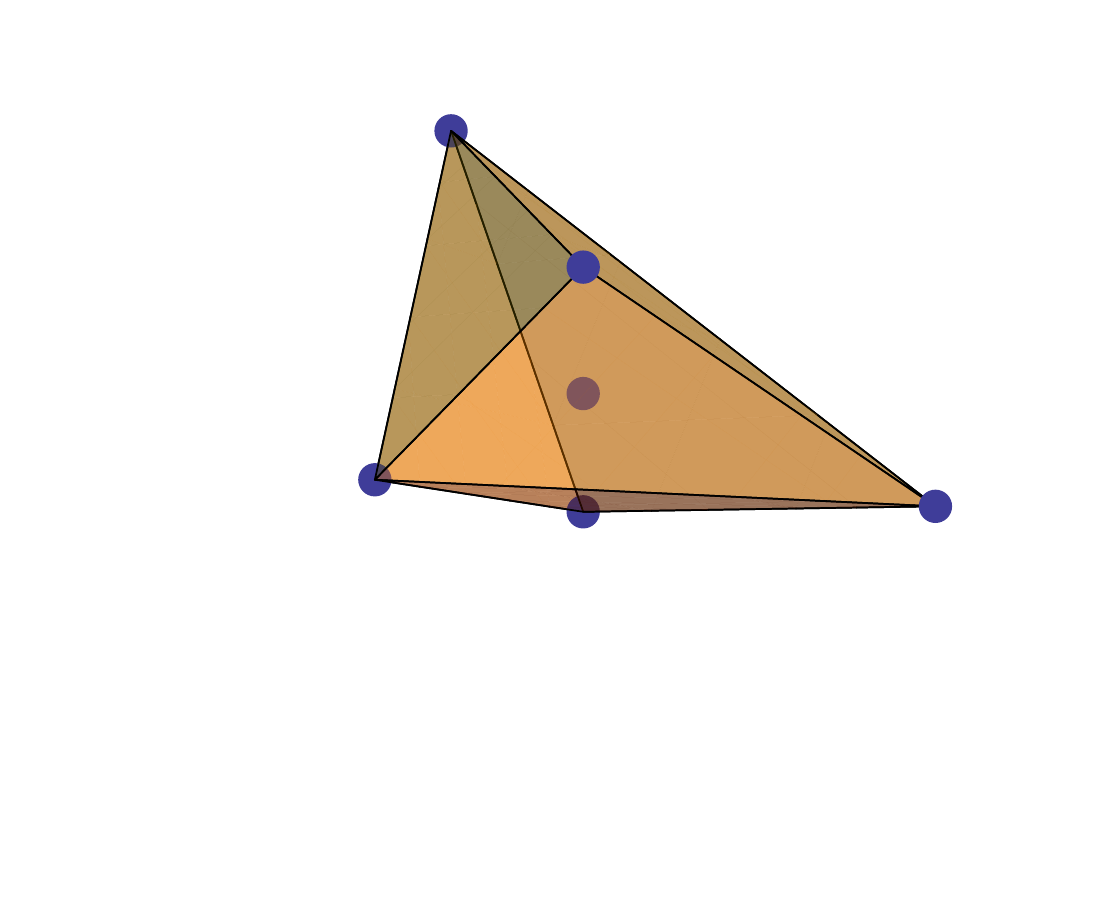}
\hfill
  \includegraphics[totalheight=5cm]{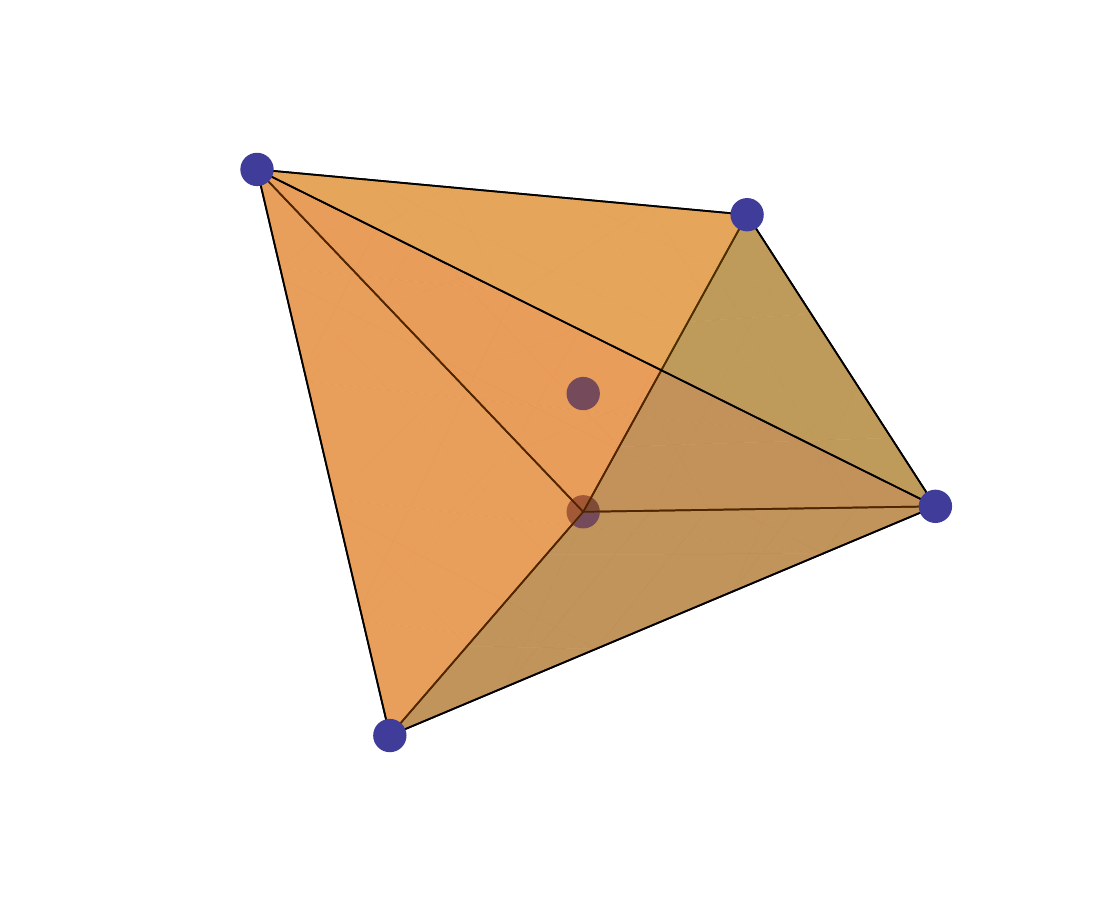}\\
 $\cB_2$ \hspace{5cm}
$\cB_3$
 \caption{The toric diagrams of (top left to bottom right) $\BP^3, \cB_1, \cB_2$ and $\cB_3$.}
  \label{f:last4fanos}
\end{center}
\end{figure}

Despite a study of all of the tilings with less than 10 nodes, it has not been possible to identify any tilings that could correspond to $\BP^3, \cB_1, \cB_2$ or $\cB_3$. Toric diagrams corresponding to these varieties are listed in \fref{f:last4fanos}. It is possible that there cannot exist a consistent CS gauge theory on M2-branes probing certain toric CY 4-folds. Another possibility is that such theories do not admit a brane tiling description.

We know that for $(3+1)$-dimensional gauge theories living on D3-branes, there is at least one theory that corresponds to a toric CY 3-fold. A way of constructing a gauge theory dual for every toric CY 4-fold is not known. The study of the fano varieties has highlighted this problem. Further investigation into this matter and the construction of an improved inverse algorithm for M2-brane theories is of great importance and should be studied in the future.

\chapter{Counting Children of a Brane Tiling}
\label{Ch:Children}
In this chapter, we will discuss some of the ideas that have helped us to count `reducible' tilings.

In section \ref{s:doubling}, the generation of tilings that correspond to theories living on D3-branes was discussed. The tilings generated were said to be `irreducible', that is they had no double-(or multi-)bonds. It was mentioned that it is possible to recover `reducible' tilings by adding multi-bonds to tilings that are irreducible. Tilings that are formed by adding multi-bonds to an irreducible `parent' tiling are known as `children'. One can count the children that can be obtained from adding multi-bonds to an irreducible `parent' tiling.

In this chapter we shall see that by using a tiling's symmetry group, it is possible to count the number of children of a parent tiling.

\Section{Counting children of the 1 hexagon tiling}

\begin{figure}
\begin{center}
\includegraphics[totalheight=3cm]{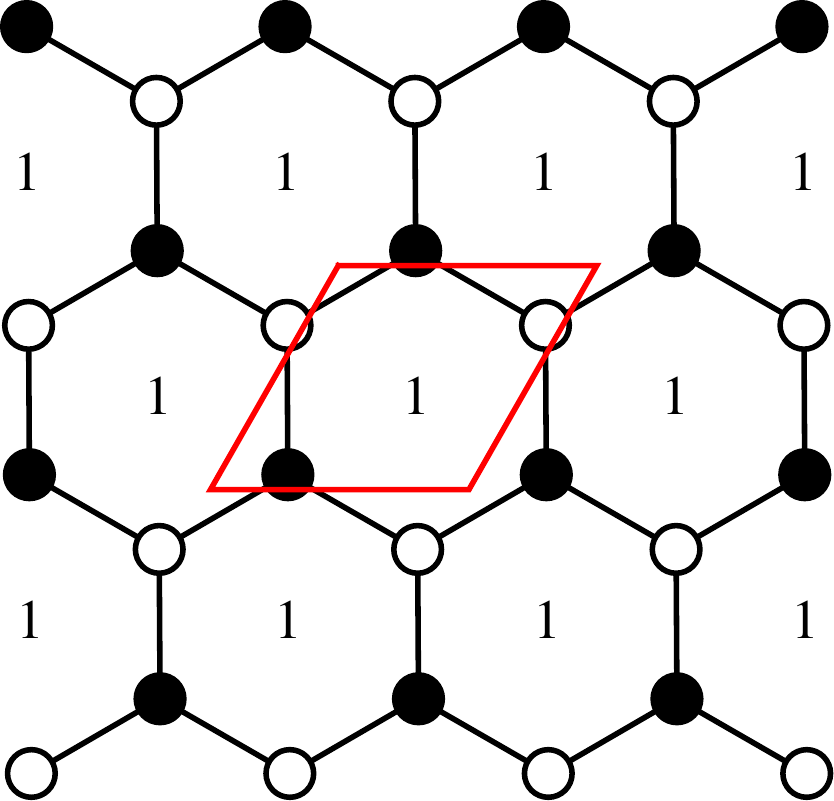}
\hspace{0.5cm}
\includegraphics[width=3cm]{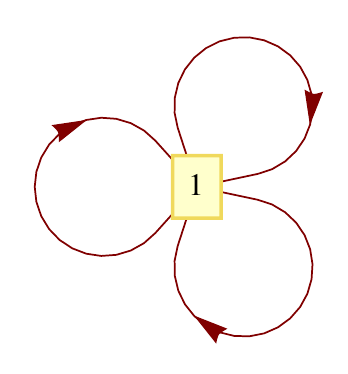}
\end{center}
\caption{The 1 hexagon tiling and its quiver}
\label{f:1hextileandquiver}
\end{figure}

Let us first consider the problem of how to count the children of the 1 hexagon model. The quiver and tiling of the 1 hexagon model are given in \fref{f:1hextileandquiver}

The first step in this counting problem is to find the symmetry group of the brane tiling. This group can be thought of as the permutation group of the edges (or fields) which keep the tiling invariant. In the case of the 1 hexagon tiling, the group is generated by two elements: rotating the tiling by $120^\mathrm{o}$ and a vertical reflection. The symmetry group of the tiling is therefore $S_3$ which corresponds to the permutation of the three edges in the tiling in all possible ways. Unsurprisingly this is also a symmetry of the quiver in \fref{f:1hextileandquiver}.

Now let us consider the problem of counting the children of the 1 hexagon tiling that have $i$ additional fields. As there is a full $S_3$ symmetry group on the tiling, this problem is equivalent to counting the number of ways it is possible to partition the number $i$ into into 3 sets.

Let us explicitly use this method to count the children of the 1 hexagon tiling with at most 2 additional fields. There is only a single way of splitting 1 into 3 partitions:
\beq
1 = 1 + 0 + 0 \equiv 0+1+0 \equiv 0+0+1
\label{e:partitionsof1}
\eeq
Therefore there is only one child of the 1 hexagon tiling with 1 additional field. This tiling is given in \tref{t:1HexModels}.

Now let us consider how many children of the 1 hexagon tiling there are with 2 additional fields. This time there are 2 ways of splitting 2 into 3 partitions:
\bea
2 &=& 2+0+0 \equiv 0+2+0 \equiv 0+0+2 \nn \\
2&=& 1+1+0 \equiv 1+0+1 \equiv 0+1+1
\label{e:partitionsof2}
\eea
and so there are two children of the 1 hexagon tiling with 2 additional fields. These two tilings are given in \tref{t:1HexModels}.

\begin{table}[h!]
\begin{center}
\begin{tabular}{cc|cc}
 Tiling & Quiver &  Tiling & Quiver\\
\hline 
 \includegraphics[totalheight=2.5cm]{Images/1Hex/Tiling-1Hex.pdf}
 &
\includegraphics[width=2.5cm]{Images/1Hex/Quiver-1Hex.pdf}
&
 \includegraphics[totalheight=2.5cm]{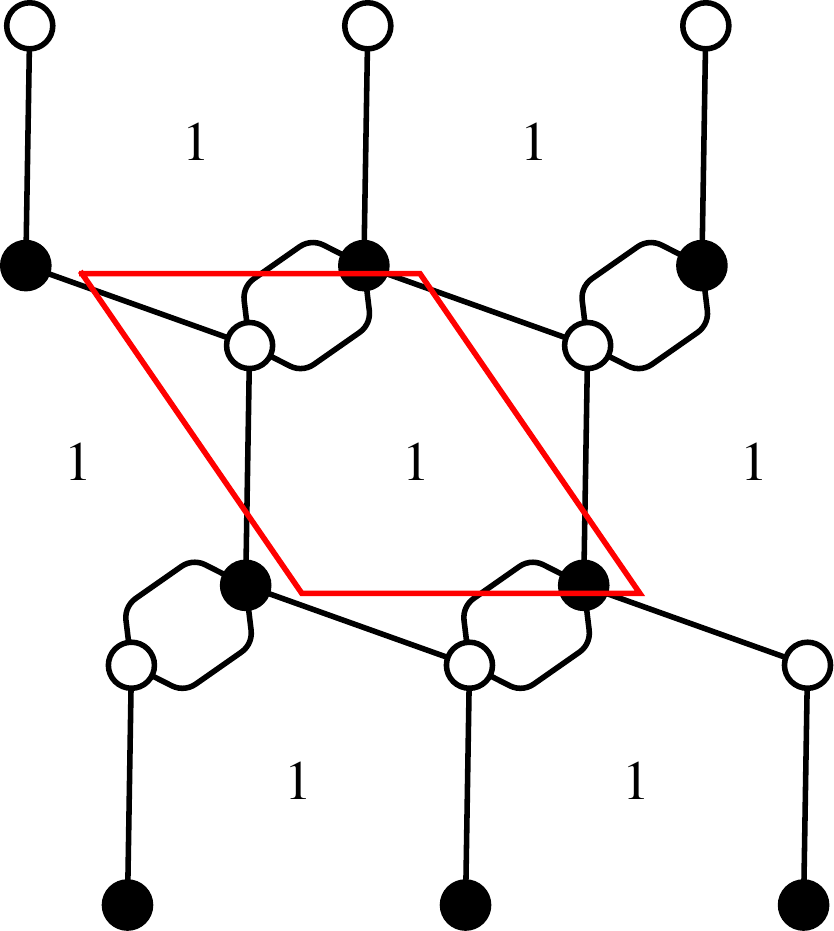}
 &
\includegraphics[width=2.5cm]{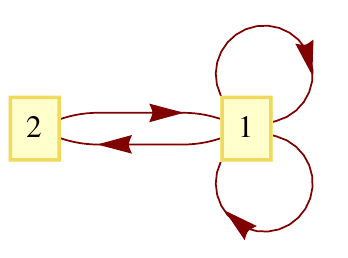}
\\
\hline
 \includegraphics[totalheight=2.5cm]{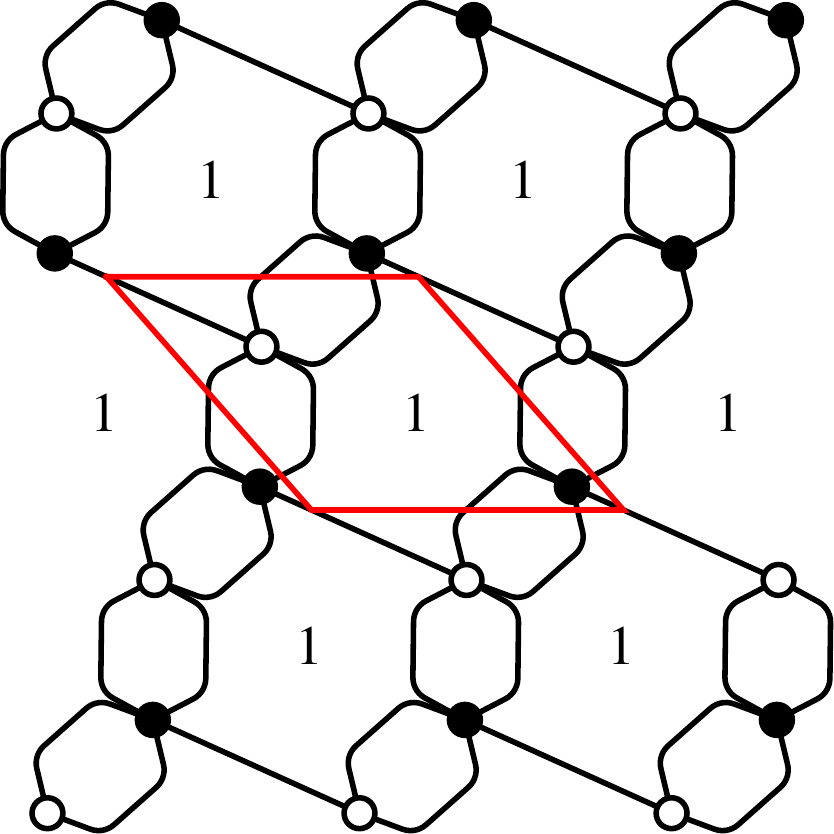}
 &
\includegraphics[width=2.5cm]{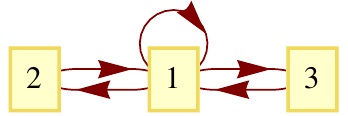}
&
 \includegraphics[totalheight=2.5cm]{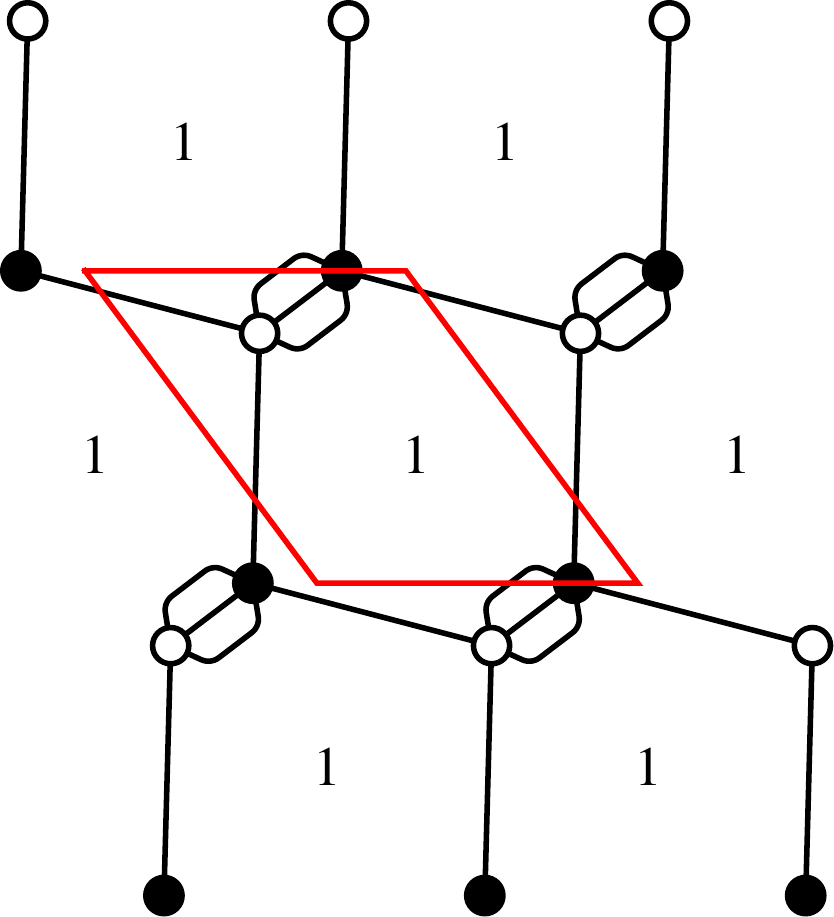}
 &
\includegraphics[width=2.5cm]{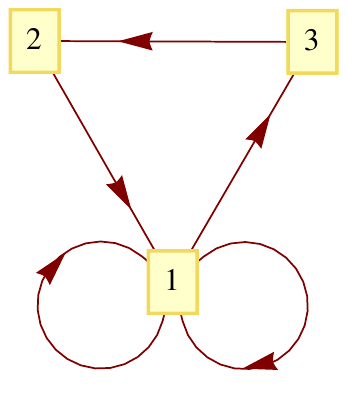}
\\
\end{tabular}
\end{center}
\caption{The 1 Hexagon Tiling and its children with at most two additional edges}
\label{t:1HexModels}
\end{table}

\subsection{Counting children using Hilbert series}

Counting the number of ways of partitioning an integer is an elementary combinatorial problem with a known solution which can be cast naturally in the language of Hilbert series. The coefficient of $t^i \nu^j$ of $g( \nu, t)$ in \eqref{eq:c3count} counts the number of ways of dividing an integer $i$ into $j$ partitions.
\beq 
g( \nu, t) = \prod_{i=0}^\infty \frac{1}{1-\nu t^i } 
\label{eq:c3count}
\eeq
Therefore the children of the 1 hexagon tiling are counted by $\mathrm{Coeff} \left( g( \nu ,t) ; \nu^3 \right)$. The coefficient of $t^k$ in the power series is equal to the number of children with $k$ additional fields \eqref{eq:c3modelcount}.

\beq
\mathrm{Coeff} \left( g( \nu ,t) ; \nu^3 \right) = \frac{1}{(1-t)(1-t^2)(1-t^3)} = 1 + t + 2 t^2 + 3 t^3 + 4 t^4 +  \ldots
\label{eq:c3modelcount}
\eeq

\subsection{Counting children using a Molien formula}

A second method of counting the children of the 1 hexagon model involves using a discrete Molien formula. This function counts the homogeneous polynomials of a given degree that are invariants a group. The key observation is that there is a one to one correspondence between these polynomials and partitions of an integer. Let us illustrate this using the 1 hexagon tiling as an example.

Suppose we have three variables $x_1$, $x_2$ and $x_3$ with an $S_3$ symmetry acting on them. One of the generators of the group acts on the variables as
\beq
x_1 \rightarrow x_2 \rightarrow x_3 \rightarrow x_1
\eeq
and the action of the second is
\beq
x_1 \leftrightarrow x_2
\eeq

We can build exactly 1 polynomial of degree 1 that is invariant under this $S_3$ symmetry, namely
\beq
x_1 + x_2 + x_3
\label{e:polyinvar1}
\eeq
There are 2 invariant polynomials of order 2
\bea
 & x_1^2 + x_2^2 + x_3^2&  \nn \\ &x_1 x_2 + x_2 x_3 + x_3 x_1&
\label{e:polyinvar2}
\eea
We can see that there is a correspondence between the polynomials above and the ways in which integers were split into partitions in \eqref{e:partitionsof1} and \eqref{e:partitionsof2}. Powers of variables in each term of the invariant polynomials correspond to the ways of partitioning i.e.
\beq
x_1^a x_2^b x_3^c \rightarrow a+b+c
\eeq

\subsubsection{An explicit Molien function}

There is an explicit formula for a generating function which counts these homogeneous polynomials. This function can be written in the form \cite{Forger:Molien}:
\beq
\frac{1}{G} \sum_{g \in G} \frac{1}{\mathrm{det}\left( \mathbb{I} - t g \right) }
\label{e:dmolien}
\eeq
Where $g \in G$ is a matrix representation of the group we are finding invariants of. Explicitly $g$ is a matrix such that 
\bea g_{ij} =
\left\{
 \begin{array}{l} 
 1 \quad \text{if} \; g \; \text{takes field} \; i \; \text{to field} \; j \\ 
 0 \quad \text{otherwise} 
 \end{array} 
 \right.
\eea
The explicit matrix representation used for $S_3$ is given in Table \ref{t:matrixs3}.

  \begin{table}
  \begin{center}
 \begin{tabular}{c|c|c}
 I & (23) & (12) 
 \\ 
 \( \left(
\begin{array}{ccc}
 1 & 0 & 0 \\
 0 & 1 & 0 \\
 0 & 0 & 1
\end{array}
\right)\)
&
\(\left(
\begin{array}{ccc}
 1 & 0 & 0 \\
 0 & 0 & 1 \\
 0 & 1 & 0
\end{array}
\right)\)
&
\(\left(
\begin{array}{ccc}
 0 & 1 & 0 \\
 1 & 0 & 0 \\
 0 & 0 & 1
\end{array}
\right)\) \\ \hline
 (132) & (123) &(13) \\
\(\left(
\begin{array}{ccc}
 0 & 1 & 0 \\
 0 & 0 & 1 \\
 1 & 0 & 0
\end{array}
\right)\)
&
\(\left(
\begin{array}{ccc}
 0 & 0 & 1 \\
 1 & 0 & 0 \\
 0 & 1 & 0
\end{array}
\right)\)
&
\(\left(
\begin{array}{ccc}
 0 & 0 & 1 \\
 0 & 1 & 0 \\
 1 & 0 & 0
\end{array}
\right)\)
\end{tabular}
\end{center}
\caption{The explicit matrix representation of $S_3$ used to count children of the 1 Hexagon tiling.}
\label{t:matrixs3}
\end{table}
 
By using the formula given in \eqref{e:dmolien} it has been possible to compute a generating function that counts children of the 1-hexagon model \eqref{e:c3molien}. We can see this exactly matches the sum we calculated previously. The term $t$ corresponds to the fact that there is only one polynomial of degree 1 and $2t^2$ corresponds to the two polynomials of degree 2.
\beq
\frac{1}{(1-t)(1-t^2)(1-t^3)} = 1 + t + 2 t^2 + 3 t^3 + 4 t^4 + 5 t^5 + 7 t^6 + 8 t^7 + 10 t^8 + 
 12 t^9 
 + \ldots
 \label{e:c3molien}
 \eeq

\begin{figure}[h!]
 \centerline{ \includegraphics[totalheight=5cm]{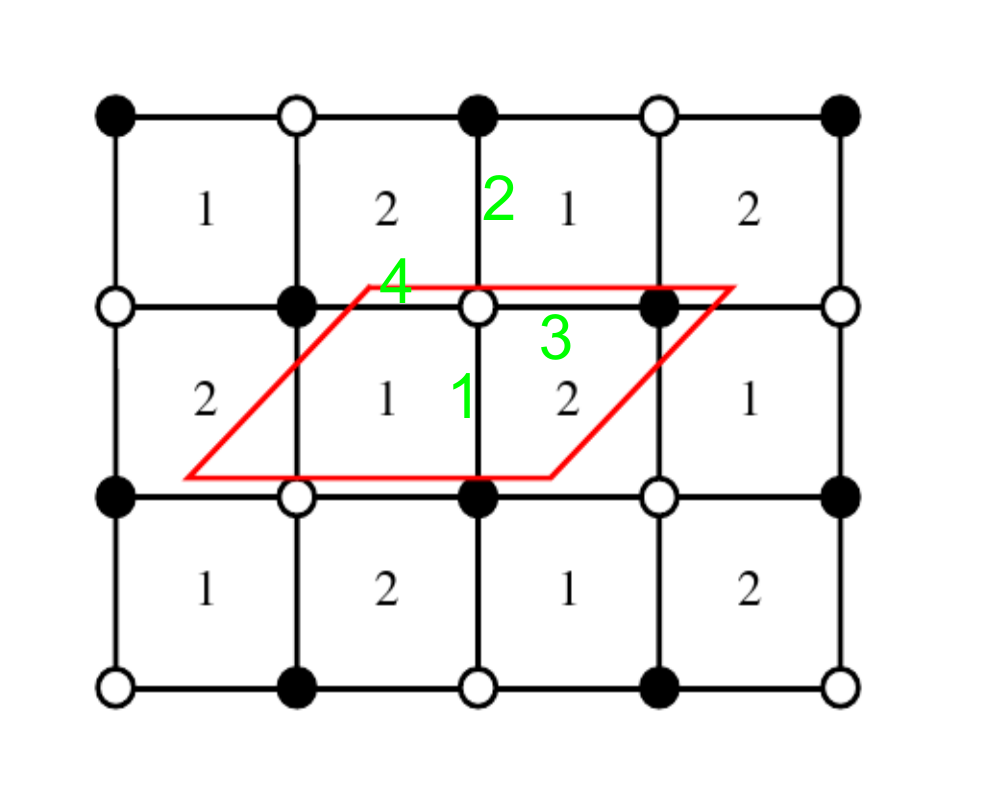}}
  \caption{The 2 Square Tiling (Fields Shown in Green)}
 \label{f:conifoldtile}
 \end{figure}

\Section{Counting children of the two square tiling} 
\label{s:conidoublings}
 Let us now attempt to count the children of the two square tiling. The 2 square tiling is given in \fref{f:conifoldtile}.\\

From analysing the two square tiling we find that the symmetry group that keeps the tiling unaltered has 2 generators. These are a reflection (corresponding to the permutation of fields $(1,2)$) and a rotation by $90^o$ (corresponding to the permutation of fields $(1,3,2,4)$).The symmetry group of the tiling is therefore the symmetry group of the square -- $D_4$.\\
 
 We can now use the Molien formula in \eqref{e:dmolien} to count the number of children of the 2 square tiling \eqref{e:conicount}. We can verify the first few terms of the partition function using Table \ref{t:2SquareModels}. There is obviously only one conifold with no doublings, and one with 1 doubling. The $3t^2$ term corresponds to a single `triple-bond' tiling and two tilings with two double bonds.\\
 \beq
 \frac{1-t^6}{(1-t) (1-t^2)^2 (1-t^3)(1-t^4) } = 1+t+3 t^2+4 t^3+8 t^4+10 t^5+16 t^6+20 t^7
+\ldots 
 \label{e:conicount}
 \eeq

\begin{table}[h!]
\begin{center}
\begin{tabular}{cc|cc}
 Tiling & Quiver & Tiling & Quiver\\
\hline \hline
 \includegraphics[totalheight=2.5cm]{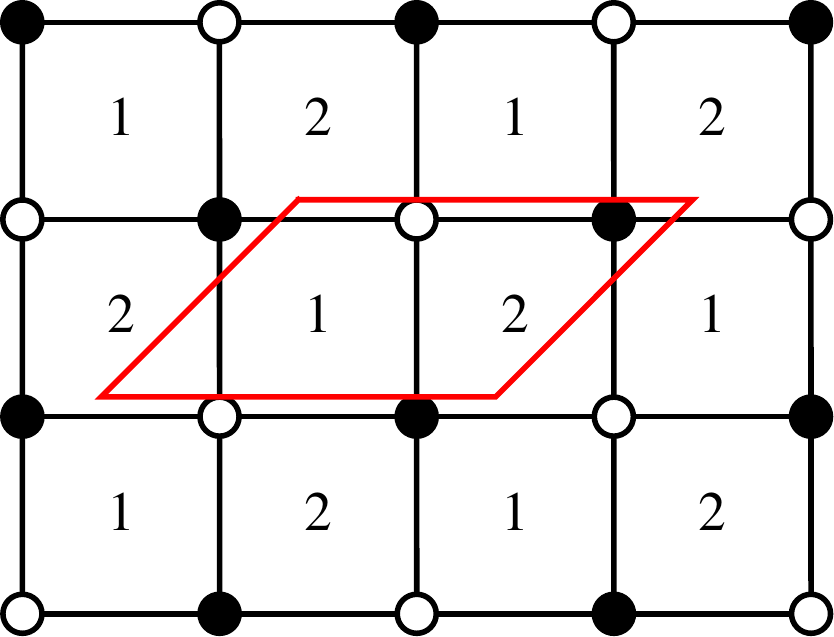}
 &
\includegraphics[width=2.5cm]{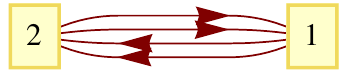}
&
 \includegraphics[totalheight=2.5cm]{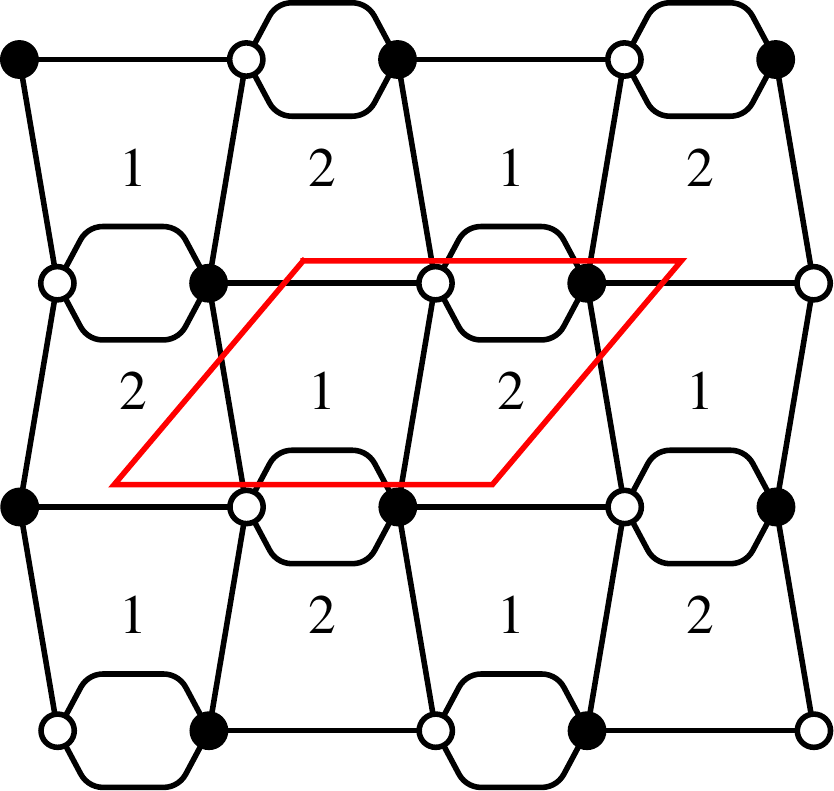}
 &
\includegraphics[width=2.5cm]{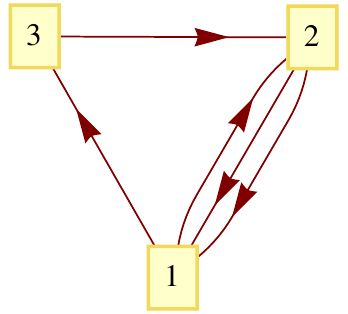}
\\
\hline
 \includegraphics[totalheight=2.5cm]{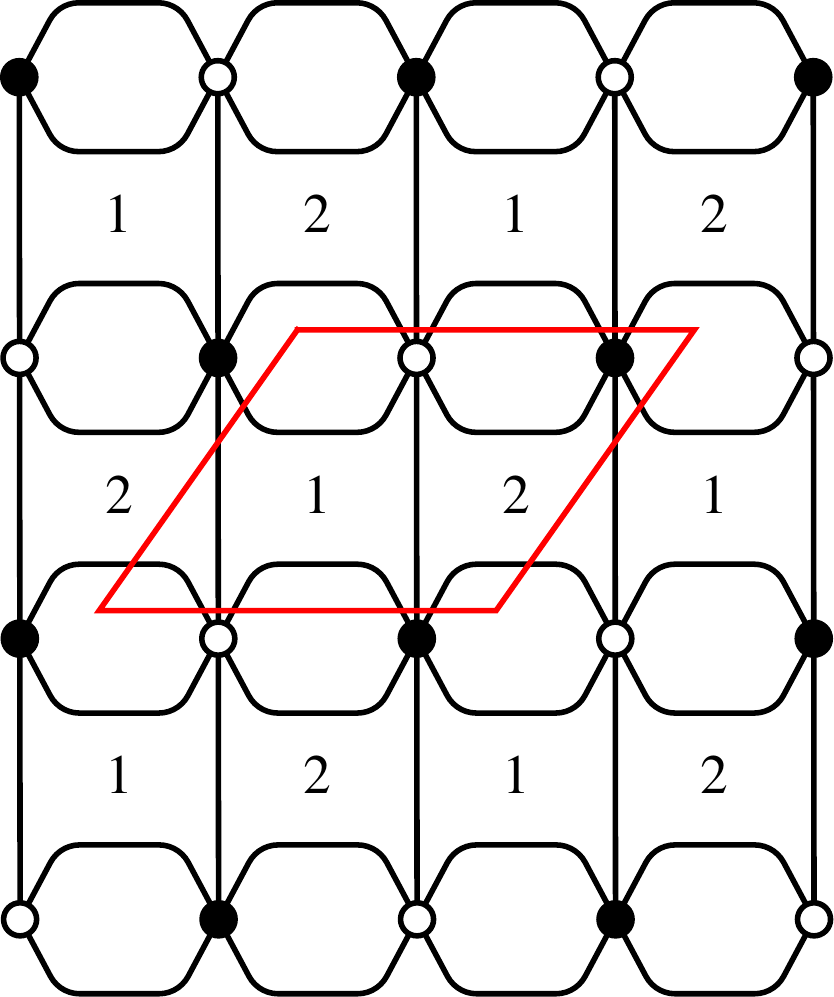}
 &
\includegraphics[width=2.5cm]{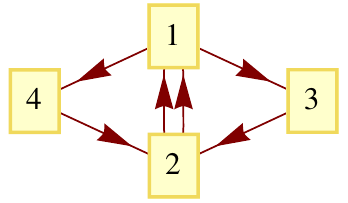}
&
 \includegraphics[totalheight=2.5cm]{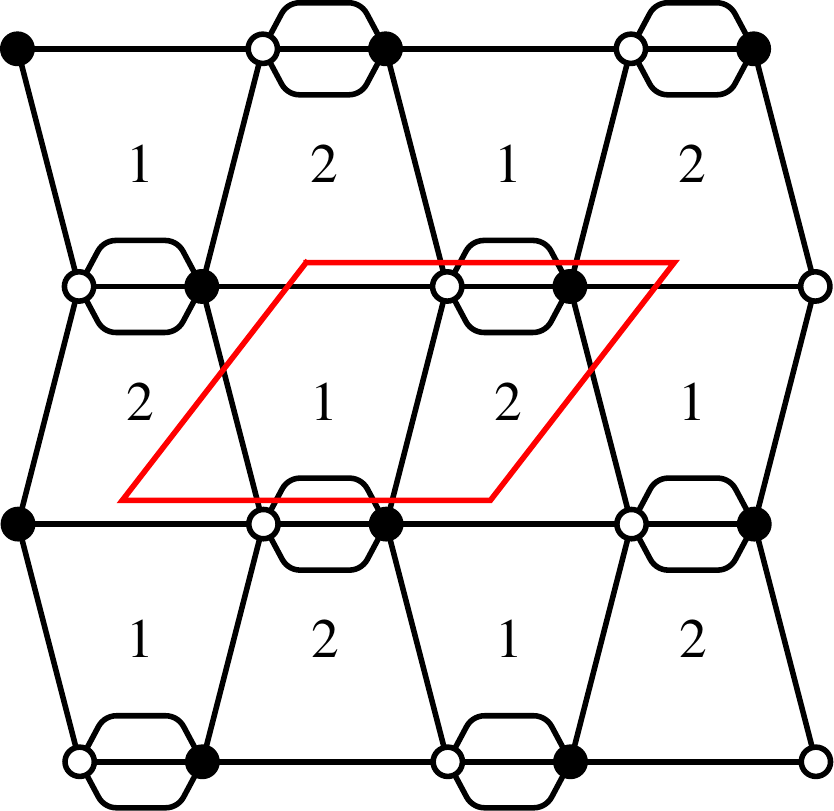}
 &
\includegraphics[width=2.5cm]{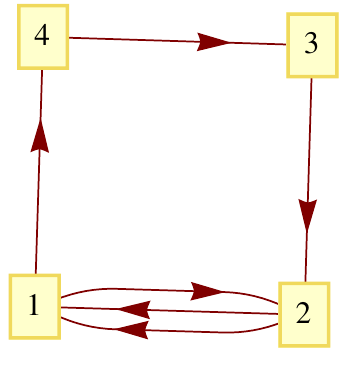}
\\
\hline
 \includegraphics[totalheight=2.5cm]{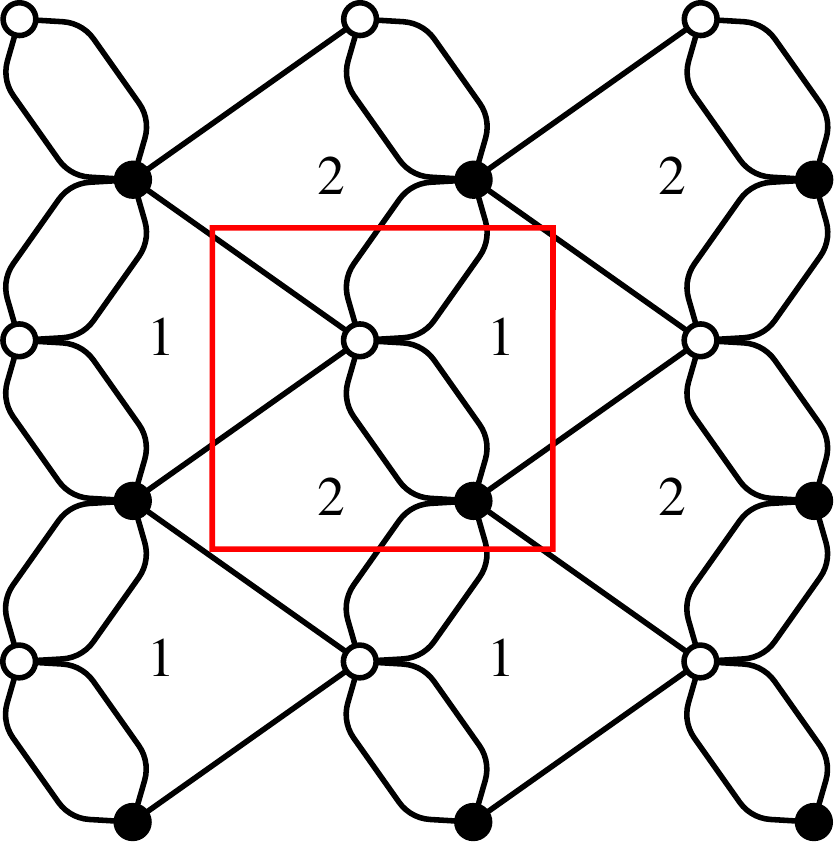}
 &
\includegraphics[width=2.5cm]{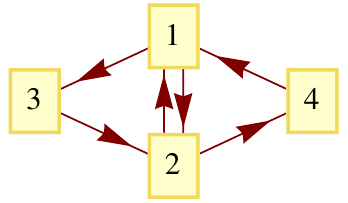} &&
\\
\end{tabular}
\end{center}
\caption{The 2 Square Tiling and its children with at most two additional edges}
\label{t:2SquareModels}
\end{table}

 \Section{Counting children of the 2 hexagon tiling} 
 
\begin{figure}[ht]
 \centerline{ \includegraphics[totalheight=3cm]{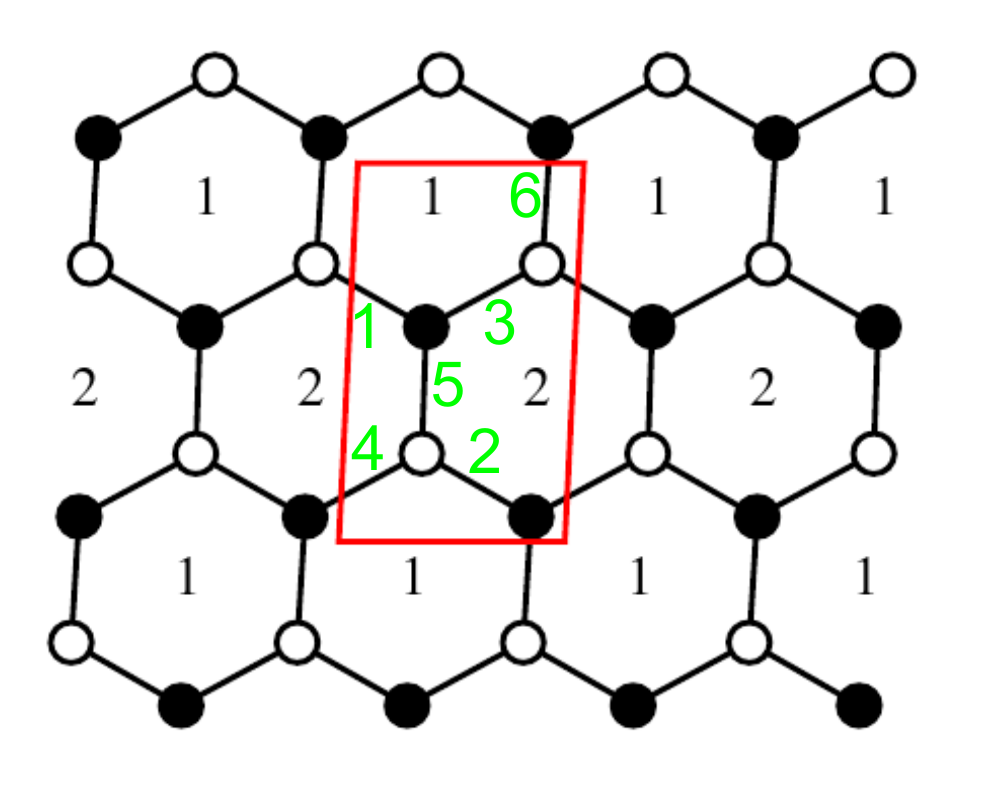}}
  \caption{The 2 Hexagon Tiling (Fields Labeled in Green)}
 \label{f:c3z2fields}
 \end{figure}

We will now attempt to count the children of the 2 hexagon tiling. The tiling has 3 different $C_2$ symmetries which can be seen in \fref{f:c3z2fields}. The first corresponds to the permutation of edges $(56)$, the second to $(13)(24)$ and the third to $(14)(23)$. It is clear that these generate a subgroup of $S_6$ and using \texttt{GAP4} \cite{GAP4} we find this subgroup to be $C_2 \times C_2 \times C_2$.\\

 As with the one hexagon model, we have used the discrete Molien formula \eqref{e:dmolien} to count children. The generating function that counts the children of the 2 hexagon tiling is given in \eqref{eq:c3z2countdoublings}
 \beq
\frac{1-t^6}{(1-t)^2(1-t^2)^4 (1-t^3)} =  1 + 2 t + 7 t^2 + 13 t^3 + 29 t^4 + 49 t^5 + 89 t^6 + 139 t^7 + 
 \ldots
 \label{eq:c3z2countdoublings}
\eeq

 The tilings of the 2 hexagon model and its children with at most 2 additional edges are given in \tref{t:2HexModels}. We can match these children with terms in the above generating function. The $2t$ term corresponds to the two children with 1 additional edge and the $7t^2$ term corresponds to the 7 children with 2 additional edges.

\begin{figure}[h!]
 \epsfxsize = 4cm
 \centerline{ \includegraphics[totalheight=3cm]{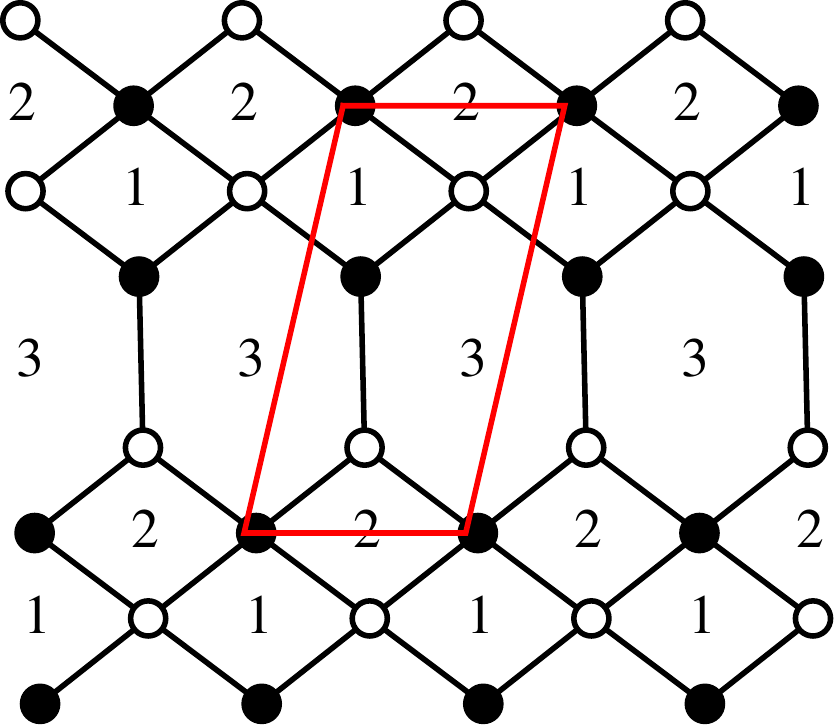}}
  \caption{The Suspended Pinch Point (SPP) Tiling}
 \label{f:SPPOnly}
 \end{figure}

\begin{table}[h]
\begin{center}
\begin{tabular}{c|c||c|c}
 Tiling  & Quiver & Tiling &Quiver\\
\hline \hline
 \includegraphics[totalheight=2cm]{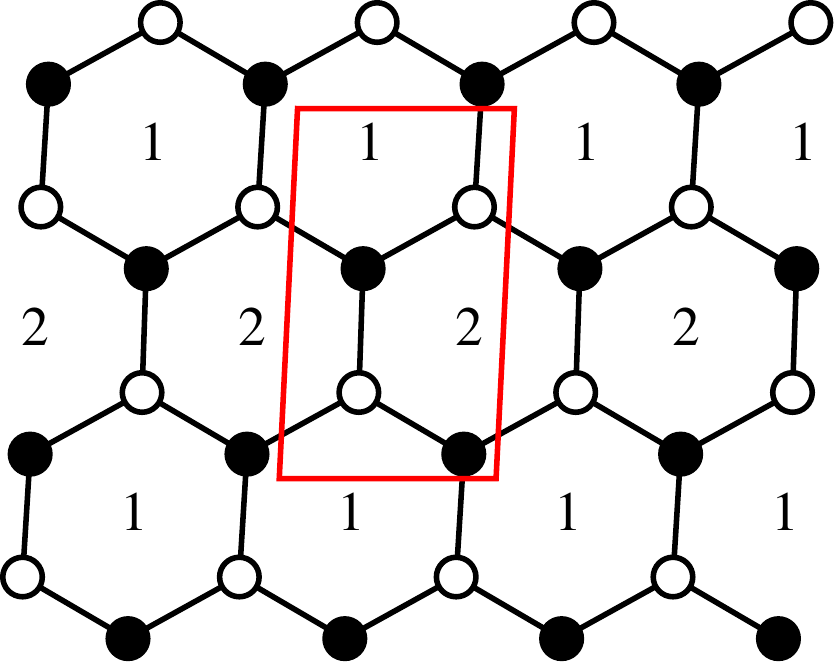}
 &
\includegraphics[width=2cm]{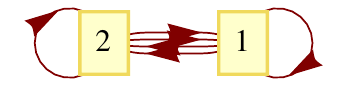}
&
 \includegraphics[totalheight=2cm]{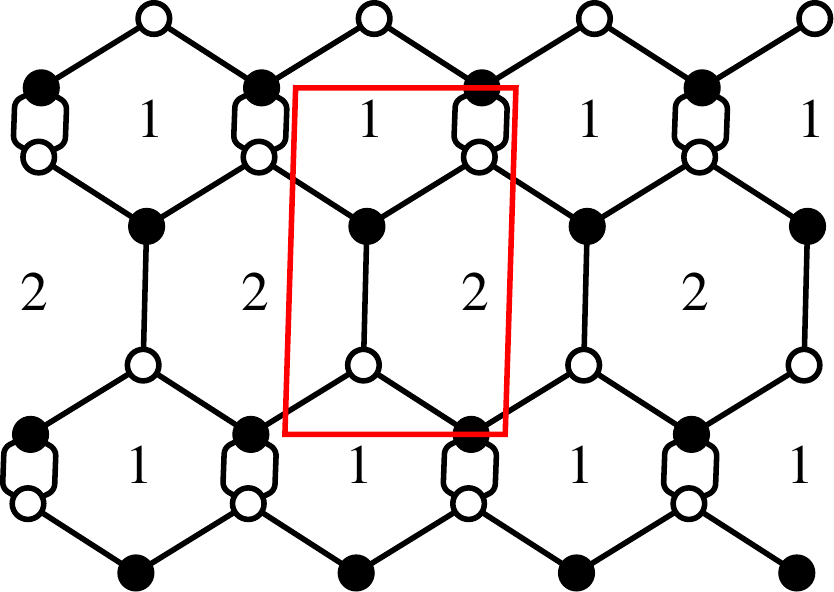}
 &
\includegraphics[width=2cm]{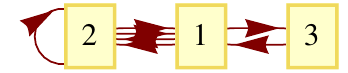}
\\
\hline
 \includegraphics[totalheight=2cm]{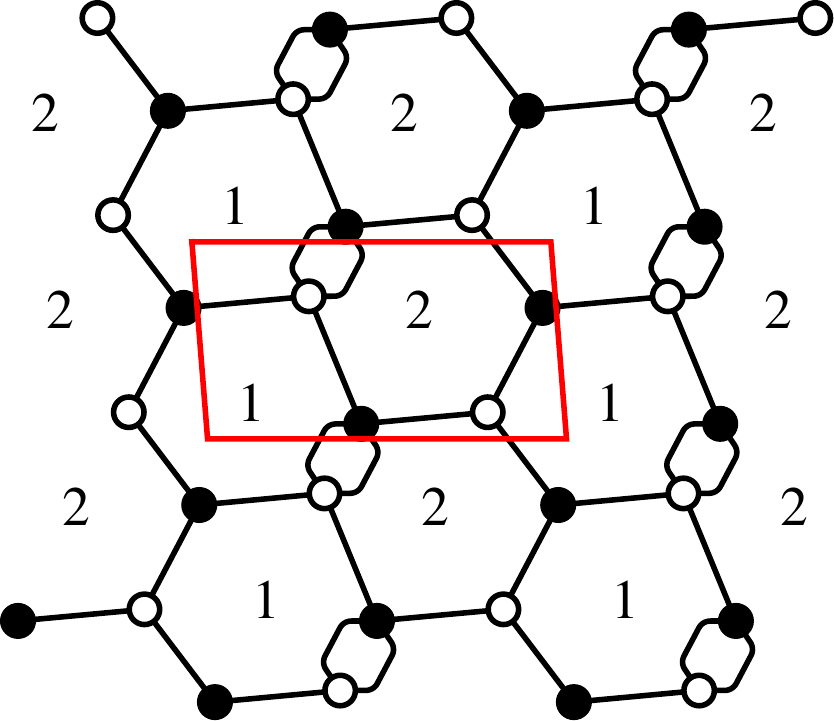}
 &
\includegraphics[width=2cm]{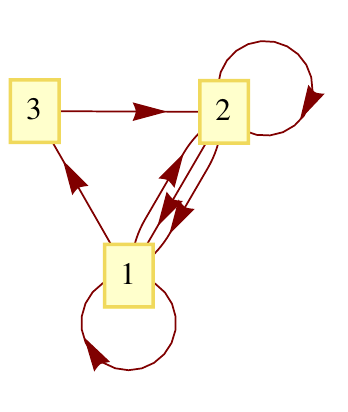}
&
 \includegraphics[totalheight=2cm]{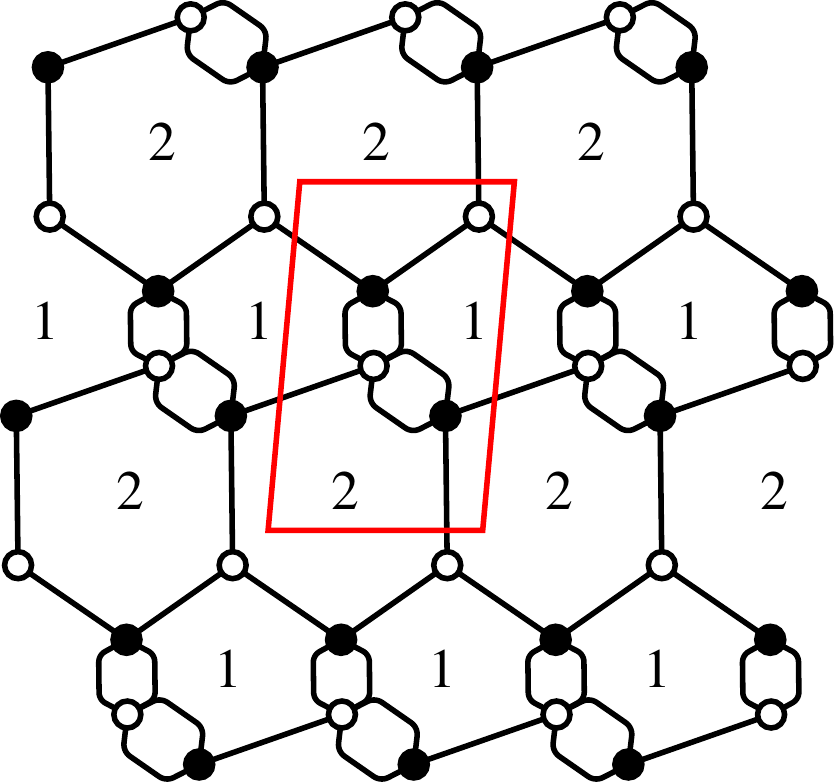}
 &
\includegraphics[width=2cm]{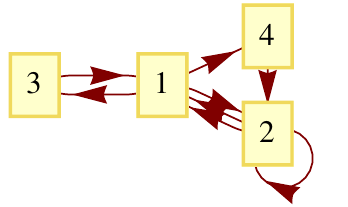}
\\
\hline
 \includegraphics[totalheight=2cm]{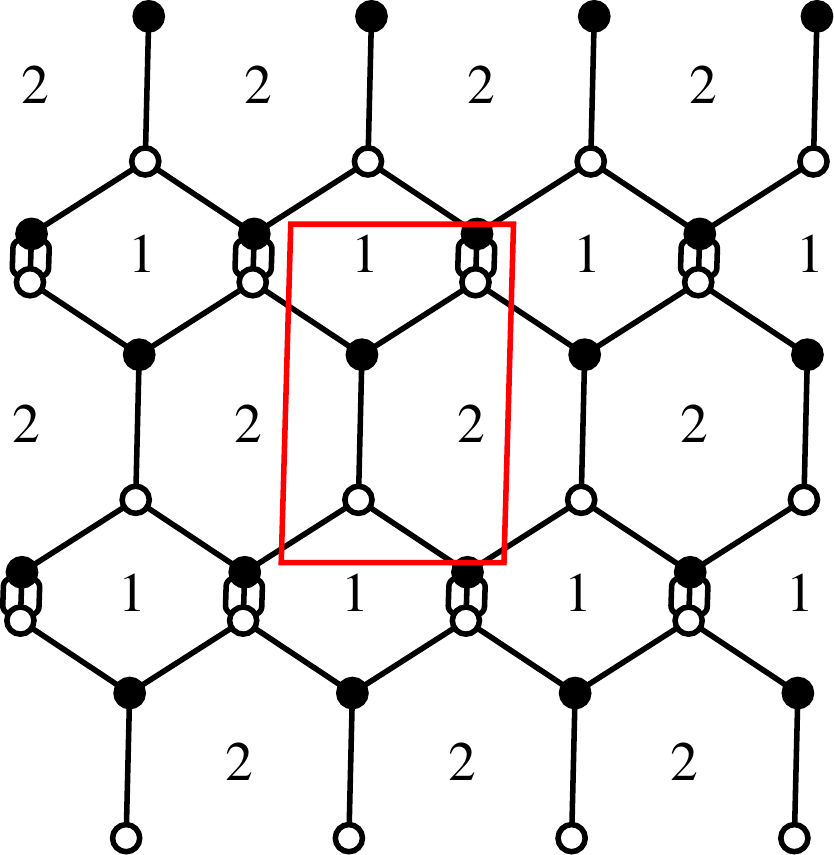}
 &
\includegraphics[width=2cm]{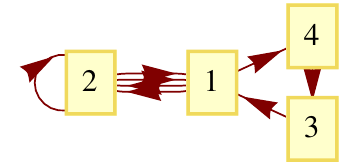}
&
 \includegraphics[totalheight=2cm]{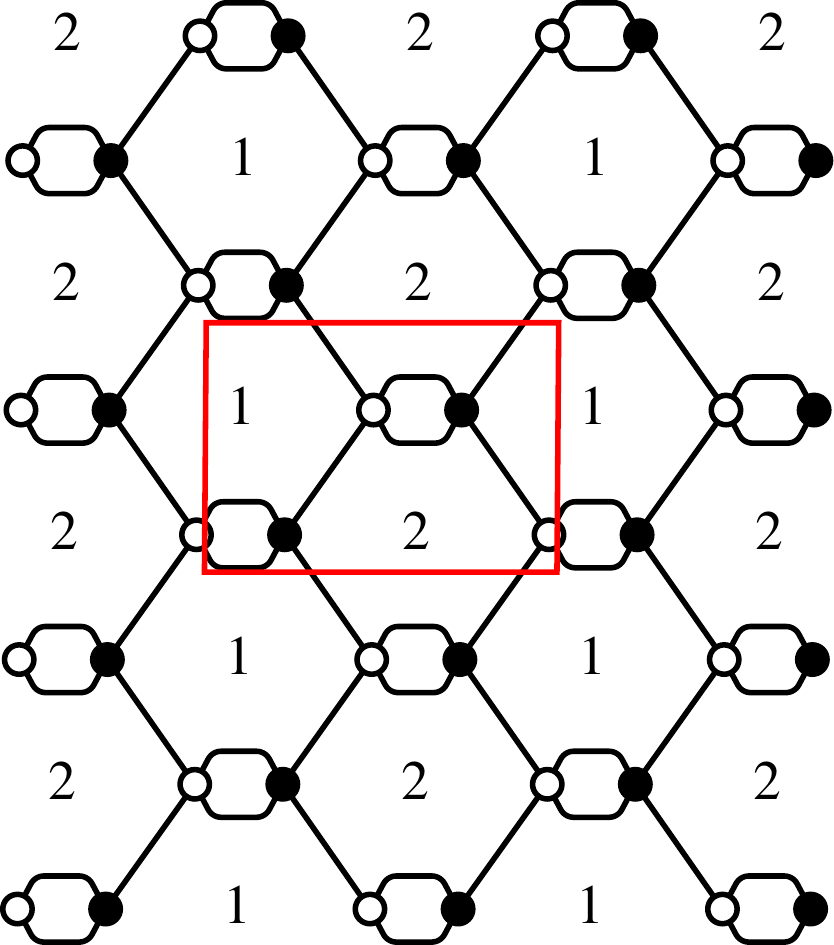}
 &
\includegraphics[width=2cm]{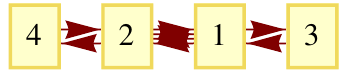}
\\
\hline
 \includegraphics[totalheight=2cm]{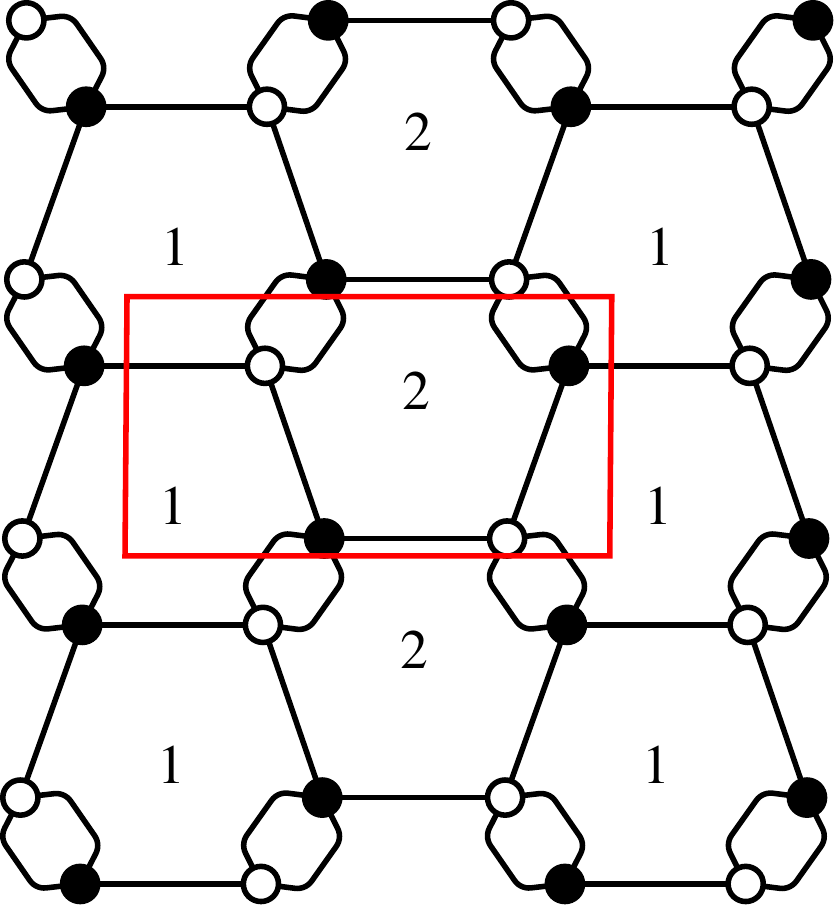}
 &
\includegraphics[width=2cm]{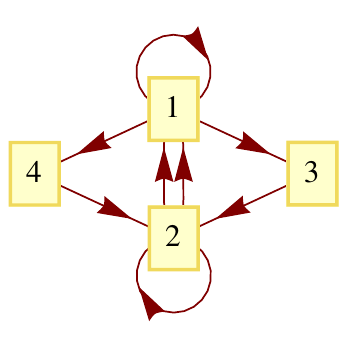}
&
 \includegraphics[totalheight=2cm]{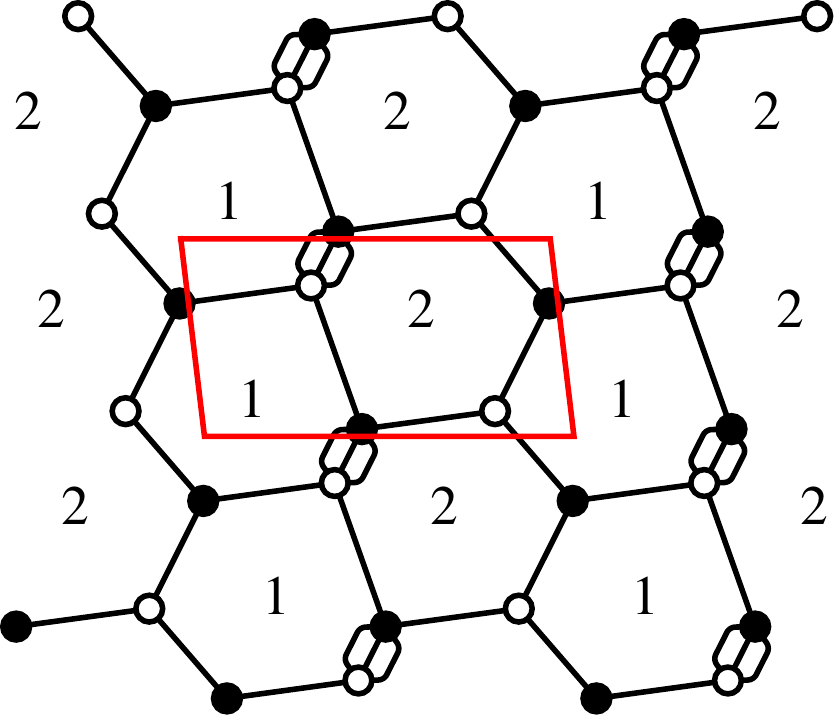}
 &
\includegraphics[width=2cm]{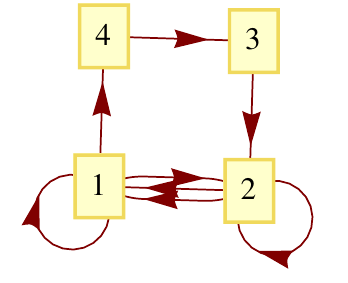}
\\
\hline
 \includegraphics[totalheight=2cm]{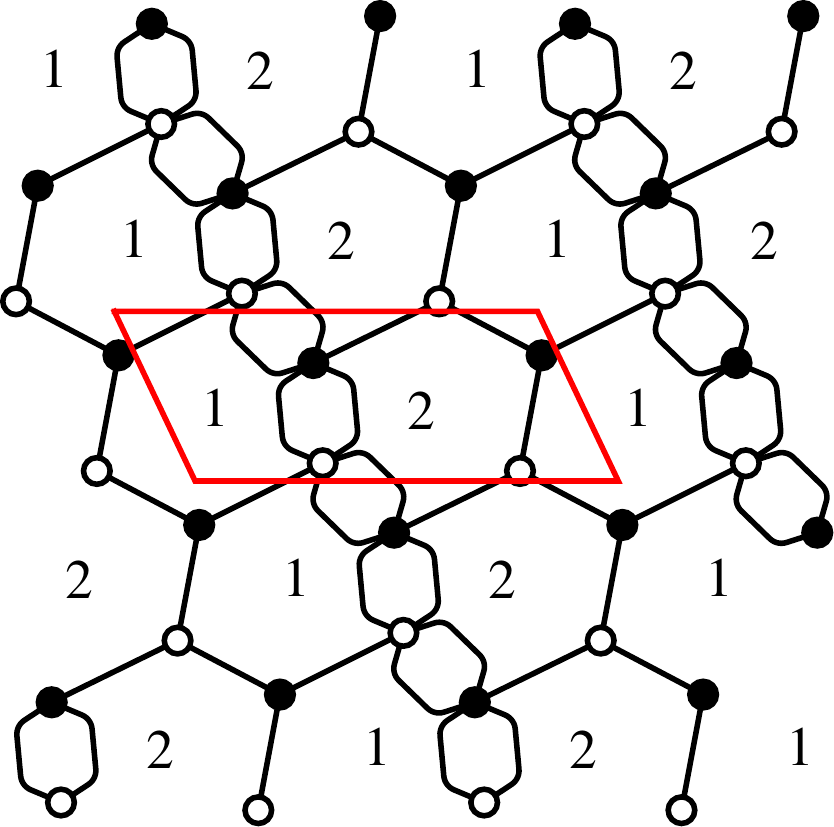}
 &
\includegraphics[width=2cm]{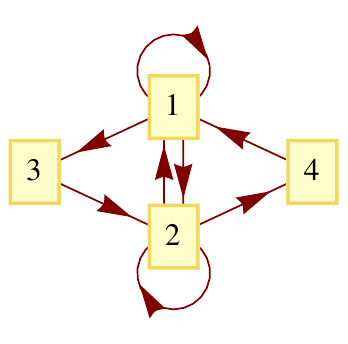}
&
 \includegraphics[totalheight=2cm]{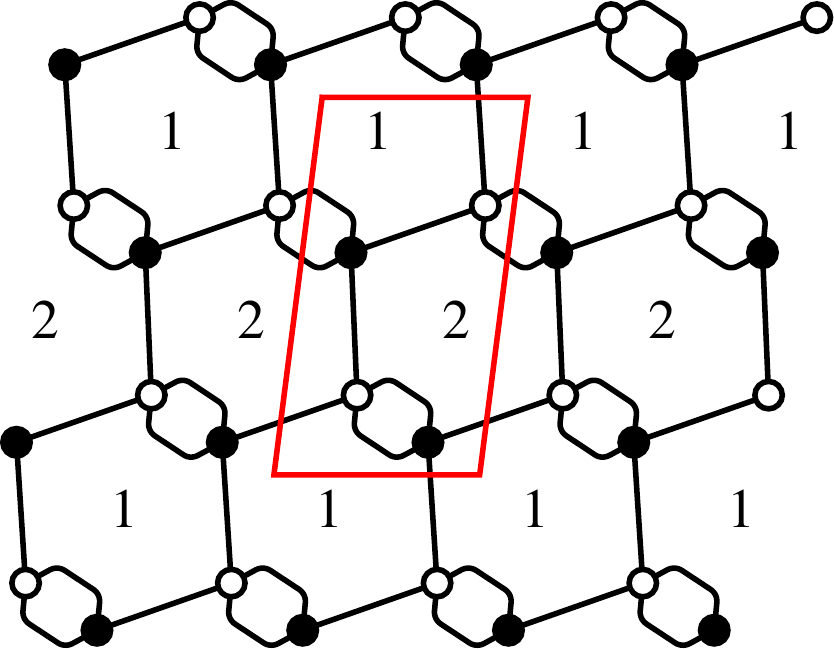}
 &
\includegraphics[width=2cm]{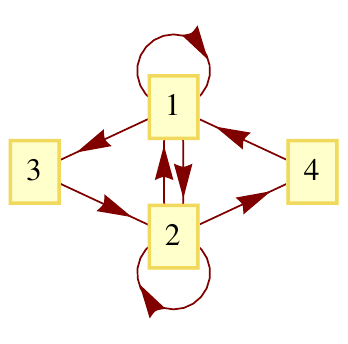}
             \end{tabular}
      \end{center}      
             \caption{The 2 Hexagon Tiling and its children with at most two additional edges}
\label{t:2HexModels}
 \end{table}
  \Section{Counting children of the 1 hexagon 2 square (or Suspended Pinch Point) tiling} 
 Let us now attempt to count the children of the 1 hexagon and 2 square tiling given in \fref{f:SPPOnly}. This model is also known as the Suspended Pinch Point (or SPP for short).

From analysing the SPP tiling we find that the symmetry group that keeps the tiling unaltered has 2 generators. The first corresponds to a horizontal reflection, the second to a rotation. The symmetry of the tiling has been found to be $C_2 \times C_2 $.

 We can use the Molien formula in \eqref{e:dmolien} to count the number of children of the SPP tiling \eqref{e:sppcount}. We can verify the first few terms of the partition function using Table \ref{t:SPPModels}. We can see that the 3 children with 1 additional edge correspond to the $3t$ term and the 11 children with 2 additional edges correspond to the $11t^2$ term.

  \beq
   \frac{1-t+2t^2}{(1-t)^4(1-t^2)^3} = 1+3 t+11 t^2+27 t^3+65 t^4+133 t^5+261 t^6+469 t^7+812 t^8+1330 t^9 + \ldots
   \label{e:sppcount}
   \eeq
 
\begin{table}[h]
          \begin{center}
\begin{tabular}{c|c||c|c}
 Tiling  & Quiver & Tiling &Quiver\\
\hline \hline
 \includegraphics[totalheight=2cm]{Images/SPP/Tiling-SPP.pdf}
 &
\includegraphics[width=2cm]{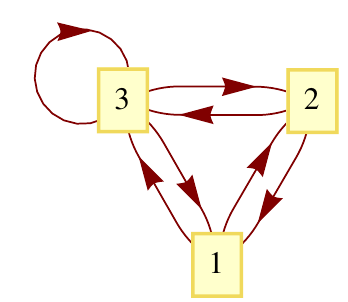}
&
 \includegraphics[totalheight=2cm]{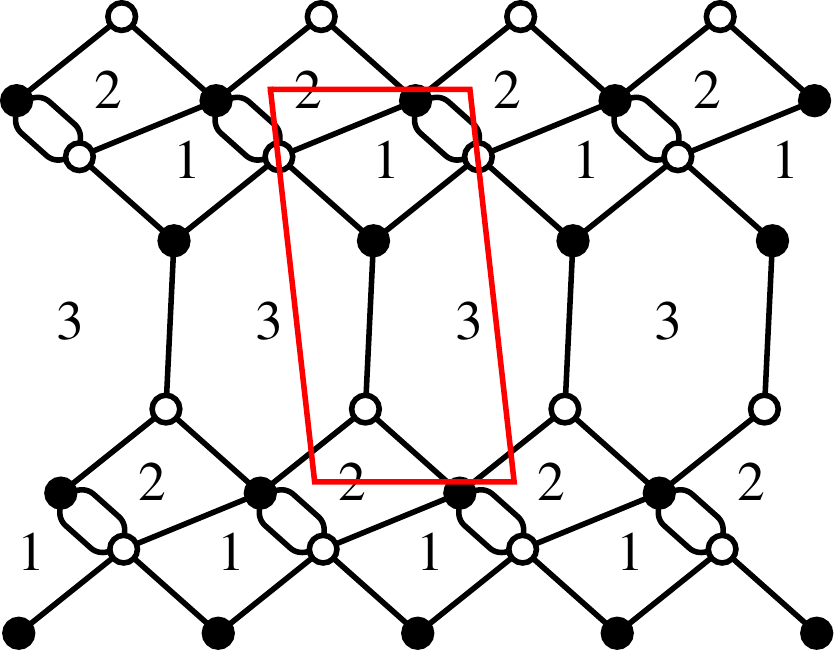}
 &
\includegraphics[width=2cm]{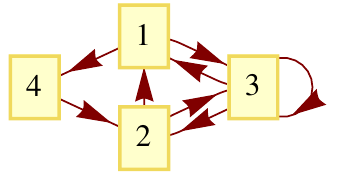}
\\ \hline
 \includegraphics[totalheight=2cm]{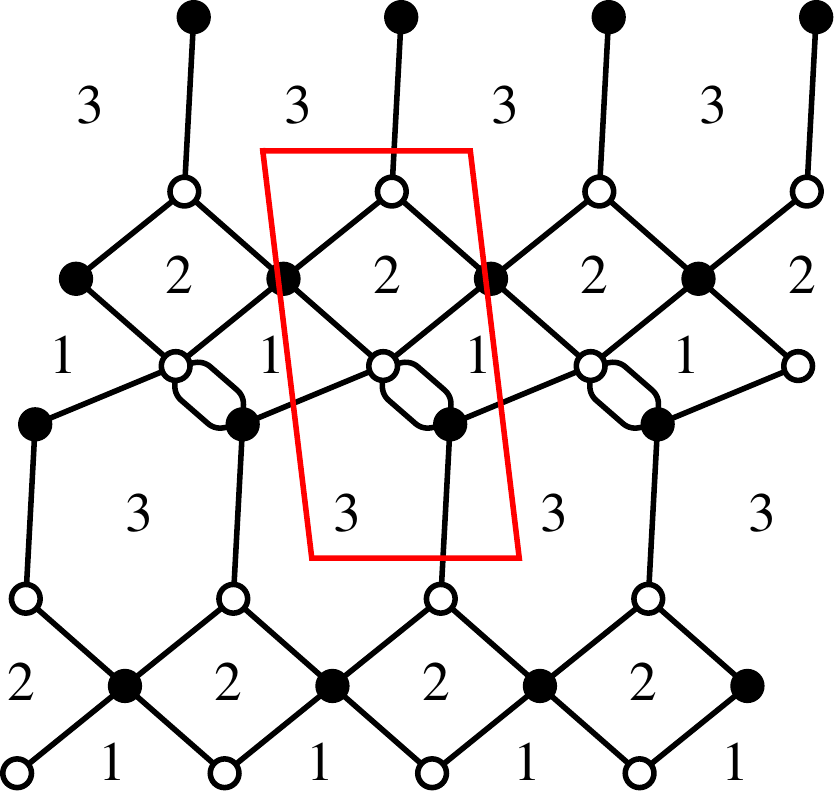}
 &
\includegraphics[width=2cm]{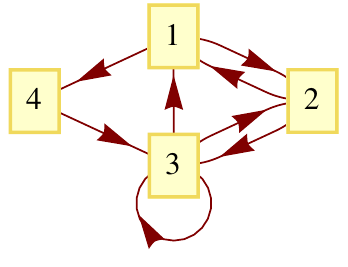}
&
 \includegraphics[totalheight=2cm]{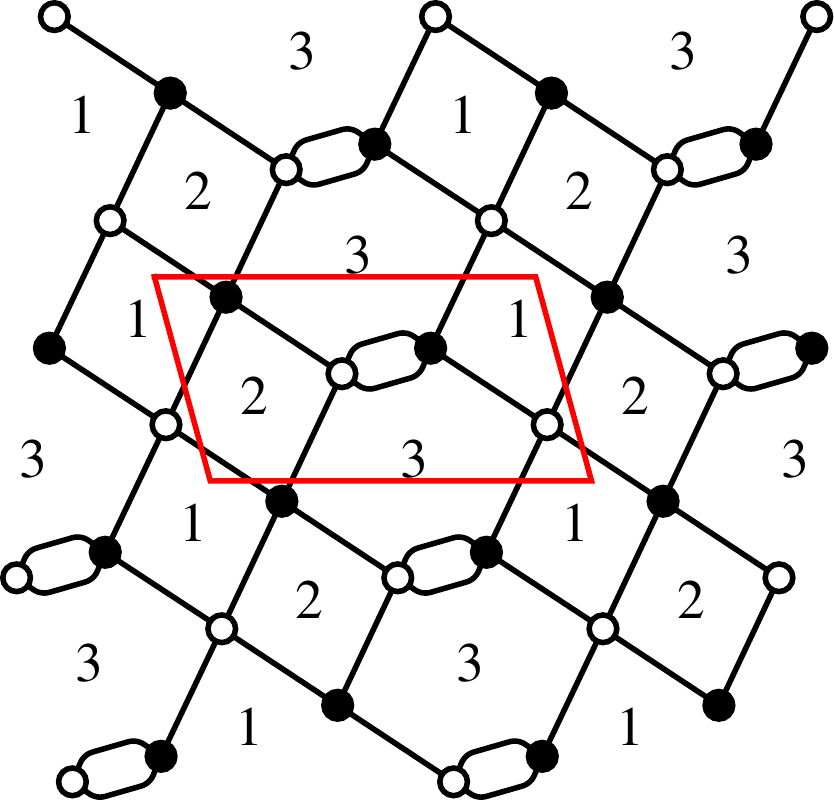}
 &
\includegraphics[width=2cm]{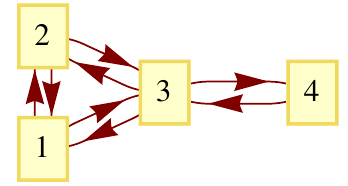}
\\ \hline \includegraphics[totalheight=2cm]{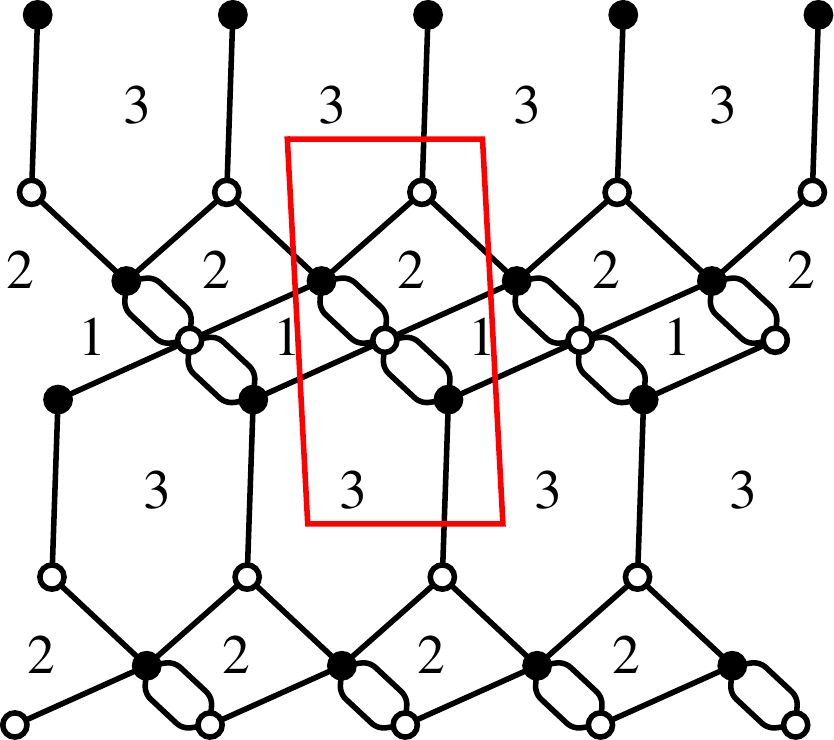}
 &
\includegraphics[width=2cm]{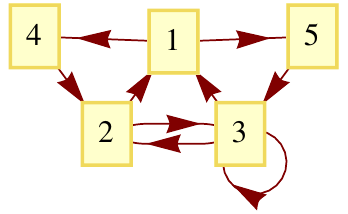}
&
 \includegraphics[totalheight=2cm]{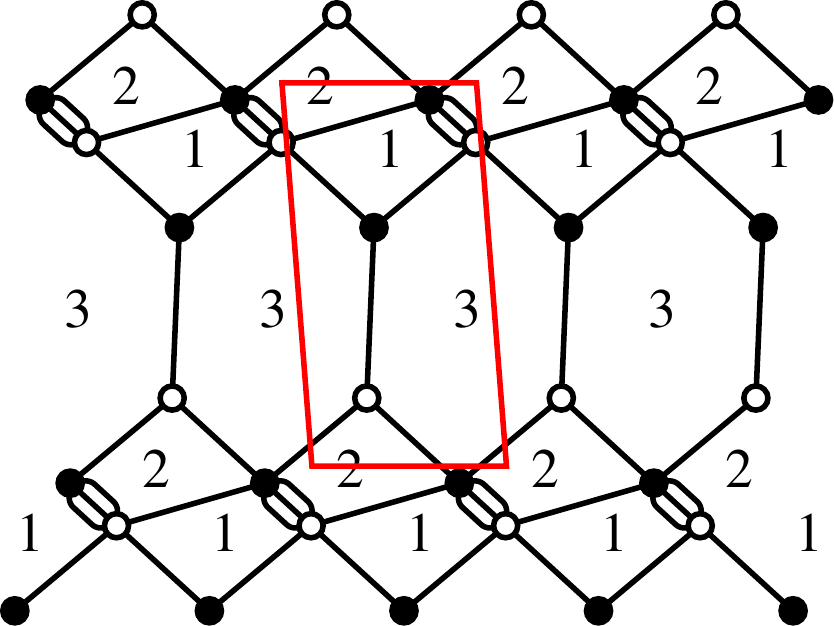}
 &
\includegraphics[width=2cm]{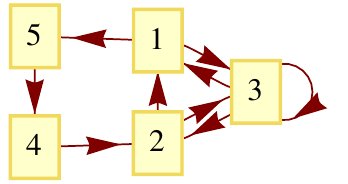}
\\ \hline \includegraphics[totalheight=2cm]{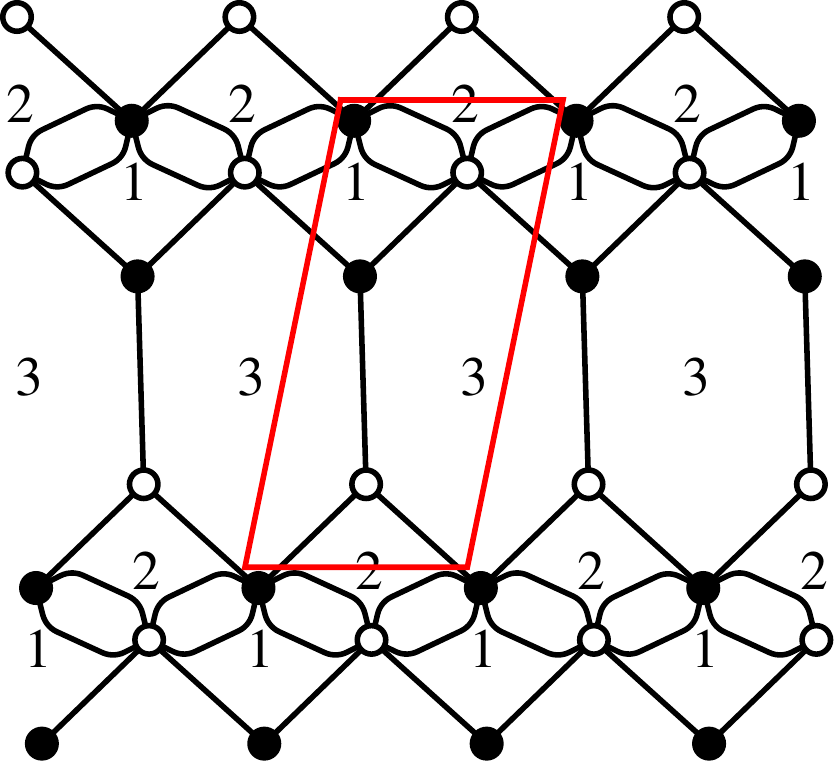}
 &
\includegraphics[width=2cm]{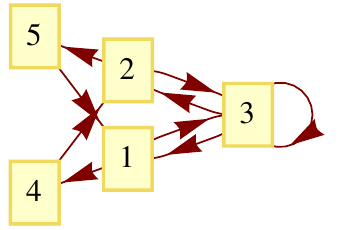}
&
 \includegraphics[totalheight=2cm]{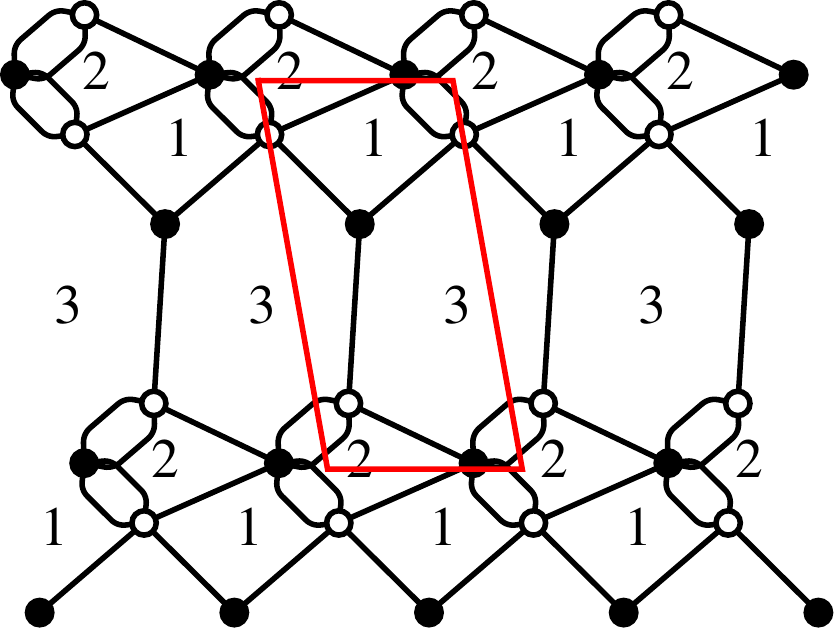}
 &
\includegraphics[width=2cm]{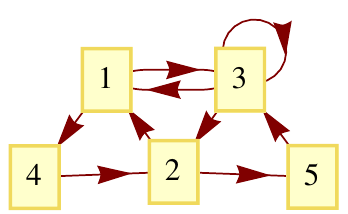}
\\ \hline \includegraphics[totalheight=2cm]{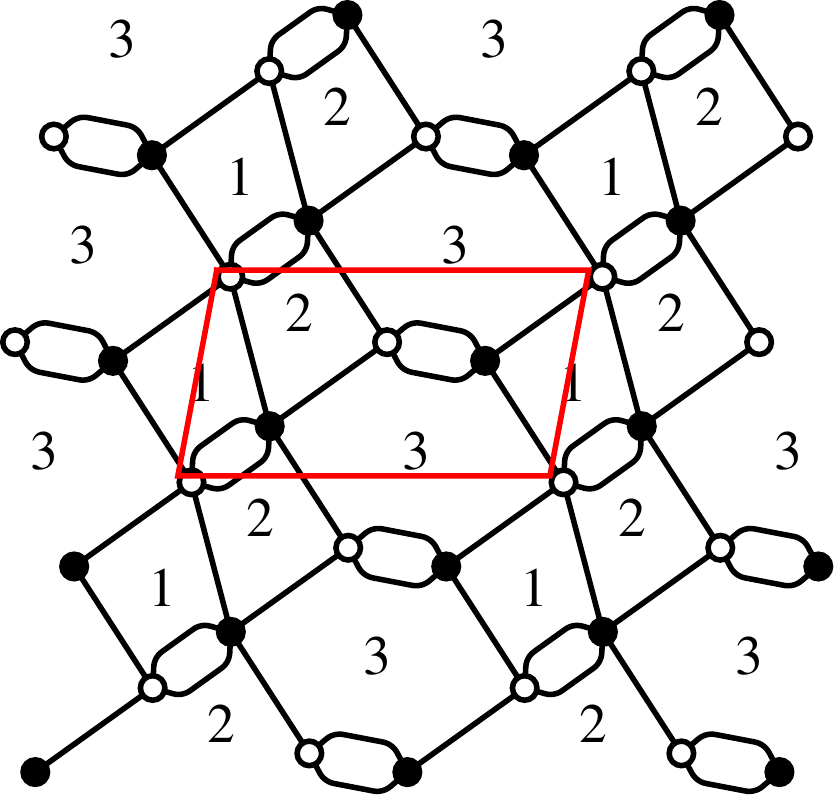}
 &
\includegraphics[width=2cm]{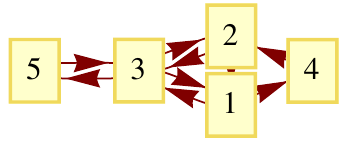}
&
 \includegraphics[totalheight=2cm]{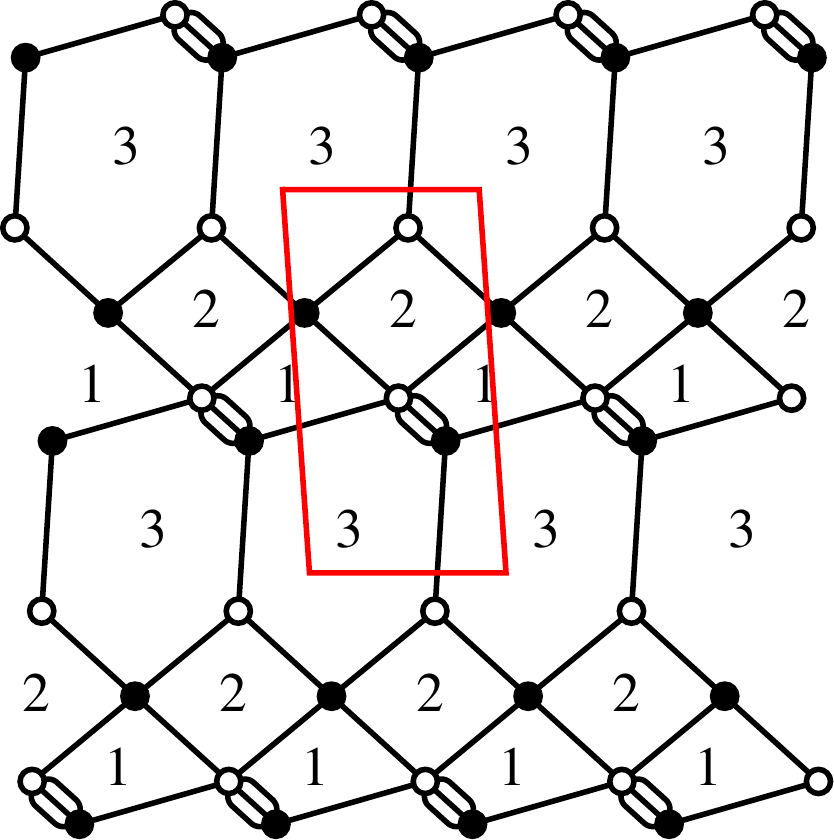}
 &
\includegraphics[width=2cm]{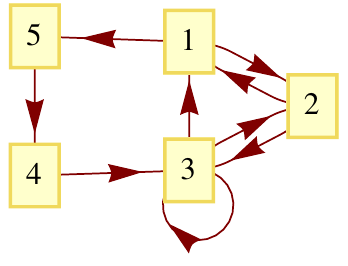}
\\ \hline \includegraphics[totalheight=2cm]{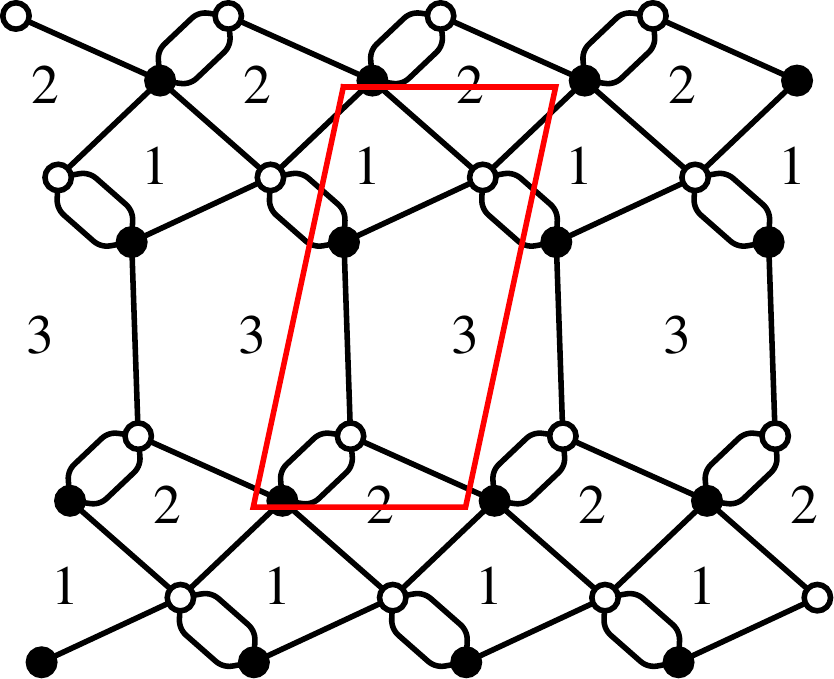}
 &
\includegraphics[width=2cm]{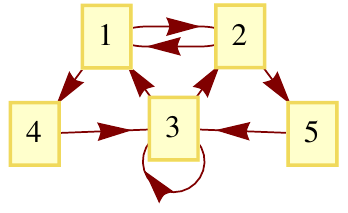}
&
 \includegraphics[totalheight=2cm]{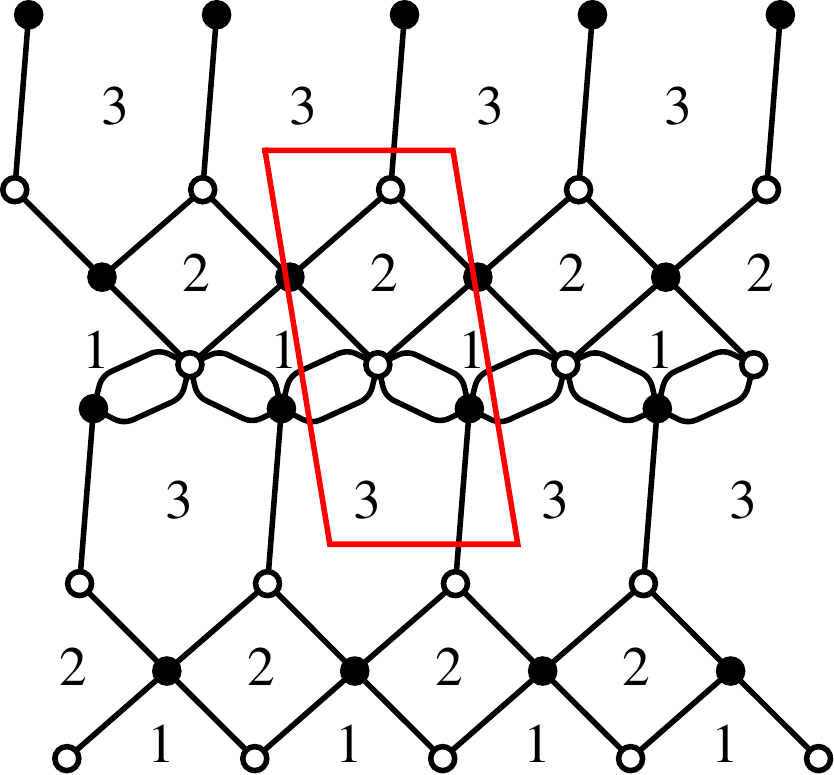}
 &
\includegraphics[width=2cm]{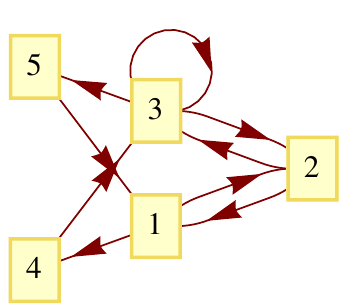}
\\ \hline \includegraphics[totalheight=2cm]{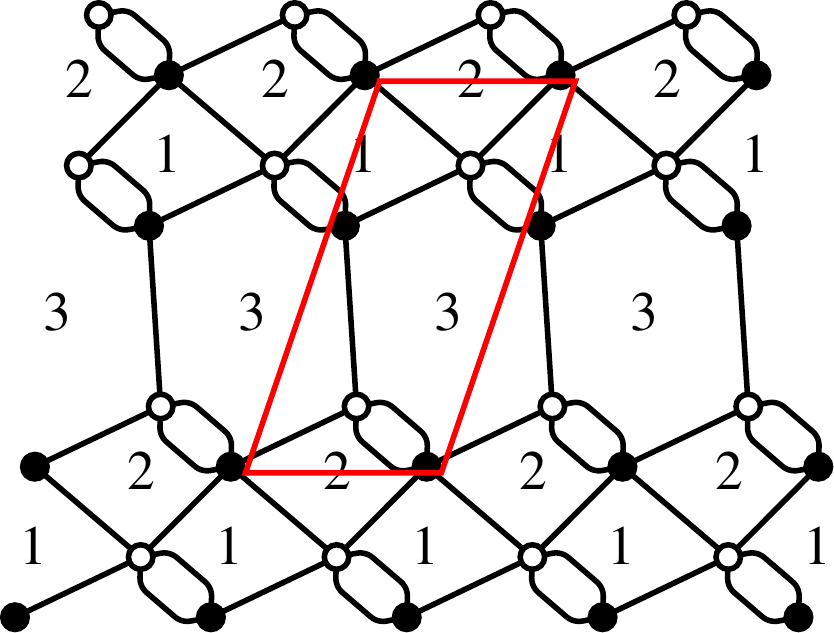}
 &
\includegraphics[width=2cm]{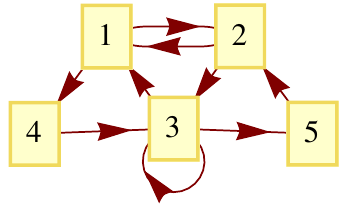}
&
 \includegraphics[totalheight=2cm]{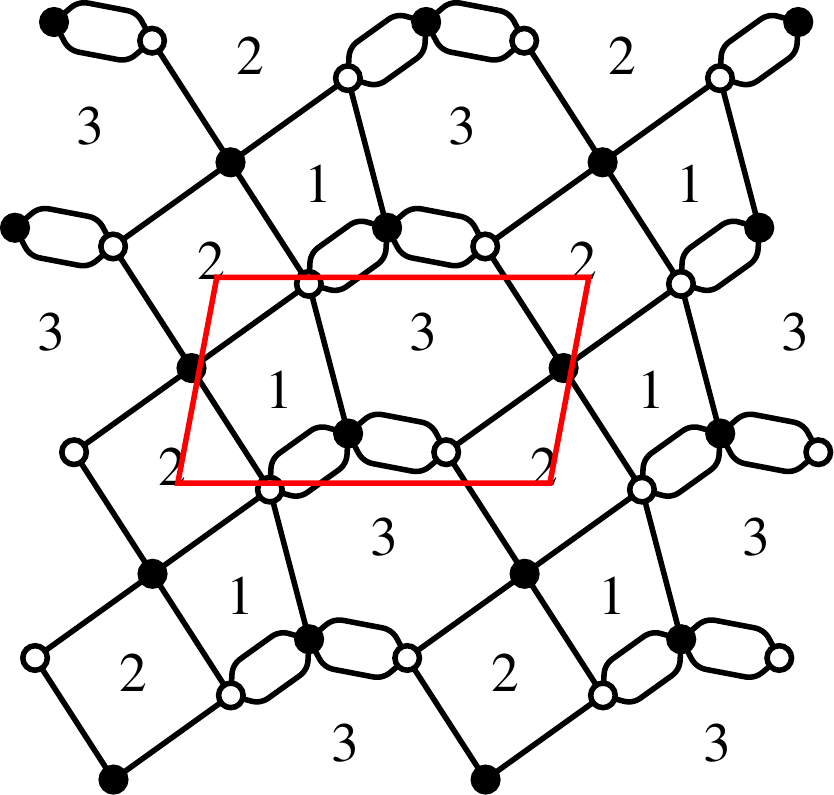}
 &
\includegraphics[width=2cm]{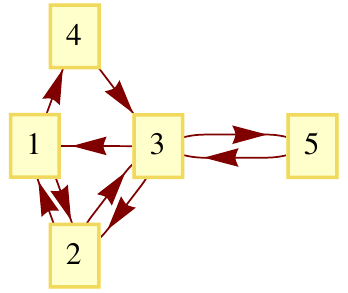}
\\ \hline \includegraphics[totalheight=2cm]{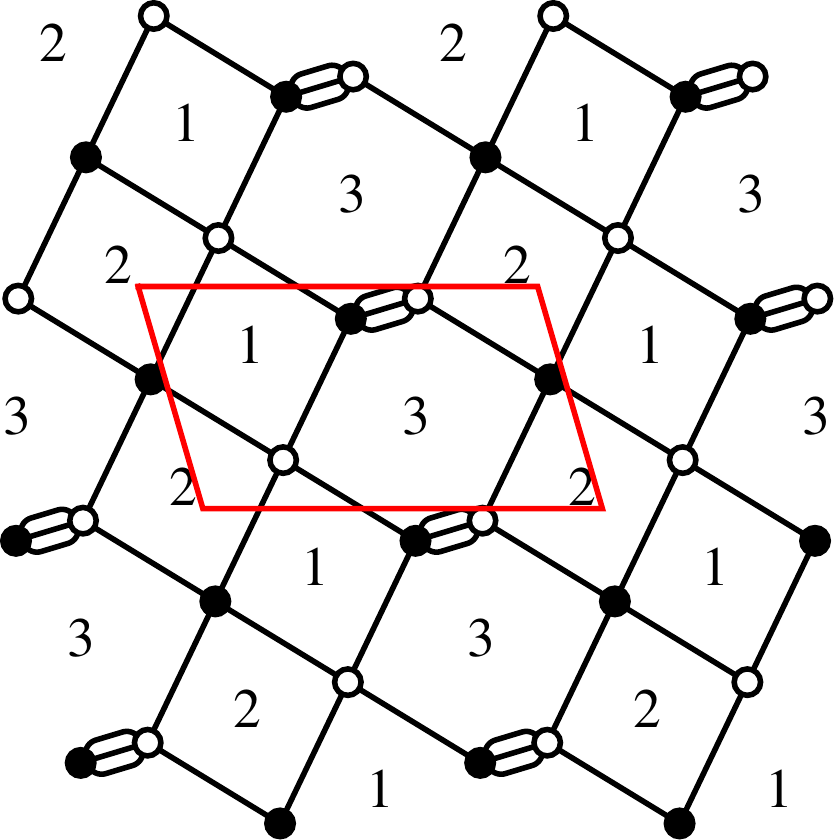}
 &
\includegraphics[width=2cm]{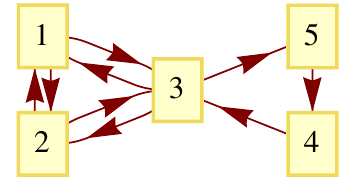}
&&
             \end{tabular}
\end{center}
             \caption{The SPP tiling and its children with at most two additional edges}      
\label{t:SPPModels}
 \end{table}  

\Section{Further work}

It is possible to count the children of any brane tiling using the tilings symmetry and the discrete Molien function. All that one needs to do is follow the procedure outlined below:

\begin{itemize}
 \item Identify the symmetry of the parent tiling.
\item Find the action of the symmetry on the $n$ edges of the tiling.
\item Write every element of the symmetry group as an $n \times n$ matrix, just as was done for $S_3$ in \tref{t:matrixs3}.
\item Use the discrete Molien formula which was given in \eqref{e:dmolien} to compute the generating function which counts the children of the parent tiling.
\end{itemize}

\chapter{Conclusion and Outlook}
Let us now conclude and discuss directions for future research.

\Section{Classification of brane tilings}

In chapter \ref{ch:btandd3}, the concept of a brane tiling was introduced and it was shown how it has been possible to generate all tilings with at most 8 superpotential terms. As has been mentioned already, these tilings can be found in Appendix \ref{a:tilings}. Tilings which fail the 3+1 dimensional consistency condition (see section \ref{ss:consist}) have been included as they are thought to be useful for defining Chern--Simons theories which can be used as world--volume theories of M2-branes.

The fact that we have been able to generate so many tilings shows the strength of the classification algorithm that has been developed. In total the algorithm has allowed us to generate close to 400 tilings using an ordinary desktop computer. Sadly we have failed to generate all tilings with 10 superpotential terms.

It might be possible to generate all brane tilings with 10 (and possibly more) superpotential terms by using an alternative tiling generation algorithm. While the method discussed earlier in this thesis involved generating quiver gauge theories, the new algorithm would involve adding edges to a template tiling.

\begin{figure}[h!]
\begin{center}
\includegraphics[width=5cm]{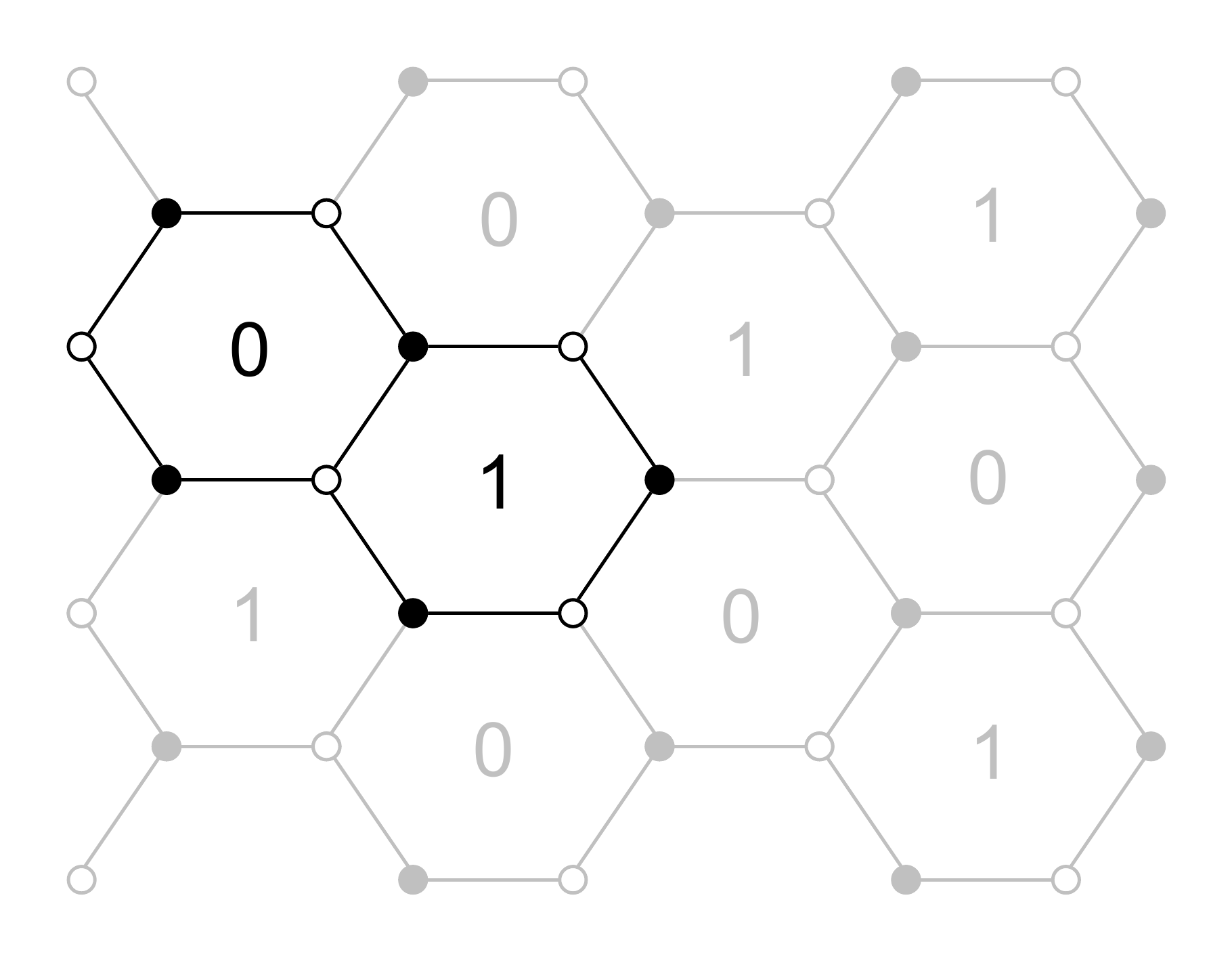}
\hspace{1cm}
\includegraphics[width=5cm]{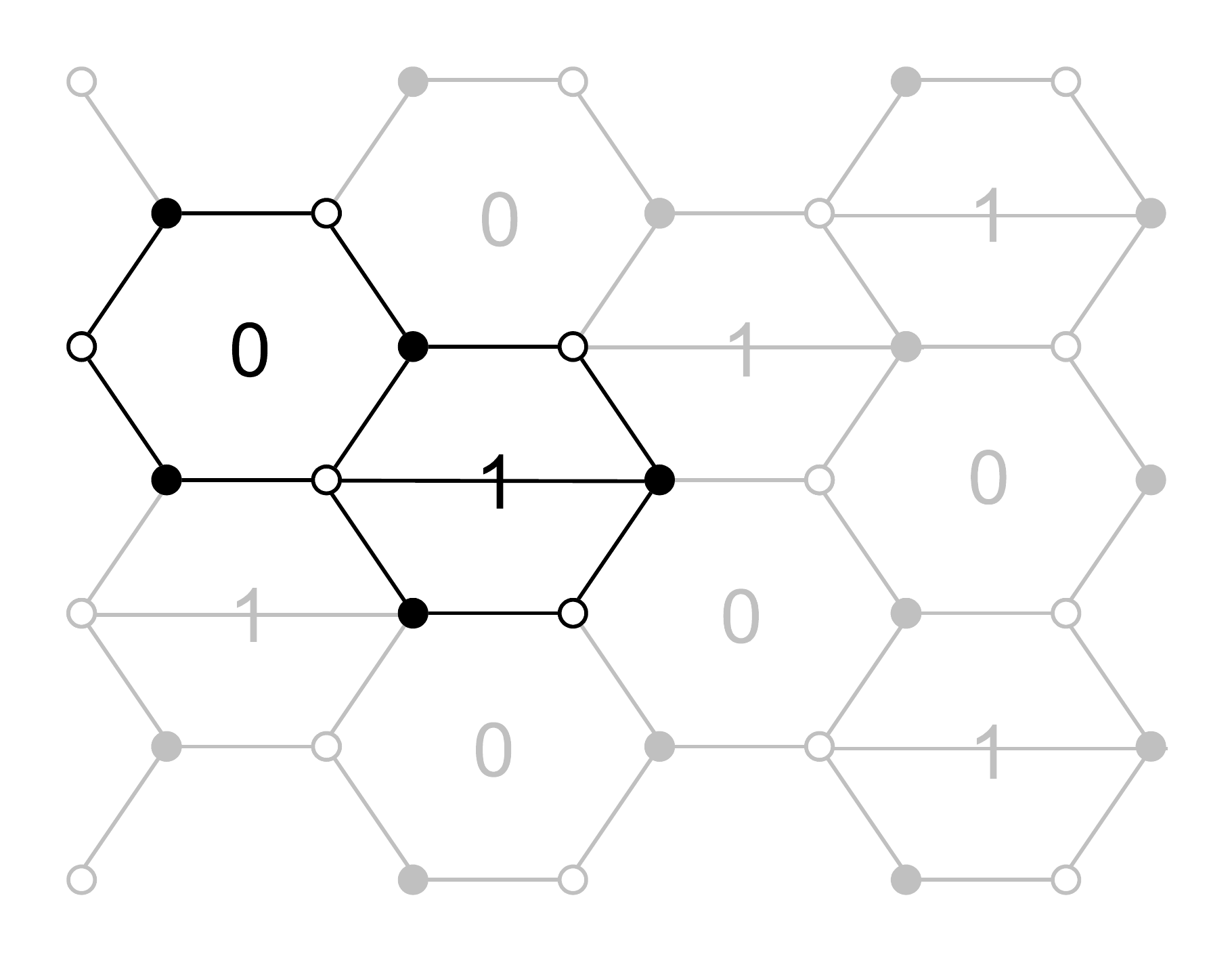}
\end{center}
\caption{Adding a diagonal edge to the 2 hexagon tiling}
\label{f:AddingFieldTo2Hex}
\end{figure}

It is possible to generate all of the tilings with four superpotential terms by considering the two hexagon tiling as a template. We can start with this template tiling and add an edge as a diagonal to a hexagon. This process is demonstrated in \fref{f:AddingFieldTo2Hex}. It has been found that all tilings with 4 superpotential terms can be reproduced by adding edges across the faces of the two-hexagon model. We find that there are two ways of adding a diagonal to one
of the hexagons, which give the models with three gauge groups -- (2.2) and (2.3) (see
Appendix \ref{a:tilings}). If we add a 2nd diagonal to the tilings we find the remaining three tilings
with four gauge groups. This procedure of finding the tilings by adding diagonals also
works in the trivial case of two superpotential terms. The conifold model can be thought of as the one-hexagon tiling with a diagonal.

We may wonder whether all tilings with 6 superpotential terms can be generated by adding diagonals to one of the 3 hexagon `template' tilings. Unfortunately this is not possible as there is a tiling with an octagonal face. It may be possible that we could generate all of these tilings by adding diagonals to a different template and it could be interesting to look into this idea further. The hope is that this idea could allow us to generate more complex tilings without the need for greater computational power.

\Section{Counting Orbifolds}

In chapter \ref{ch:orbs}, three different methods of counting abelian CY orbifolds of $\BC^3$ were discussed. The first method was to encode the action of an abelian group on $\BC^3$ using a set of 3-vectors. The second method was to use toric data (triangles on a $\BZ^2$ lattice) to count the orbifolds.
The third and final method discussed was to count brane tilings formed from only hexagons. These three different methods have given an identical counting of abelian CY orbifolds of $\BC^3$ of order 50 or less.

A generating function that counts the orbifolds of $\BC^3$ was given. The key to understanding this function is by using the cycle index and Burnside's lemma \cite{HananyOrlando10}. It has been found that these tools can be used to find generating functions that count orbifolds of the conifold, orbifolds of $L^{aba}$ and also orbifolds of higher dimensional spaces such as $\BC^4$, $\BC^5$ and $\BC^6$ \cite{RakNew}.

A question that remains unanswered is whether it is possible to find a generating function that counts brane tilings according the number of superpotential terms. The success with generating functions for orbifolds is certainly a step towards this goal, although it is unclear how to make further progress in this direction.

\Section{Brane Tilings and M2-branes}

The relationship between brane tilings and M2-branes was discussed in chapter \ref{ch:phases}. The forward algorithm for M2-branes was explained and implemented for a few simple tilings. Two different tilings (with CS levels) were found to correspond to M2-branes in flat space. As the theories have the same mesonic moduli space, we call them `toric dual'. Three phases of \conxc were investigated.

In section \ref{Sec:Higgs} connections between different M2-brane theories were established via the Higgs mechanism. In particular, the three phases of \conxc were Higgsed to the phases of $\BC^4$. The Higgs mechanism shows one of the strengths of the brane tiling. Giving a VEV to a field reduces to the simple operation of removing an edge from the tiling.

Chern--Simons theories that correspond to 14 of the smooth toric fano 3-folds were found in chapter \ref{Ch:Fanos}. These theories were constructed by using an inverse algorithm for M2-branes which relies on the projection of 3--dimensional toric data to 2--dimensional toric data and then forming CS theories from tilings that correspond to the 2--dimensional toric data when viewed as D3-brane theories.

The current inverse algorithm for M2-branes can be used for simple M2-brane models quite easily, however it does have issues. The first problem is that the algorithm involves projections of 3--dimensional toric data in all possible ways and this quickly becomes computationally expensive. A second failure is that there is no guarantee that there is a CS theory with a tiling description that corresponds to some toric CY 4-fold. We found no Chern--Simons theories corresponding to Fanos 4, 35, 36 and 37 which highlights this issue. 

A direction for future research would be to find exactly the class of toric CY 4--folds that correspond to brane tilings with CS levels. Currently the only way of checking for a model is by using the projection method given here. A second direction would be to investigate further M2-branes probing toric fano 4 ($\BC^4 / \BZ_4$). It would be interesting to see whether one could form a worldvolume theory of M2-branes that has this geometry as its mesonic moduli space. It is expected that this theory would have no brane tiling description. It is not even clear whether this theory would be a quiver Chern--Simons theory.

\Section{Counting Children of Brane Tilings}

In chapter \ref{Ch:Children} we discussed how it is possible to count the children of an `irreducible' parent tiling. We illustrated how it is possible to use the symmetry of a parent tiling to form a generating function that counts its children according to the number of fields that are added to the tiling. Explicit generating functions that count children of the 1 hexagon tiling, the 2 square tiling, the 2 hexagon tiling and the SPP (1 hexagon, 2 square) tiling were given.

It is definitely possible to develop the idea of counting children of a parent tiling further. Firstly, it is theoretically possible to count the children of every tiling given in Appendix \ref{a:tilings}. In order to do this efficiently, a better way of finding the symmetry of a generic brane tiling should be developed. Currently this is done by eye and by identifying the group using GAP \cite{GAP4}.

A second way in which it might be possible to extend the idea of counting children is to find generating functions that count children of the orbifolds of some base tiling. A first task would be to count the children of all tilings that can be formed using only hexagons and to see whether there is any pattern in this series of generating functions.
\chapter*{Acknowledgements}
This content of this thesis simply could not have been produced without the help of some very kind and exceptionally talented individuals.

Firstly I would like to thank my supervisor Amihay Hanany who has given me great guidance over the past few years. It has also been a pleasure working with Noppadol $(\Omega)$ Mekareeya, Jurgis Pasukonis, Rak-Kyeong Seong and Giuseppe Torri and I would like to wish them all every success in the future. I would like to mention that the results published in `On the Classification of Brane Tilings' \cite{Davey:2009bp} were used by Jurgis in order to fulfill the requirements for the degree of Master of Science in Theoretical Physics of Imperial College London in the Summer of 2009.

I would like to say how much of a pleasure it has been being a member of the Theoretical Physics group at Imperial College. I'd like to thank everyone in the group for allowing me to have such an enjoyable time at the college.

There are many people in the wider Theoretical Physics community that I would like to thank. I'd like to explicitly thank Yang-Hui He and Bogdan Stefanski for help with an application for an EPSRC postdoctoral position.

I am also very grateful for the Science and Technology Facilities Council (STFC) studentship which I have held for almost 4 years.
\appendix
\chapter{A Catalogue of Brane Tilings}
\label{a:tilings}
In this appendix, we present all brane tilings with at most 8 superpotential terms. The brane tilings are presented along with the toric diagram and an identification number (\#). For the tilings with 2 and 4 superpotential terms the common name of the 3+1 dimensional theory that the tiling corresponds to as well as the quiver diagram are presented.

\Section{Tilings with two superpotential terms}

\begin{table}[ht]
\begin{center}

\end{center}
\caption{Tilings with 2 superpotential terms}
\label{t:tilings2}
\end{table}


\Section{Tilings with four superpotential terms}


\begin{table}[h]

\begin{center}

%

\end{center}

\caption{Tilings with 4 superpotential terms and 2 gauge groups}
\label{t:tilings4-2}
\end{table}

\FloatBarrier

\begin{table}[ht!]

\begin{center}
%
\end{center}

\caption{Tilings with 4 superpotential terms and 3 or 4 gauge groups}
\label{t:tilings4-4}
\end{table}


\Section{Tilings with six superpotential terms}
\begin{table}[h]
\begin{center}
%
\end{center}
\caption{Tilings with 6 superpotential terms (1 of 2)}
\label{t:tilings6-1}
\end{table}
\begin{table}[h]
\begin{center}
%
\end{center}
\caption{Tilings with 6 superpotential terms (2 of 2)}
\label{t:tilings6-2}
\end{table}

\FloatBarrier
\Section{Tilings with eight superpotential terms}
\begin{table}[h]
\begin{center}
%
\end{center}
\caption{Tilings with 8 superpotential terms (1 of 17)}
\label{t:tilings8-1}
\end{table}
\begin{table}[h]
\begin{center}
%
\end{center}
\caption{Tilings with 8 superpotential terms (2 of 17)}
\label{t:tilings8-2}
\end{table}
\begin{table}[h]
\begin{center}
%
\end{center}
\caption{Tilings with 8 superpotential terms (8 of 17)}
\label{t:tilings8-8}
\end{table}
\FloatBarrier
\begin{table}[h]
\begin{center}
%
\\
\end{tabular}
\end{center}
\caption{Tilings with 8 superpotential terms (17 of 17)}
\label{t:tilings8-17}
\end{table}

\chapter{Brane tilings Corresponding to the Fano 3-folds}
\label{fanoapp}
In this appendix we give further details of the theories corresponding to 14 of the 18 smooth toric fano 3 folds. Toric data along with a tiling and set of Chern-Simons levels are presented for each fano 3-fold. The forward algorithm is applied to each model to show that its moduli space corresponds a fano variety. The symmetry of the mesonic moduli space of each model is investigated. A full analysis of the moduli space of each theory can be found in ``M2-Branes and Fano 3-folds'' \cite{Davey:Fano}.

\begin{figure}[ht]
\begin{center}
  \includegraphics[totalheight=3cm]{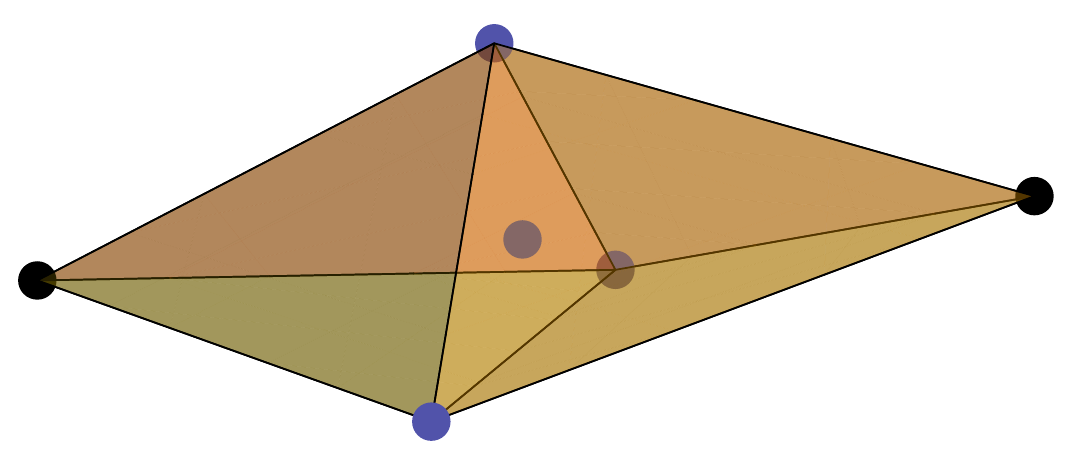}
 \caption{The toric diagram of the $\cB_4$ (Toric Fano 24).}
  \label{f:tdtoricfano24}
\end{center}
\end{figure}

\Section{$\cB_4$ (Toric Fano 24)}
The toric data of $\cB_4$ (Toric Fano 24) is given in \eqref{e:gtfano24}. The toric diagram of the variety is displayed in \fref{f:tdtoricfano24}.
\beq
G_t =\left(
\begin{array}{cccccc}
 1 &-1 & 0 & 0 & 0 & 0 \\
       0 & 1 &-1 & 0 & 0 & 0 \\
       0 & 0 & 0 & 1 & -1 & 0 
\end{array} \right)
\label{e:gtfano24}
\eeq

\begin{figure}[ht]
\includegraphics[totalheight=5cm]{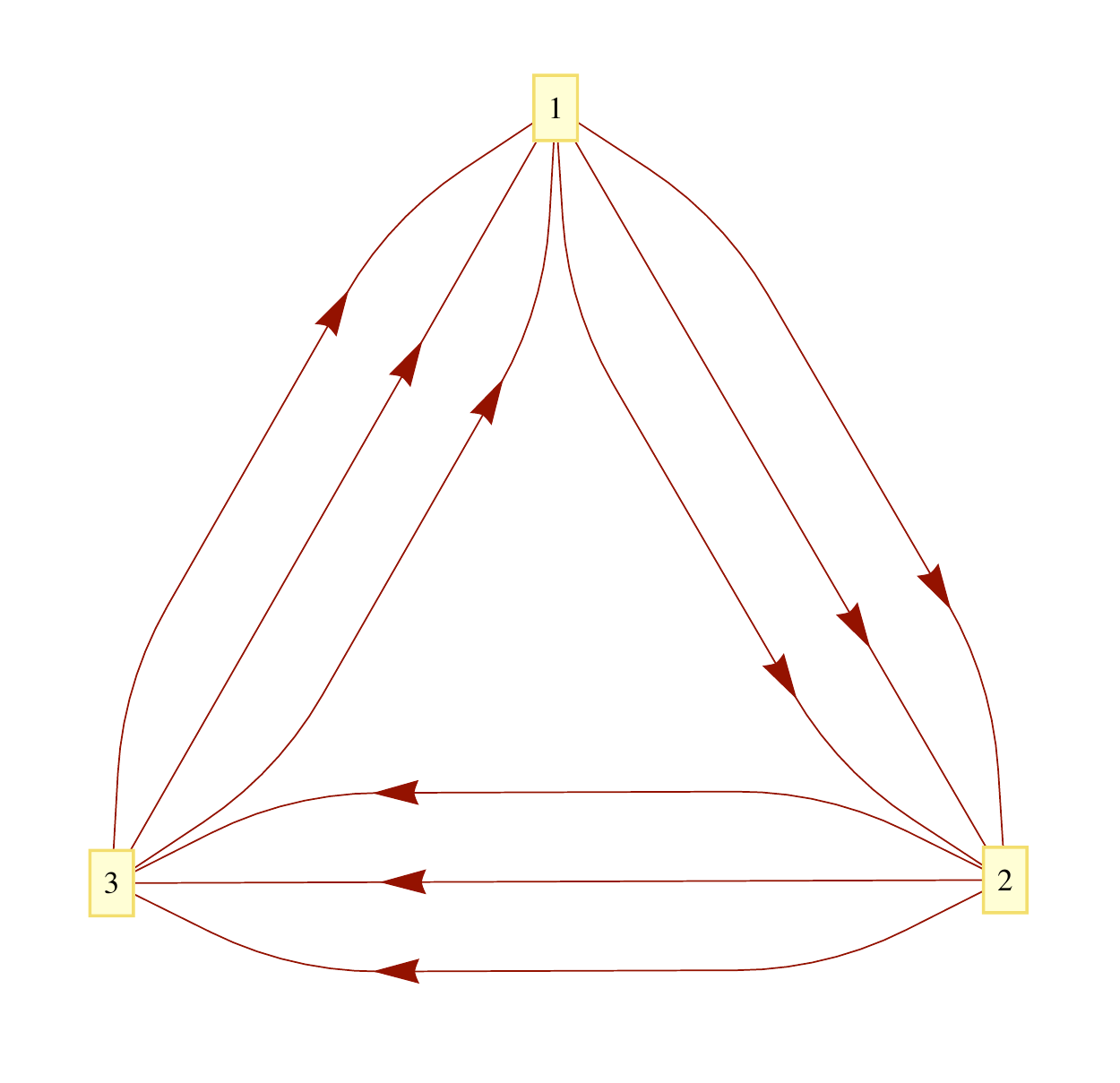}
\hfill
\includegraphics[totalheight=5cm]{FinalFano/tiling24.pdf}
\caption{(i) Quiver diagram of the $M^{1,1,1}$  theory.\ (ii) Tiling of the $M^{1,1,1}$  theory.}
  \label{f:m111}
\end{figure}

A gauge theory description of M2-branes placed at the tip of the cone over $\cB_4$ (also known as $M^{1,1,1}$) has been found. 
The $M^{1,1,1}$ theory \cite{Hanany:2008cd, Martelli:2008si, HananyZaff08,Davey:2009sr, Hanany:2009vx,Davey:2009qx,Davey:2009bp,Davey:2009et, Petrini:2009ur, Fabbri:1999hw} has 3 gauge groups and 9 chiral multiplets, which we shall call $X_{12}^i, X_{23}^i, X_{31}^i$ (with $i=1,2,3$). The quiver diagram and tiling are given in \fref{f:m111}.  
Note that in $3+1$ dimensions, this tiling corresponds to the gauge theory living on D3-branes probing the cone over the $dP_0$ surface.
The superpotential is given by
\bea
W= \tr \left( \epsilon_{ijk} X^i_{12} X^j_{23} X^k_{31} \right)~. 
\eea
The CS levels are $\vec{k} = (1, -2, 1) $.

\paragraph{The Kasteleyn matrix.} The Chern-Simons levels $k_a$ for gauge groups can be written in terms of the integers $n_i$ that correspond to Chern-Simons variables for fields. The two variables are related by the incidence matrix $k_a = \sum_i d_{ai} n_i$ \cite{Davey:2009sr}. In this case $n^i_{jk}$ are related to the levels $k_a$ by
\bea
\begin{array}{rcl}
\text{Gauge group 1~:} \qquad k_1   &=   n^{1}_{12} + n^{2}_{12} + n^{3}_{12} - n^{1}_{31} - n^{2}_{31} - n^{3}_{31}~, \nn \\
\text{Gauge group 2~:} \qquad k_2   &=   n^{1}_{23} + n^{2}_{23} + n^{3}_{23} - n^{1}_{12} - n^{2}_{12} - n^{3}_{12}~, \nn \\
\text{Gauge group 3~:} \qquad k_3   &=   n^{1}_{31} + n^{2}_{31} + n^{3}_{31} - n^{1}_{23} - n^{2}_{23} - n^{3}_{23}~.  
\label{e:kafano24}
\end{array}
\eea  
We will choose $n^i_{jk}$ to be
\bea
n^{1}_{12}=-n^{1}_{23}= 1,\qquad n^i_{jk}=0 \;\;\text{otherwise}~.
\eea
The Kasteleyn matrix is calculated as follows. Since the fundamental domain contains 3 pairs of black and white nodes, the Kasteleyn matrix is a $3 \times 3$ matrix: 
\bea
K =   \left(
\begin{array}{c|ccc}
& w_1 & w_2 & w_3\\
\hline
b_1 & z^{n^{1}_{31}} &  z^{n^{3}_{12}} & y z^{n^{2}_{23}} \\
b_2 &  \frac{1}{x} z^{n^{3}_{23}} & z^{n^{2}_{31}} & z^{n^{1}_{12}} \\
b_3 &  z^{n^{2}_{12}} &  \frac{x}{y} z^{n^{1}_{23}} & z^{n^{3}_{31}} \end{array}
\right) ~.
\label{e:kastfano24}
\eea
The permanent of the Kasteleyn matrix is given by
\bea
\mathrm{perm}(K) &=&   x y^{-1} z^{(n^{1}_{12} + n^{1}_{23} + n^{1}_{31})}
+ y z^{(n^{2}_{12} + n^{2}_{23} + n^{2}_{31})}
+ x^{-1} z^{(n^{3}_{12} + n^{3}_{23} + n^{3}_{31})} \nn \\
&+& z^{(n^{1}_{12} + n^{2}_{12} + n^{3}_{12})}
+ z^{(n^{1}_{23} + n^{2}_{23} + n^{3}_{23})}
+ z^{(n^{1}_{31} + n^{2}_{31} + n^{3}_{31})}\nn \\&=&
 x y^{-1}+ y+ x^{-1}+z + z^{-1}+1\;\nn \\
&& \text{(for $n^{1}_{12} = - n^{1}_{23} = 1,~ n^i_{jk}=0 \;
\text{otherwise}$)~.}
\label{e:charpolyfano24}
\eea
The powers of $x, y$ and $z$ in each term of \eref{e:charpolyfano24} give the coordinates of each point in the toric diagram. These points are collected in the columns of the following $G_K$ matrix: 
\bea
G_K = \left(
\begin{array}{cccccc}
  1 & 0 & -1 & 0 &  0 & 0 \\
 -1 & 1 &  0 & 0 &  0 & 0\\
  0 & 0 &  0 & 1 & -1 & 0
\end{array}
\right)~.
\eea
This matrix can be related to the $G_t$ matrix which was given in \eqref{e:gtfano24} by a series of permutations of rows and columns, and so both $G_t$ and $G_K$ correspond to the same variety. The simple roots of $SU(3)$ are found in the first 3 columns of $G_K$ and the simple roots of $SU(2)$ can be found in the 4th and 5th columns. This is consistent with the global symmetry of the geometry which was given in \tref{t:fanotable}.

\begin{figure}[ht]
\begin{center}
  \includegraphics[totalheight=3cm]{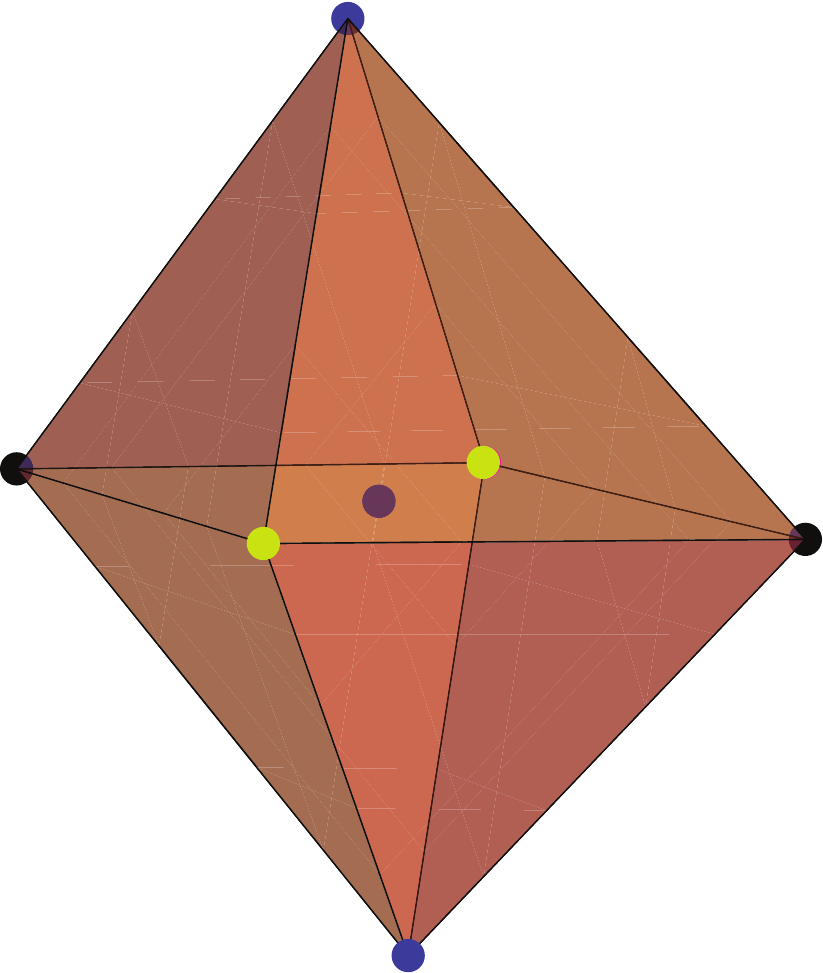}
 \caption{The toric diagram of the $\cC_3$ (Toric Fano 62).}
  \label{f:tdtoricfano62}
\end{center}
\end{figure}

\Section{$\cC_3$ (Toric Fano 62)}
The toric data of $\cC_3$ (Toric Fano 62) is given in \eqref{e:gtfano62}. The toric diagram of the variety is displayed in \fref{f:tdtoricfano62}.
\beq
G_t =\left(
\begin{array}{ccccccc}
 1 &-1 & 0 & 0 & 0 & 0 & 0 \\
 0 & 0 & 1 &-1 & 0 & 0 & 0 \\
 0 & 0 & 0 & 0 & 1 &-1 & 0 
\end{array} \right)
\label{e:gtfano62}
\eeq

\begin{figure}[ht]
\begin{center}
  \includegraphics[totalheight=5.7cm]{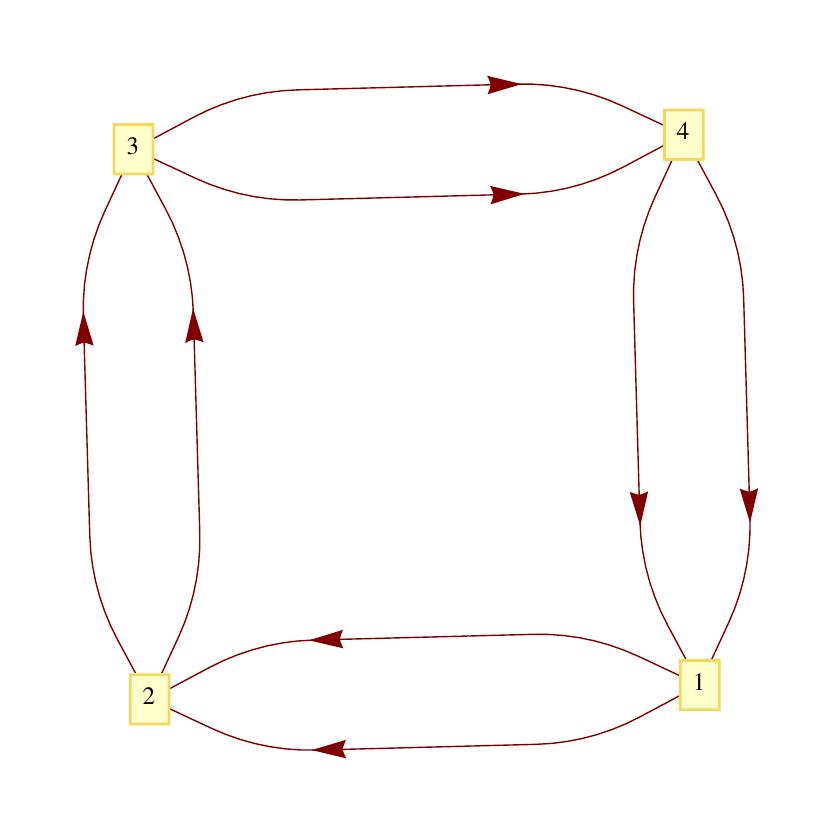}
\hfill
  \includegraphics[totalheight=5cm]{FinalFano/tiling62I.pdf}
 \caption{(i) Quiver for Phase I of $Q^{1,1,1}/\BZ_2$. \ (ii) Tiling for Phase I of $Q^{1,1,1}/\BZ_2$.}
  \label{f:phase1f0}
\end{center}
\end{figure}

A CS theory corresponding to this fano variety was introduced in \cite{Hanany:2008cd, HananyZaff08} as a modified $\BF_0$ theory. We shall consider a phase of the theory that has 4 gauge groups and has bi-fundamental fields $X_{12}^i$, $X_{23}^i$, $X_{34}^i$ and $X_{41}^i$ (with $i=1,2$). This theory is sometimes known as Phase I of $Q^{1,1,1}/\BZ_2$. The superpotential of this theory is given by
\bea
W = \epsilon_{ij} \epsilon_{pq} \tr(X_{12}^i X_{23}^p X_{34}^j X_{41}^q)~. \label{suppotph1q111z2}
\eea
The quiver diagram and tiling are shown in Figure \ref{f:phase1f0}.  
The fields are assigned to the edges in the tiling according to \fref{f:phase1f0} (ii).
Note that, in 3+1 dimensions, this quiver and this tiling correspond to Phase I of the $\BF_0$ theory \cite{ master3, Masterspace, MMasterspace, Forcella:2009bv, chiralops}.
The CS levels are chosen to be $\vec{k} = (1,-1,-1,1)$.

\paragraph{The Kasteleyn matrix.} The CS levels can be written in terms of the integers $n^i_{jk}$ as:
\bea \label{e:csfano62phI}
\begin{array}{rcl}
\text{Gauge group 1~:} \qquad k_1 &=  n^{1}_{12} + n^{2}_{12} - n^{1}_{41} - n^{2}_{41} ~, \\
\text{Gauge group 2~:} \qquad k_2 &=  n^{1}_{23} + n^{2}_{23} - n^{1}_{12} - n^{2}_{12} ~, \\
\text{Gauge group 3~:} \qquad k_3 &=  n^{1}_{34} + n^{2}_{34} - n^{1}_{23} - n^{2}_{23} ~, \\
\text{Gauge group 4~:} \qquad k_4 &=  n^{1}_{41} + n^{2}_{41} - n^{1}_{34} - n^{2}_{34} ~.
\end{array}
\eea  

The Kasteleyn matrix can be computed from the tiling. The fundamental domain contains two black nodes and two white nodes and, therefore, the Kasteleyn matrix is a $2\times 2$ matrix:
\be
K =   \left(
\begin{array}{c|cc}
& w_1 & w_2 \\
\hline
b_1 & z^{n^{2}_{12}} +  \frac{1}{x} z^{n^{1}_{34}} &\ z^{n^{2}_{41}} +  \frac{1}{y} z^{n^{1}_{23}}   \\
b_2 & z^{n^{1}_{41}} +  y z^{n^{2}_{23}} &\ z^{n^{1}_{12}} + x z^{n^{2}_{34}}  
\end{array}
\right) ~. \nn
\ee
The permanent of this matrix is given by
\bea 
\perm~K &=&  x z^{(n^{2}_{12} + n^{2}_{34})} +  x^{-1} z^{(n^{1}_{12} + n^{1}_{34})} + y z^{(n^{2}_{23} + n^{2}_{41})} + y^{-1} z^{(n^{1}_{23} + n^{1}_{41})}\nn \\
&&+ z^{(n^{1}_{12} + n^{2}_{12})} + z^{(n^{1}_{34} + n^{2}_{34})} +  z^{(n^{1}_{23} + n^{2}_{23})} + z^{(n^{1}_{41} + n^{2}_{41})} \nn \\ 
&=&  x  +  x^{-1} +  y  + y^{-1} + z+ z^{-1} +  2 \nn \\
&&\qquad \text{(for $n^{2}_{12} = - n^{2}_{34} = 1,~ n^i_{jk}=0 \; \text{otherwise}$)} ~. 
\label{permKph1f0}
\eea

The powers of $x, y$ and $z$ in each term of \eref{permKph1f0} give the coordinates of each point in the toric diagram. These points are collected in the columns of the following $G_K$ matrix: 
\bea
G_K = \left(
 \begin{array}{cccccccc}
 1 &-1 & 0 & 0 & 0 & 0 & 0 & 0 \\
 0 & 0 & 1 &-1 & 0 & 0 & 0 & 0 \\
 0 & 0 & 0 & 0 & 1 &-1 & 0 & 0
\end{array}
\right)
\eea
This matrix identical to the $G_t$ matrix which was given in \eqref{e:gtfano62} if the double multiplicity at the origin is ignored (which is thought to be unimportant), and so both $G_t$ and $G_K$ correspond to the same variety. The simple roots of $SU(2)$ are found in three different pairs of columns so the non abelian part of the global symmetry is identified as being $SU(2)^3$ and is consistent with the symmetry given in \tref{t:fanotable}.

\Section{$\cC_4$ (Toric Fano 123)}

\begin{figure}[ht]
\begin{center}
  \includegraphics[totalheight=3cm]{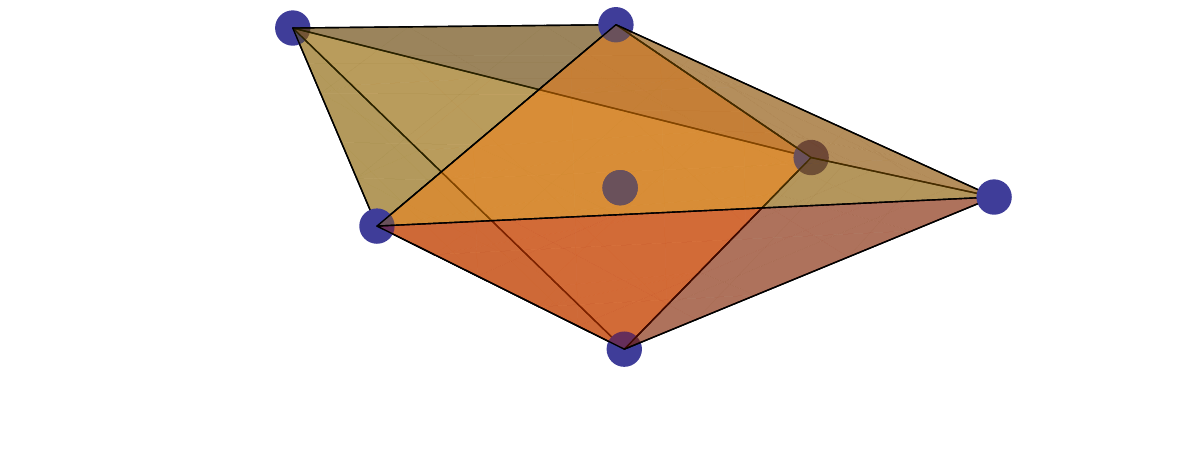}
 \caption{The toric diagram of $\cC_4$ (Toric Fano 123).}
  \label{f:tdtoricfano123}
\end{center}
\end{figure}

The toric data of $\cC_4$ (Toric Fano 123) is given in \eqref{e:gtfano123} and the toric diagram of the variety is displayed in \fref{f:tdtoricfano123}. This geometry is also known as $dP_1 \times \BP^1$.
\beq
G_t =\left(
\begin{array}{ccccccc}
 1 &-1 & 0 & 0 & 0 & 0 & 0 \\
 0 & 0 & 1 &-1 & 0 & 0 & 0 \\
 0 & 0 & 1 & 1 &-1 & 0 & 0 
\end{array} \right)
\label{e:gtfano123}
\eeq

\begin{figure}[ht]
\begin{center}
 \includegraphics[totalheight=5cm]{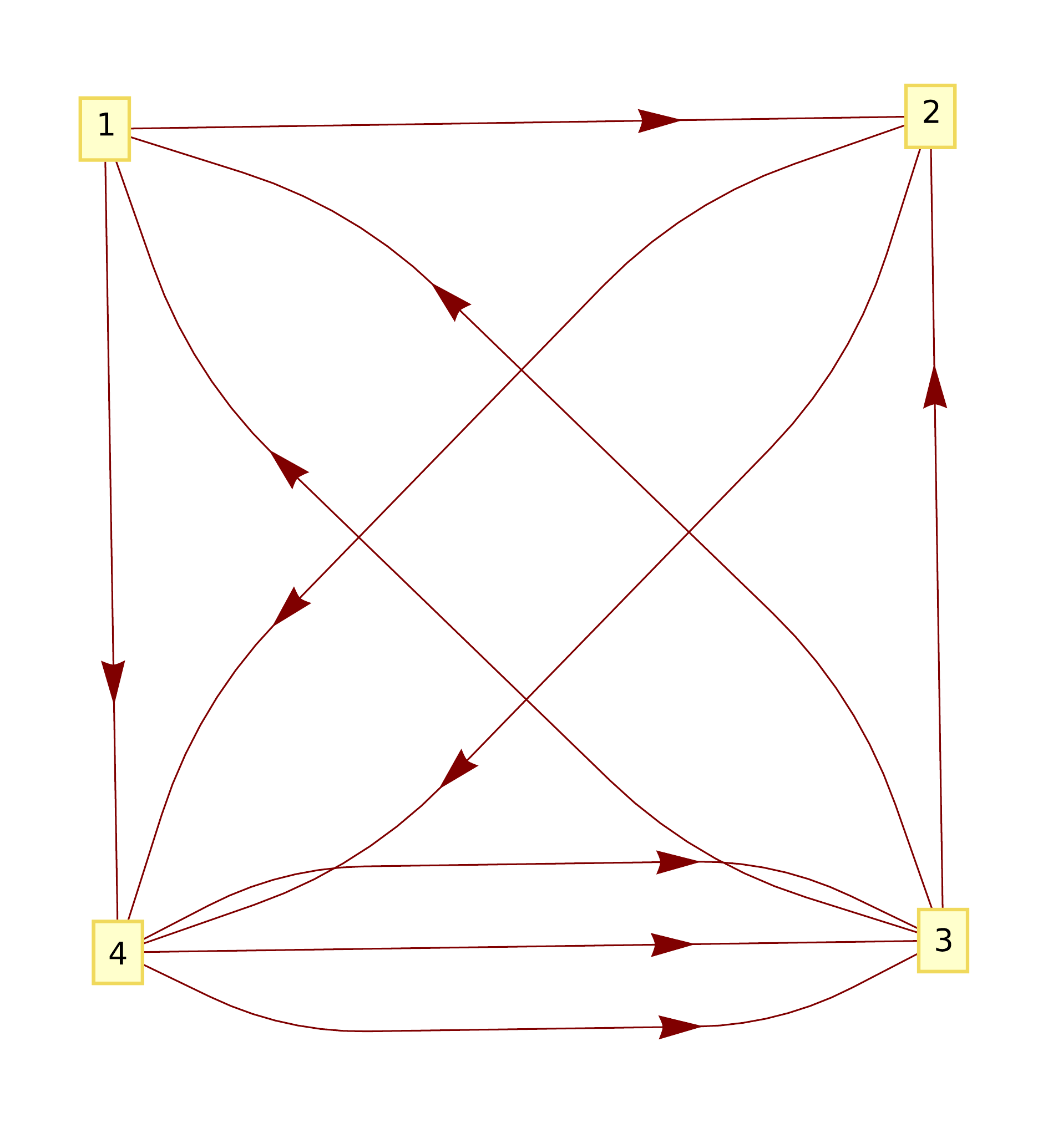}
 \hfill
 \includegraphics[totalheight=5cm]{FinalFano/tiling123.pdf}
 \caption{(i) Quiver diagram of the $dP_1 \times \BP^1$ theory.\ (ii) Tiling of the $dP_1 \times \BP^1$ theory.}
  \label{f:tqfano123}
\end{center}
\end{figure}

The CS theory we have found corresponding to the fano has 4 gauge groups and chiral fields $X_{14}$, $X_{12}$, $X_{32}$, $X^i_{43}$, $X^j_{24}$ and $X^j_{31}$ (with $i=1,2,3$ and $j=1,2$).
The quiver diagram and the tiling are presented in Figure \ref{f:tqfano123}.
Note that in $3+1$ dimensions this tiling corresponds to the gauge theory on D3-branes probing a cone over the $dP_1$ surface.
The superpotential can be read off from the tiling and can be written as:
\bea
W = \tr \left[ \epsilon_{ij} \left( X_{14}X^{i}_{43}X^{j}_{31} + X_{32}X^{i}_{24}X^{j}_{43} - X_{12}X^{i}_{24}X^{3}_{43}X^{j}_{31}\right) \right]~.
\label{e:sptoric123}
\eea

The CS levels are chosen to be $\vec{k} = (1,1,-1,-1)$

\paragraph{The Kasteleyn matrix.} The Chern-Simons levels can be parametrized as follows:
\bea
\begin{array}{ll}
\text{Gauge group 1~:} \qquad k_1  &=   n_{12} + n_{14} - n^{1}_{31} - n^{2}_{31}~, \nn \\
\text{Gauge group 2~:} \qquad k_2  &=   n^{1}_{24} + n^{2}_{24} - n_{12} - n_{32}~, \nn \\
\text{Gauge group 3~:} \qquad k_3  &=   n_{32} + n^{1}_{31} + n^{2}_{31} - n^{1}_{43} - n^{2}_{43} - n^{3}_{43} ~, \nn \\
\text{Gauge group 4~:} \qquad k_4  &=   n^{1}_{43} + n^{2}_{43} + n^{3}_{43} - n^{1}_{24} - n^{2}_{24} - n_{14}~. \nn \\

\label{e:kafano123}
\end{array}
\eea
Let us choose:
\bea
n^1_{24} = - n^1_{31} = 1,\quad n^i_{jk}=0 \; \text{otherwise}~.
\eea
The Kasteleyn matrix $K$ can be computed for this model. The fundamental domain contains three black and three white nodes, hence $K$ is a $3\times 3$ matrix:
\bea
K =   \left(
\begin{array}{c|ccc}
& b_1 & b_2 & b_3\\
\hline
w_1 &  z^{n_{14}} &  z^{n^1_{43}} &  \frac{x}{y} z^{n^2_{31}} \\
w_2 &  y z^{n^1_{31}} &  z^{n^2_{24}} &  z^{n^3_{43}} +  x z^{n_{12}} \\
w_3 &  z^{n^2_{43}} &  \frac{1}{x} z^{n_{32}} &  z^{n^1_{24}} \end{array}
\right) ~.
\label{e:kastfano123}
\eea
The permanent of this matrix is given by:
\bea
\perm~K &=& 
   z^{(n^1_{24} + n^{2}_{24} + n_{14})} + z^{(n^{1}_{31} + n^2_{31} + n_{32})} +  y z^{(n^{1}_{31} + n^{1}_{24} + n^{1}_{43})} \nn \\
&+&  x y^{-1} z^{(n^{2}_{31}+ n^{2}_{24} + n^{2}_{43})} +  x z^{(n^1_{43} + n^{2}_{43} + n_{12})}  +  x^{-1} z^{(n^{3}_{43} + n_{14} + n_{32})} \nn \\
&+&  z^{(n^1_{43} + n^{2}_{43} + n^{3}_{43})} +  z^{(n_{12} + n_{14}+ n_{32})}\nn \\
&=& z  + z^{-1} + y + x y^{-1} + x + x^{-1} + 2\nn\\ 
&& \text{(for $n^1_{24} = - n^1_{31} = 1,\quad n^i_{jk}=0\; ~ \text{otherwise}$)} ~.
\label{e:permKfano123}
\eea
The coordinates of the toric diagram are collected in the columns of the following matrix:
\bea
G_K = \left(
\begin{array}{cccccccc}
  0 & 0 & 0 &  1 & 1 & -1 & 0 & 0\\
  0 & 0 & 1 & -1 & 0 &  0 & 0 & 0\\
  1 &-1 & 0 &  0 & 0 &  0 & 0 & 0
\end{array}
\right)~.
\eea
Multiplying on the left by {\footnotesize $\left( \begin{array}{ccc} 0&0&1\\0&1&0\\1&0&0 \end{array} \right) \in GL(3, \BZ)$} we can find the equivalent toric data:
\bea
G'_K = \left(
\begin{array}{cccccccc}
  1 & -1 &  0 &  0 & 0 &  0 & 0 & 0\\
  0 &  0 &  1 & -1 & 0 &  0 & 0 & 0\\
  0 &  0 &  0 &  1 & 1 & -1 & 0 & 0
\end{array}
\right)~.
\eea
One can observe that the first two rows contain the weights of two $SU(2)$s. This is consistent with the non-abelian symmetry for the model which was quoted earlier in \tref{t:fanotable}. The $G'_K$ matrix is also identical (up to multiplicities of toric points) to the $G_t$ matrix in \eqref{e:gtfano123}.

\Section{$\cC_5$ (Toric Fano 68)}

\begin{figure}[ht]
\begin{center}
  \includegraphics[totalheight=3cm]{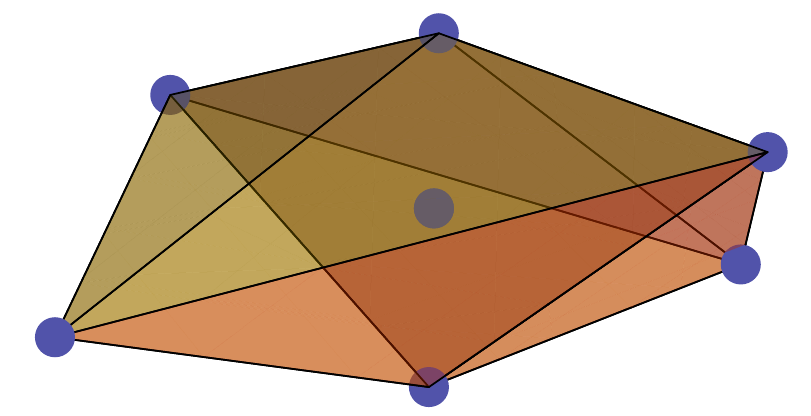}
 \caption{The toric diagram of $\cC_5$ (Toric Fano 68).}
  \label{f:tdtoricfano68}
\end{center}
\end{figure}

The toric data of $\cC_5$ (Toric Fano 68) is given in \eqref{e:gtfano68} and the toric diagram of the variety is displayed in \fref{f:tdtoricfano68}.
\beq
G_t =\left(
\begin{array}{ccccccc}
1 &-1 & 0 &  0 & 0 & 0 & 0 \\
       0 & 0 & 1 & -1 & 0 & 0 & 0 \\
       0 & 1 & 0 & -1 & 1 &-1 & 0
\end{array} \right)
\label{e:gtfano68}
\eeq

\begin{figure}[ht]
\begin{center}
  \includegraphics[totalheight=5.7cm]{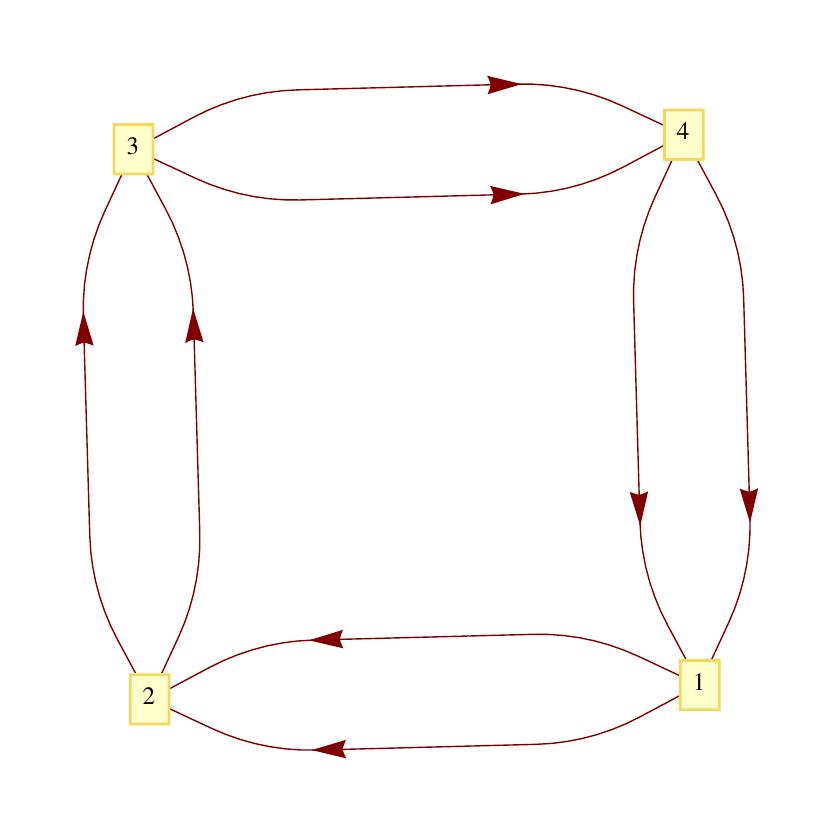}
\hfill
  \includegraphics[totalheight=5cm]{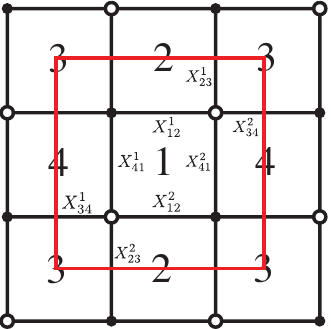}
 \caption{Quiver and Tiling for $\cC_5$ (Toric Fano 68)}
  \label{f:phase1f0-2}
\end{center}
\end{figure}

The CS theory we have found corresponding to the fano has a quiver diagram, superpotential and tiling that are identical to those used in the discussion of Fano 62 (also known as $\CC_\mathrm{3}$ or Phase I of $Q^{1,1,1}/\BZ_2$). For convenience the quiver and tiling are given again in \fref{f:phase1f0-2} and the superpotential is given in \eref{suppotph1q111z2-2}. The CS levels chosen this time are $\vec{k} = (1,-2,1,0)$.

\bea
W = \epsilon_{ij} \epsilon_{pq} \tr(X_{12}^i X_{23}^p X_{34}^j X_{41}^q)~. \label{suppotph1q111z2-2}
\eea

\paragraph{The Kasteleyn matrix.} The CS levels can be written in terms of the integers $n^i_{jk}$ as:
\bea \label{e:csfano62phI-2}
\begin{array}{rcl}
\text{Gauge group 1~:} \qquad k_1 &=  n^{1}_{12} + n^{2}_{12} - n^{1}_{41} - n^{2}_{41} ~, \\
\text{Gauge group 2~:} \qquad k_2 &=  n^{1}_{23} + n^{2}_{23} - n^{1}_{12} - n^{2}_{12} ~, \\
\text{Gauge group 3~:} \qquad k_3 &=  n^{1}_{34} + n^{2}_{34} - n^{1}_{23} - n^{2}_{23} ~, \\
\text{Gauge group 4~:} \qquad k_4 &=  n^{1}_{41} + n^{2}_{41} - n^{1}_{34} - n^{2}_{34} ~.
\end{array}
\eea  
In particular, for this model the following choice is made:
\bea
n^{1}_{12} = - n^{2}_{23} = 1,\quad n^i_{jk}=0 \; \text{otherwise}~.
\eea
The Kasteleyn matrix $K$ can be computed for this model. The fundamental domain contains two black nodes and two white nodes, which implies that $K$ is a $2\times 2$ matrix\footnote{Although the tiling of this model is identical to that of the first phase of $Q^{1,1,1}/\BZ_2$, a different weight assignment is used in the Kasteleyn matrix. This choice will make the non-abelian factors of the global symmetry more apparent in the $G_K$ matrix.}:
\be
K= \left(
\begin{array}{c|cc}
& w_1 & w_2 \\
\hline
b_1 &  z^{n^{2}_{12}} +  x z^{n^{1}_{34}} &\  z^{n^{2}_{41}} +  y z^{n^{1}_{23}}   \\
b_2 &  z^{n^{1}_{41}} +  \frac{1}{y} z^{n^{2}_{23}} &\  z^{n^{1}_{12}} + \frac{1}{x} z^{n^{2}_{34}}  
\end{array}
\right) ~. \label{e:kastfano68ph1}
\ee
The permanent of this Kasteleyn matrix can be written as:
\bea
\mathrm{perm}~K 
&=&  x z^{(n^{1}_{12} + n^{1}_{34})} +  x^{-1} z^{(n^{2}_{12} + n^{2}_{34})} +   y z^{(n^{1}_{23} + n^{1}_{41})} +  y^{-1} z^{(n^{2}_{23} + n^{2}_{41})}\nn \\
&+& z^{( n^{1}_{12} + n^{2}_{12})}+  z^{( n^{1}_{23} + n^{2}_{23} )} + z^{(n^{1}_{34} + n^{2}_{34})} + z^{( n^{1}_{41} + n^{2}_{41})}\nn \\
&=&   x + x^{-1} z + y + y^{-1} z^{-1} +  z + z^{-1} + 2\nn \\
&& \text{(for $n^{2}_{12} = - n^{2}_{23} = 1,\quad n^i_{jk}=0\; \text{otherwise}$)} ~. 
\label{e:permKfano68ph1}
\eea
The coordinates of the toric diagram are collected in the columns of the following matrix:
\bea
G_K = \left(
\begin{array}{cccccccc}
   1 & -1 &  0 &  0 & 0 &  0 & 0 & 0 \\
   0 &  0 &  1 & -1 & 0 &  0 & 0 & 0 \\
   0 &  1 &  0 & -1 & 1 & -1 & 0 & 0
\end{array}
\right)~.
\eea
Note that the first two rows of the $G_K$ matrix contain the weights of two $SU(2)$ groups; this implies that the non-abelian part of the mesonic symmetry is $SU(2)\times SU(2)$ which is consistent with \tref{t:fanotable}. $G_K$ is identical to the $G_t$ matrix in \fref{e:gtfano68} (up to the multiplicity of the internal point).

\Section{$\cC_1$ (Toric Fano 105)}

\begin{figure}[ht]
\begin{center}
  \includegraphics[totalheight=3cm]{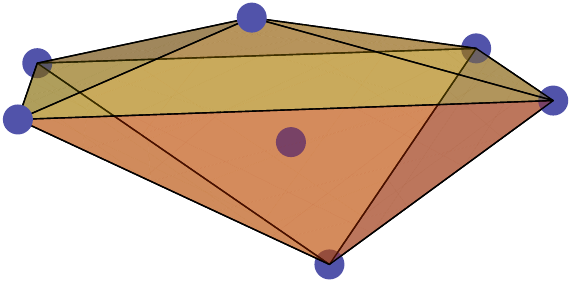}
 \caption{The toric diagram of $\cC_1$ (Toric Fano 105).}
  \label{f:tdtoricfano105}
\end{center}
\end{figure}

The toric data of $\cC_1$ (Toric Fano 105) is given in \eqref{e:gtfano105} and the toric diagram of the variety is displayed in \fref{f:tdtoricfano105}.
\beq
G_t =\left(
\begin{array}{ccccccc}
1 &-1 & 0 &  0 & 0 & 0 & 0 \\
 0 & 0 & 1 & -1 & 0 & 0 & 0 \\
 0 & 1 & 0 &  1 &-1 & 1 &  0
\end{array} \right)
\label{e:gtfano105}
\eeq

\begin{figure}[ht]
\begin{center}
  \includegraphics[totalheight=3cm]{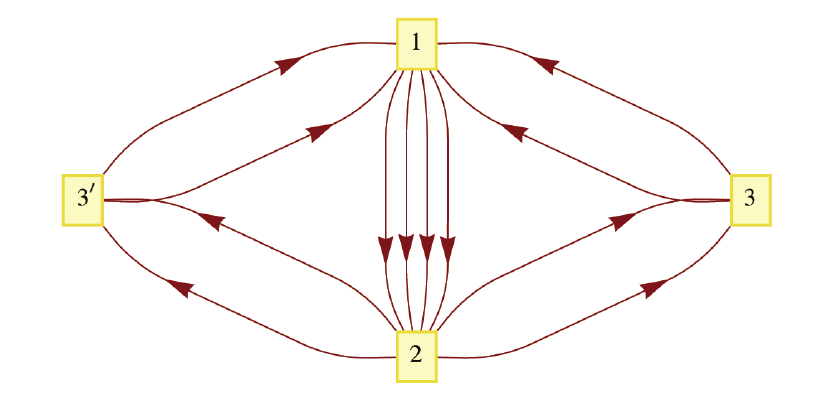}
\hfill
  \includegraphics[totalheight=6cm]{FinalFano/tiling62II.pdf}
 \caption{(i) Quiver diagram for the $\cC_1$ theory.\ (ii) Tiling for the $\cC_1$ theory.}
  \label{f:tqtoricfano105}
\end{center}
\end{figure}

A CS theory has been found that corresponds to this fano variety. It has 4 gauge groups and 12 chiral fields, which are denoted by $X_{12}^{ij}$, $X_{23}^i$, $X_{23'}^i$, $X_{31}^i$ and $X^{i}_{3'1}$ (with $i,j=1,2$). The quiver diagram and tiling are given in Figure \ref{f:tqtoricfano105}. We pick the CS levels to be $\vec{k} = (2,0,-1,-1)$.  The superpotential of this model is shown in (\ref{e:spotfano62phII}).

\bea
W &=& \epsilon_{ij}\epsilon_{kl} \tr(X^{ik}_{12}X^{l}_{23}X^{j}_{31}) - \epsilon_{ij}\epsilon_{kl} \tr(X^{ki}_{12}X^{l}_{23'} X^{j}_{3'1})~.
\label{e:spotfano62phII}
\eea

\paragraph{The Kasteleyn matrix.} The Chern-Simons levels for this model can be parametrized in terms of integers as shown in \ref{e:csfano62phII}.

\bea
\text{Gauge group 1~:} \qquad k_1    &=&   n^{11}_{12} + n^{12}_{12} + n^{21}_{12} + n^{22}_{12} - n^{1}_{31} - n^{2}_{31} - n^{1}_{3'1} - n^{2}_{3'1}  \nn  \\
\text{Gauge group 2~:} \qquad k_2    &=&   n^{1}_{23} + n^{2}_{23} + n^{1}_{23'} + n^{2}_{23'} - n^{11}_{12} - n^{12}_{12} - n^{21}_{12} - n^{22}_{12}  \nn \\
\text{Gauge group 3~:} \qquad k_3    &=&   n^{1}_{31} + n^{2}_{31} - n^{1}_{23} - n^{2}_{23}  \nn \\
\text{Gauge group $3'$~:}\qquad k_{3'} &=&   n^{1}_{3'1} + n^{2}_{3'1} - n^{1}_{23'} - n^{2}_{23'}  
\label{e:csfano62phII}
\eea  

Let us choose:
\bea
n^{12}_{12} = n^{21}_{12} = n^2_{23'} = -n^{22}_{12} = -n^{2}_{31} = 1,\quad n^i_{kl}=n^{ij}_{kl}=0 \; \text{otherwise}~.
\eea
The Kasteleyn matrix for this model can be written as:
\be
K= \left(
\begin{array}{c|cccc}
& w_1 & w_2 & w_3 & w_4\\
\hline
b_1 &  y z^{n^{2}_{23}}  &\  \frac{1}{x}  z^{n^{1}_{31}}      &\           0           &\  z^{n^{21}_{12}} \\
b_2 &  x z^{n^{2}_{31}}  &\   \frac{1}{y} z^{n^{1}_{23}}      &\  z^{n^{12}_{12}}      &\       0            \\
b_3 &         0          &\   z^{n^{22}_{12}}                &\   z^{n^{1}_{3'1}}      &\ z^{n^{1}_{23'} } \\
b_4 &    z^{n^{11}_{12}} &\       0                           &\   z^{n^2_{23'} }      &\  z^{n^{2}_{3'1}}
\end{array}
\right) ~. \label{e:kastfano105}
\ee
The permanent of the Kasteleyn matrix is
\bea
\perm~K &=&  y  z^{(n^{2}_{23} + n^{2}_{3'1} + n^{12}_{12} + n^{22}_{12})} +  \frac{1}{y} z^{(n^{1}_{23} + n^{1}_{3'1} + n^{21}_{12} + n^{11}_{12})} + x z^{(n^{2}_{23'} + n^{2}_{31} + n^{21}_{12} + n^{22}_{12})} \nn \\
&+&  \frac{1}{x} z^{(n^{1}_{23'} + n^{1}_{31} + n^{11}_{12} + n^{12}_{12})} + z^{(n^{1}_{31} + n^{2}_{31} + n^{1}_{3'1} + n^{2}_{3'1})} +  z^{(n^{1}_{23'} + n^{2}_{23'} + n^{1}_{23} + n^{2}_{23})}\nn \\
&+&  z^{(n^{11}_{12} + n^{21}_{12} + n^{12}_{12} + n^{22}_{12})} +  z^{(n^{1}_{23} + n^{2}_{23} + n^{1}_{3'1} + n^{2}_{3'1})} + z^{(n^{1}_{31} + n^{2}_{31} + n^{1}_{23'} + n^{2}_{23'})}\nn \\
&=&  y + y^{-1} z + x + x^{-1} z  + z^{-1} + 2z + 2 \nn \\
&& \text{(for $n^{12}_{12} = n^{21}_{12} = n^2_{23'} = -n^{22}_{12} = -n^{2}_{31} = 1,$}\nn \\ &&\text{$n^i_{kl}=n^{ij}_{kl}=0 \; ~ \text{otherwise}$)} ~.\nn \\
\label{e:permKfano105}
\eea

The coordinates of the toric diagram are given by the powers of each monomial in \eref{e:permKfano105} and can be encoded in columns of the following matrix:
\bea
G_K = \left(
\begin{array}{ccccccccc}
   1 & -1 &  0 &  0 & 0 &  0 & 0 & 0 & 0\\
   0 &  0 &  1 & -1 & 0 &  0 & 0 & 0 & 0\\
   0 &  1 &  0 &  1 &-1 &  1 & 1 & 0 & 0
\end{array}
\right).~
\eea
The first and second rows of the $G_K$ matrix correspond to powers of $y$ and $x$ in \eqref{e:permKfano105} respectively. The two simple roots of $SU(2)$ which are visible in the first 4 columns of $G_K$ are consistent with the non-abelian part of the mesonic symmetry being $SU(2)\times SU(2)$. The $G_K$ matrix above is equal to the $G_t$ matrix for this fano, up to multiplicity of toric points.

\Section{$\cC_2$ (Toric Fano 136)}

\begin{figure}[ht]
\begin{center}
  \includegraphics[totalheight=2cm]{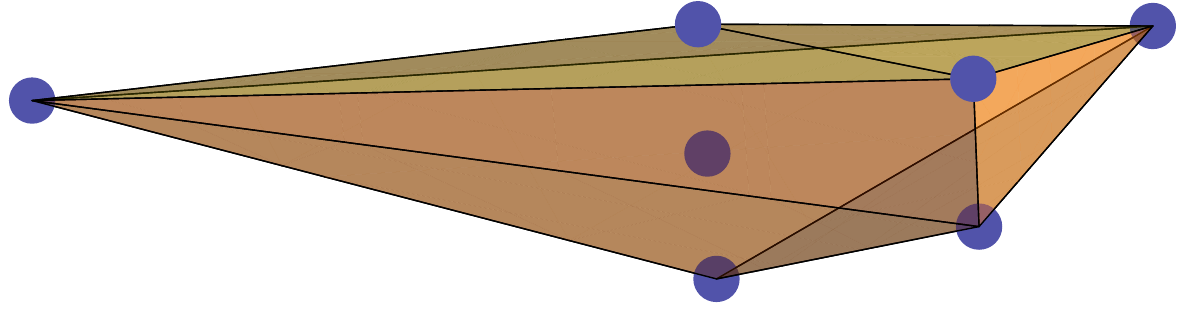}
 \caption{The toric diagram of $\cC_2$ (Toric Fano 136).}
  \label{f:tdtoricfano136}
\end{center}
\end{figure}

The toric data of $\cC_2$ (Toric Fano 136) is given in \eqref{e:gtfano136} and the toric diagram of the variety is displayed in \fref{f:tdtoricfano136}.
\beq
G_t =\left(
\begin{array}{ccccccc}
 1 &-1 & 0 & 0 & 0 & 0 & 0 \\
 0 & 1 &-1 &-1 & 0 & 0 & 0 \\
 0 & 0 & 1 & 2 &-1 & 1 & 0
\end{array} \right)
\label{e:gtfano136}
\eeq

\begin{figure}[ht]
\begin{center}
\includegraphics[totalheight=3cm]{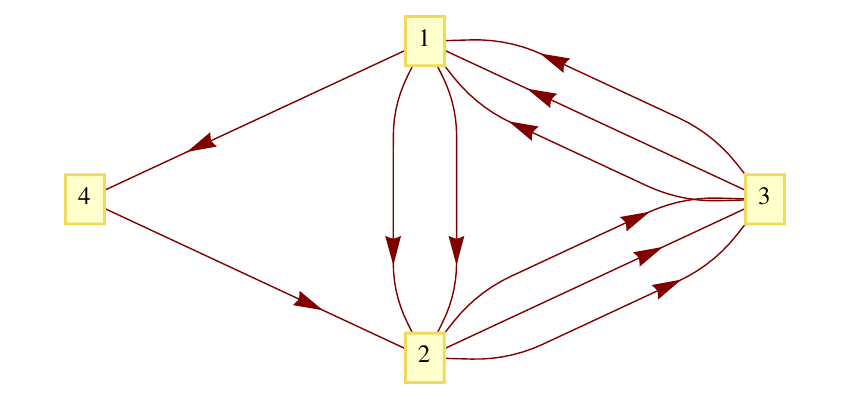}
\hfill
\includegraphics[totalheight=4cm]{FinalFano/tiling136.pdf}
 \caption{(i) Quiver diagram of the $\cC_2$ model.\ (ii) Tiling of the $\cC_2$ model.}
  \label{f:tqfano136}
\end{center}
\end{figure}

A CS theory corresponding to this fano variety has been found. This theory has 4 gauge groups and chiral fields $X^i_{23}$, $X^i_{31}$ (with $i=1,2,3$), $X^j_{12}$ (with $j = 1,2$), $X_{14}$ and $X_{42}$. The tiling and the quiver diagram are presented in Figure \ref{f:tqfano136}.  Note that the former can be obtained by adding a `double bond' to the 3 hexagon tiling.  The superpotential of this model can be written as
\beq
W = \epsilon_{ij} \tr (X^{i}_{31}X^{j}_{12}X^3_{23})+\epsilon_{ij} \tr (X^{i}_{12}X^{j}_{23}X^3_{31})+\epsilon_{ij}\tr (X^{i}_{23}X^{j}_{31}X_{14}X_{42})
\label{e:spotfano136}
\eeq
and we choose CS levels to be $\vec{k} = (-1,2,0,-1)$.

\paragraph{The Kasteleyn matrix.} The Chern-Simons levels can be parametrized in terms of integers $n^i_{jk}$ and $n_{jk}$ as follows:
\bea
\begin{array}{ll}
\text{Gauge group 1~:} \qquad k_1  &=   n_{14} + n^{1}_{12} + n^{2}_{12} - n^{1}_{31} - n^{2}_{31} - n^{3}_{31} ~, \nn \\
\text{Gauge group 2~:} \qquad k_2  &=   n^{1}_{23} + n^{2}_{23} + n^{3}_{23} - n^{1}_{12} - n^{2}_{12} - n_{42}  ~, \nn \\
\text{Gauge group 3~:} \qquad k_3  &=   n^{1}_{31} + n^{2}_{31} + n^{3}_{31} - n^{1}_{23} - n^{2}_{23} - n^{3}_{23}  ~, \nn \\
\text{Gauge group 4~:} \qquad k_4  &=   n_{42} - n_{14}  ~.
\label{e:kafano136}
\end{array}
\eea
Let us choose
\bea
n^2_{31} = n^3_{23} = - n_{42} = 1,\quad n^i_{jk}=n_{jk}=0 \; \text{otherwise}~.
\eea
which is consistent with our earlier choice of $\vec{k}$. The Kasteleyn matrix for this model can be computed. Since the fundamental domain contains six nodes in total, $K$ is a $3\times 3$ matrix:
\bea
K =   \left(
\begin{array}{c|ccc}
& b_1 & b_2 & b_3\\
\hline
w_1 &  z^{n_{42}} +  z^{n_{14}} &  z^{n^2_{23}} &  \frac{y}{x} z^{n^1_{31}} \\
w_2 &  x z^{n^2_{31}} & z^{n^1_{12}} & z^{n^3_{23}} \\
w_3 &  z^{n^1_{23}} &  \frac{1}{y} z^{n^3_{31}} &  z^{n^2_{12}} \end{array}
\right) ~. \label{e:kastfano136}
\eea
The permanent of this matrix is given by
\bea
\perm~K &=& 
 x z^{(n^{2}_{31} + n^{2}_{12} + n^{2}_{23})} +  x^{-1} y z^{(n^{1}_{31} + n^{1}_{12} + n^{1}_{23})} +   y^{-1} z^{(n^{3}_{31} + n_{42} + n^{3}_{23})}\nn \\
&+&  y^{-1} z^{(n^{3}_{31} + n_{14} + n^3_{23})} +  z^{(n^{1}_{12} + n^{2}_{12} + n_{42})}+  z^{(n^{1}_{31} + n^{2}_{31} + n^{3}_{31})}\nn \\
&+&  z^{(n^{1}_{23} + n^{2}_{23} + n^{3}_{23})} +  z^{(n^{1}_{12} + n^{2}_{12} + n_{14})} \nn \\
&=&  x z+ x^{-1} y +y^{-1} + y^{-1} z + z^{-1} +  2z + 1\nn\\ 
&& \text{(for $n^2_{31} = n^3_{23} = - n_{42} = 1,$} \nn \\
&& \text{$ n^i_{jk}=n_{jk}=0 \; ~ \text{otherwise}$)} ~.
\label{e:permKfano136}
\eea
The powers of $x, y$ and $z$ in each term of \eref{e:permKfano136} give the coordinates of each point in the toric diagram. These these points can be written as columns of the following matrix: 
\bea
 \left(
\begin{array}{cccccccc}
   1 & -1 &  0 &  0 & 0 & 0 & 0 & 0 \\
   0 &  1 & -1 & -1 & 0 & 0 & 0 & 0 \\
   1 &  0 &  0 &  1 &-1 & 1 & 1 & 0 
\end{array}
\right)~. \nn
\eea
By multiplying the matrix above on the left by {\footnotesize $\left( \begin{array}{ccc} 1&0&0\\0&1&0\\-1&-1&1 \end{array} \right) \in GL(3, \BZ)$}, the following matrix is obtained:
\bea
G_K = \left(
\begin{array}{cccccccc}
   1 & -1 &  0 &  0 & 0 & 0 & 0 & 0 \\
   0 &  1 & -1 & -1 & 0 & 0 & 0 & 0 \\
   0 &  0 &  1 &  2 &-1 & 1 & 1 & 0 
\end{array}
\right)~.
\eea
The first row of this matrix contains the weights of the fundamental representation of $SU(2)$, which implies that the non-abelian part of the mesonic symmetry contains one $SU(2)$ factor. This is consistent with \tref{t:fanotable}. $G_K$ matches the $G_t$ matrix in \eqref{e:gtfano136} up to multiplicity of toric points.

\Section{$\cD_1$ (Toric Fano 131)}

\begin{figure}[ht]
\begin{center}
  \includegraphics[totalheight=2.5cm]{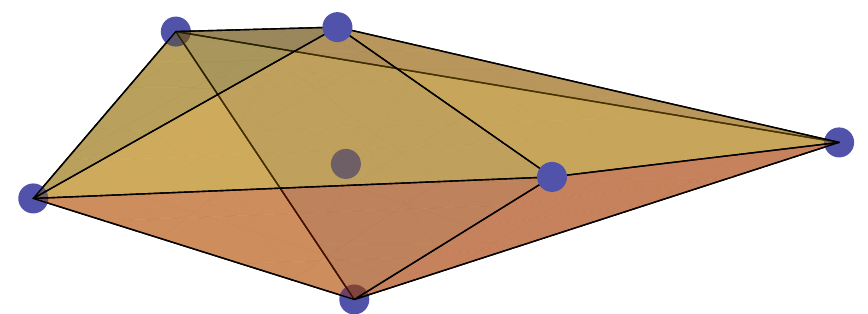}
 \caption{The toric diagram of $\cD_1$ (Toric Fano 131).}
  \label{f:tdtoricfano131}
\end{center}
\end{figure}

The toric data of $\cD_1$ (Toric Fano 131) is given in \eqref{e:gtfano131} and the toric diagram of the variety is displayed in \fref{f:tdtoricfano131}.
\beq
G_t =\left(
\begin{array}{ccccccc}
 1 &-1 & 0 & 0 & 0 & 0 & 0 \\
 0 & 1 & 1 &-1 & 0 & 0 & 0 \\
 0 & 0 & 0 & 1 &-1 & 1 & 0
\end{array} \right)
\label{e:gtfano131}
\eeq

\begin{figure}[ht]
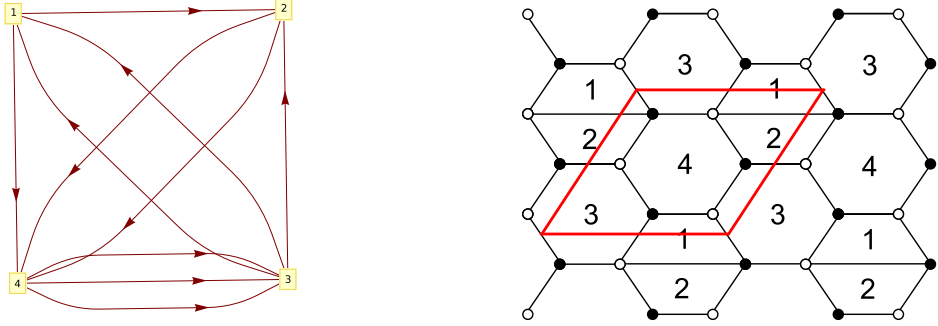

\begin{center}
 \includegraphics[totalheight=5cm]{FinalFano/quiver123.pdf}
 \hfill
 \includegraphics[totalheight=5cm]{FinalFano/tiling123.pdf}
 \caption{Quiver and Tiling for the Chern Simons theory corresponding to $\cD_1$ (Toric Fano 131).}
  \label{f:tqfano123-2}
\end{center}
\end{figure}

The CS theory corresponding to this fano variety has 4 gauge groups and chiral fields $X_{13}$, $X_{12}$, $X_{42}$, $X^i_{34}$, $X^j_{23}$ and $X^j_{41}$ (with $i=1,2,3$ and $j=1,2$).
 The tiling and the quiver diagram can be found in \fref{f:tqfano123-2}. The CS levels are $\vec{k} = (-1,-1,0,2)$.
The superpotential of the theory can be found in \eqref{e:sptoric123-2}.

\bea
W = \tr \left[ \epsilon_{ij} \left( X_{14}X^{i}_{43}X^{j}_{31} + X_{32}X^{i}_{24}X^{j}_{43} - X_{12}X^{i}_{24}X^{3}_{43}X^{j}_{31}\right) \right]~.
\label{e:sptoric123-2}
\eea

\paragraph{The Kasteleyn matrix.} The Chern-Simons levels $\vec{k}$ can be written in terms of the integers $n^i_{jk}$ and $n_{jk}$ as shown below

\bea
\begin{array}{ll}
\text{Gauge group 1~:} \qquad k_1  &=   n_{12} + n_{13} - n^{1}_{41} - n^{2}_{41}~, \nn \\
\text{Gauge group 2~:} \qquad k_2  &=   n^{1}_{23} + n^{2}_{23} - n_{12} - n_{42}~, \nn \\
\text{Gauge group 3~:} \qquad k_3  &=   n^{1}_{34} + n^{2}_{34} + n^{3}_{34} - n^{1}_{23} - n^{2}_{23} - n_{13}~, \nn \\
\text{Gauge group 4~:} \qquad k_4  &=   n_{42} + n^{1}_{41} + n^{2}_{41} - n^{1}_{34} - n^{2}_{34} - n^{3}_{34} ~.
\label{e:kafano131}
\end{array}
\eea

For this theory, let us choose:
\bea
n^1_{34} = n_{13} = - n^1_{41} = - n_{12} =1,\quad n^i_{jk}=n_{jk}=0 \; \text{otherwise}~.
\eea
The fundamental domain contains three pairs of black and white nodes, and so the Kasteleyn matrix $K$ is a $3\times 3$ matrix\footnote{Note that, in order to make the non-abelian mesonic symmetry more apparent in the $G_K$ matrix, the weight assignment is different to \ref{e:kastfano123}}:
\bea
K =   \left(
\begin{array}{c|ccc}
& b_1 & b_2 & b_3\\
\hline
w_1 & z^{n_{13}} & z^{n^1_{34}} & \frac{y}{x} z^{n^2_{41}} \\
w_2 & x z^{n^1_{41}} & z^{n^2_{23}} & z^{n^3_{34}} + y z^{n_{12}} \\
w_3 & z^{n^2_{34}} & \frac{1}{y} z^{n_{42}} & z^{n^1_{23}} \end{array}
\right) ~.
\label{e:kastfano131}
\eea
The permanent of this matrix is given by
\bea
\perm~K &=& x z^{(n^1_{41} + n^1_{23} + n^1_{34} )} +  x^{-1} y z^{(n^2_{41} + n^2_{23} + n^2_{34})} +  y z^{(n^1_{34} + n^2_{34} + n_{12})}\nn \\
&+&  y^{-1} z^{(n^3_{34} + n_{42} + n_{13})} +  z^{(n^1_{41} + n^2_{41} + n_{42})} +  z^{(n^1_{23} + n^2_{23} + n_{13})}\nn \\
&+&  z^{(n^1_{34} + n^2_{34} + n^3_{34})} +  z^{(n_{12} + n_{42} + n_{13})}\nn \\
&=&  x + x^{-1} y + y  + y^{-1} z + z^{-1} + 2z + 1\nn\\ 
&&  \text{(for $n^1_{34} = n_{13} = - n^1_{41} = - n_{12} =1,\quad n^i_{jk}=n_{jk}=0 \; ~ \text{otherwise}$)} ~.\nn \\
\label{e:permKfano131}
\eea

 The coordinates of the toric diagram are collected in the columns of the following matrix:
\bea
G_K = \left(
\begin{array}{cccccccc}
  1 &-1 & 0 &  0 & 0 &  0 & 0 & 0\\
  0 & 1 & 1 & -1 & 0 &  0 & 0 & 0\\
  0 & 0 & 0 &  1 &-1 &  1 & 1 & 0
\end{array}
\right)~.
\eea
The first row of this matrix contains weights of the fundamental representation of $SU(2)$ which matches the non abelian symmetry of the mesonic moduli space which was given in \tref{t:fanotable}. The $G_K$ matrix also matches the $G_t$ matrix which was given in \eqref{e:gtfano131} up to toric multiplicity.

\Section{$\cD_2$ (Toric Fano 139)}

\begin{figure}[ht]
\begin{center}
  \includegraphics[totalheight=2.5cm]{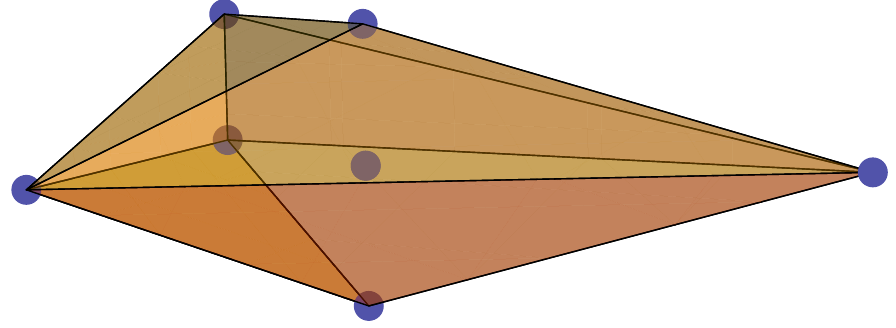}
 \caption{The toric diagram of $\cD_2$ (Toric Fano 139).}
  \label{f:tdtoricfano139}
\end{center}
\end{figure}

The toric data of $\cD_2$ (Toric Fano 139) is given in \eqref{e:gtfano139} and the toric diagram of the variety is displayed in \fref{f:tdtoricfano139}.
\beq
G_t =\left(
\begin{array}{ccccccc}
 1 &-1 & 0 & 0 & 0 & 0 & 0 \\
 0 & 1 &-1 &-1 & 0 & 0 & 0 \\
 0 & 0 & 0 & 1 & 1 &-1 & 0
\end{array} \right)
\label{e:gtfano139}
\eeq

\begin{figure}[ht]
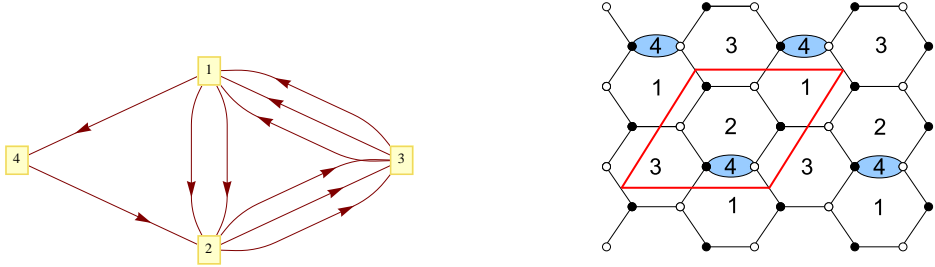

\begin{center}
\includegraphics[totalheight=3cm]{FinalFano/quiver136.pdf}
\hfill
\includegraphics[totalheight=4cm]{FinalFano/tiling136.pdf}
 \caption{Quiver and Tiling of the $\cD_2$ model (Toric Fano 139)}
  \label{f:tqfano139}
\end{center}
\end{figure}

The CS theory corresponding to this fano variety has 4 gauge groups and chiral fields $X^i_{23}$, $X^i_{31}$ (with $i=1,2,3$), $X^j_{12}$ (with $j = 1,2$), $X_{14}$ and $X_{42}$. The tiling and the quiver diagram are identical to those of the $\cC_2$ theory. For convenience they are given again in \fref{f:tqfano139}. The CS levels of this theory are $\vec{k} = (-1,1,1,-1)$.  The superpotential is given in \eref{e:spotfano136-2}.

\beq
W = \epsilon_{ij} \tr (X^{i}_{31}X^{j}_{12}X^3_{23})+\epsilon_{ij} \tr (X^{i}_{12}X^{j}_{23}X^3_{31})+\epsilon_{ij}\tr (X^{i}_{23}X^{j}_{31}X_{14}X_{42})
\label{e:spotfano136-2}
\eeq

\paragraph{The Kasteleyn matrix.} The Chern-Simons levels can be parametrized in terms of integers as according to (\ref{e:kafano139}).

\bea
\begin{array}{ll}
\text{Gauge group 1~:} \qquad k_1  &=   n_{14} + n^{1}_{12} + n^{2}_{12} - n^{1}_{31} - n^{2}_{31} - n^{3}_{31} ~, \nn \\
\text{Gauge group 2~:} \qquad k_2  &=   n^{1}_{23} + n^{2}_{23} + n^{3}_{23} - n^{1}_{12} - n^{2}_{12} - n_{42}  ~, \nn \\
\text{Gauge group 3~:} \qquad k_3  &=   n^{1}_{31} + n^{2}_{31} + n^{3}_{31} - n^{1}_{23} - n^{2}_{23} - n^{3}_{23}  ~, \nn \\
\text{Gauge group 4~:} \qquad k_4  &=   n_{42} - n_{14}  ~.
\label{e:kafano139}
\end{array}
\eea

For this model let us choose:
\bea
n^3_{31} = - n_{42} = 1,\quad n^i_{jk}=n_{jk}=0 \; \text{otherwise}~.
\eea
The Kasteleyn matrix $K$ for this model can be calculated. The fundamental domain contains six nodes in total, hence $K$ is a $3\times 3$ matrix:
\bea
K =   \left(
\begin{array}{c|ccc}
& b_1 & b_2 & b_3\\
\hline
w_1 & z^{n_{42}} + z^{n_{14}} &  z^{n^2_{23}} & \frac{y}{x} z^{n^1_{31}} \\
w_2 & x z^{n^2_{31}} & z^{n^1_{12}} & z^{n^3_{23}} \\
w_3 & z^{n^1_{23}} & \frac{1}{y} z^{n^3_{31}} & z^{n^2_{12}} \end{array}
\right) ~. \label{e:kastfano139}
\eea
The permanent of this matrix is given by:
\bea
\perm~K &=& 
 x z^{(n^{2}_{12} + n^{2}_{23} + n^{2}_{31})} +  x^{-1} y z^{(n^{1}_{12} + n^{1}_{23} + n^{1}_{31})} +  y^{-1} z^{(n^{3}_{23} + n^{3}_{31} + n_{42})}\nn \\
&+&  y^{-1} z^{(n^3_{23} + n^{3}_{31} + n_{14})} +  z^{(n^{1}_{31} + n^{2}_{31} + n^{3}_{31})}+  z^{(n^{1}_{12} + n^{2}_{12} + n_{42})}\nn \\
&+&  z^{(n^{1}_{12} + n^{2}_{12} + n_{14})} +  z^{(n^{1}_{23} + n^{2}_{23} + n^{3}_{23})} \nn \\
&=& x + x^{-1} y + y^{-1} +  y^{-1} z + z + z^{-1} + 2\nn\\ 
&&  \text{(for $n^3_{31} = - n_{42} = 1,\quad n^i_{jk}=n_{jk}=0 \; ~ \text{otherwise}$)} ~.
\label{e:permKfano139}
\eea

The coordinates of the toric diagram are collected in the columns of the following matrix:
\bea
G_K = \left(
\begin{array}{cccccccc}
   1 & -1 &  0 &  0 & 0 & 0 & 0 & 0 \\
   0 &  1 & -1 & -1 & 0 & 0 & 0 & 0 \\
   0 &  0 &  0 &  1 & 1 &-1 & 0 & 0 
\end{array}
\right)~.
\eea
The first row of this matrix contains the weights of the fundamental representation of $SU(2)$, which implies that the non-abelian part of the mesonic symmetry is $SU(2)$. The $G_K$ above is identical to $G_t$ in \eqref{e:gtfano139} up to multiplicity of toric points.

\Section{$\cE_1 $(Toric Fano 218)}

\begin{figure}[ht]
\begin{center}
  \includegraphics[totalheight=2.5cm]{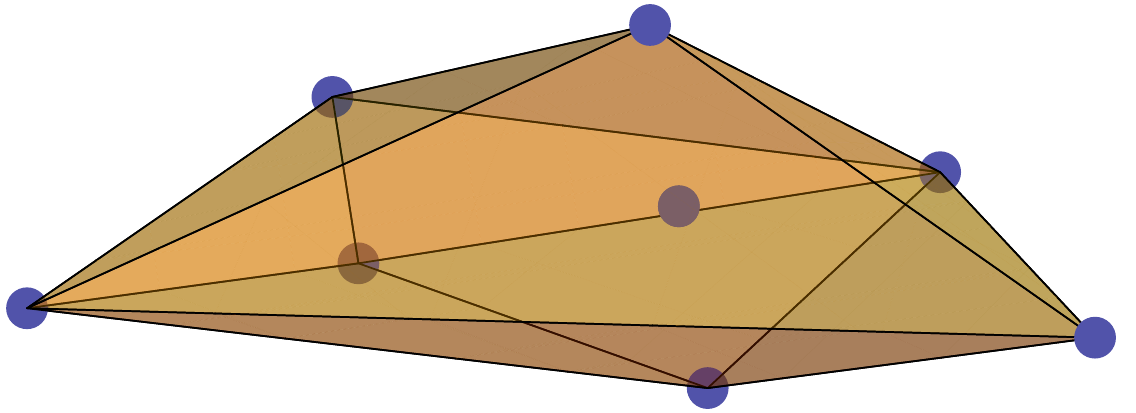}
 \caption{The toric diagram of $\cE_1$ (Toric Fano 218).}
  \label{f:tdtoricfano218}
\end{center}
\end{figure}

The toric data of $\cE_1$ (Toric Fano 218) is given in \eqref{e:gtfano218} and the toric diagram of the variety is displayed in \fref{f:tdtoricfano218}.
\beq
G_t =\left(
\begin{array}{cccccccc}
1 &-1 & 0 & 0 & 0 & 0 & 0 & 0 \\
      0 & 1 & 1 & 1 &-1 & 0 & 0 & 0 \\
      0 & 0 & 0 & 1 &-1 &-1 & 1 & 0
\end{array} \right)
\label{e:gtfano218}
\eeq

\begin{figure}[ht]
\begin{center}
\includegraphics[totalheight=3.5cm]{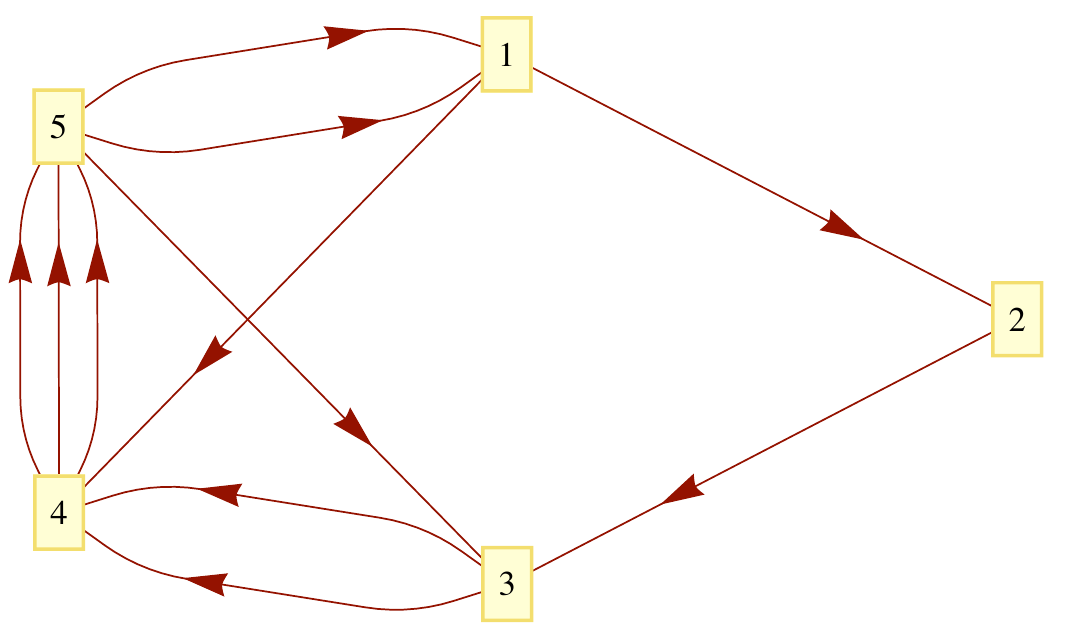}
\hfill
\includegraphics[totalheight=4.5cm]{FinalFano/tiling218.pdf}
 \caption{(i) Quiver diagram of the $\cE_1$  model. (ii) Tiling of the $\cE_1$ model.}
  \label{f:tqfano218}
\end{center}
\end{figure}

The CS theory corresponding to this fano variety has 5 gauge groups and chiral superfields $X^i_{45}$ (with $i=1,2,3$), $X^j_{51}$, $X^j_{34}$ (with $j=1,2$), $X_{14}$, $X_{12}$, $X_{53}$ and $X_{23}$.
The tiling and quiver of this theory are shown in Figure \ref{f:tqfano218}.  The superpotential can be read off from the tiling:

\bea
W = \tr \left[ \epsilon_{ij} \left( X^i_{51}X_{12}X_{23}X^j_{34} X^3_{45} +  X_{53}X^i_{34}X^j_{45} + X_{14} X^i_{45}X^j_{51} \right) \right]~.
\label{e:spfano218}
\eea

We choose the CS levels to be $\vec{k}=(1,-1,0,-1,1)$

\paragraph{The Kasteleyn matrix.} The CS levels can be parametrized in terms of integers $n^i_{jk}$ and $n_{jk}$ as follows:
\bea
\begin{array}{ll}
\text{Gauge group 1~:} \qquad k_1 &=   n_{12} + n_{14} - n^{1}_{51} - n^{2}_{51} ~,  \nn \\
\text{Gauge group 2~:} \qquad k_2 &=   n_{23} - n_{12} ~,  \nn \\
\text{Gauge group 3~:} \qquad k_3 &=   n^{1}_{34} + n^{2}_{34} - n_{23} - n_{53} ~,  \nn \\
\text{Gauge group 4~:} \qquad k_4 &=   n^{1}_{45} + n^{2}_{45} + n^{3}_{45} - n^{1}_{34} - n^{2}_{34} - n_{14} ~,  \nn \\
\text{Gauge group 5~:} \qquad k_5 &=   n_{53} + n^{1}_{51} + n^{2}_{51} - n^{1}_{45} - n^{2}_{45} - n^{3}_{45} ~.
\label{e:kafano218}
\end{array}
\eea  
Let us choose
\bea
n_{12}=-n^3_{45}= 1,\qquad n^i_{jk}=n_{jk}=0 \;\text{otherwise}~.
\eea
which is consistent with the earlier choice of $\vec{k}$. The fundamental domain contains three pairs of black and white nodes and, therefore, the Kasteleyn matrix is a $3\times 3$ matrix:
\bea
K =   \left(
\begin{array}{c|ccc}
& b_1 & b_2 & b_3\\
\hline
w_1 & z^{n_{14}} & z^{n^1_{45}} & \frac{y}{x} z^{n^2_{51}} \\
w_2 & x z^{n^1_{51}} & z^{n^2_{34}} & z^{n^3_{45}} + y z^{n_{12}} + y z^{n_{23}} \\
w_3 & z^{n^2_{45}} & \frac{1}{y} z^{n_{53}} & z^{n^1_{34}} \end{array}
\right) ~.
\label{e:kastfano218}
\eea
The permanent of the Kasteleyn matrix is given by:
\bea
\mathrm{perm}(K) &=&  x z^{(n^1_{34} + n^1_{45} + n^1_{51})} +  x^{-1} y z^{(n^2_{34} + n^2_{45} + n^2_{51})} +  y z^{(n^1_{45} + n^2_{45} + n_{23})}\nn \\
&+&   y z^{(n^1_{45} + n^2_{45} + n_{12})} +  y^{-1} z^{(n^3_{45} + n_{53} + n_{14})} + z^{(n^1_{45} + n^2_{45} + n^3_{45})}\nn \\
&+& z^{(n_{53} + n_{14} + n_{12})} + z^{(n_{53} + n_{14} + n_{23})} + z^{(n^1_{51} + n^2_{51} + n_{53})} + z^{(n^1_{34} + n^2_{34} + n_{14})}\nn \\
&=&  x+ x^{-1} y + y + y z + y^{-1} z^{-1} + z^{-1} + z + 3\nn \\
&& \; \text{(for $n_{12}=-n^3_{45}= 1,\qquad n^i_{jk}=n_{jk}=0 \; \text{otherwise}$)}.
\label{e:charpolyfano218}
\eea

The powers of $x,y$ and $z$ in each term of \eref{e:charpolyfano218} give the coordinates of the toric diagram. They are collected in the columns of the following matrix:
\bea
G_K = \left(
\begin{array}{cccccccccc}
  1 & -1 & 0 & 0 &  0 &  0 & 0 & 0 & 0 & 0 \\
  0 &  1 & 1 & 1 & -1 &  0 & 0 & 0 & 0 & 0 \\
  0 &  0 & 0 & 1 & -1 & -1 & 1 & 0 & 0 & 0 
\end{array}
\right)~.
\label{e:Gkfano218}
\eea
Since the first row contains the weights of the fundamental representation of $SU(2)$, the mesonic symmetry contains $SU(2)$. The $G_K$ above is identical to $G_t$ in \eqref{e:gtfano218} up to multiplicity of toric points.

\Section{$\cE_2$ (Toric Fano 275)}

\begin{figure}[ht]
\begin{center}
  \includegraphics[totalheight=3cm]{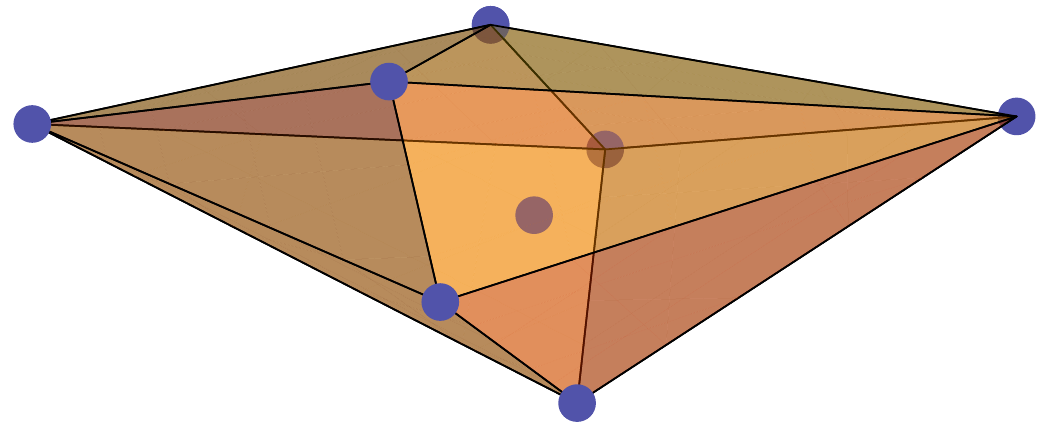}
 \caption{The toric diagram of $\cE_2$ (Toric Fano 275).}
  \label{f:tdtoricfano275}
\end{center}
\end{figure}

The toric data of $\cE_2$ (Toric Fano 275) is given in \eqref{e:gtfano275} and the toric diagram of the variety is displayed in \fref{f:tdtoricfano275}.
\beq
G_t =\left(
\begin{array}{cccccccc}
 1 &-1 & 0 & 0 & 0 & 0 & 0 & 0 \\
 0 & 0 & 1 &-1 &-1 & 0 & 0 & 0 \\
 0 & 1 & 0 & 0 & 1 &-1 & 1 & 0
\end{array} \right)
\label{e:gtfano275}
\eeq

\begin{figure}[ht]
\begin{center}
\includegraphics[totalheight=3.5cm]{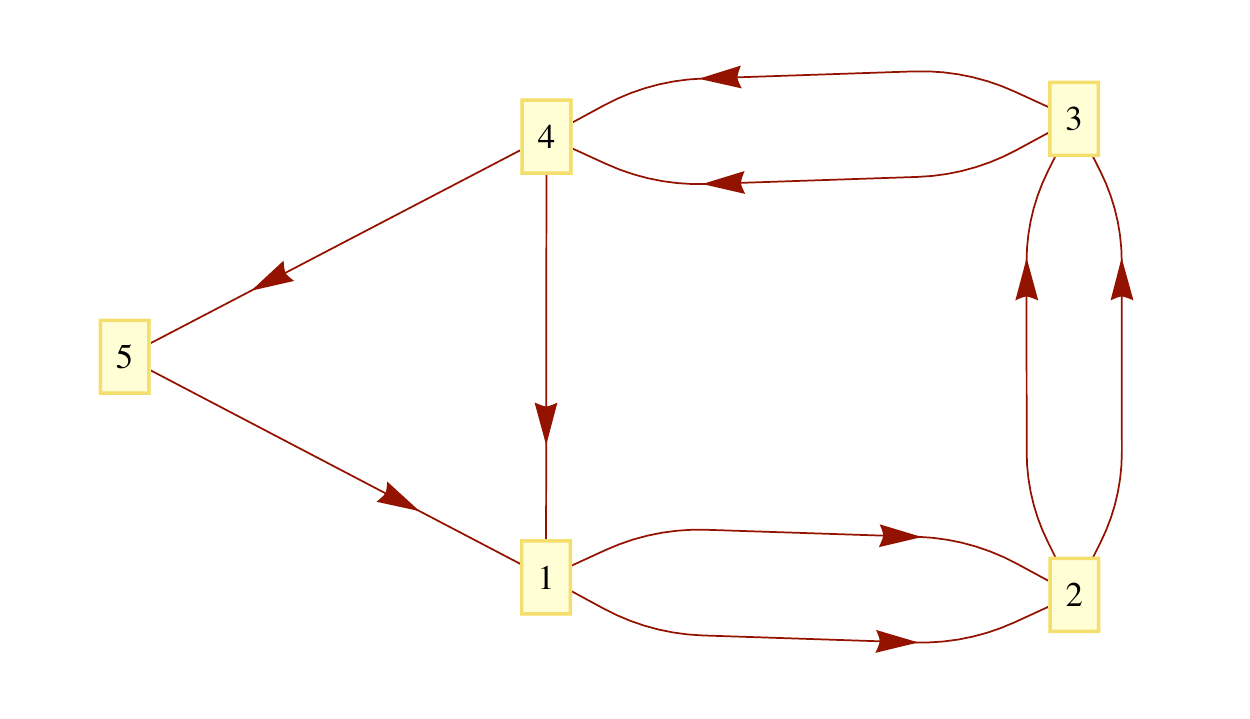}
\hfill
\includegraphics[totalheight=4.5cm]{FinalFano/tiling275.pdf}
  \caption{(i) Quiver of the $\cE_2$ model. \quad (ii) Tiling of the $\cE_2$ model.}
  \label{t:fano275tileandquiver}
  \end{center}
\end{figure}

The CS theory corresponding to this fano variety has 5 gauge groups and bi-fundamental fields $X_{34}^i$, $X_{12}^i$, $X_{23}^i$, $X_{41}$, $X_{51}$, $X_{45}$ (with $i=1,2$). The quiver diagram and tiling are drawn in Figure \ref{t:fano275tileandquiver}. 

The superpotential of the theory is given by
\bea
W = \tr \left[ \epsilon_{ij} ( X_{45} X_{51} X^i_{12}X^1_{23} X^j_{34} -  X_{41} X^i_{12} X^2_{23} X^j_{34}) \right]~.
\label{e:spotfano275}
\eea
The CS levels are chosen to be $\vec{k} = (1,0,-1,-1,1)$.

\paragraph{The Kasteleyn matrix.}   The Chern-Simons levels can be parametrized as follows:
\bea
\begin{array}{ll}
\text{Gauge group 1:} \qquad  k_1 &=   n^{1}_{12} + n^{2}_{12} - n_{41} - n_{51} ~, \nn \\
\text{Gauge group 2:} \qquad  k_2 &=   n^{1}_{23} + n^{2}_{23} - n^{1}_{12} - n^{2}_{12} ~, \nn \\
\text{Gauge group 3:} \qquad  k_3 &=   n^{1}_{34} + n^{2}_{34} - n^{1}_{23} - n^{2}_{23} ~,   \nn \\
\text{Gauge group 4:} \qquad  k_4 &=   n_{41} + n_{45} - n^{1}_{34} - n^{2}_{34} ~, \nn \\
\text{Gauge group 5:} \qquad  k_5 &=   n_{51} - n_{45} ~.
\label{e:kafano275}
\end{array}
\eea
Let us choose 
\bea
n^{1}_{12} = n^{2}_{23} = -n_{45} = 1,~ n^i_{jk}=n_{jk}=0 \; \text{otherwise}~.
\eea
which is consistent with our previous choice of $\vec{k}$. The fundamental domain contains two pairs of black and white nodes and, therefore, the Kasteleyn matrix is a $2 \times 2$ matrix: 
\bea
K= \left(
\begin{array}{c|cc}
& w_1 & w_2 \\
\hline
b_1 & z^{n^{1}_{34}} + x z^{n^{2}_{12}} &\ z^{n^{1}_{23}} + \frac{1}{y} z^{n_{45}} + \frac{1}{y} z^{n_{51}}   \\
b_2 & z^{n^{2}_{23}} + y z^{n_{41}} &\ z^{n^{2}_{34}} + \frac{1}{x} z^{n^{1}_{12}}
\end{array}
\right)~.
\label{e:kastfano275}
\eea
The permanent of this matrix is given by:
\bea
\perm(K) &=&  x z^{(n^2_{12} + n^2_{34})} +  x^{-1}z^{(n^1_{12} + n^1_{34})} +  y z^{(n^1_{23} + n_{41})} +  y^{-1} z^{(n^2_{23} + n_{45})}\nn \\
&+& y^{-1} z^{(n^2_{23} + n_{51})} +  z^{(n_{41} + n_{45})} + z^{(n^1_{23} + n^2_{23})} + z^{(n^1_{12} + n^2_{12})}\nn \\
&+&z^{(n^1_{34} + n^2_{34})} + z^{(n_{41} + n_{51})}\nn \\
&=& x + x^{-1} z +y + y^{-1} + y^{-1} z +z^{-1} + 2z + 2\nn \\
&& \text{(for $n^{1}_{12} = n^{2}_{23} = -n_{45} = 1$} \nn \\
&& \text{$n^i_{jk}=n_{jk}=0\; \text{otherwise}$)} ~.
\label{e:charpolyfano275}
\eea

 The powers of $x, y$ and $z$ in each term of \eref{e:charpolyfano275} give the coordinates of each point in the toric diagram. These points can be collected in the columns of the following $G_K$ matrix: 
\beq
G_K = \left(
\begin{array}{cccccccccc}
 1 &-1 & 0 & 0 & 0 & 0 & 0 & 0 & 0 & 0 \\
 0 & 0 & 1 &-1 &-1 & 0 & 0 & 0 & 0 & 0 \\
 0 & 1 & 0 & 0 & 1 &-1 & 1 & 1 & 0 & 0
\end{array} \right)
\eeq
Since the first row contains the weights of the fundamental representation of $SU(2)$, the mesonic symmetry contains $SU(2)$. The $G_K$ above is identical to $G_t$ in \eqref{e:gtfano275} up to multiplicity of toric points.

\Section{$\cE_3$ (Toric Fano 266)}

\begin{figure}[ht]
\begin{center}
  \includegraphics[totalheight=3.5cm]{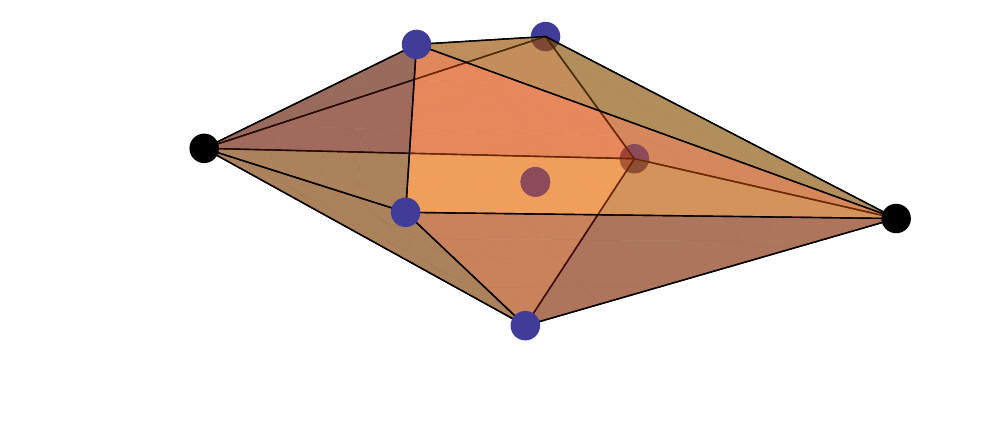}
 \caption{The toric diagram of $\cE_3$ (Toric Fano 266).}
  \label{f:tdtoricfano266}
\end{center}
\end{figure}

The toric data of $\cE_3$ (Toric Fano 266) is given in \eqref{e:gtfano266} and the toric diagram of the variety is displayed in \fref{f:tdtoricfano266}.
\beq
G_t =\left(
\begin{array}{cccccccc}
 1 &-1 & 0 & 0 & 0 & 0 & 0 & 0 \\
 0 & 0 & 1 &-1 & 1 & 0 & 0 & 0 \\
 0 & 0 & 0 & 0 & 1 & 1 &-1 & 0
\end{array} \right)
\label{e:gtfano266}
\eeq

\begin{figure}[ht]
\begin{center}
\includegraphics[totalheight=3.5cm]{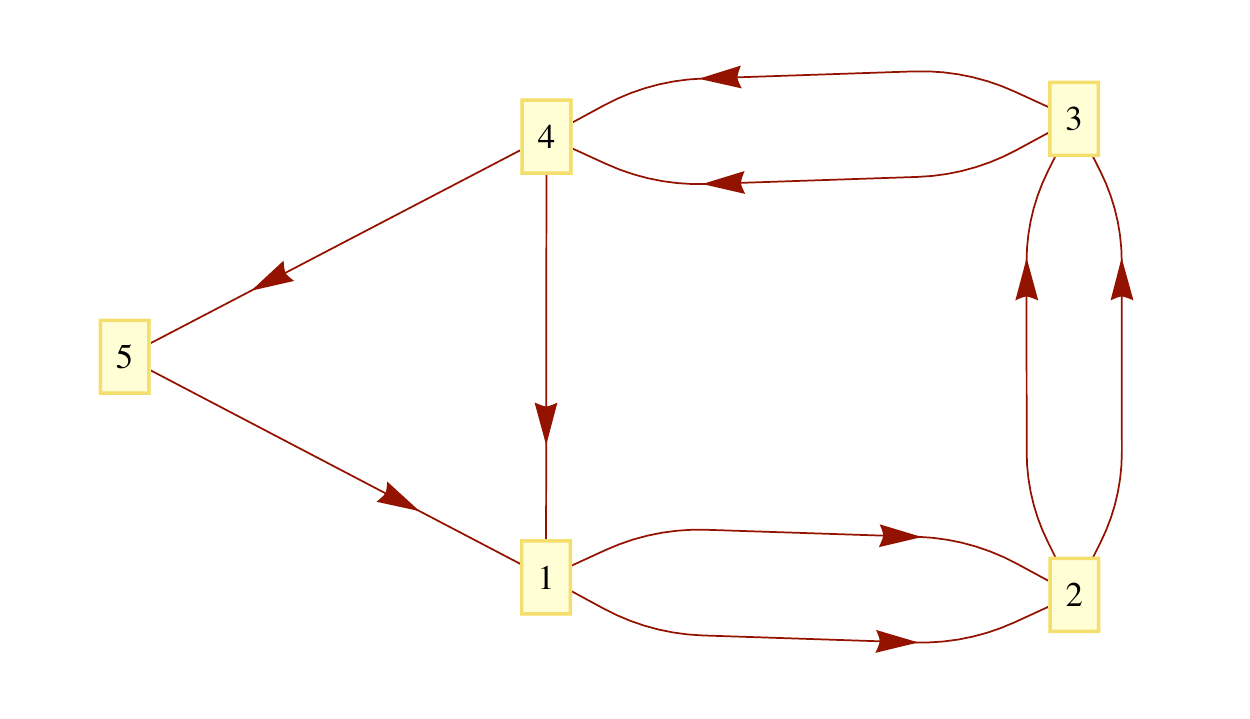}
\hfill
\includegraphics[totalheight=4.5cm]{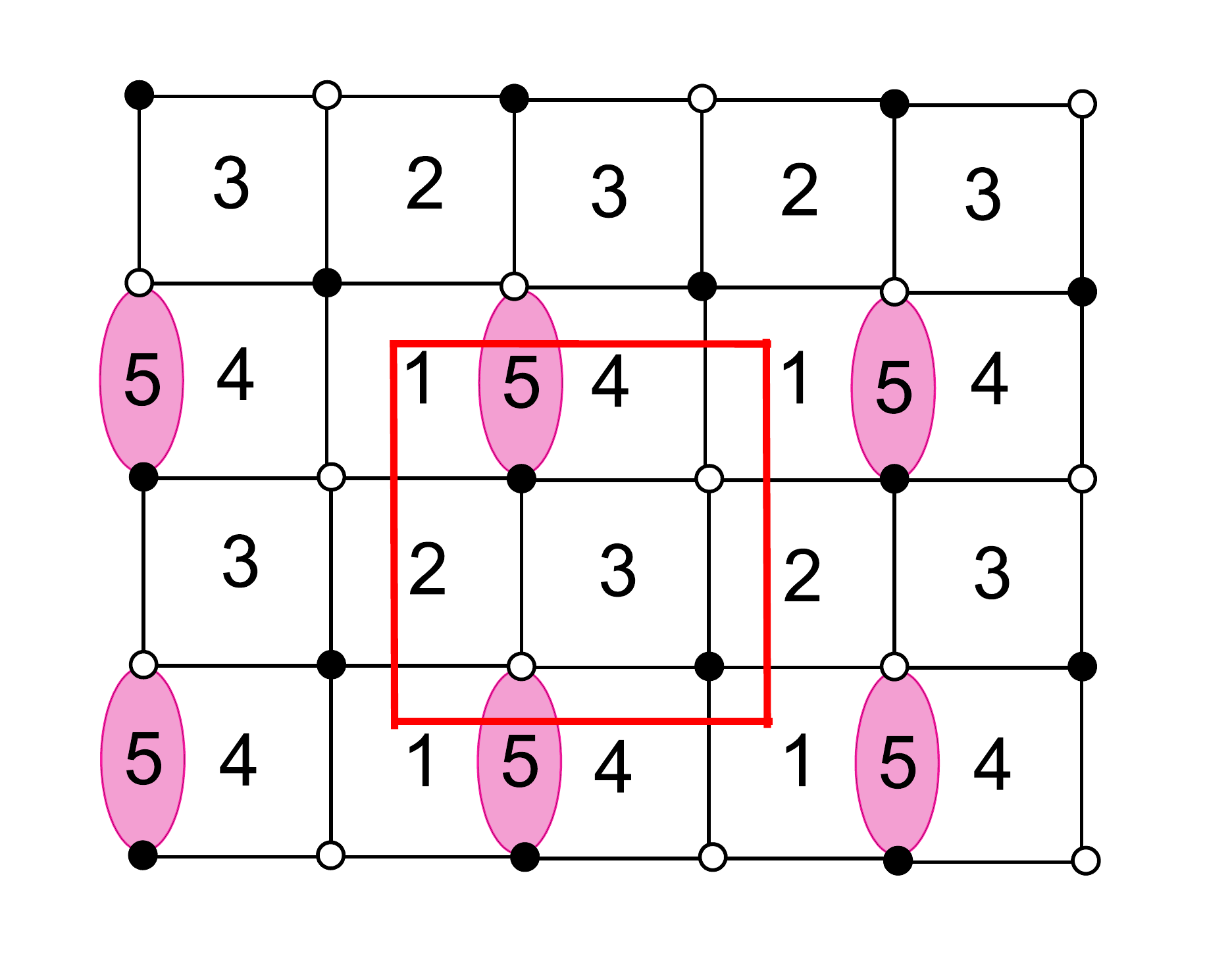}
  \caption{(i) Quiver of the $\cE_3$ model. \quad (ii) Tiling of the $\cE_3$ model.}
  \label{t:fano275tileandquiver-2}
  \end{center}
\end{figure}

The CS theory corresponding to this fano variety is the same as the one used to describe the $\cE_2$ geometry. It has 5 gauge groups and bi-fundamental fields $X_{34}^i$, $X_{12}^i$, $X_{23}^i$, $X_{41}$, $X_{51}$, $X_{45}$ (with $i=1,2$). The quiver diagram and tiling are drawn in Figure \ref{t:fano275tileandquiver-2}. 

The superpotential of the theory is given by
\bea
W = \tr \left[ \epsilon_{ij} ( X_{45} X_{51} X^i_{12}X^1_{23} X^j_{34} -  X_{41} X^i_{12} X^2_{23} X^j_{34}) \right]~.
\label{e:spotfano275-2}
\eea

We choose the CS levels to be $\vec{k} = (1,1,-1,0,-1)$.

\paragraph{The Kasteleyn matrix.}   A parametrization of the Chern-Simons levels in terms of the integers $n^i_{jk}$ and $n_{jk}$ is given by:
\bea
\begin{array}{ll}
\text{Gauge group 1:} \qquad  k_1 &=   n^{1}_{12} + n^{2}_{12} - n_{41} - n_{51} ~, \nn \\
\text{Gauge group 2:} \qquad  k_2 &=   n^{1}_{23} + n^{2}_{23} - n^{1}_{12} - n^{2}_{12} ~, \nn \\
\text{Gauge group 3:} \qquad  k_3 &=   n^{1}_{34} + n^{2}_{34} - n^{1}_{23} - n^{2}_{23} ~,   \nn \\
\text{Gauge group 4:} \qquad  k_4 &=   n_{41} + n_{45} - n^{1}_{34} - n^{2}_{34} ~, \nn \\
\text{Gauge group 5:} \qquad  k_5 &=   n_{51} - n_{45} ~.
\label{e:kafano266}
\end{array}
\eea

We will choose 
\bea
n^{2}_{23} =-n_{51} =   1,~ n^i_{jk}=n_{jk}=0 \; \text{otherwise}~.
\eea
The Kasteleyn matrix can now be constructed. The fundamental domain contains two black 
nodes and two white nodes and, therefore, the Kasteleyn matrix is a $2 \times 2$ matrix\footnote{The weight assignment is different from that chosen in \ref{e:kastfano275}. This will make the non-abelian part of the mesonic symmetry more evident in the $G_K$ matrix}: 

\bea
K= \left(
\begin{array}{c|cc}
& w_1 & w_2 \\
\hline
b_1 & z^{n^{1}_{34}} + \frac{1}{x} z^{n^{2}_{12}} &\ z^{n^{1}_{23}} + y z^{n_{45}} +  y z^{n_{51}}     \\
b_2 & z^{n^{2}_{23}} + \frac{1}{y} z^{n_{41}} &\ z^{n^{2}_{34}} + x z^{n^{1}_{12}}
\end{array}
\right)~.
\label{e:kastfano266}
\eea
The permanent of this matrix is given by
\bea
\perm(K) &=&  x {z^{(n^{1}_{12} + n^{1}_{34})}} +  x^{-1} z^{(n^{2}_{12} + n^{2}_{34})}+  y z^{(n^{2}_{23} + n_{51})}+  y^{-1} {z^{(n^1_{23} + n_{41})}}\nn \\
&+&  y z^{(n^2_{23} + n_{45})}+  z^{(n^{1}_{23} + n^{2}_{23})}+  z^{(n_{41} + n_{51})}+  z^{(n^{1}_{12} + n^{2}_{12})}\nn \\
&+& z^{(n^{1}_{34} + n^{2}_{34})}+  z^{(n_{45} + n_{41})}\nn \\
&=& x + x^{-1} +y + y^{-1}+ y z+  z + z^{-1}+ 3\qquad \nn \\
&&\text{(for $n^{2}_{23} =-n_{51} =   1,~ n^i_{jk}=n_{jk}=0\; \text{otherwise}$)} ~. \nn \\ 
\label{e:charpolyfano266}
\eea

The powers of $x, y$ and $z$ in each term of the permanent of the Kasteleyn matrix give the coordinates of each point in the toric diagram. The coordinates of each point in the toric diagram form columns of the following $G_K$ matrix: 
\bea
G_K = \left(
\begin{array}{cccccccccc}
 1 &-1 & 0 & 0 & 0 & 0 & 0 & 0 & 0 & 0 \\
 0 & 0 & 1 &-1 & 1 & 0 & 0 & 0 & 0 & 0 \\
 0 & 0 & 0 & 0 & 1 & 1 &-1 & 0 & 0 & 0
\end{array}
\right) 
\eea
The first row of the above matrix contains the weights of the fundamental representation of $SU(2)$. Therefore, the mesonic moduli space contains an $SU(2)$ symmetry. The $G_K$ above is identical to $G_t$ in \eqref{e:gtfano266} up to multiplicity of toric points.

\Section{$\cE_4$ (Toric Fano 271)}

\begin{figure}[ht]
\begin{center}
  \includegraphics[totalheight=3.5cm]{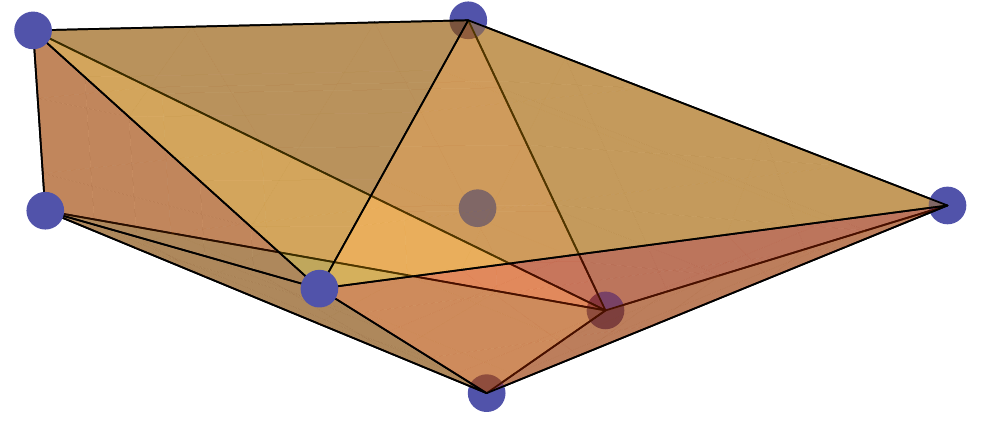}
 \caption{The toric diagram of $\cE_4$ (Toric Fano 271).}
  \label{f:tdtoricfano271}
\end{center}
\end{figure}

The toric data of $\cE_4$ (Toric Fano 271) is given in \eqref{e:gtfano271} and the toric diagram of the variety is displayed in \fref{f:tdtoricfano271}.
\beq
G_t =\left(
\begin{array}{cccccccc}
1 &-1 & 0 & 0 & 0 & 0 & 0 & 0 \\
 0 & 1 & 1 &-1 &-1 & 0 & 0 & 0 \\
 0 & 0 & 0 & 0 & 1 & 1 &-1 & 0
\end{array} \right)
\label{e:gtfano271}
\eeq

\begin{figure}[ht]
\begin{center}
\includegraphics[totalheight=3.5cm]{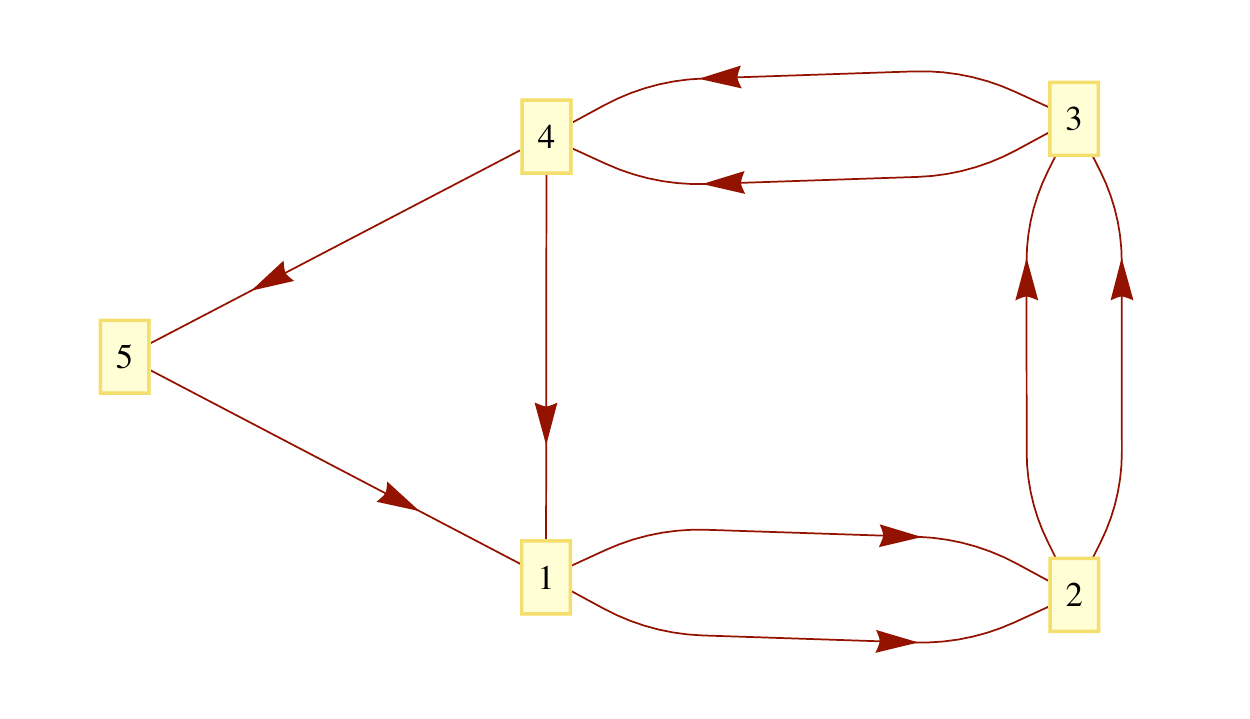}
\hfill
\includegraphics[totalheight=4.5cm]{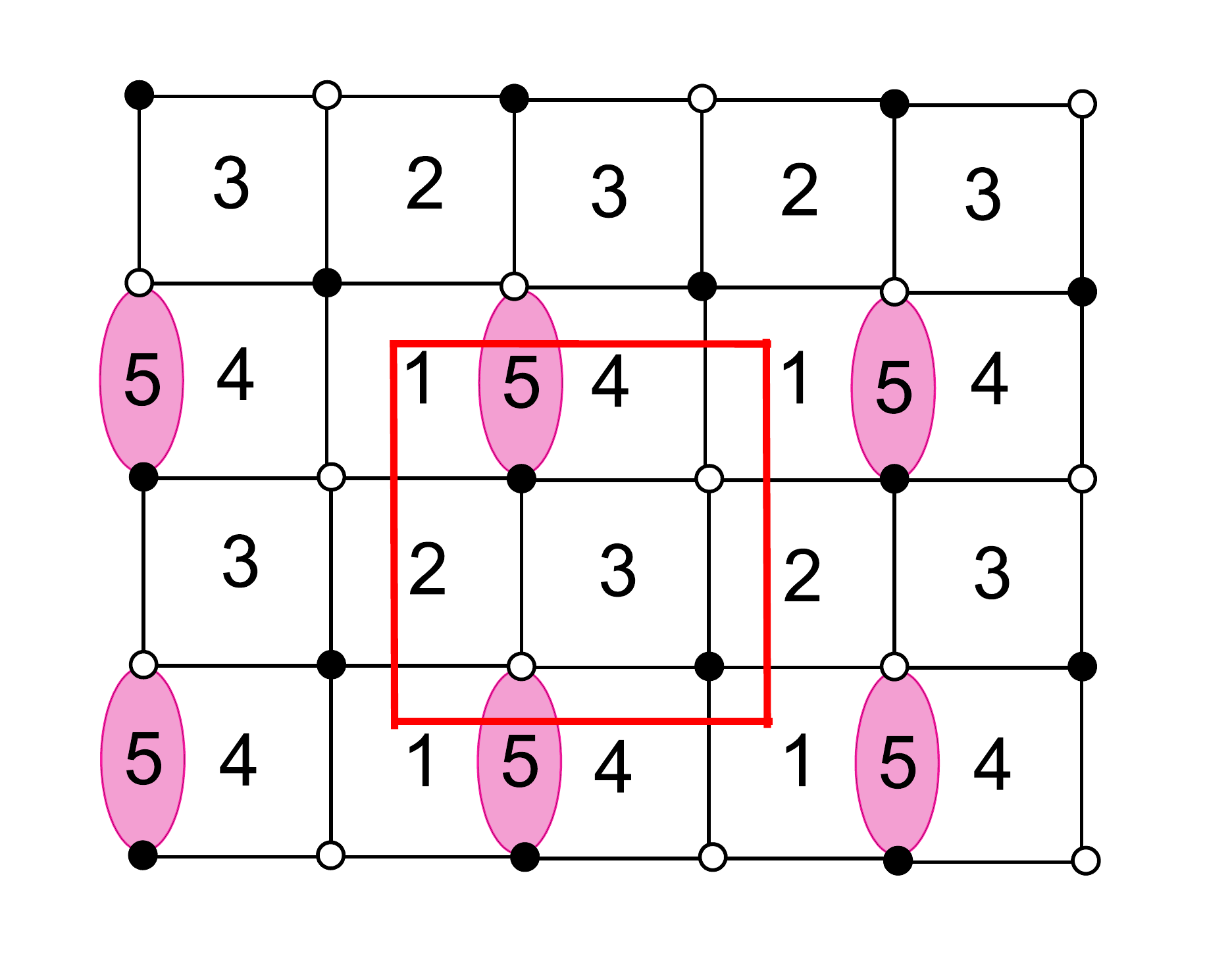}
  \caption{(i) Quiver of the $\cE_3$ model. \quad (ii) Tiling of the $\cE_3$ model.}
  \label{t:fano275tileandquiver-3}
  \end{center}
\end{figure}

The CS theory corresponding to this fano variety is the same as the one used to describe the $\cE_2$ geometry. It has 9 chiral fields: $X^i_{12}$, $X^i_{23}$, $X^i_{41}$ (with $i=1,2$), $X_{35}$, $X_{54}$ and $X_{34}$. The quiver diagram and the tiling are given in \ref{t:fano275tileandquiver-3}. The superpotential can be read from \eqref{e:spotfano275}. For this model, we choose the CS levels to be $\vec{k}=(1,-1,0,-1,1)$.

The superpotential of the theory is given by
\bea
W = \tr \left[ \epsilon_{ij} ( X_{45} X_{51} X^i_{12}X^1_{23} X^j_{34} -  X_{41} X^i_{12} X^2_{23} X^j_{34}) \right]~.
\label{e:spotfano275-3}
\eea

\paragraph{The Kasteleyn matrix.} The Chern-Simons levels for this model can be written as:
\bea
\begin{array}{ll}
\text{Gauge group 1:} \qquad  k_1 &=   n^{1}_{12} + n^{2}_{12} - n_{41} - n_{51} ~, \nn \\
\text{Gauge group 2:} \qquad  k_2 &=   n^{1}_{23} + n^{2}_{23} - n^{1}_{12} - n^{2}_{12} ~, \nn \\
\text{Gauge group 3:} \qquad  k_3 &=   n^{1}_{34} + n^{2}_{34} - n^{1}_{23} - n^{2}_{23} ~,   \nn \\
\text{Gauge group 4:} \qquad  k_4 &=   n_{41} + n_{45} - n^{1}_{34} - n^{2}_{34} ~, \nn \\
\text{Gauge group 5:} \qquad  k_5 &=   n_{51} - n_{45} ~,
\label{e:kafano271}
\end{array}
\eea

We choose:
\bea
n^2_{12}=-n_{45}= 1,\qquad n^i_{jk}=n_{jk}=0 \;\;\text{otherwise}~.
\eea
which is consistent with our previous choice of $\vec{k}$. The fundamental domain of the tiling contains two white nodes and two black nodes, thus the Kasteleyn matrix is a $2 \times 2$ matrix and can be written as:
\bea
K =   \left(
\begin{array}{c|cc}
& w_1 & w_2\\
\hline
b_1 & z^{n^{1}_{34}} + x z^{n^{2}_{12}} &\ z^{n^{1}_{23}} + y z^{n_{45}} + y z^{n_{51}}   \\
b_2 & z^{n^{2}_{23}} + \frac{1}{y} z^{n_{41}} &\ z^{n^{2}_{34}} + \frac{1}{x} z^{n^{1}_{12}}
\end{array}
\right) ~.
\label{e:kastfano271}
\eea
The permanent of the Kasteleyn matrix is equal to:
\bea
\mathrm{perm}(K) &=&  x z^{(n^2_{12} + n^2_{34})}+  x^{-1} z^{(n^1_{12} + n^1_{34})}+  z^{(n^1_{12} + n^2_{12})}+  z^{(n_{41} + n_{45})}\nn \\
&+&  y z^{(n^2_{23} + n_{45})}+  y z^{(n_{51} + n^2_{23})} +  y^{-1} z^{(n_{41} + n^1_{23})}+  z^{(n_{51} + n_{41})}\nn \\
&+&  z^{(n^1_{23} + n^2_{23})}+  z^{(n^1_{34} + n^2_{34})}\nn \\
&=&  x + x^{-1} z+  z+  z^{-1} + y z^{-1} + y + y^{-1} + 3 \nn \\
&&\text{(for $n^2_{12}=-n_{45}= 1,\qquad n^i_{jk}=n_{jk}=0 \; \text{otherwise}$)}.
\label{e:charpolyfano271}
\eea

The powers of $x$, $y$ and $z$ in each of the terms in \eref{e:charpolyfano271} are the coordinates of the toric diagram in the following matrix:
\bea
\left(
\begin{array}{cccccccccc}
  1 & -1 &  0 &  0 & 0 & 0 &  0 & 0 & 0 & 0 \\
  0 &  0 &  0 &  0 & 1 & 1 & -1 & 0 & 0 & 0 \\
  0 &  1 &  1 & -1 &-1 & 0 &  0 & 0 & 0 & 0 
\end{array}
\right)~.
\label{e:Gkfano271}
\eea
Multiplying this matrix on the left by {\footnotesize $\left( \begin{array}{ccc} 1&0&0\\0&0&1\\0&1&0 \end{array} \right) \in GL(3, \BZ)$} gives us
\bea
G_K = \left(
\begin{array}{cccccccccc}
  1 & -1 &  0 &  0 & 0 & 0 &  0 & 0 & 0 & 0 \\
  0 &  1 &  1 & -1 &-1 & 0 &  0 & 0 & 0 & 0 \\
  0 &  0 &  0 &  0 & 1 & 1 & -1 & 0 & 0 & 0 
\end{array}
\right)~.
\label{e:Gkpfano271}
\eea
The first row of \eqref{e:Gkpfano271} contains the weights of the fundamental representation of $SU(2)$. Therefore the mesonic symmetry contains $SU(2)$. The $G_K$ matrix above is identical to $G_t$ in \eqref{e:gtfano271} up to multiplicity of toric points.

\Section{$\cF_2$ (Toric Fano 369)}

\begin{figure}[ht]
\begin{center}
  \includegraphics[totalheight=3cm]{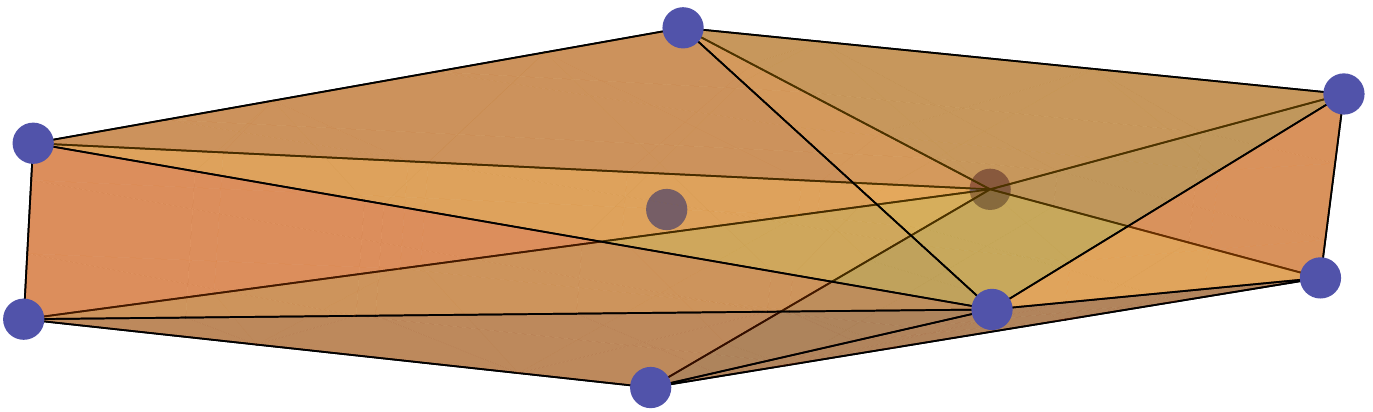}
 \caption{The toric diagram of $\cF_2$ (Toric Fano 369).}
  \label{f:tdtoricfano369}
\end{center}
\end{figure}

The toric data of $\cF_2$ (Toric Fano 369) is given in \eqref{e:gtfano369} and the toric diagram of the variety is displayed in \fref{f:tdtoricfano369}.
\beq
G_t =\left(
\begin{array}{ccccccccc}
 1 &-1 & 0 & 0 & 0 & 0 & 0 & 0 & 0 \\
 0 & 1 & 1 &-1 & 1 &-1 & 0 & 0 & 0 \\
 0 & 0 & 0 & 0 & 1 &-1 &-1 & 1 & 0
\end{array} \right)
\label{e:gtfano369}
\eeq

\begin{figure}[ht]
\begin{center}
\includegraphics[totalheight=2.5cm]{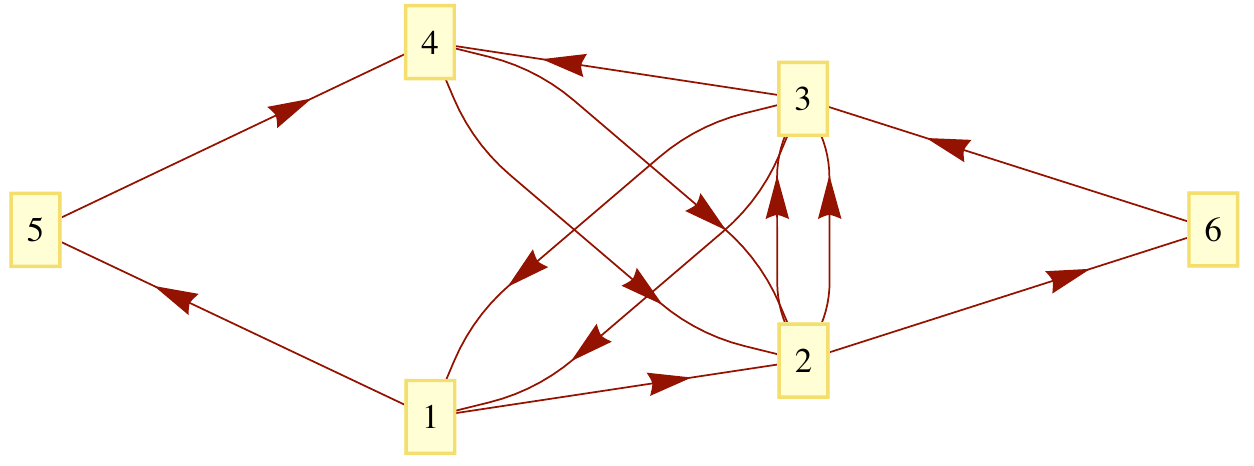}
\hfill
\includegraphics[totalheight=4cm]{FinalFano/tiling369.pdf}
\caption{(i) Quiver of the $\cF_2$ model. \quad (ii) Tiling of the $\cF_2$ model.}
  \label{f:tqfano369}
\end{center}
\end{figure}

The CS theory corresponding to this fano variety has 6 gauge groups and chiral fields $X^{i}_{23}, X^{i}_{31}, X^{i}_{42}$ (with $i=1,2$), $X_{12}, X_{34}, X_{26},X_{63}, X_{15}$ and $X_{54}$. The quiver diagram and the tiling of this model are presented in Figure \ref{f:tqfano369}.  The superpotential of this model can be read off from the tiling and can be written as:
\beq
W = \tr \left[\epsilon_{ij} \left(X_{12}X^i_{23}X^j_{31} + X_{34}X^i_{42}X^j_{23} + X_{26}X_{63}X^{i}_{31}X_{15}X_{54}X^j_{42} \right) \right].
\label{spfano369}
\eeq
The CS levels are chosen to be $\vec{k}=(0,-1,0,-1,1,1)$.

\paragraph{The Kasteleyn matrix.} The Chern-Simons levels can be parametrized in terms of the integers $n^i_{jk}$ or $n_{jk}$ as follows:
\bea
\begin{array}{ll}
\text{Gauge group 1~:} \qquad k_1 &=   n_{12} + n_{15} - n^{1}_{31} - n^{2}_{31}  ~,  \nn \\
\text{Gauge group 2~:} \qquad k_2 &=   n_{26} + n^{1}_{23} + n^{2}_{23} - n^{1}_{42} - n^{2}_{42} - n_{12}  ~,  \nn \\
\text{Gauge group 3~:} \qquad k_3 &=   n_{34} + n^{1}_{31} + n^{2}_{31} - n^{1}_{23} - n^{2}_{23} - n_{63} ~,  \nn \\
\text{Gauge group 4~:} \qquad k_4 &=   n^{1}_{42} + n^{2}_{42} - n_{34} - n_{54} ~,  \nn \\
\text{Gauge group 5~:} \qquad k_5 &=   n_{54} - n_{15} ~,  \nn \\
\text{Gauge group 6~:} \qquad k_6 &=   n_{63} - n_{26} ~.
\label{e:kafano369}
\end{array}
\eea  
We will choose
\bea
n_{54}=-n_{26}= 1,\qquad n^i_{jk}=n_{jk}=0 \;\;\text{otherwise}~.
\eea
Which is consistent with our previous choice of $\vec{k}$. Since the fundamental domain contains 3 pairs of black and white nodes, the Kasteleyn matrix of this model is a $3\times 3$ matrix:
\bea
K =   \left(
\begin{array}{c|ccc}
& b_1 & b_2 & b_3\\
\hline
w_1 & z^{n_{12}} & \frac{y}{x} z^{n^2_{31}} & z^{n^1_{23}} \\
w_2 & z^{n^2_{23}} & z^{n^1_{42}} & \frac{1}{y} z^{n_{34}} \\
w_3 & x z^{n^1_{31}} \quad & z^{n_{63}} + z^{n_{26}} + y z^{n_{15}} + y z^{n_{54}} \quad & z^{n^2_{42}} \end{array}
\right) ~.
\label{e:kastfano369}
\eea
The permanent of the Kasteleyn matrix is
\bea
\mathrm{perm}(K) &=&  x z^{(n^1_{23} + n^1_{42} + n^1_{31})} +  x^{-1} y z^{(n^2_{23} + n^2_{42} + n^2_{31})}+   y  z^{(n^1_{23} + n^2_{23} + n_{15})}\nn \\
&+&  y^{-1} z^{(n_{63} + n_{12} + n_{34})} +  y z^{(n^1_{23} + n^2_{23} + n_{54})} +  y^{-1} z^{(n_{26} + n_{12} + n_{34})} \nn \\
&+& z^{(n^1_{23} + n^2_{23} + n_{26} )}+ z^{(n_{54} + n_{12} + n_{34})}+ z^{(n^1_{23} + n^2_{23} + n_{63})} \nn \\
&+&  z^{(n_{15} + n_{12} + n_{34})} + z^{(n^1_{42} + n^2_{42} + n_{12})}+  z^{(n^1_{31} + n^2_{31} + n_{34})}\nn \\
&=& x + x^{-1} y + y + y^{-1} + y z + y^{-1} z^{-1} + z^{-1} + z + 4\nn \\
&& \; \text{(for $n_{54}=-n_{26}= 1,\; n^i_{jk}=n_{jk}=0 \; \text{otherwise}$)}~.
\label{e:charpolyfano369}
\eea
The powers of $x, y$ and $z$ of each term in \eref{e:charpolyfano369} give the coordinates of the toric diagram:
\bea
G_K = \left(
\begin{array}{cccccccccccc}
  1 & -1 & 0 & 0 & 0 &  0 & 0 & 0 & 0 & 0 & 0 & 0 \\
  0 &  1 & 1 &-1 & 1 & -1 & 0 & 0 & 0 & 0 & 0 & 0 \\
  0 &  0 & 0 & 0 & 1 & -1 &-1 & 1 & 0 & 0 & 0 & 0 
\end{array}
\right)~.
\label{e:Gkfano369}
\eea
The first row contains the powers of the weights of the fundamental representation of $SU(2)$. Thus, the mesonic symmetry of this model contains $SU(2)$. The $G_K$ matrix above is identical to $G_t$ in \eqref{e:gtfano369} up to multiplicity of toric points.

\Section{$\cF_1$ (Toric Fano 324)}

\begin{figure}[ht]
\begin{center}
  \includegraphics[totalheight=3cm]{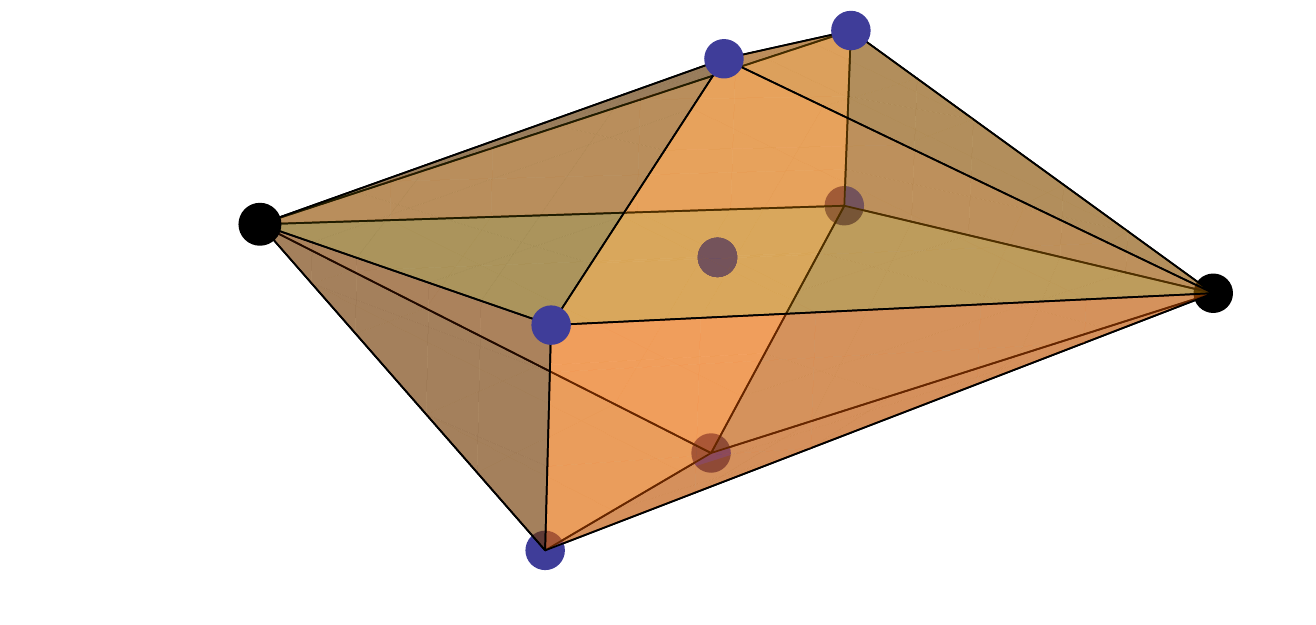}
 \caption{The toric diagram of $\cF_1$ (Toric Fano 324).}
  \label{f:tdtoricfano324}
\end{center}
\end{figure}

The toric data of $\cF_1$ (Toric Fano 324) is given in \eqref{e:gtfano324} and the toric diagram of the variety is displayed in \fref{f:tdtoricfano324}.
\beq
G_t =\left(
\begin{array}{ccccccccc}
  1 & -1 & 0 &  0 & 0 & 0 &  0 &  0 & 0  \\
  0 &  0 & 1 & -1 & 1 &-1 &  0 &  0 & 0  \\
  0 &  0 & 0 &  0 & 1 &-1 & -1 &  1 & 0 
\end{array} \right)
\label{e:gtfano324}
\eeq

\begin{figure}[ht]
\begin{center}
\includegraphics[totalheight=3.4cm]{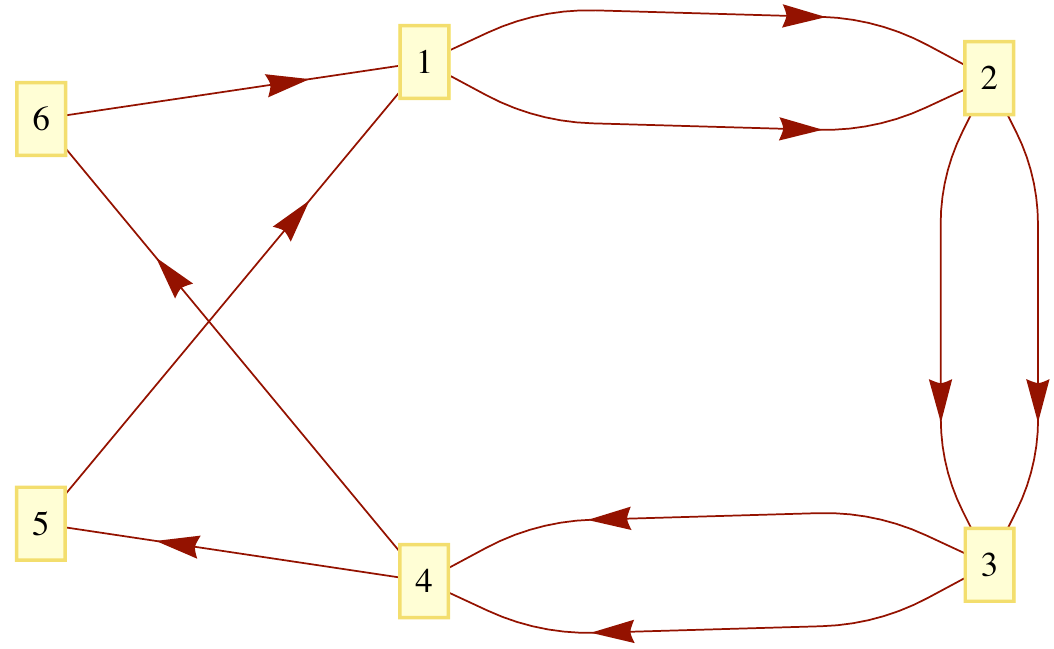}
\hfill
\includegraphics[totalheight=4.5cm]{FinalFano/tiling324.pdf}
 \caption{(i) Quiver of the $\cF_1$  model. \qquad (ii) Tiling of the $\cF_1$  model.}
  \label{f:fano324tileandquiver}
  \end{center}
\end{figure}

The CS theory corresponding to this fano variety has 6 gauge groups and 10 chiral fields: $X^i_{12}, X^i_{23}, X^i_{34}$ (with $i=1,2$), $X_{46},X_{61}, X_{45}$ and $X_{51}$. The quiver diagram and tiling are presented in Figure \ref{f:fano324tileandquiver}.
The superpotential can be read off from the tiling as
\bea
W = \tr \left[ \epsilon_{ij}  \left(X^i_{12}X^1_{23}X^j_{34}X_{45}X_{51} - X^j_{12}X^2_{23}X^i_{34}X_{46}X_{61}\right) \right]~.
\label{e:spotfano324}
\eea
We will choose the CS levels to be $\vec{k}=(0,0,0,0,-1,1)$.

\paragraph{The Kasteleyn matrix.} The Chern-Simons levels of this model can be written in terms of the integers $n^i_{jk}$ and $n_{jk}$ as:
\bea
\begin{array}{ll}
\text{Gauge group 1:} \qquad  k_1 &=   n^{1}_{12} + n^{2}_{12} - n_{51} - n_{61} ~, \nn \\
\text{Gauge group 2:} \qquad  k_2 &=   n^{1}_{23} + n^{2}_{23} - n^{1}_{12} - n^{2}_{12} ~, \nn \\
\text{Gauge group 3:} \qquad  k_3 &=   n^{1}_{34} + n^{2}_{34} - n^{1}_{23} - n^{2}_{23} ~, \nn \\
\text{Gauge group 3:} \qquad  k_4 &=   n_{45} + n_{46} - n^{1}_{34} - n^{2}_{34} ~, \nn \\
\text{Gauge group 4:} \qquad  k_5 &=   n_{51} - n_{45} ~, \nn \\
\text{Gauge group 5:} \qquad  k_6 &=   n_{61} - n_{46} ~.
\label{e:kafano324}
\end{array}
\eea
Let us choose 
\bea
n_{45} =-n_{46} =   1,~ n^i_{jk}=n_{jk}=0 \; \text{otherwise}~.
\eea
Which is consistent with our earlier choice of $\vec{k}$. The fundamental domain contains two pairs of white and black nodes, and so the Kasteleyn matrix is a $2 \times 2$ matrix:
\bea
K= \left(
\begin{array}{c|cc}
& w_1 & w_2 \\
\hline
b_1 & z^{n^{2}_{34}} + x z^{n^{1}_{12}} &\ z^{n^{2}_{23}} + \frac{1}{y} z^{n_{46}}+ \frac{1}{y} z^{n_{61}}   \\
b_2 & z^{n^{1}_{23}} + y z^{n_{45}} + y z^{n_{51}} &\ z^{n^{1}_{34}} + \frac{1}{x} z^{n^{2}_{12}}  
\end{array}
\right)~.
\label{e:kastfano324}
\eea
The permanent of the Kasteleyn matrix can be written as
\bea
\perm~K &=&  x z^{(n^1_{12} + n^1_{34})} +  x^{-1} z^{(n^2_{12} + n^2_{34})} +  y z^{(n_{51} + n^2_{23})}+  y^{-1} z^{(n_{61} + n^1_{23})} \nn \\ 
&+&  y z^{(n_{45} + n^2_{23})} +  y^{-1} z^{(n^1_{23} + n_{46})} +  z^{(n_{51} + n_{46})}+  z^{(n_{61} + n_{45})}\nn \\  
&+& z^{(n^1_{12} + n^2_{12})} + z^{(n^1_{34} + n^2_{34})} +  z^{(n_{61} + n_{51})} + z^{(n^1_{23} + n^2_{23})} + z^{(n_{46} + n_{45} )}\nn \\
&=& x + x^{-1} + y + y^{-1} + y z+ y^{-1} z^{-1} + z^{-1} + z + 5\nn \\
&& \text{(for $n_{45} =-n_{46} =   1,~ n^i_{jk}=n_{jk}=0 \; \text{otherwise}$)} ~. 
\label{e:charpolyfano324}
\eea

The powers of $x$, $y$ and $z$ in each of the terms of \eref{e:charpolyfano324} give the coordinates of the toric diagram and can be collected in the columns of the $G_K$ matrix:
\bea
G_K = \left(
\begin{array}{ccccccccccccc}
  1 & -1 & 0 &  0 & 0 & 0 &  0 &  0 & 0 & 0 & 0 & 0 & 0 \\
  0 &  0 & 1 & -1 & 1 &-1 &  0 &  0 & 0 & 0 & 0 & 0 & 0 \\
  0 &  0 & 0 &  0 & 1 &-1 & -1 &  1 & 0 & 0 & 0 & 0 & 0
\end{array}
\right)~.
\label{e:Gkfano324}
\eea
The first row contains the weights of the fundamental representation of $SU(2)$ and so the mesonic symmetry contains $SU(2)$. The $G_K$ matrix above is identical to $G_t$ in \eqref{e:gtfano324} up to multiplicity of toric points.


\end{document}